\documentclass[oldversion]{aa}
\usepackage{epsfig}
\usepackage{natbib}
\usepackage{lscape}
\usepackage{wasysym}
\bibpunct{(}{)}{;}{a}{}{,}
\begin{document}

\title{The HARPS search for southern extrasolar planets\thanks{Based on 
observations collected at the La Silla Parana Observatory,
ESO (Chile) with the HARPS spectrograph at the 3.6-m telescope (ESO runs ID 
72.C-0488, 082.C-0212, and 085.C-0063).  }}

\subtitle{XXV. Results from the metal-poor sample}

\author{
  N.C. Santos\inst{1,2,3} \and
  M. Mayor\inst{3} \and
 X. Bonfils\inst{3,4} 
 X. Dumusque\inst{1,3} \and
 F. Bouchy\inst{5} \and
 P. Figueira\inst{1} \and
 C. Lovis\inst{3} \and	
 C. Melo\inst{6} \and
 F. Pepe\inst{3} \and
 D. Queloz\inst{3} \and
 D. S\'egransan\inst{3} \and
 S. G. Sousa\inst{1} \and
 S. Udry\inst{3}
  }

\institute{
    Centro de Astrof{\'\i}sica, Universidade do Porto, Rua das Estrelas, 
    4150-762 Porto, Portugal
    \and
    Departamento de F{\'\i}sica e Astronomia, Faculdade de Ci\^encias, Universidade do Porto, Portugal
    \and
    Observatoire de Gen\`eve, Universit\'e de Gen\`eve, 51 ch. des Maillettes, 1290 Sauverny, Switzerland
     \and 
     Laboratoire d'Astrophysique, Observatoire de Grenoble, Universit\'e J. Fourier, CNRS (UMR5571), BP 53, F-38041 Grenoble, Cedex 9, France
    \and
    Institut d'Astrophysique de Paris, UMR7095 CNRS, Universit\'e Pierre \& Marie Curie, 98bis Bd Arago, 75014 Paris, France
    \and
    European Southern Observatory, Casilla 19001, Santiago 19, Chile
}

%    Physikalisches Institut, University of Bern, Sidlerstrasse 5, 3012 Bern, Switzerland
%    \and

\date{Accepted for publications}

\abstract{
Searching for extrasolar planets around stars of different metallicity may provide strong constraints to the models of 
planet formation and evolution. 
In this paper we present the overall results of a HARPS (a high-precision spectrograph mostly dedicated to deriving
precise radial velocities) program to search for planets orbiting 
a sample of 104 metal-poor stars (selected [Fe/H] below $-$0.5). Radial velocity time series of each star are presented and searched for signals using several statistical diagnostics. Stars with detected signals are presented, including 3 attributed to the presence of previously announced giant planets
orbiting the stars HD171028, HD181720, and HD190984. Several binary stars and at least one case of a coherent signal caused by activity-related phenomena are presented. One 
very promising new, possible giant planet orbiting the star HD\,107094 is discussed, and the results are analyzed in light of the metallicity-giant planet correlation. We conclude that the frequency of giant planets orbiting metal-poor stars may be 
higher than previously thought, probably reflecting the higher precision of the HARPS survey.  In the metallicity domain of our sample, we also find evidence that the frequency of planets is a steeply rising function of the stellar metal content, as found for higher metallicity stars. 
  \keywords{planets and satellites: formation --
  	    Stars: abundances --
	    planetary systems --
	    Techniques: spectroscopic --
	    Techniques: radial velocities
	    }}

\authorrunning{Santos et al.}
%\titlerunning{}
\maketitle

\section{Introduction}

The past 15 years have seen a plethora of exoplanet discoveries, with more than 450 announcements so far\footnote{See updated table at http://www.exoplanet.eu}. Mostly thanks to the development of the radial-velocity technique, some of the discovered planets have been found to have masses as low as $\sim$2 times the mass of our Earth \citep[][]{Mayor-2009}. The holy grail of exoplanet searches, namely the first detection of an exo-Earth, is closer than ever. 

{
The increasing number of detected planets is also providing strong 
constraints for the models of planet formation and evolution \citep[][]{Udry-2007,Ida-2004a,Mordasini-2009a}. 
Two main models of giant planet formation have been discussed in the literature. On the one hand,
the core-accretion model tells us that giant planets can be formed by the accretion of solids in a protoplanetary disk,
building up a $\sim$10\,M$_\oplus$ core followed by rapid agglomeration of gas \citep[][]{Pollack-1996,Ida-2004a,Mordasini-2009a}.
On the other hand, the ``disk instability'' model suggests that giant planets may be the outcome of direct gravitational
instability of the gas \citep[][]{Boss-1997,Mayer-2002}. The two models predict, however, very different outcomes
for the planet formation process as a function of the chemical composition of the disk (i.e. density of solids) and stellar mass \citep[][]{Matsuo-2007}.
In particular, while the core accretion model predicts a strong metallicity-planet correlation and a higher prevalence of giant planets around
higher mass solar-type stars \citep[][]{Mordasini-2009b,Laughlin-2004}, such trends are less evident 
for the disk instability process \citep[][]{Boss-2002,Boss-2006}. 

Concerning stellar metallicity, the first large, uniform spectroscopic studies comparing large samples of stars with and without planets have confirmed former suspicions \citep[][]{Gonzalez-1997} that stars hosting giant planets are more metal-rich (on average) than single, field stars \citep[][]{Santos-2001,Santos-2004b,Fischer-2005}. 
Stellar mass is also suspected to have a strong influence on the planet formation process \citep[][]{Johnson-2007,Lovis-2007}, but
a clear understanding of this picture is required. This need has inspired the construction of specific samples to search for planets 
around stars of different mass/evolutionary states \citep[e.g.][]{Bonfils-2005b,Setiawan-2005,Johnson-2007,Lovis-2007} 
and different chemical compositions \citep[e.g.][]{Tinney-2003,Fischer-2005b,DaSilva-2006,Santos-2007,Sozzetti-2009}.
}

The HARPS guaranteed time observations (GTO) program started to follow several different samples
of solar-type stars in October 2003 \citep[][]{Mayor-2003b}. The unparalleled long-term precision
of HARPS allowed discovery of several planets among the targets, including the large majority
of the known planets with masses near the mass of Neptune or 
below \citep[e.g.][]{Santos-2004a,Lovis-2006,Mayor-2009}. 

\begin{figure*}[t!]
\center
\resizebox{8.0cm}{!}{\includegraphics{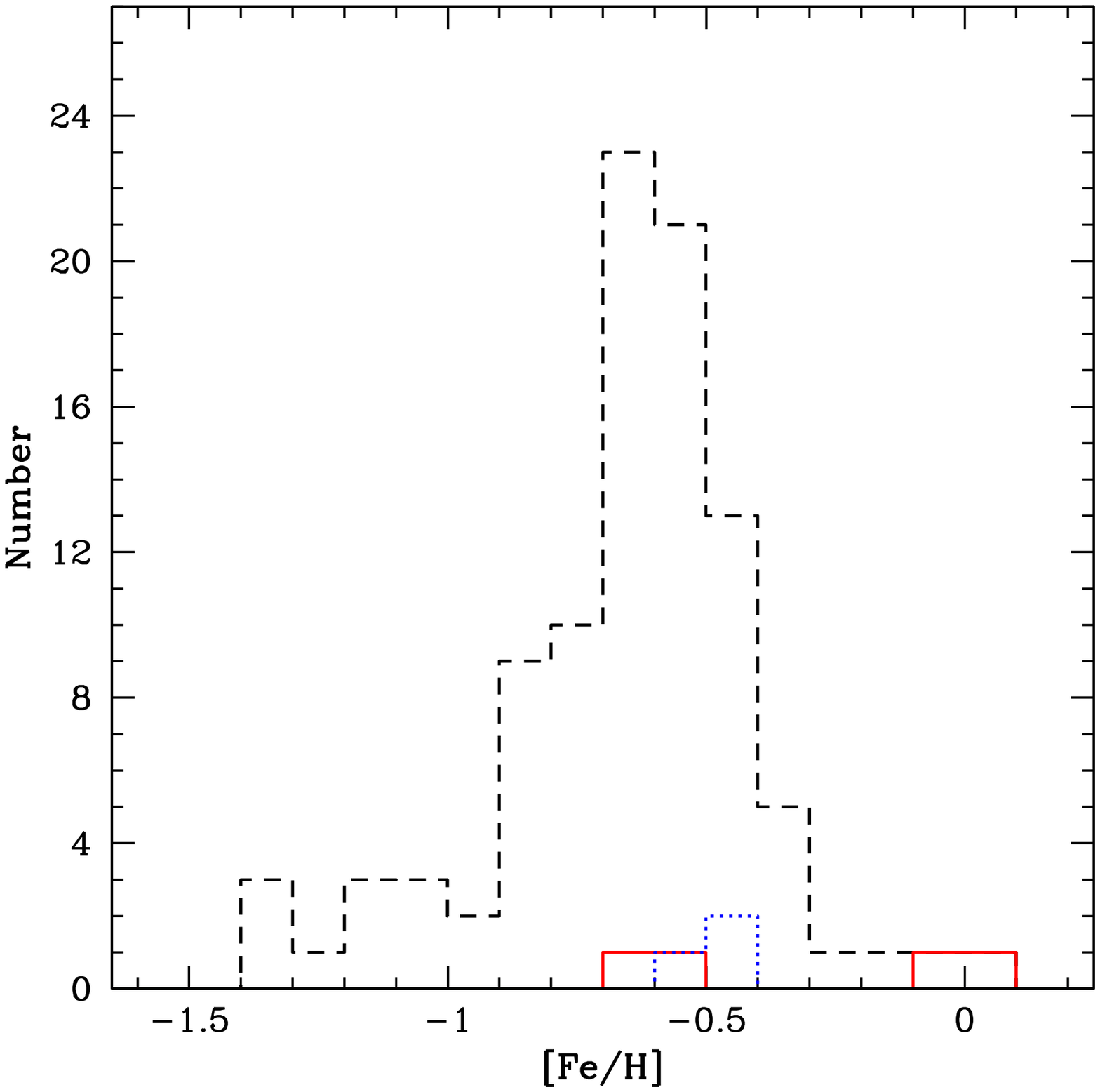}}
\resizebox{8.0cm}{!}{\includegraphics{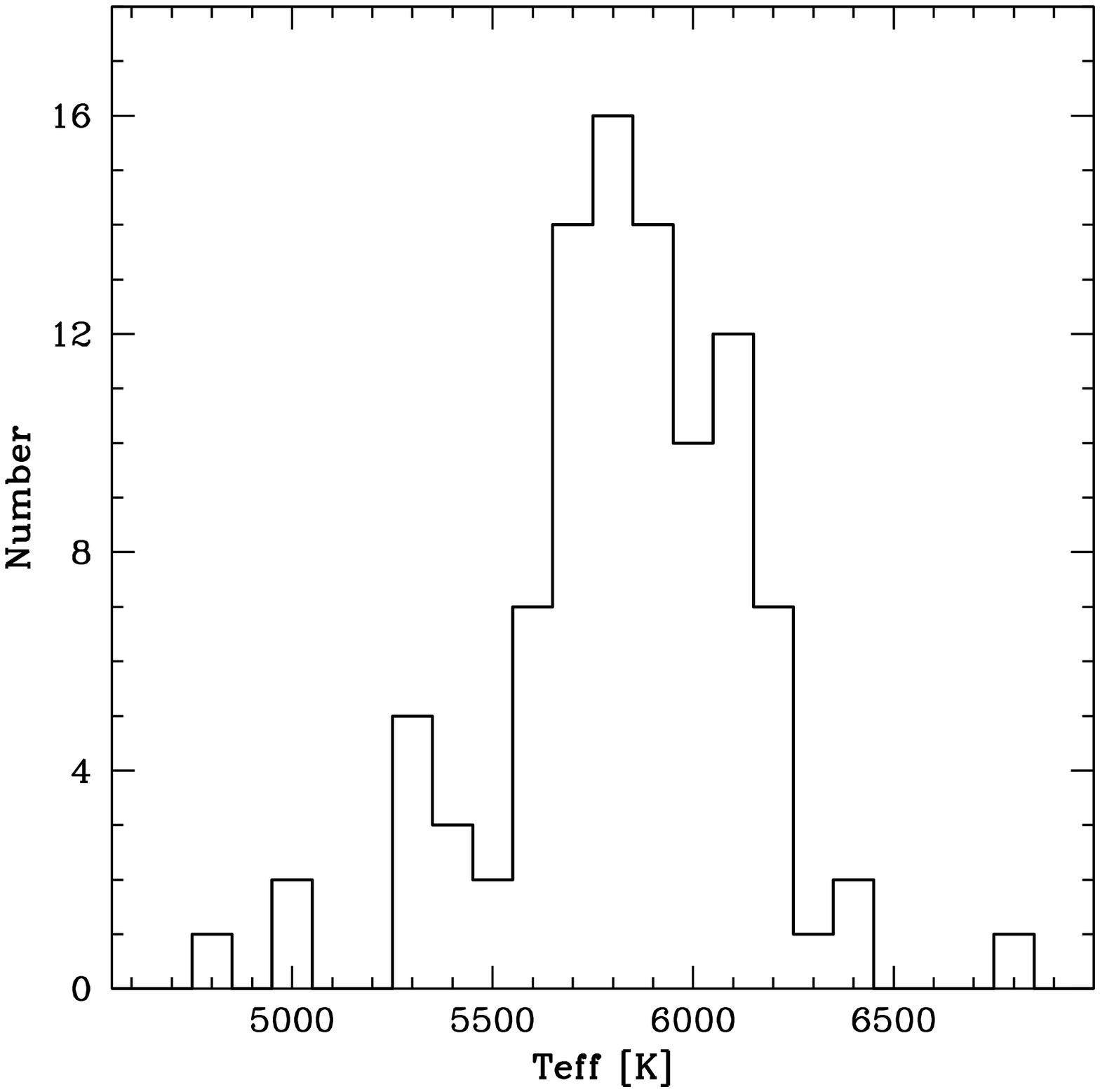}} \\
\resizebox{8.0cm}{!}{\includegraphics{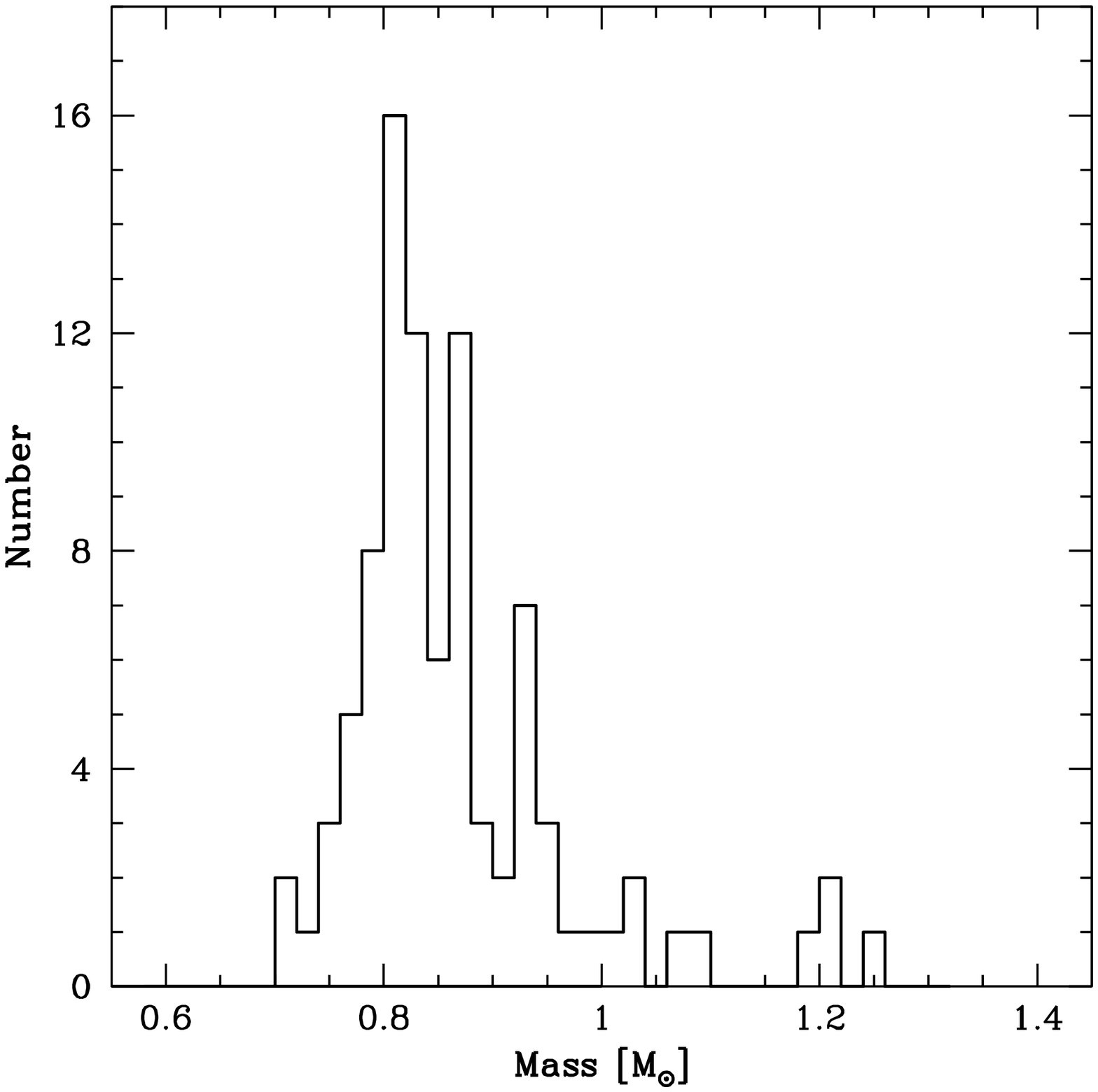}}
\resizebox{8.0cm}{!}{\includegraphics{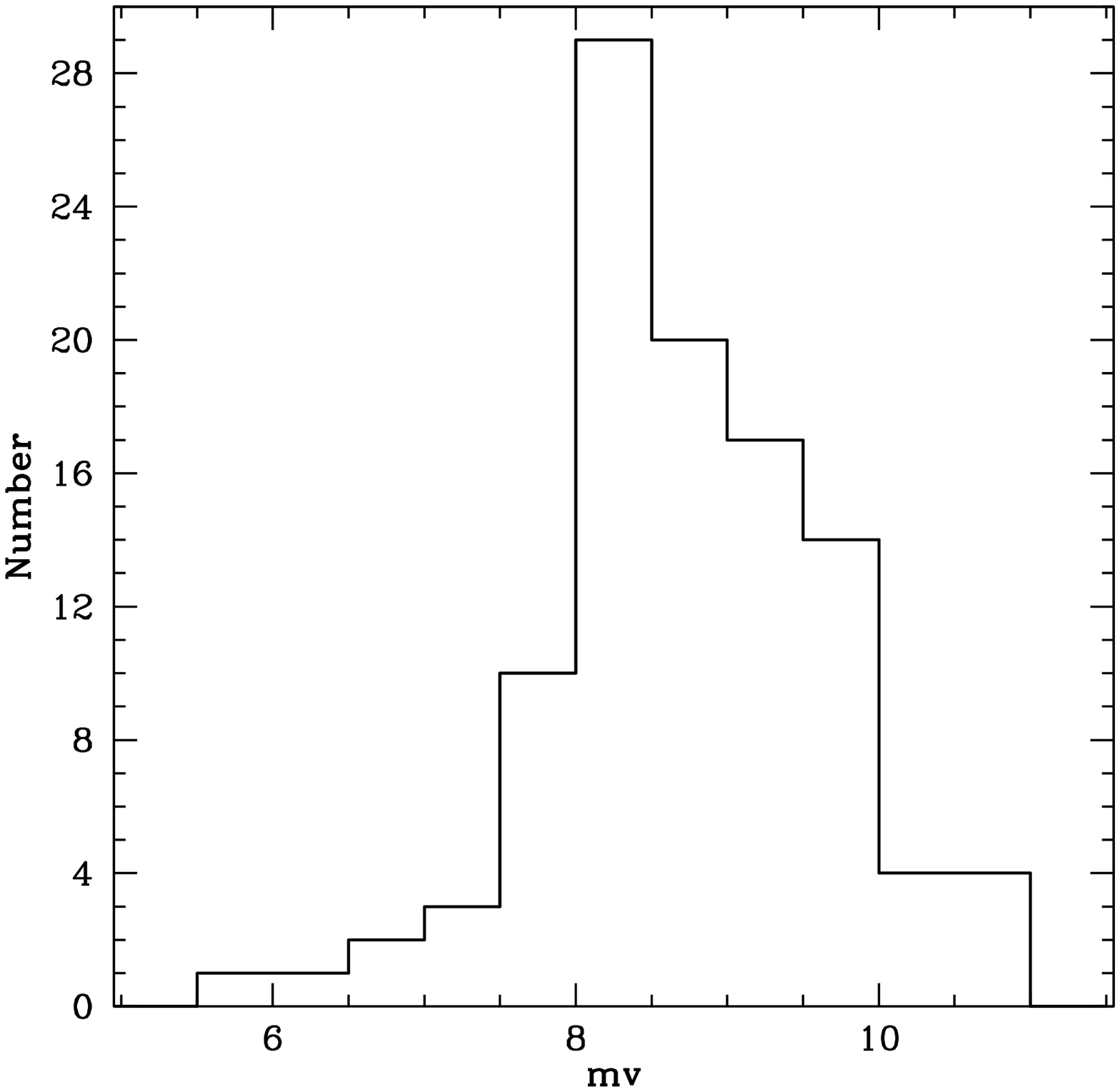}}
\caption{Distributions of metallicity (top left), effective temperature (top right), stellar mass (bottom left), and visual magnitude (bottom right) for all the stars in our sample.}
\label{fig:param}
\end{figure*}

One of the HARPS GTO sub-samples was built to explore how frequently giant planets orbit metal-poor stars to determine 
the metallicity limit below which no giant planets can be observed. In this paper we present the overall results of this program.
In Sects.\,\ref{sec:sample} and \ref{sec:observations} we present our initial sample and observations. In Sect.\,\ref{sec:cleaning}
we show how the sample was cleaned {\it a posteriori} of the presence of stellar binaries and fast-rotating, active stars. 
The orbital parameters for some of the binaries in the sample are also listed. The remaining stars are then discussed in 
 Sect.\,\ref{sec:results}.  The three planets already announced and one more promising candidate are presented.
We conclude in Sect.\,\ref{sec:conclusions}.

\section{The HARPS metal-poor sample}
\label{sec:sample}

\begin{figure}[b]
\resizebox{\hsize}{!}{\includegraphics[bb= 30 430 580 690]{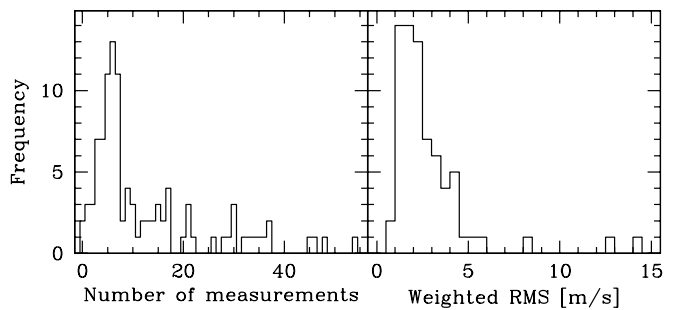}}
\caption{{\it Left}: Histogram of the number of radial-velocity measurements for the stars in our sample. {\it Right:} Distribution of weighted rms values for our stars with at least 5 radial-velocity measurements. For clarity, the distribution is only shown for rms values below 15\,m\,s$^{-1}$.}
\label{fig:mesures}
\end{figure}

To explore the low-metallicity tail of the planet-host stars' distribution, 
one of the samples studied within the HARPS GTO program followed 104 
metal-poor or mild metal-poor solar-type stars. This sample was chosen based on a preliminary version of
the large catalog of \citet[][]{Nordstrom-2004}. From this catalog, we took all
late-F, G, and K stars ($b-y>$0.330) south of $+$10$^{\mathrm{o}}$ of declination that 
have a visual V magnitude brighter than 12. From these, we then excluded all identified visual
and spectroscopic binaries, all stars suspected to be giants, and all those 
with measured projected rotational velocity $v\,\sin{i}$ above $\sim$6.0\,km\,s$^{-1}$ 
\citep[to indirectly exclude the most active stars, for which it is not possible to achieve a high 
radial-velocity precision -- ][]{Saar-1997,Santos-2000a,Paulson-2002}. Finally, we considered 
only those targets with estimated photometric [Fe/H] between $-$0.5 and $-1.5$. As
seen below, a detailed spectroscopic analysis has later shown that some of the
targets fall significantly out of this metallicity interval (see Fig.\,\ref{fig:param}). 

The final 104 stars in the sample (Table\,\ref{tab:targets}) have their V magnitudes between 5.9 and 10.9, 
distributed around an average value of 8.7. After a 15-minute exposure,
these magnitudes allow us to obtain an S/N high enough to derive radial-velocities
with a precision better than 1\,m\,s$^{-1}$ for the majority of the targets.
We note, however, that not all of these were interesting targets for a high-precision radial velocity
planet search program (Sect.\,\ref{sec:cleaning}).

\begin{table}[b]
\caption{List of targets, coordinates (equinox 2000.0), and magnitudes. Full table available online.}
\label{tab:targets}
\begin{tabular}{llll}
\hline\hline
\noalign{\smallskip}
Star & alpha & delta & mv\\
\hline
HD224817 & 00:00:58.2 & $-$11:49:25 & 8.4\\
HD967      & 00:14:04.4 & $-$11:18:41 & 8.4\\
HD11397 & 01:51:40.5 & $-$16:19:03 & 9.0\\
HD16784 & 02:40:38.7 & $-$30:08:07 & 8.0\\
HD17548 & 02:48:51.8 & $-$01:30:34 & 8.2\\
... & ... & ...& ...\\
\hline
\noalign{\smallskip}
\end{tabular}
\end{table}

Precise spectroscopic atmospheric parameters and masses for the targets were derived 
in a separate paper \citep[][]{Sousa-2010}. In Fig.\,\ref{fig:param} we plot the distributions
of metallicities, temperatures, stellar masses, and V magnitudes for the sample. In the upper-left panel, the dashed line represents the whole sample, 
while the filled and dotted lines represent the distributions for stars with high amplitude long term 
radial velocity trends (discussed in Sect.\,\ref{sec:drifts}) and stars with known planets (Sect.\,\ref{sec:planets}), respectively.
Overall, the metallicity distribution peaks around $-$0.65\,dex and has a strong fall above $-$0.4\,dex. The sample includes, however, several stars 
with spectroscopic metallicity above $\sim$$-$0.5\,dex,
even though the initial cut, based on photometric metallicity estimates, was set at $-$0.5\,dex. 
Some of these outliers can be explained by the error bars, both in the spectroscopic and (initial) 
photometric metallicities (the sample had been cut using the latter). The most extreme cases
(HD\,62849, HD\,123517, HD\,144589, and CD$-$231087, all with [Fe/H] above $-$0.3\,dex) are less easy 
to explain, though. HD\,62849 ([Fe/H]=$-$0.17) is a close visual binary (see Sect.\,\ref{sec:bin}). We cannot exclude
light contamination from the secondary star having biased the derivation of the stellar parameters. In any case,
this star was later excluded from the sample (Sect.\,\ref{sec:cleaning}). CD$-$2310879 is the hottest star in our sample,
and it presents a high activity level (Sect.\,\ref{sec:active}). We cannot reject that the presence of active regions has induced 
a systematic error in our parameters. No clear explanation exists for the remaining two stars, namely HD\,123517 and HD\,144589. 
Interestingly, both present clear long-term trends in the radial-velocity data (see discussion in Sect.\,\ref{sec:drifts}). Their 
high metallicities may be a signature of long period giant planets.

\begin{table}[t!]
\caption{Stars excluded from the sample.}
\label{tab:clean}
\begin{tabular}{ll}
\hline\hline
\noalign{\smallskip}
Star & Comment\\
\hline
BD$-$00\,4234	& Active/Fast rotator\\
BD$-$03\,2525 & SB2\\
CD$-$45\,2997	& SB1$\dagger$\\
%CD$-$23\,10879 & Active\\
CD$-$43\,6810	& SB1\\
HD16784		& SB1\\
HD25704    	& Close visual binary\\
HD62849    	& Close visual binary\\
HD128575	& Active/Fast rotator\\
HD134113	& SB1$\dagger$\\
HD161265 	& SB2\\
HD164500 	& SB2\\
HD175179  	& Close visual binary\\
HD187151 	& SB2\\
HD197890	& Active/Fast rotator\\
HD221580	& Giant\\
HD224347	& SB1$\dagger$\\
\hline
\noalign{\smallskip}
\end{tabular}
\newline
$\dagger$ Orbital fit was possible.
\end{table}

\begin{figure}[t!]
\resizebox{\hsize}{!}{\includegraphics[bb= 25 147 580 441]{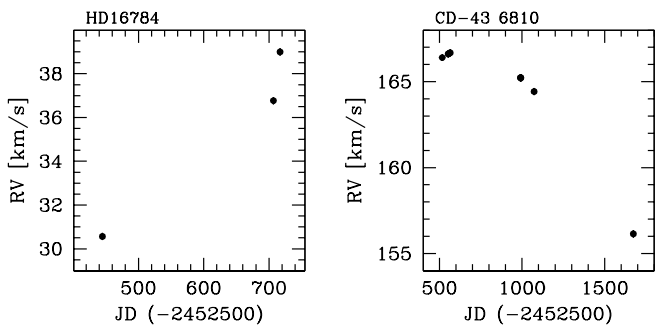}}
\caption{Radial-velocity measurements of \object{HD16784} and \object{CD-436810}, two SB1 binaries from the sample.}
\label{fig:sb1}
\end{figure}

\begin{figure}[t!]
\resizebox{7.8cm}{!}{\includegraphics{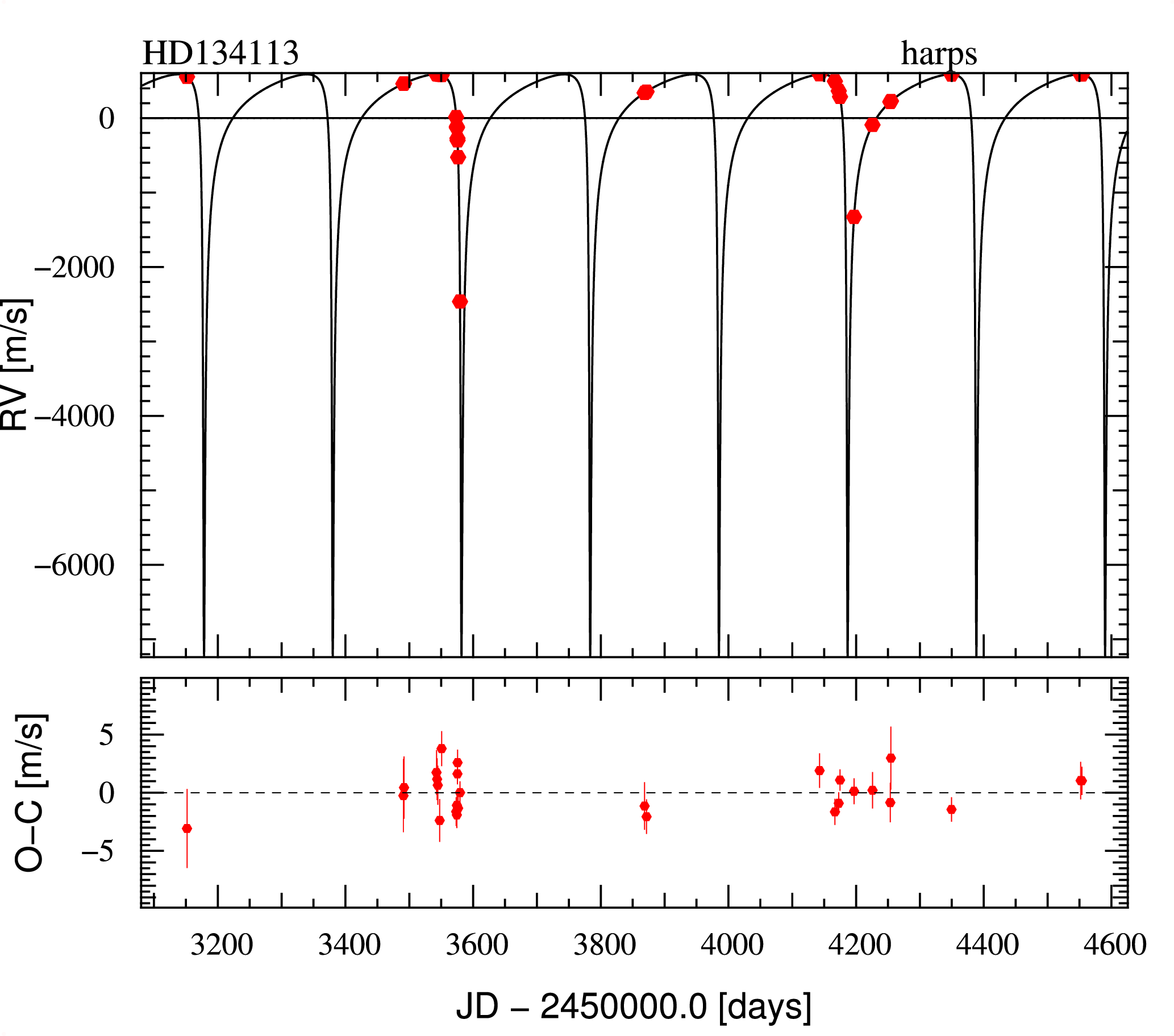}}\\
\resizebox{7.8cm}{!}{\includegraphics{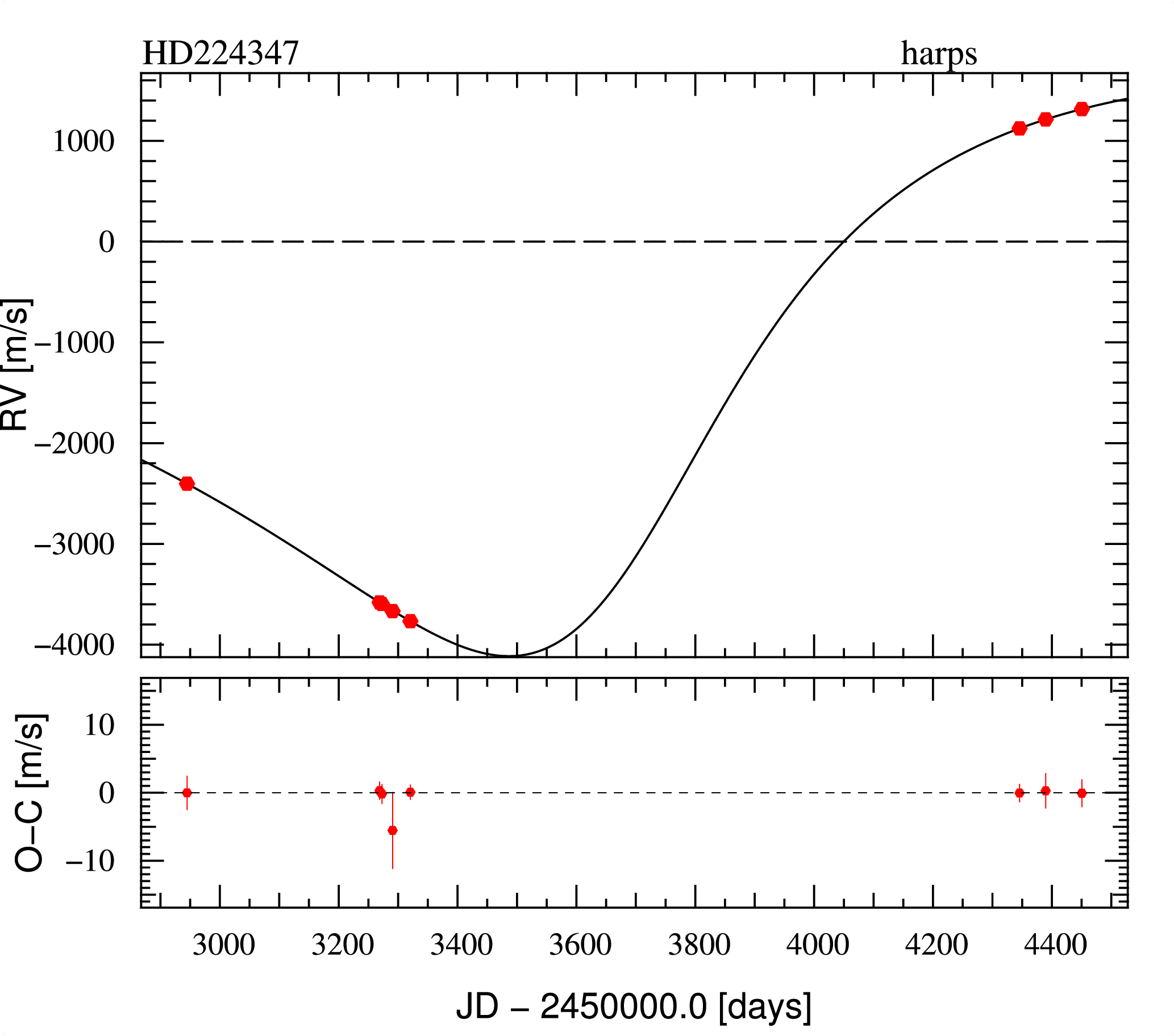}}\\
\resizebox{7.8cm}{!}{\includegraphics{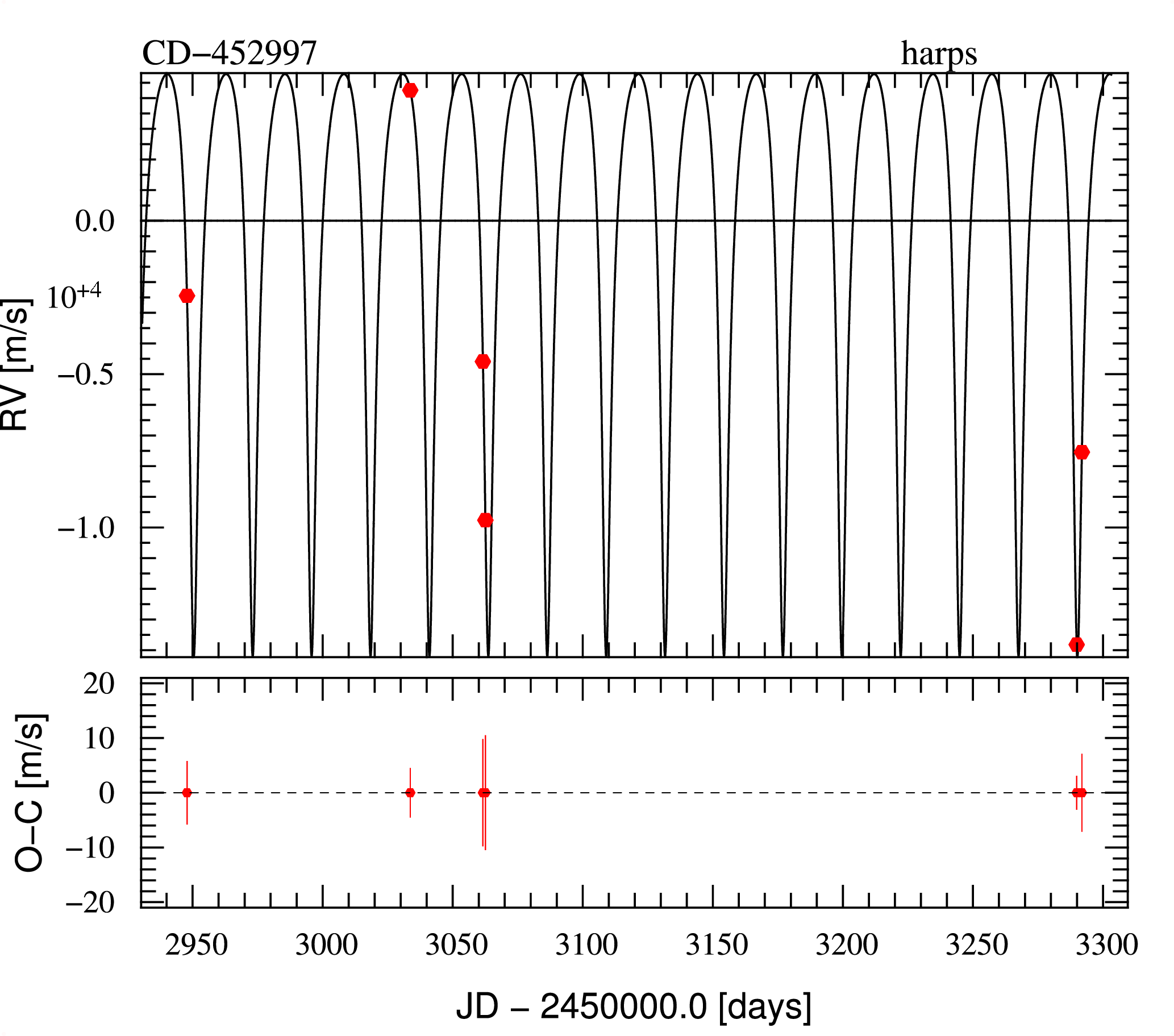}}
\caption{{\it Top}: Radial-velocity measurements of \object{HD\,134113} as a function of time, and
the best Keplerian fit to the data. The residuals of the fit are shown in the lower panel. 
{\it Center and bottom}: Same for HD224347 and CD$-$452997, respectively.}
\label{fig:orbitbin}
\end{figure}

The mass and effective temperature distributions are rather peaked, with maxima around 0.8\,M$_\odot$ and 5700\,K. 
Both distributions are roughly Gaussian shaped, although the former shows a tail towards higher masses.
All stars except HD\,221580 (Sect.\ref{sec:fastrot}) have spectroscopic parameters (temperature and surface gravity) typical 
of dwarfs or subgiants \citep[for more details see][]{Sousa-2010}.

In Table\,\ref{tab:actvsini}, we present the basic stellar parameters for the sample stars, excluding objects discussed in Sect.\,\ref{sec:cleaning}. The effective temperatures,
surface gravities, and stellar metallicities were taken from the detailed spectroscopic analysis presented in \citet[][]{Sousa-2010}. 
The B-V values were taken from the Hipparcos catalog \citep[][]{ESA-1997}, when available, and for the remaining
stars from the NOMAD catalog \citep[][]{Zacharias-2004}. 
The estimate of the projected rotational velocity $v\,\sin{i}$ was derived using a calibration based on the width of the HARPS cross-correlation function,
similar to the one presented for CORALIE in \citet[][]{Santos-2002a}. Due to different observing uncertainties, we adopted a value of $<$1\,km\,s$^{-1}$ for stars presenting an estimated $v\,\sin{i}$ below this value. The activity level of the star, denoted here by the S$_{MW}$ 
and $\log{R'_{HK}}$ values\footnote{Both are presented since the latter depends on B-V, a parameter that we do not
control.}, were derived from the analysis of the \ion{Ca}{ii} H and K lines in the HARPS spectra, using a methodology similar 
to the one presented in \citet[][]{Santos-2000a}. 

%Extreme metallicity cases:
%
%HD62849, feh=-0.17: BVIS
%HD123517; feh=+0.09: linear trend
%HD144589; feh=-0.05: linear trend
%CD-231087; feh=-0.24: very hot and very active; maybe out of method and/or activity errors

\begin{table*}[t]
\caption{List of targets with number of measurements, time span between first and last measurements, average radial velocity, stellar atmospheric parameters, B-V color, activity level, and  $v\,\sin{i}$. Full table
available online.}
\label{tab:actvsini}
\begin{tabular}{lccccccccccc}
\hline\hline
\noalign{\smallskip}
Star & N & Span  & $<$RV$>$      &  T$_{eff}$ & $\log{g}$ & [Fe/H] & S$_{MW}$ & $\sigma(S_{MW})$ & $B-V$ & $\log{R'_{HK}}$ & $v\,\sin{i}$ \\
        &     & [days] & [km\,s$^{-1}$]  & [K]             & [cgs]         &             &                     &                                    &              &                               &[km\,s$^{-1}$]\\
\hline
HD967 & 34 & 2132 & -24.3 & 5568$\pm$17 & 4.53$\pm$0.02 & -0.68$\pm$0.01 & 0.176 & 0.003 & 0.645 & -4.91 & $<$1\\
HD11397 & 33 & 2075 & 41.1 & 5564$\pm$26 & 4.46$\pm$0.04 & -0.54$\pm$0.02 & 0.184 & 0.002 & 0.693 & -4.89 & $<$1\\
HD17548 & 10 & 1504 & -15.4 & 6011$\pm$26 & 4.44$\pm$0.02 & -0.53$\pm$0.02 & 0.162 & 0.005 & 0.529 & -4.91 & $<$1\\
HD17865 & 21 & 1853 & 31.9 & 5877$\pm$24 & 4.32$\pm$0.03 & -0.57$\pm$0.02 & 0.159 & 0.002 & 0.568 & -4.96 & $<$1\\
HD22879 & 36 & 1857 & 120.4 & 5884$\pm$33 & 4.52$\pm$0.03 & -0.81$\pm$0.02 & 0.165 & 0.003 & 0.554 & -4.91 & $<$1\\
... & ... & ...& ... & ... & ...&... & ... & ...& ... & ... & ...\\
\hline
\noalign{\smallskip}
\end{tabular}
\end{table*}

\section{Observations}
\label{sec:observations}

From October 2003 to July 2010, a total of 1301 precise radial-velocity measurements were
obtained with HARPS for the stars in our program. The observations were done as part of
the HARPS GTO program, and later complemented by two separate follow-up
programs for a few stars (ESO programs ID 082.C-0212 and 085.C-0063\footnote{In this latter case only for a
star discussed in \ref{sec:ambiguous}}) aimed at searching for Neptune-like planets orbiting a sample
of moderately metal-poor stars. Most of the stars have 5 or more measurements (Fig.\,\ref{fig:mesures}).

During the first three years of the survey, the measurements were obtained with $\sim$2\,m\,s$^{-1}$ precision.
The defined exposure times were not long enough to average out the ``noise'' coming from
stellar oscillations \citep[][]{Bouchy-2005c}. Since 2006, this strategy has been revised, and the 
full precision of HARPS was exploited. Exposure times were then set to be 900\,s,
allowing the precision of the individual measurements to be increased to below 1\,m\,s$^{-1}$.
The observing strategy used does not avoid other sources of noise, including those
coming from stellar granulation and activity \citep[e.g.][]{Dumusque-2010}. As seen in Fig.\,\ref{fig:mesures}, if we consider only objects with at least 
5 measurements, most of the stars present a radial-velocity rms of $\sim$1-2.5\,m\,s$^{-1}$, as expected.
In this figure we limited the rms range to values up to $\sim$15 \,m\,s$^{-1}$ for the sake of clarity. Only 10 objects fall above this limit.

The data were reduced using the latest version of the HARPS pipeline. This includes a correction for the secular 
accelleration \citep[e.g.][]{Zechmeister-2009}, for which we used Hipparcos parallaxes and proper motions \citep[][]{ESA-1997}, whenever available.

\section{Cleaning the sample}
\label{sec:cleaning}

A few stars from our original sample were found to be unsuitable targets for a planet search survey for different reasons (binarity, activity, high rotation -- Table\,\ref{tab:clean}). 
In this section we present these targets, which will be excluded from the rest of the discussion.

\subsection{Binaries}
\label{sec:bin}

Though known binary stars were {\it a priori} excluded from the sample, a few SB1, SB2, or close visual binary systems passed through our first selection criteria. After a few measurements, these cases were usually discarded from the sample. 

% BVIS:
%HD25704 
%HD62849 
%HD175179

%SB2:
%BD-032525  (Many refs.; SB2 by inspection of spectrum and by Pourbaix (2004))
%HD161265  (2 references in literature; None on binarity; bad spectrum)
%HD164500 (8 refs. 2 RVs in Nordstrom, no reference to binarity; but SB2 confirmed by inspection of spectrum)
%HD187151 (1 single reference (Nordstrom); no RV; SB2 by inspection of spectrum)

%SB1 without orbit:
%HD16784 (many refs. Nordstrom: no ref to binarity; only 1 Nordstrom RV)
%CD-436810 (not many refs, and no ref to binarity; only 1 Nordstrom RV)

% Notas:

Three of these stars, \object{HD25704} (number of measurements, Nmes=20), \object{HD62849} (Nmes=17), and \object{HD175179} (Nmes=3), were found to be close visual binary systems, after visual inspection of the guiding camera images.
These have been at some point excluded from the sample given that the light contamination from the stellar companion in the HARPS fiber precludes any determination of precise radial velocities. This contamination is seeing dependent. These stars are not adequate for a planet search survey.

\object{BD$-$03\,2525} (Nmes=2), \object{HD161265} (Nmes=2), \object{HD164500} (Nmes=2), and \object{HD187151} (Nmes=1) were also excluded from the sample since an inspection of the cross-correlation function revealed that they are SB2 binaries. \object{BD$-$03\,2525} was also classified as a spectroscopic binary in  \citet[][]{Pourbaix-2004} (after the beginning of our survey).

Five other stars in our sample were found to be SB1 binaries. For three of these (see next section), we obtained enough HARPS points to obtain a reliable orbital solution (Sect.\,\ref{sec:binaryfits}). For the remaining two, \object{HD16784} (Nmes=3; span=272 days) and \object{CD$-$43\,6810} (Nmes=9; span=1157\,days), the reduced number of measurements or their span did not allow us to fit any Keplerian function to the data (Fig.\,\ref{fig:sb1}).  

\subsubsection{Binaries with orbital fits}
\label{sec:binaryfits}

For three of the binary stars in our sample, we gathered enough measurements to allow
derivation of at least a tentative orbital solution. 

\begin{table*}[t]
\caption[]{Elements of the fitted orbits for SB1 binaries.}
\begin{tabular}{lllll}
\hline
\hline
\noalign{\smallskip}
			&  \object{HD\,134113} & \object{HD\,224347}	    &  \object{CD$-$45\,2997}$\dagger$\\
\hline
$P$             	& 201.674$\pm$0.008             & 6380$\pm$2500	&	22.6	    & [d]\\
$T$             	& 2\,453\,984.87$\pm$0.01    & 2\,453\,665$\pm$78      &  2\,453\,176.8 &  [d]\\
$a$			& 0.66                                         & 7.0				   & 0.15 & [AU]\\
$e$             	& 0.888$\pm$0.001                 & 0.56$\pm$0.11		    & 0.50 & \\
$V_r$           	& $-$60.19$\pm$0.06             & 1.38$\pm$0.17 &  39.2 & [km\,s$^{-1}$]\\
$\omega$        	& 163.1$\pm$0.2                     & 221$\pm$4	& 172		   & [degr] \\ 
$K_1$           	& 3907$\pm$42                       & 2890$\pm$210		& 9500	  & [m\,s$^{-1}$] \\
$f_1(m)$        	& 121,138\,10$^{-6}$	     &  9\,062.746\,10$^{-6}$ &	 1\,301.588\,10$^{-6}$	   & [M$_{\odot}$]\\ 
$\sigma(O-C)$	& 1.55                                        & 	0.56		& --	   & [m\,s$^{-1}$]  \\    
$N$             	& 28                                           & 8			& 6		   &  \\
$m_2\,\sin{i}$  	& 48                                           & 210			& 92	   & [M$_{\mathrm{Jup}}$]\\
\noalign{\smallskip}
\hline
\end{tabular}
\newline
$\dagger$ Number of points is similar to number of free parameters; no errors available.
\label{tab:orbitbin}
\end{table*}

{\it \object{HD\,134113}}: Though previously known to be a spectroscopic binary \citep[][]{Latham-2002}, we missed this classification when preparing the sample. A total of 28 precise radial-velocities were thus obtained for this star. A Keplerian fit to the data confirms and refines the previously suspected solution, with P= 201.7\,days, $e$=0.89, and K=3.9\,km\,s$^{-1}$. The complete set of orbital parameters is listed in Table\,\ref{tab:orbitbin}. Considering a stellar mass of 0.88 solar masses \citep[][]{Sousa-2010}, this signal is compatible with a minimum mass companion with 48 times the mass of Jupiter, a possible brown dwarf companion.

{\it \object{HD\,224347}}: There is no previous reference in the literature of this star being a spectroscopic binary.
Our 8 precise radial velocities obtained with HARPS show, however, a clear high-amplitude signal.
The radial-velocities are well fit with a Keplerian function with period 6380\,days, eccentricity of
0.56, and K=2.9\,km\,s$^{-1}$. The large uncertainties (see Table\,\ref{tab:orbitbin}) come from our data not covering one full orbital period, and the phase space is not covered well. This orbital solution should be considered as preliminary. Considering a stellar mass of 0.95 solar masses \citep[][]{Sousa-2010}, this signal is compatible with the presence of a minimum mass companion with 210 times the mass of Jupiter orbiting HD\,224347.

{\it \object{CD$-$45\,2997}}: 6 radial-velocity measurements of this star were obtained, showing a clear high-amplitude signal in the data. Though the number of measurements is small (and comparable with the number of free parameters to fit), a tentative Keplerian fit holds a period of 22.6\,days, eccentricity of 0.50, and a semi-amplitude of 9.5\,km\,s$^{-1}$. The signal is compatible with a 92 Jupiter mass stellar companion to this 0.72 solar mass star. We note, however, that this orbital solution should be regarded with caution.

\subsection{Fast-rotating or giant stars}
\label{sec:fastrot}

{\it \object{HD221580}}: The 54 precise radial-velocity measurements of this star show a high-amplitude radial-velocity variation (total amplitude $\sim$80\,m\,s$^{-1}$, rms=19.4\,m\,s$^{-1}$). The data presents a rather complex behavior, and no clear periodicity could be found. Later detailed spectroscopic analysis has revealed that this star is a metal-poor giant, with T$_{eff}$=5322\,K, $\log{g}$=2.68, and [Fe/H]=$-$1.13 \citep[][]{Sousa-2010}. The observed ``noise'' is thus likely to be related to intrinsic stellar variability. Very similar stellar parameters were derived by \citet[][]{Gratton-2000}, who classified this star as a red horizontal branch giant. No reference to its evolutionary status is given in \citet[][]{Nordstrom-2004}.

{\it \object{HD128575}, \object{HD197890}, and \object{BD$-$00\,4234}}: These three stars were excluded from the sample since the first (and only) obtained HARPS spectrum has shown that their CCFs were very large and shallow, or were very difficult to identify. This behavior is typical of fast-rotating stars. These objects are not suitable for a precise radial velocity search for planets.

\section{Results}
\label{sec:results}

After removing the 16 targets discussed above, we are left with 88 stars that we consider suitable for a high-precision radial-velocity planet search.
From these, however, 24 objects have 5 or less measurements, making it difficult to make any firm conclusions. These are discussed separately from the others. For the remaining 64 stars (with 6 or more measurements), a more detailed analysis
is done.

\subsection{Stars with $\leq$5 measurements}

\begin{figure*}[t!]
\resizebox{5.9cm}{!}{\includegraphics[bb= 18 160 580 430]{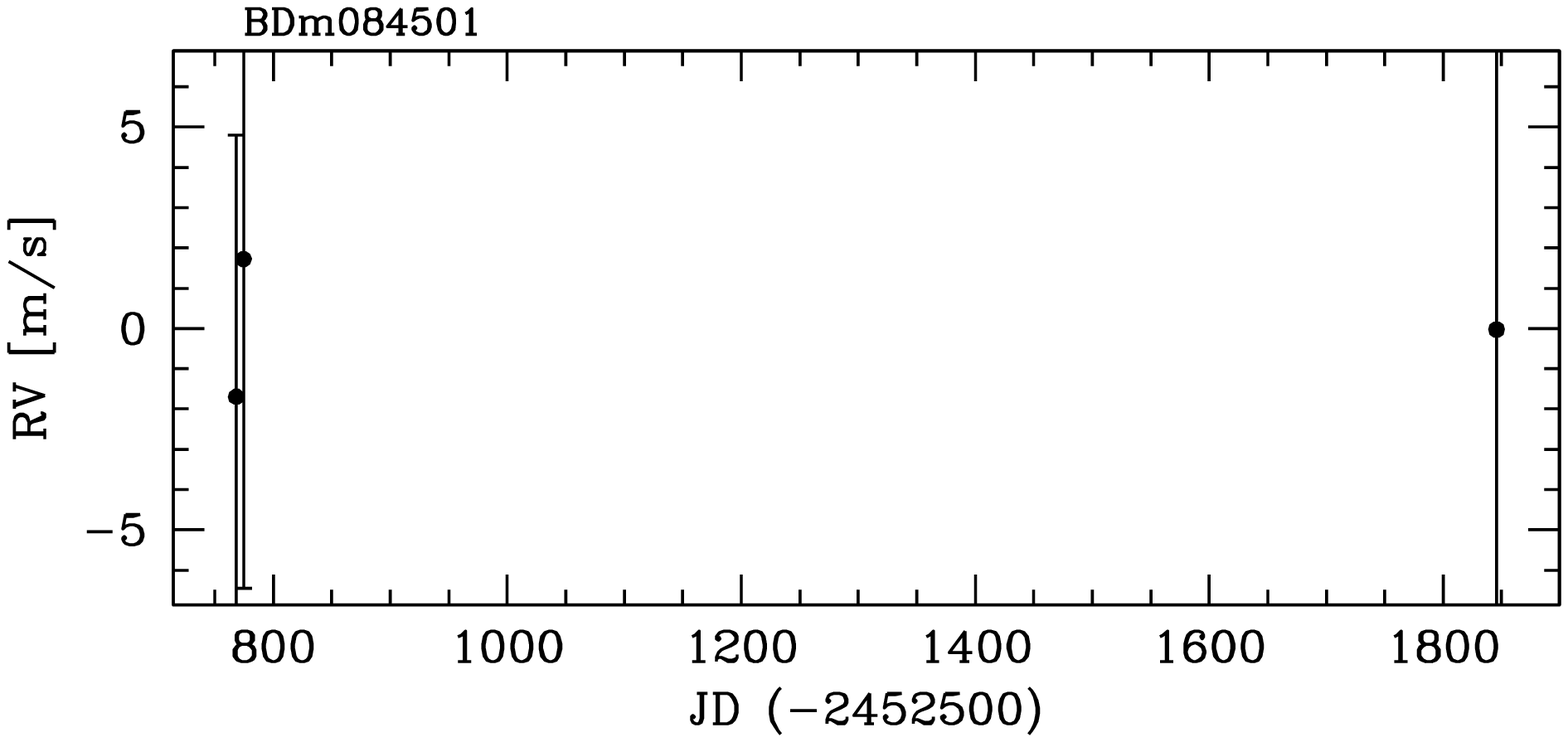}}
\resizebox{5.9cm}{!}{\includegraphics[bb= 18 160 580 430]{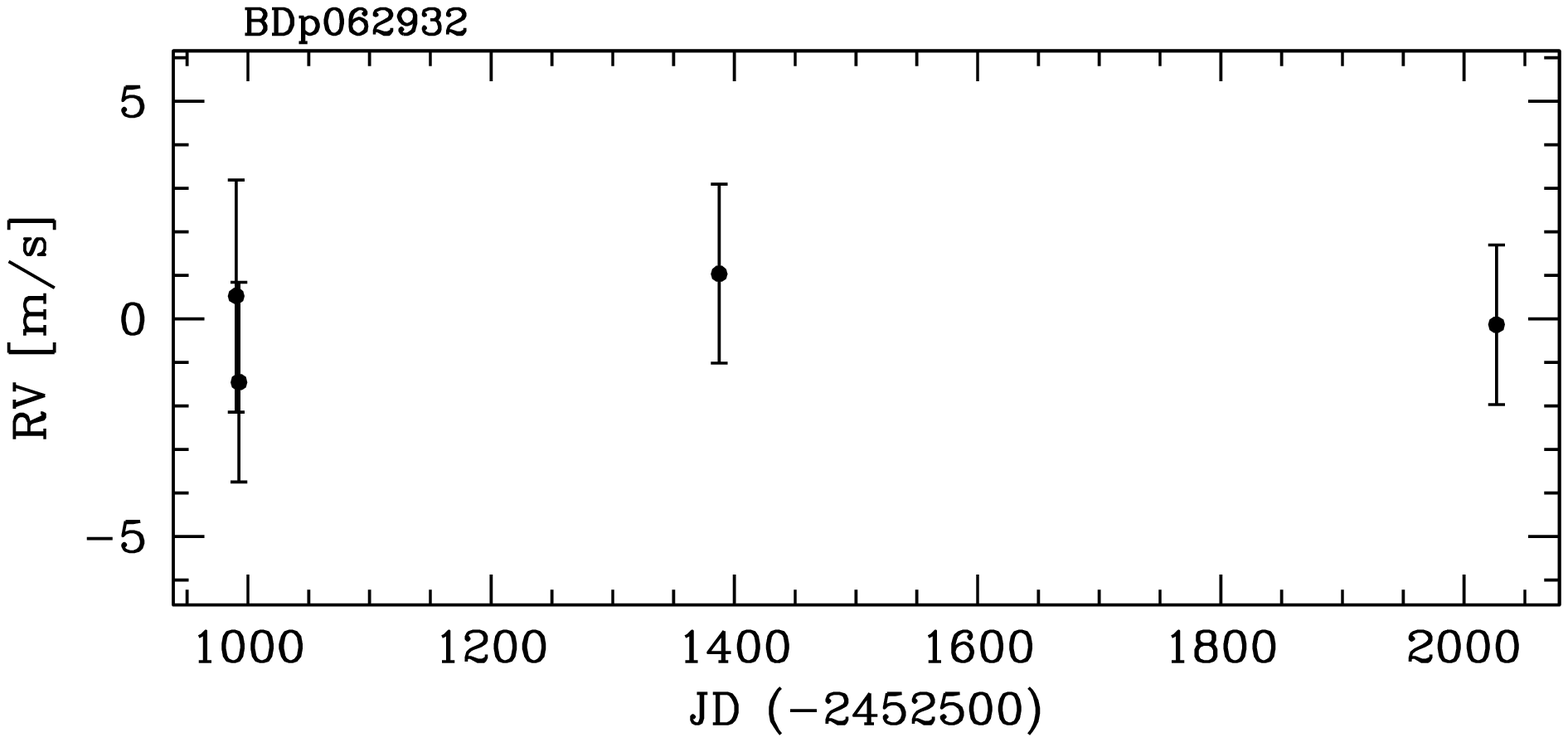}}
\resizebox{5.9cm}{!}{\includegraphics[bb= 18 160 580 430]{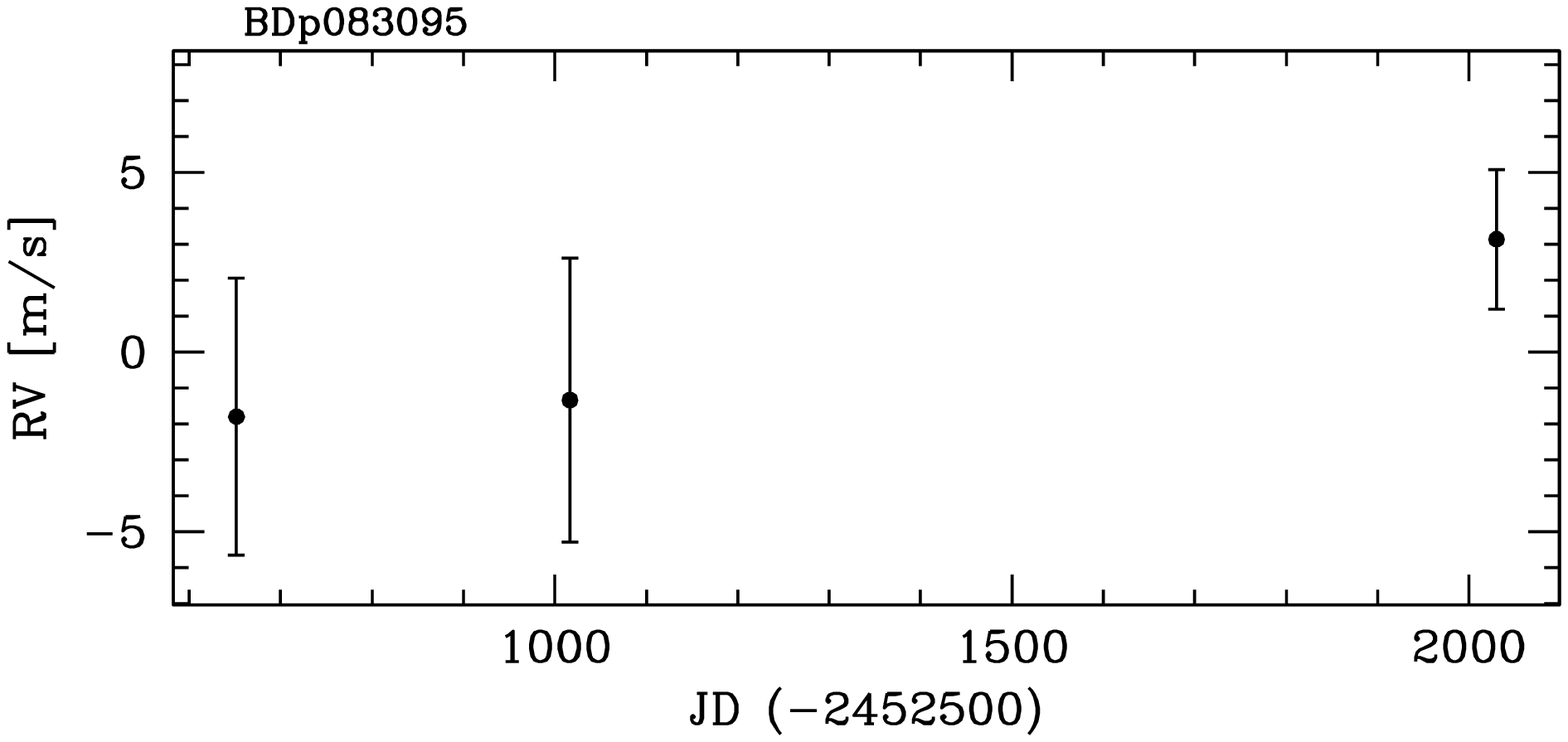}}\\
\resizebox{5.9cm}{!}{\includegraphics[bb= 18 160 580 430]{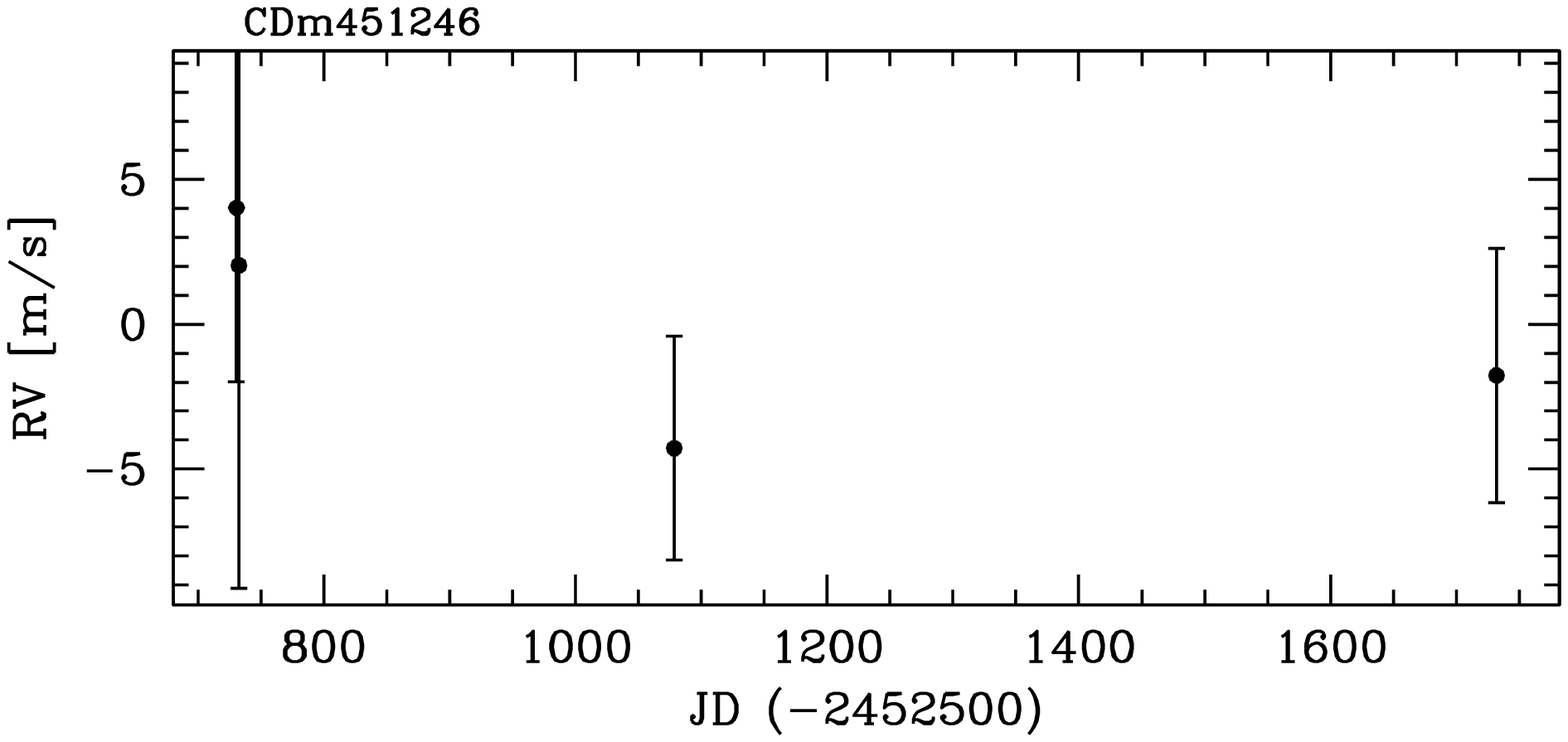}}
\resizebox{5.9cm}{!}{\includegraphics[bb= 18 160 580 430]{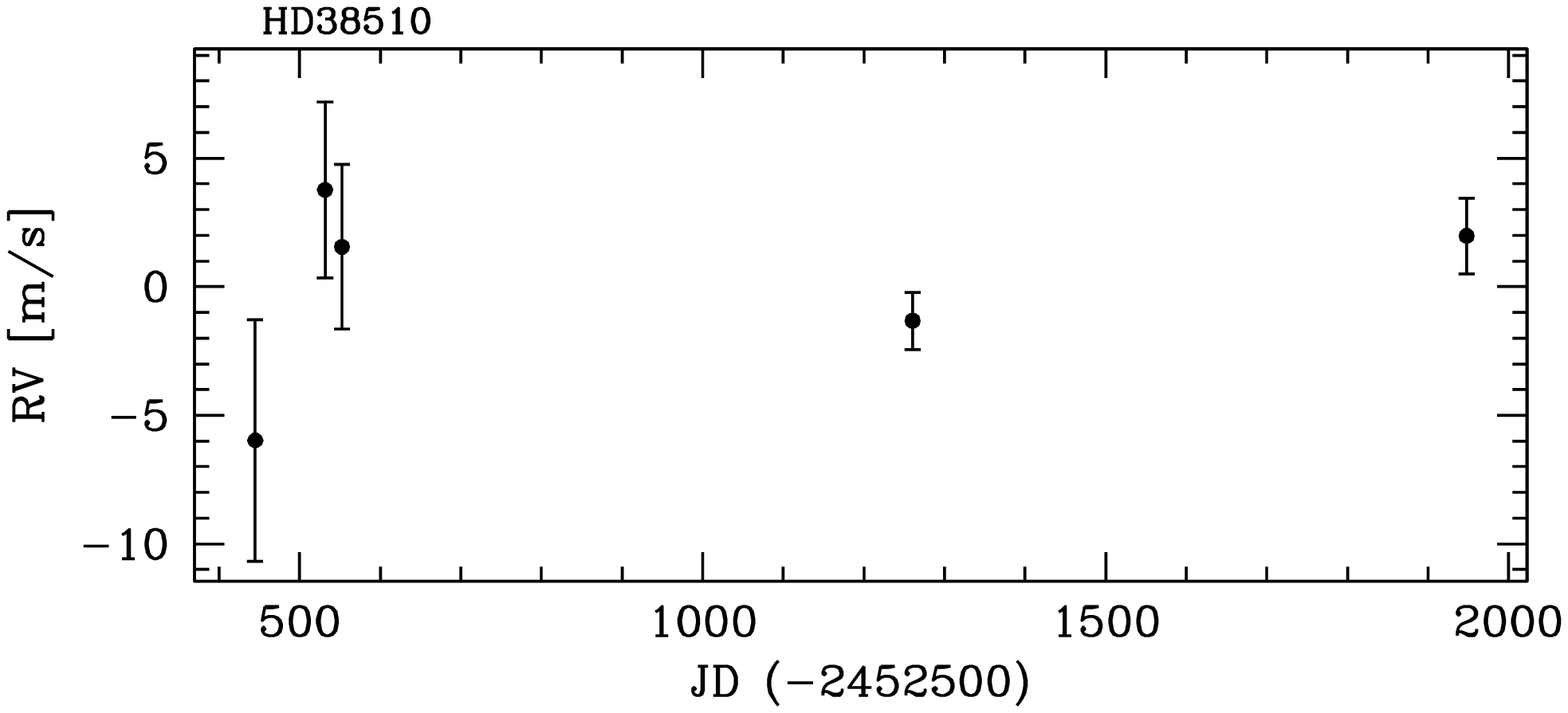}}
\resizebox{5.9cm}{!}{\includegraphics[bb= 18 160 580 430]{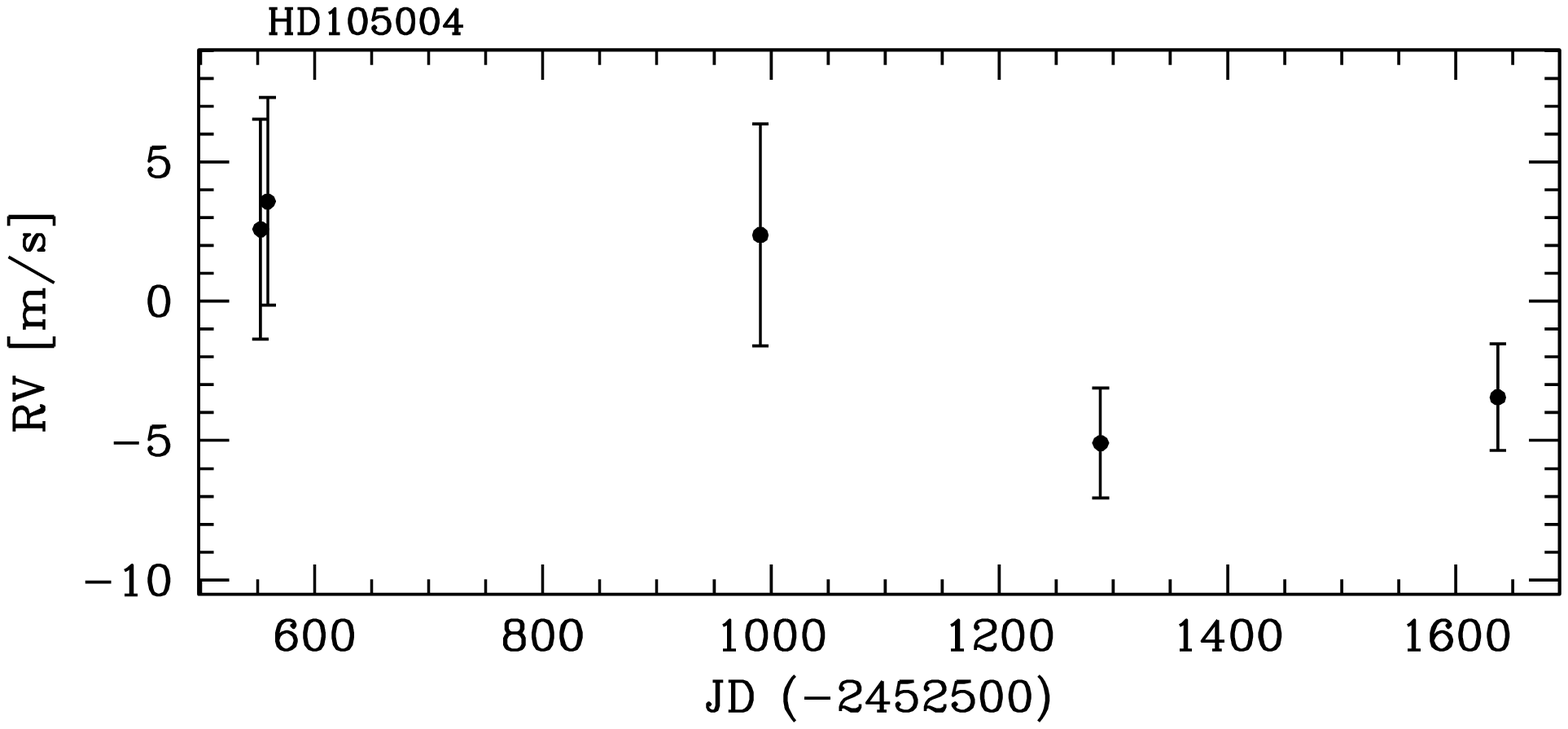}}\\
\resizebox{5.9cm}{!}{\includegraphics[bb= 18 160 580 430]{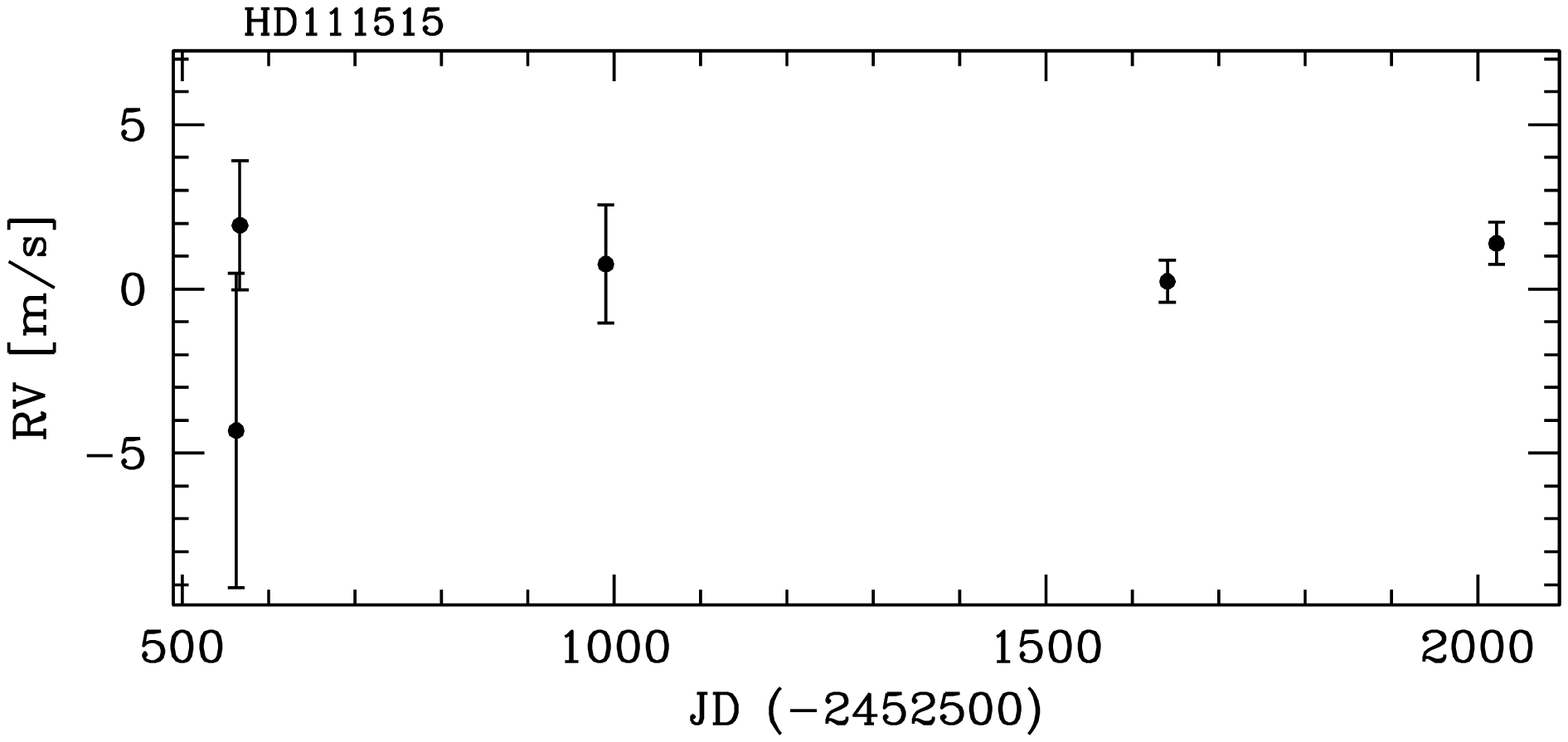}}
\resizebox{5.9cm}{!}{\includegraphics[bb= 18 160 580 430]{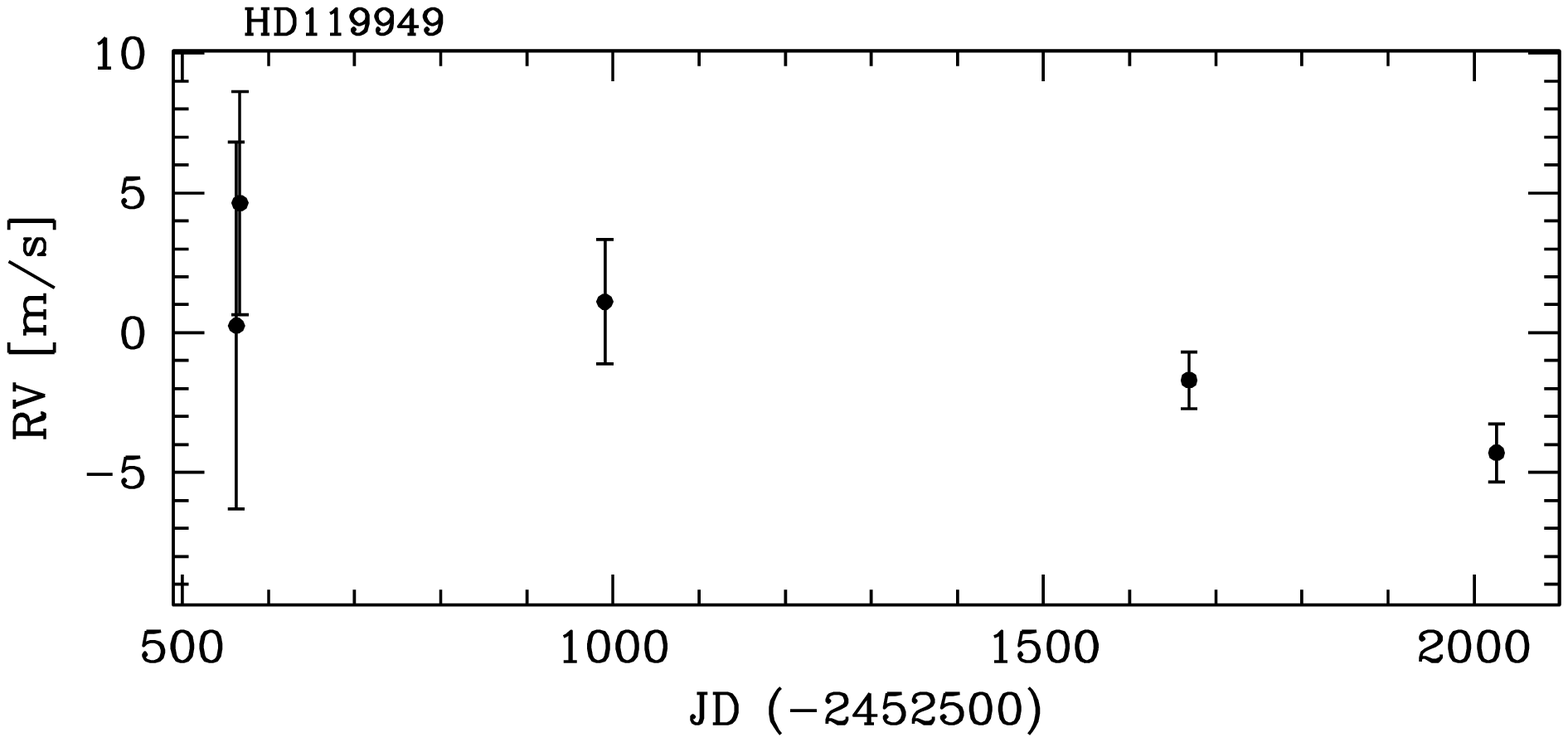}}
\resizebox{5.9cm}{!}{\includegraphics[bb= 18 160 580 430]{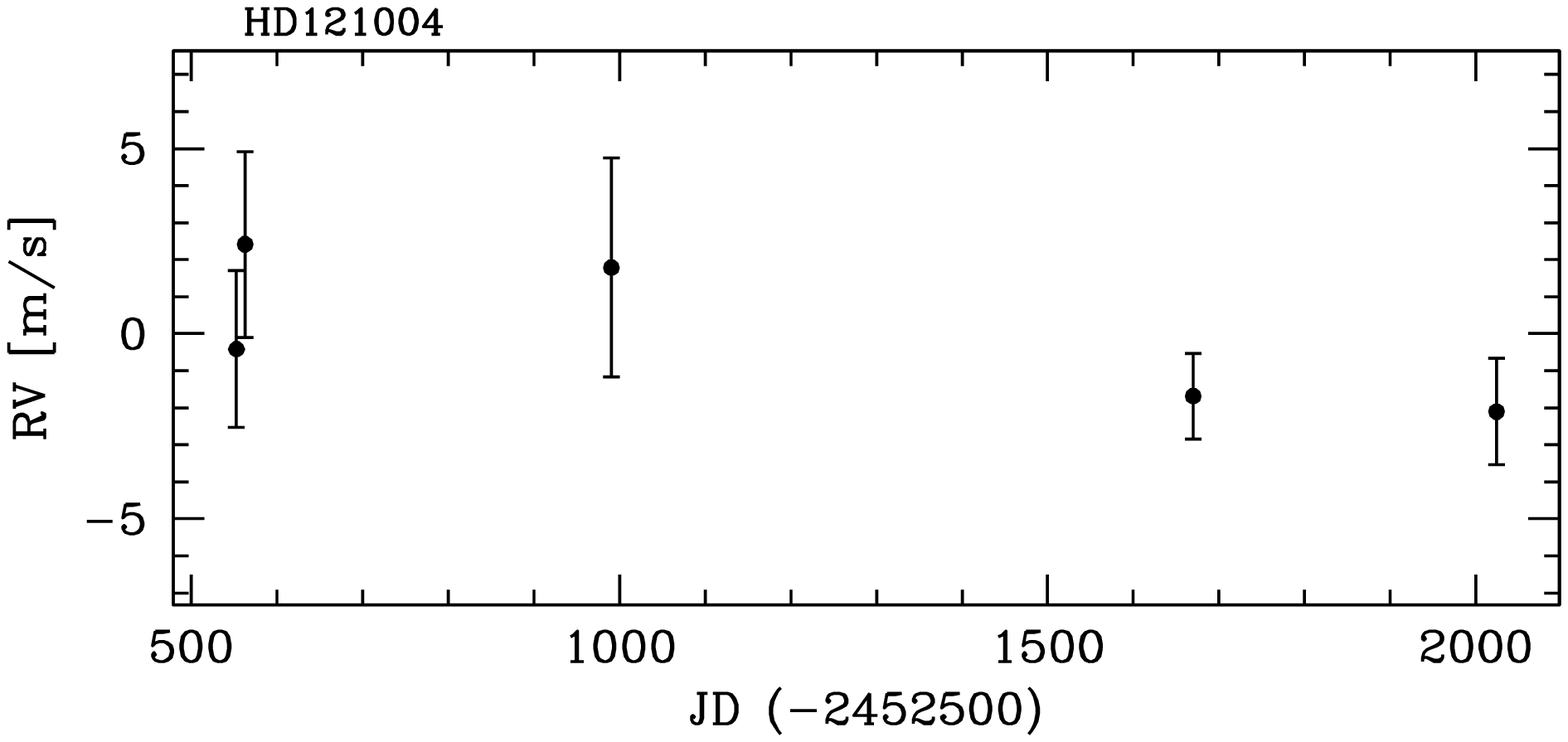}}\\
\resizebox{5.9cm}{!}{\includegraphics[bb= 18 160 580 430]{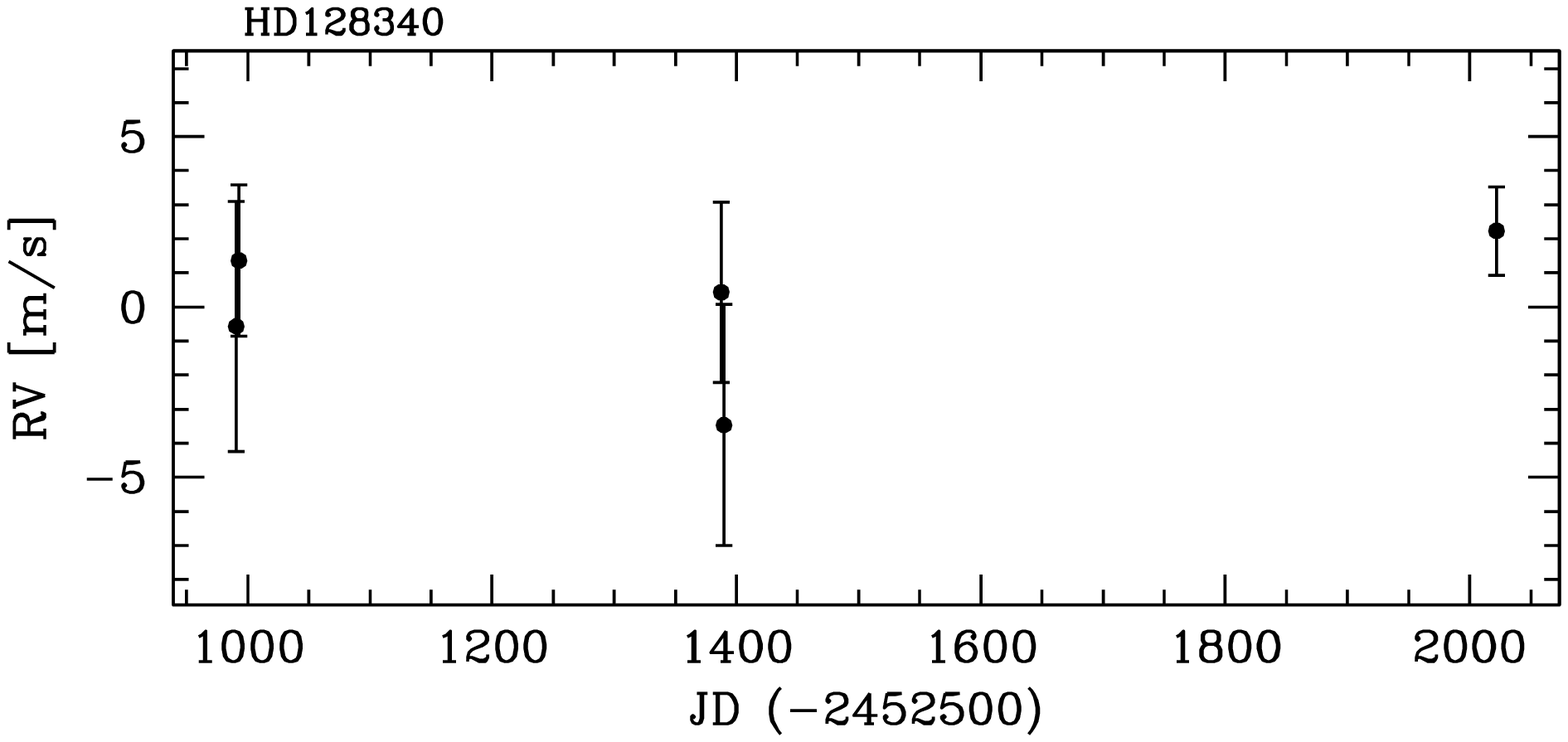}}
\resizebox{5.9cm}{!}{\includegraphics[bb= 18 160 580 430]{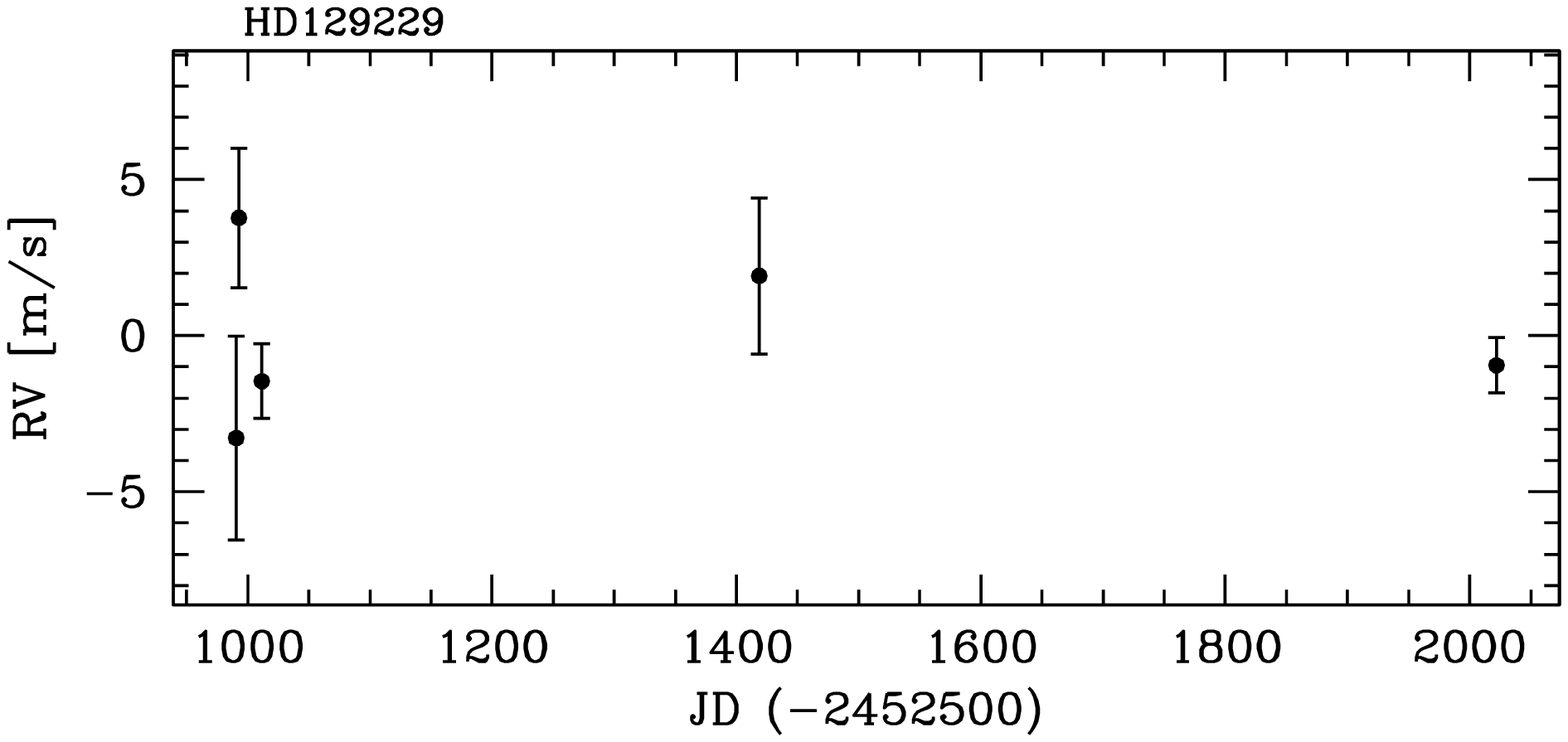}}
\resizebox{5.9cm}{!}{\includegraphics[bb= 18 160 580 430]{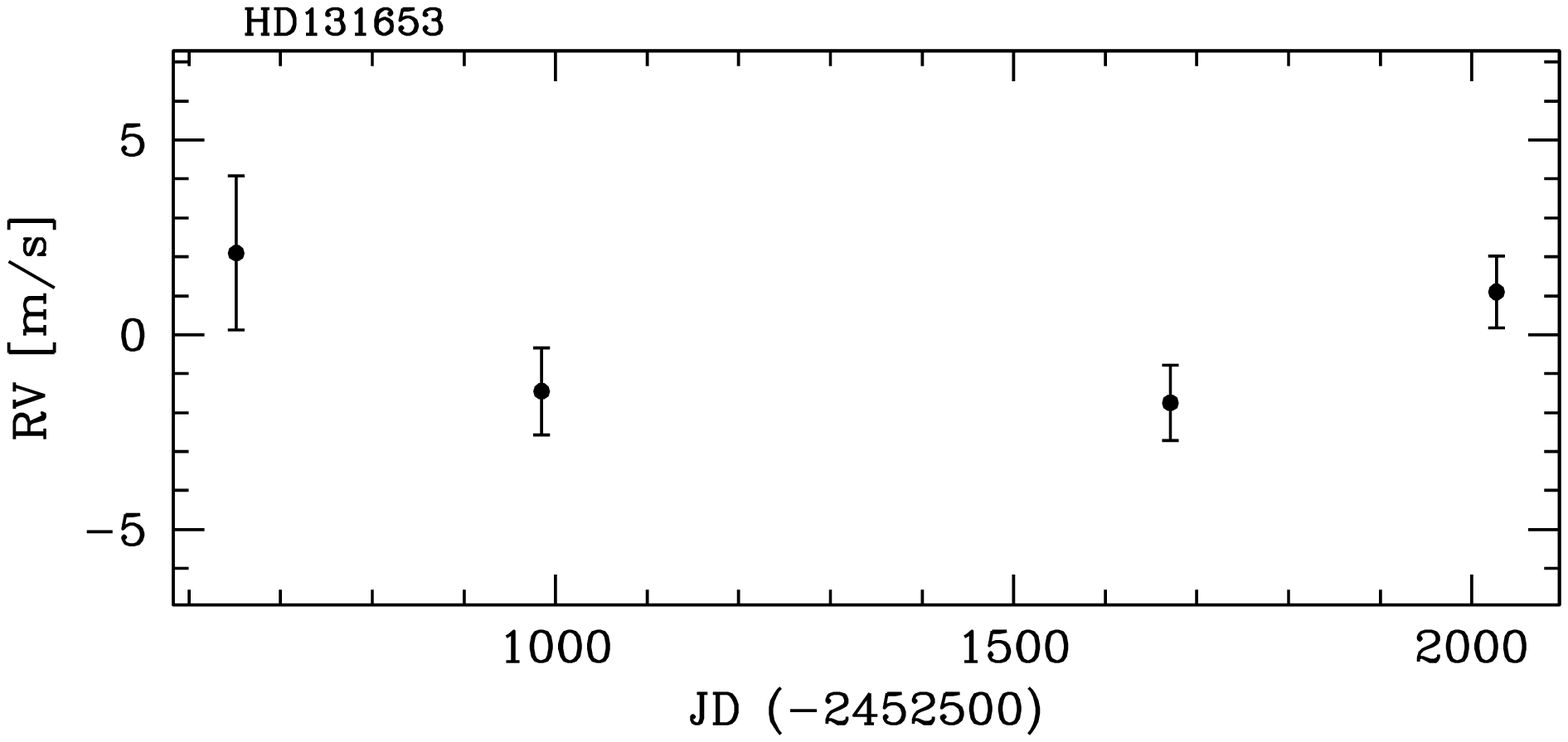}}\\
\resizebox{5.9cm}{!}{\includegraphics[bb= 18 160 580 430]{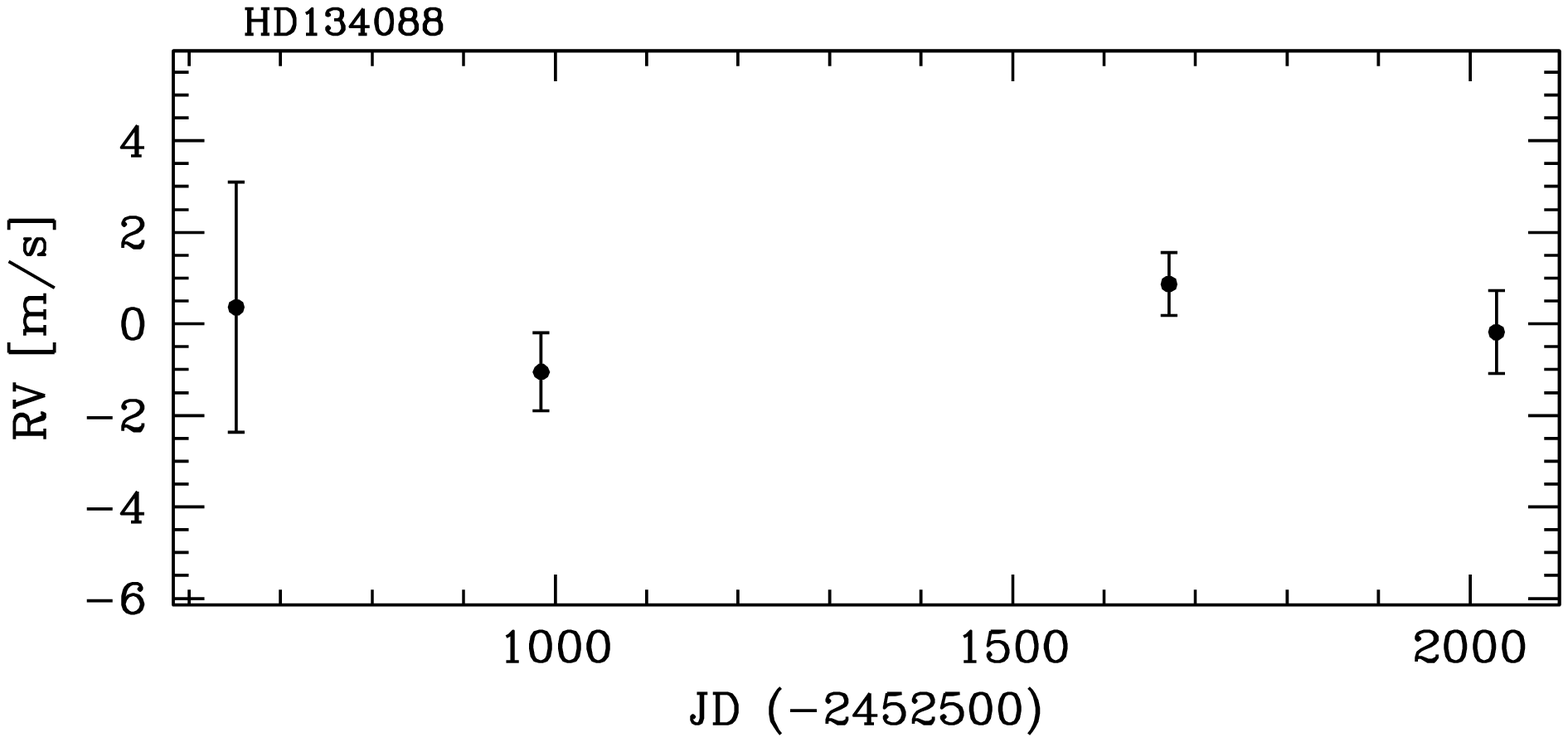}}
\resizebox{5.9cm}{!}{\includegraphics[bb= 18 160 580 430]{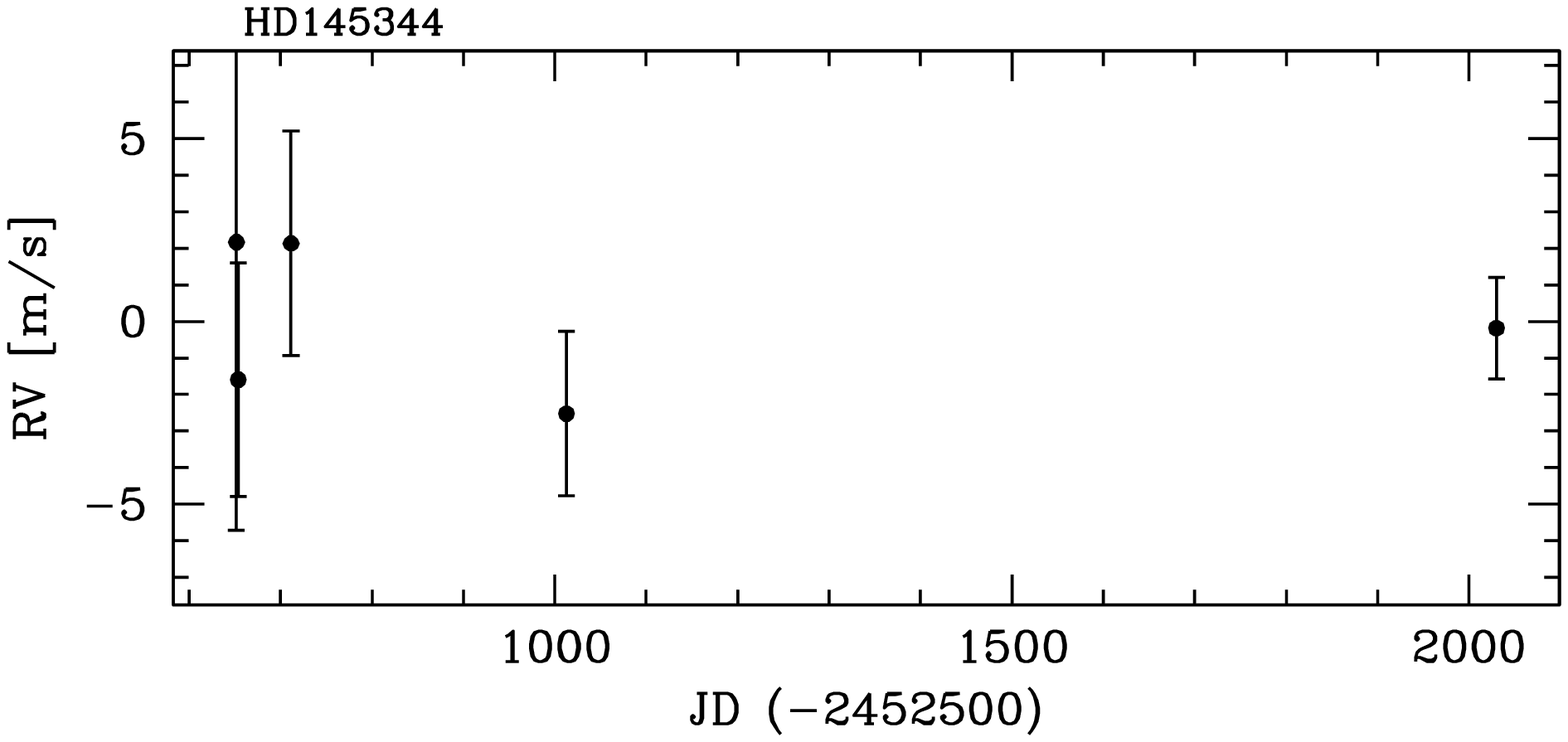}}
\resizebox{5.9cm}{!}{\includegraphics[bb= 18 160 580 430]{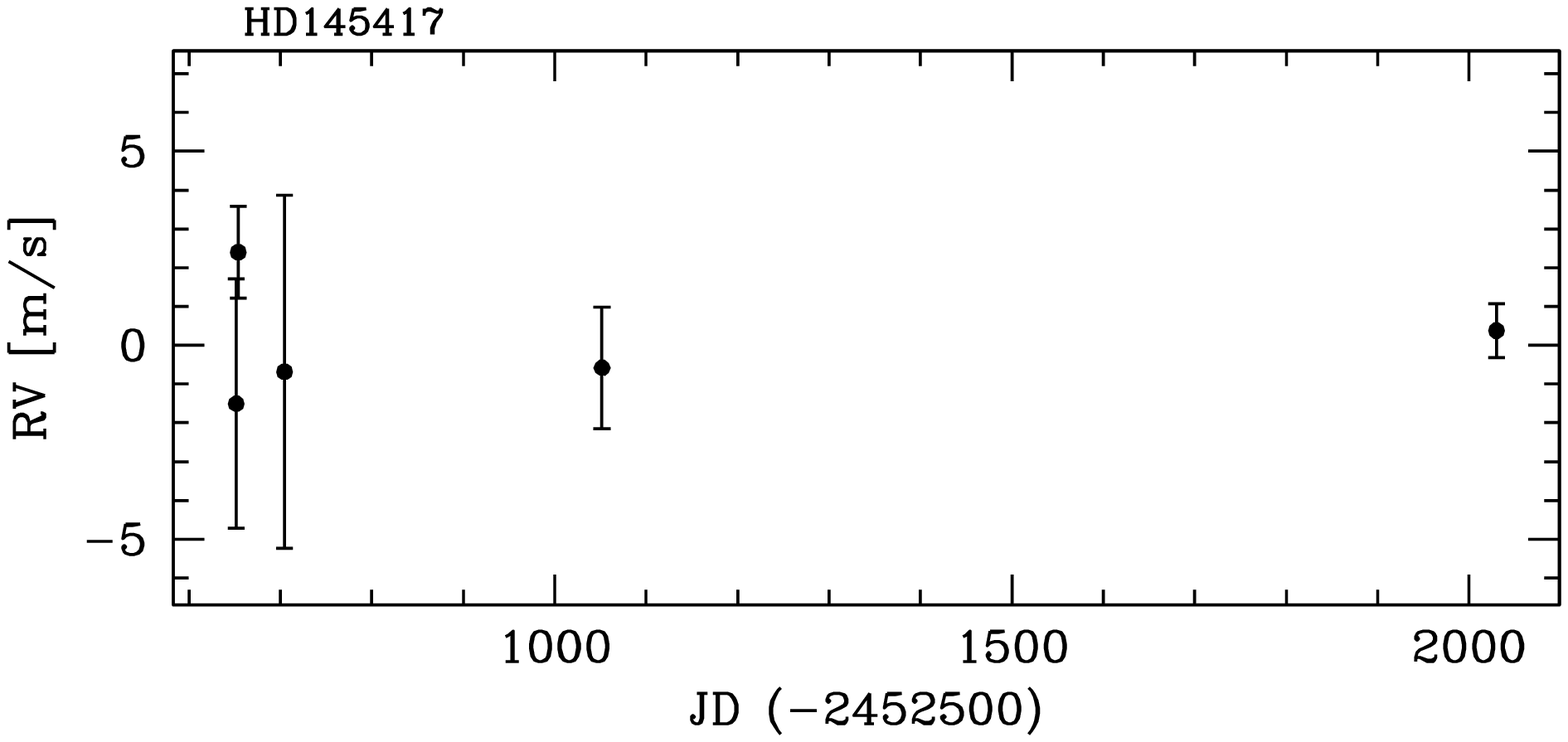}}\\
\resizebox{5.9cm}{!}{\includegraphics[bb= 18 160 580 430]{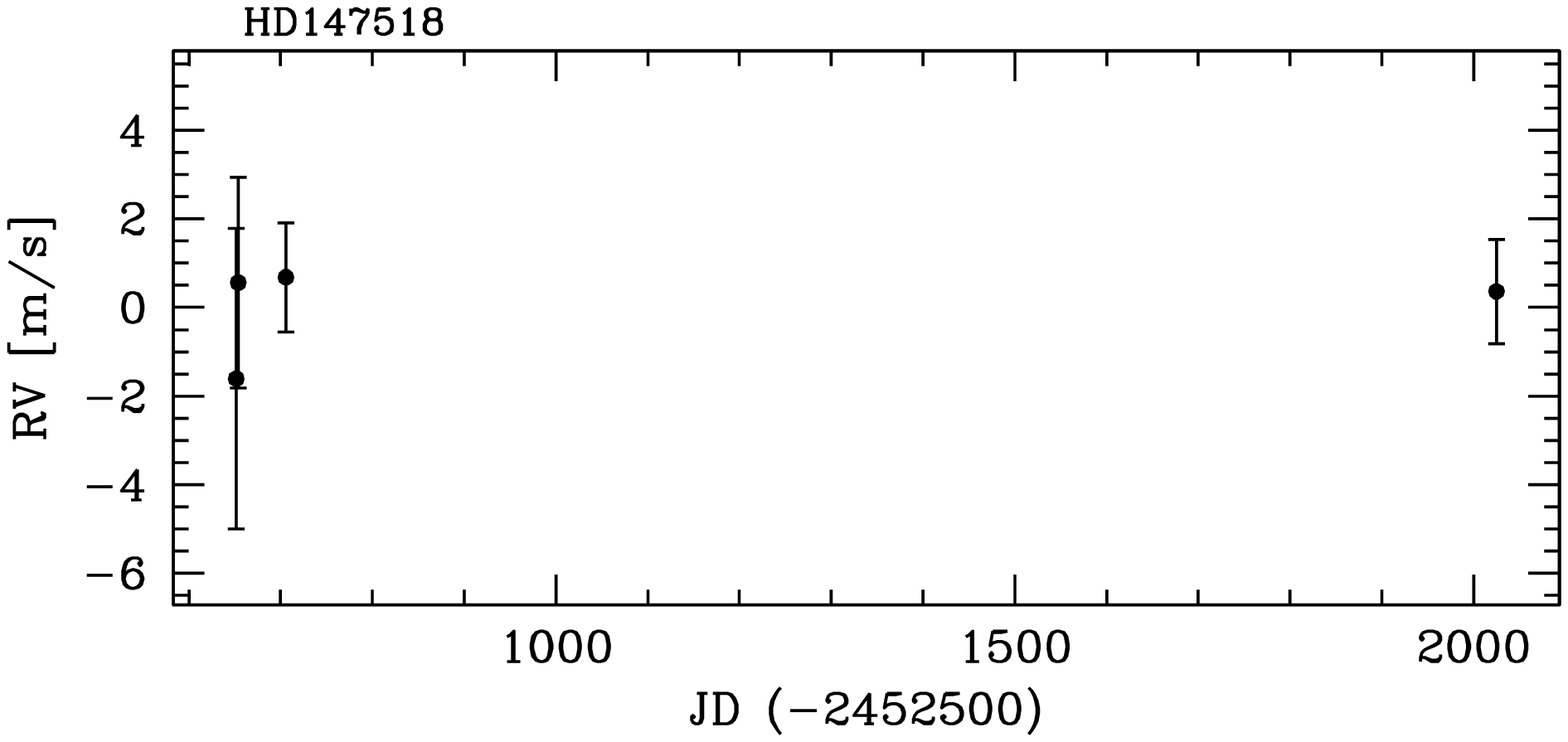}}
\resizebox{5.9cm}{!}{\includegraphics[bb= 18 160 580 430]{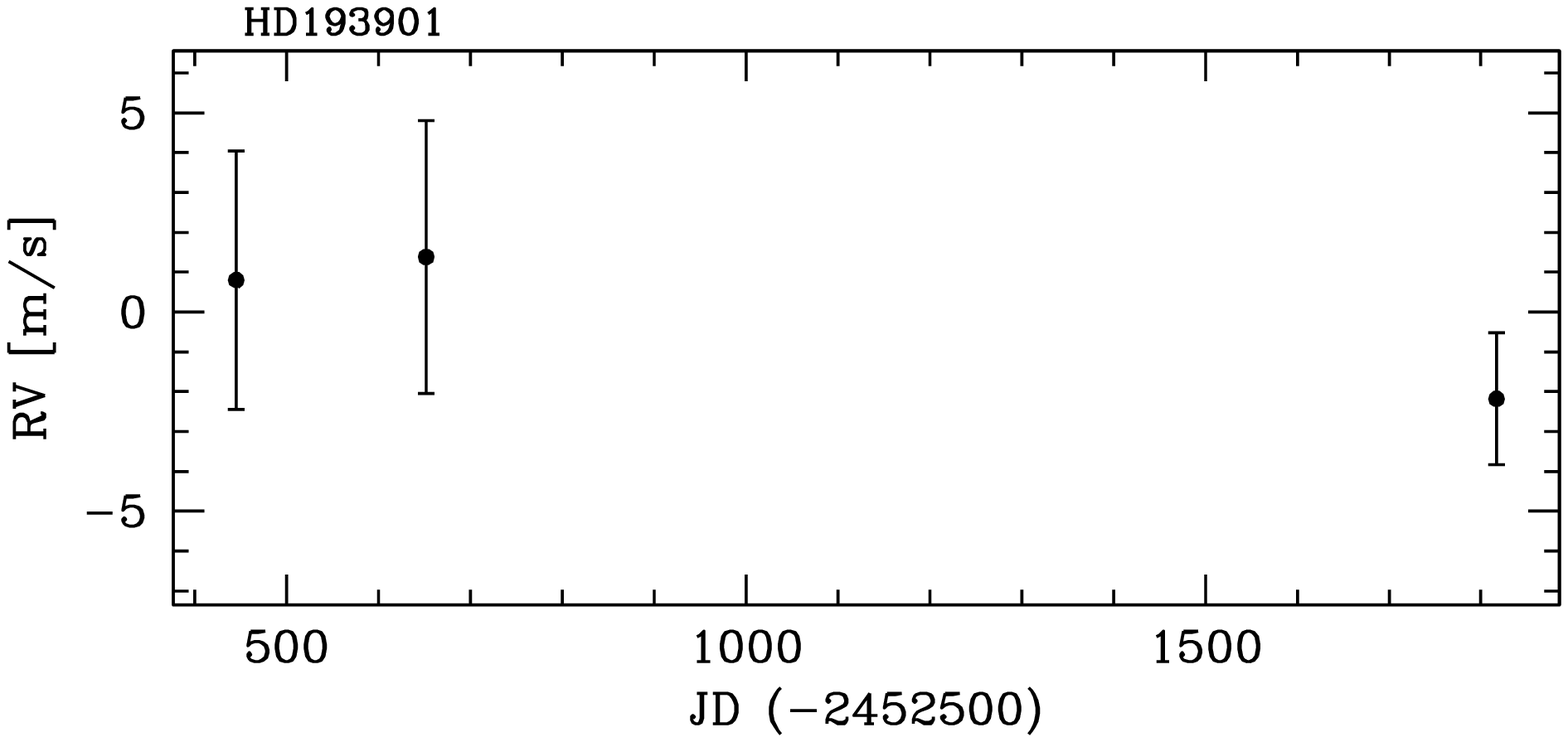}}
\resizebox{5.9cm}{!}{\includegraphics[bb= 18 160 580 430]{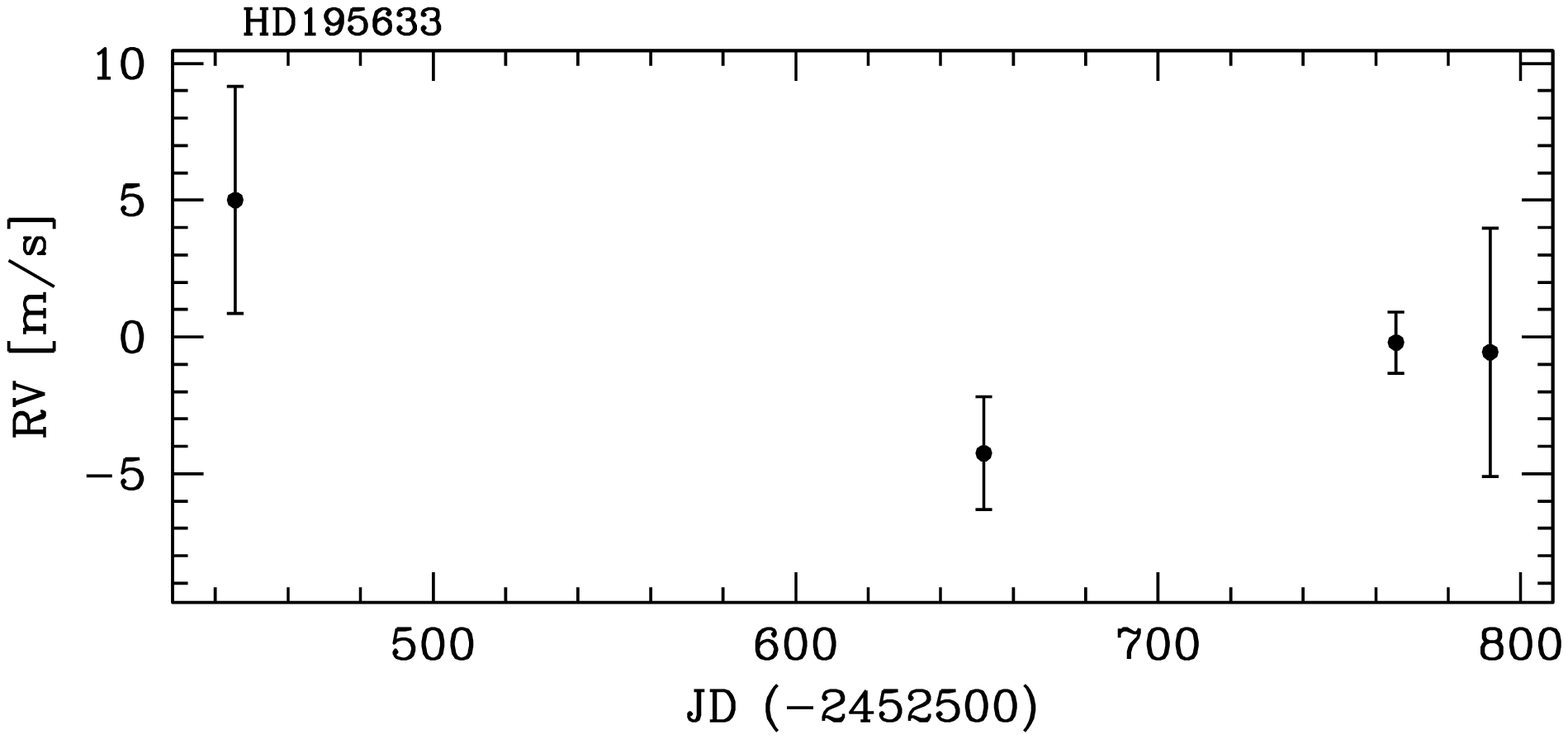}}\\
\resizebox{5.9cm}{!}{\includegraphics[bb= 18 160 580 430]{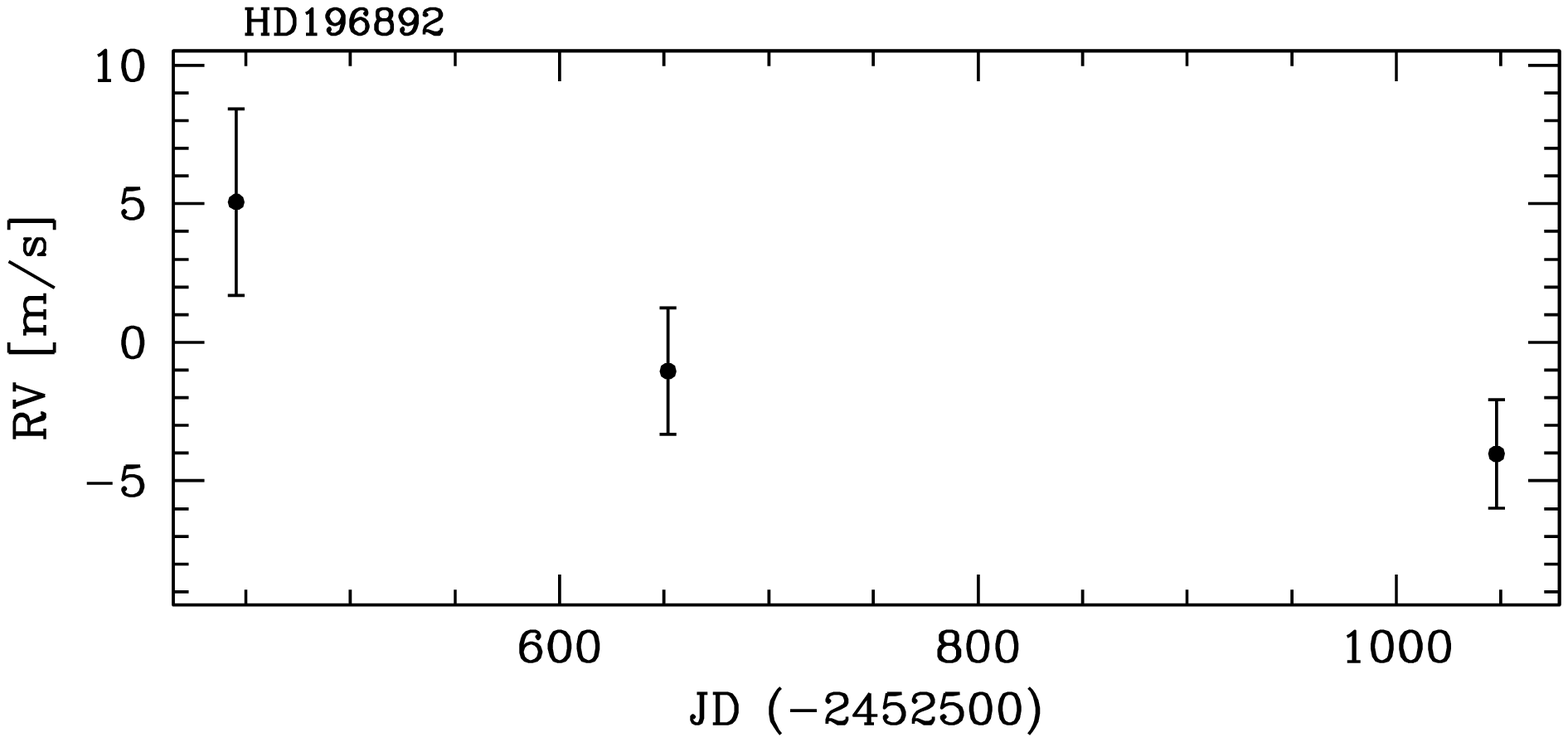}}
\resizebox{5.9cm}{!}{\includegraphics[bb= 18 160 580 430]{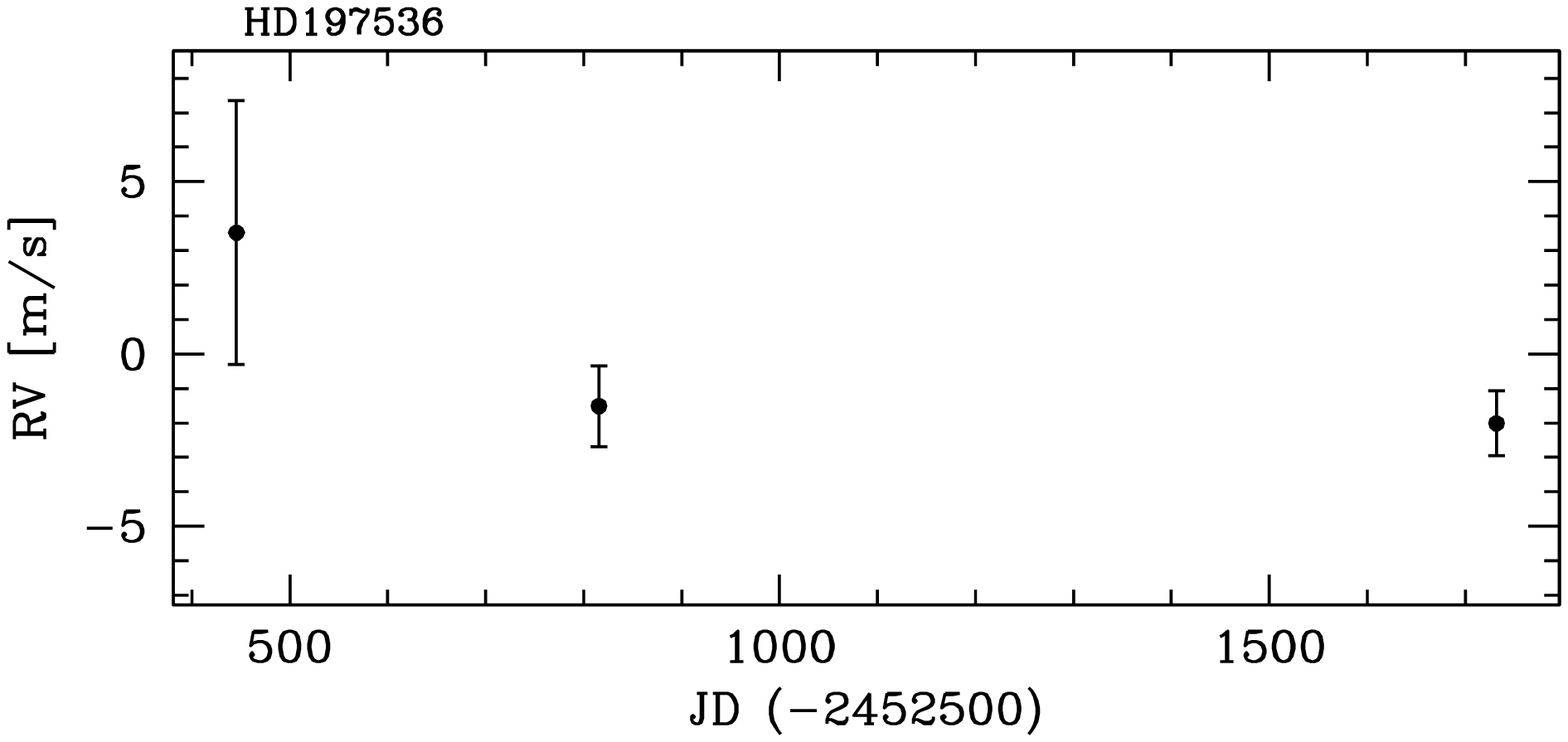}}
\resizebox{5.9cm}{!}{\includegraphics[bb= 18 160 580 430]{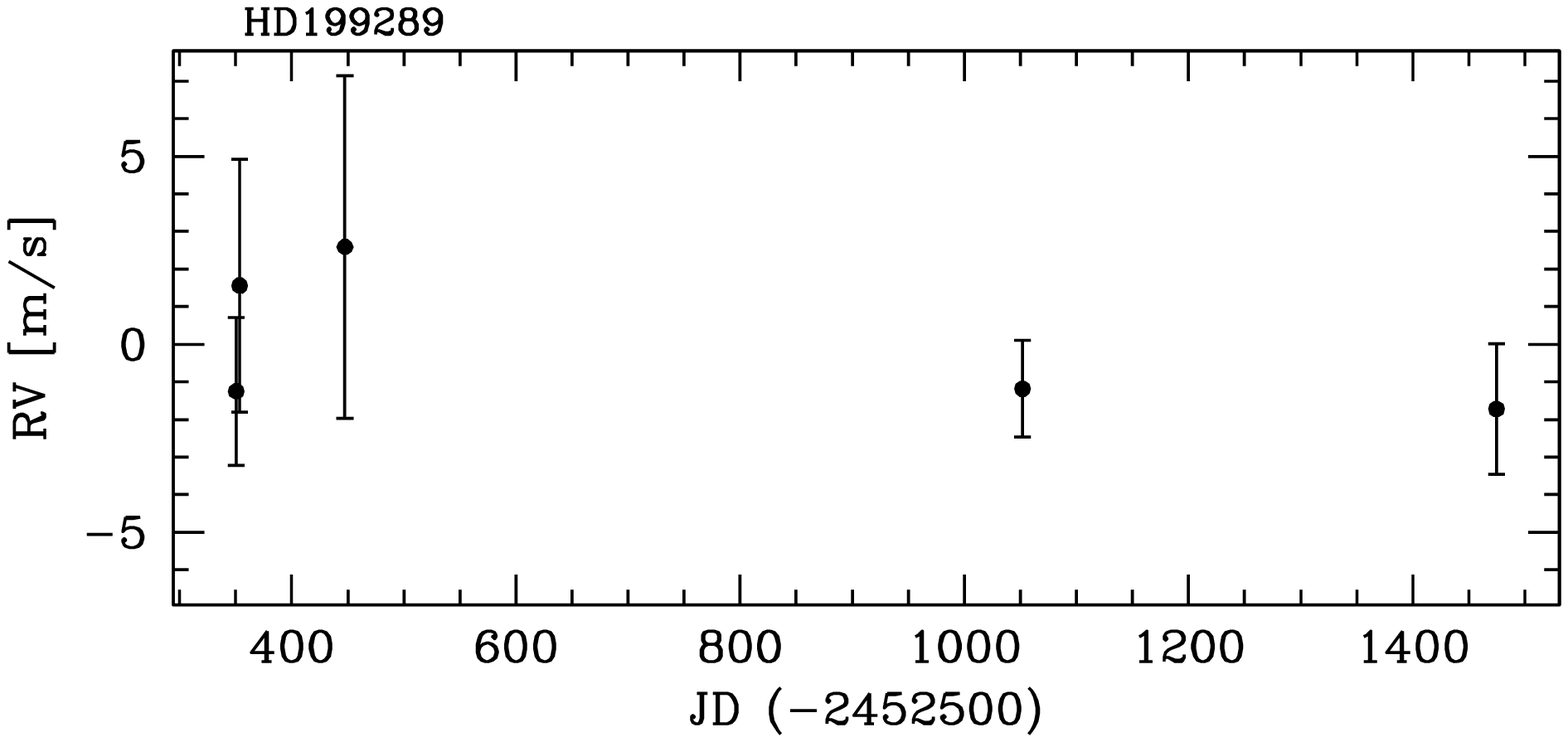}}\\
\resizebox{5.9cm}{!}{\includegraphics[bb= 18 160 580 430]{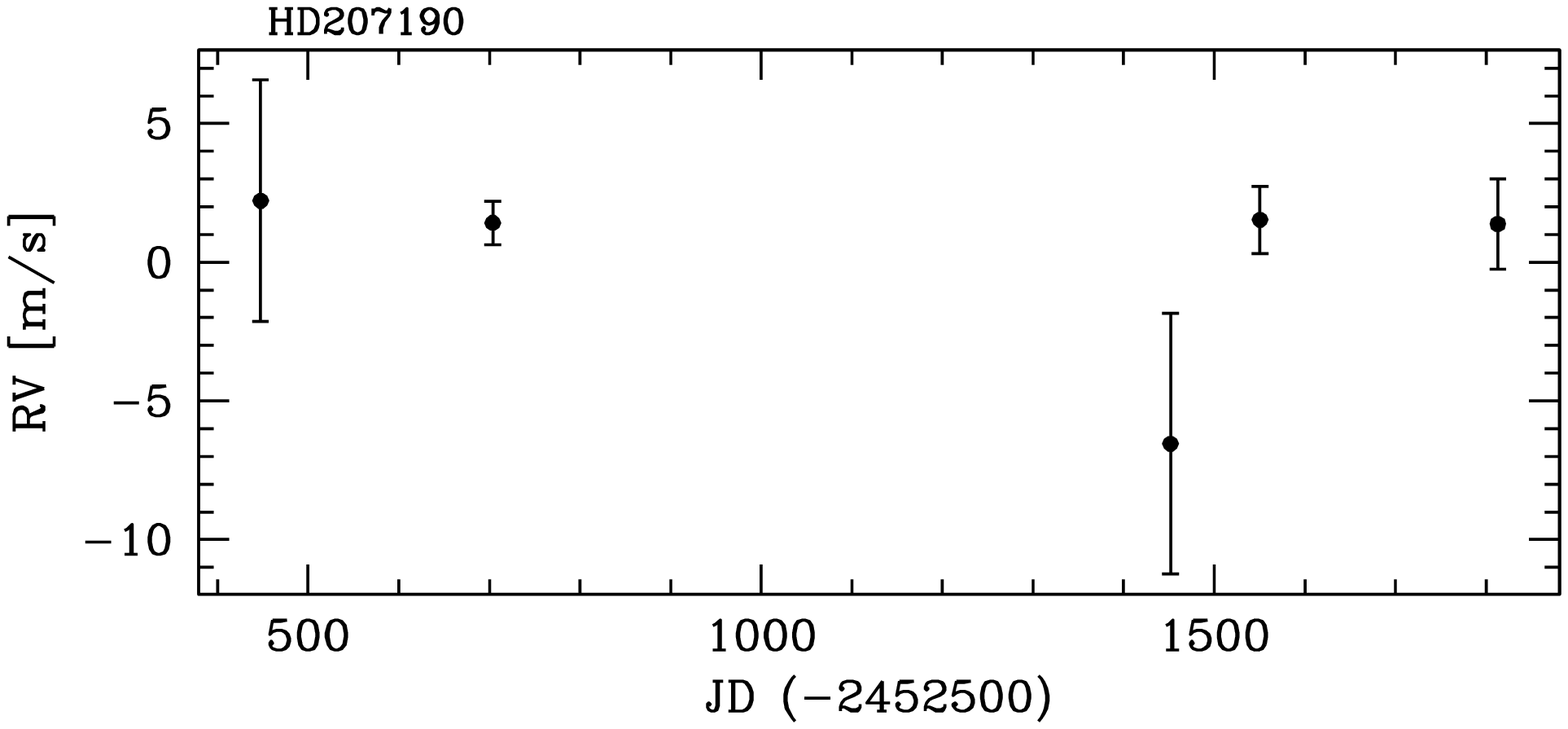}}
\resizebox{5.9cm}{!}{\includegraphics[bb= 18 160 580 430]{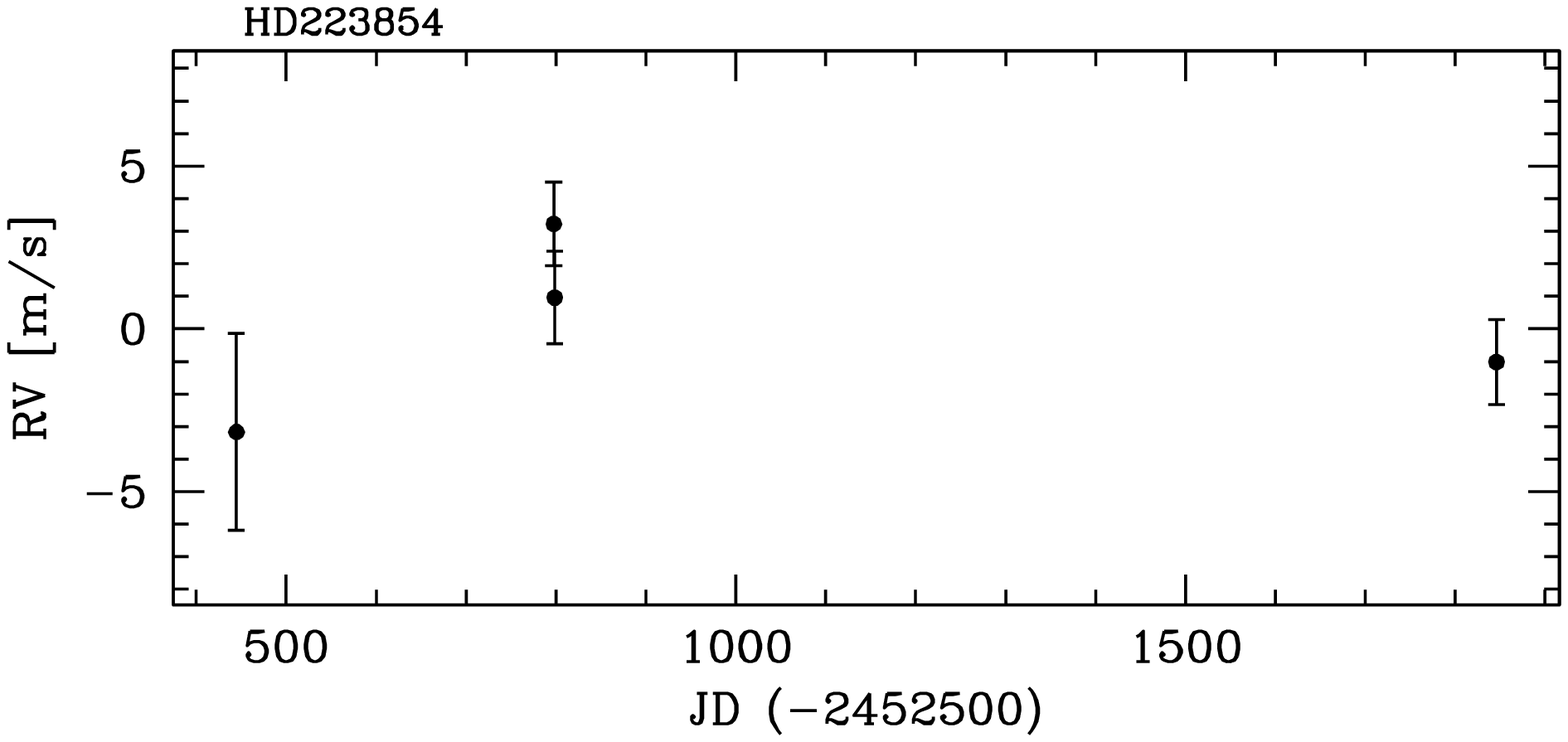}}
\caption{Radial velocity time series for stars with 2 to 5 radial velocity measurements.}
\label{fig:5mes}
\end{figure*}

Besides several of the cases discussed in Sect.\,\ref{sec:cleaning}, 24 stars in our sample were observed only 5 or less
times during the period of our survey (Table\,\ref{tab:5mes}, Fig.\,\ref{fig:5mes}; one star, with one single measurement, was 
not included in the figure). With so few measurements, 
not much information can be extracted about these targets: a Keplerian fit, for instance, has 6 free parameters. We thus 
decided to separate all the targets with this reduced number of observations from the rest of the sample.

\begin{table}[t!]
\caption{Stars with 5 or less measurements.}
\begin{tabular}{lrcccr}
\hline\hline
\noalign{\smallskip}
Star & $<$RV$>$$\dagger$  & $\sigma$(RV)$\dagger$  & $<$$\sigma_{mes}$$>$ & N & Span \\
  & [km\,s$^{-1}$]  & [m\,s$^{-1}$]  & [m\,s$^{-1}$] & & [days]\\
\hline
\object{BD$-$084501} & 84.469 &  1.7 & 7.4 &  3 &  1077\\
\object{BD$+$062932} & $-$144.809 &  1.0 & 2.2 &  4 &  1036\\
\object{BD$+$063077} & 46.843 &  -- & 7.3 &  1 &  --\\
\object{BD$+$083095} & $-$51.148 &  3.2 & 3.3 &  3 &  1378\\
\object{CD$-$451246} & 108.847 &  3.8 & 6.3 &  4 &  1001\\
\object{HD38510} & 183.828 &  2.6 & 2.8 &  5 &  1502\\
\object{HD105004} & 121.495 &  3.9 & 3.1 &  5 &  1084\\
\object{HD111515} & 2.657 &  0.9 & 2.0 &  5 &  1458\\
\object{HD119949} & $-$39.327 &  2.6 & 3.0 &  5 &  1463\\
\object{HD121004} & 245.318 &  1.7 & 2.0 &  5 &  1470\\
\object{HD128340} & 2.997 &  2.0 & 2.7 &  5 &  1031\\
\object{HD129229} & $-$15.932 &  1.9 & 2.0 &  5 &  1031\\
\object{HD131653} & $-$67.090 &  1.7 & 1.2 &  4 &  1375\\
\object{HD134088} & $-$61.111 &  1.0 & 1.3 &  4 &  1377\\
\object{HD145344} & $-$11.875 &  1.8 & 3.6 &  5 &  1378\\
\object{HD145417} & 8.799 &  1.3 & 2.2 &  5 &  1378\\
\object{HD147518} & $-$42.127 &  0.6 & 2.0 &  4 &  1373\\
\object{HD193901} & $-$171.416 &  2.2 & 2.8 &  3 &  1371\\
\object{HD195633} & $-$45.832 &  3.1 & 3.0 &  4 &  346\\
\object{HD196892} & $-$34.498 &  4.1 & 2.5 &  3 &  602\\
\object{HD197536} & $-$17.117 &  1.4 & 2.0 &  3 &  1286\\
\object{HD199289} & $-$6.144 &  1.2 & 2.6 &  5 &  1123\\
\object{HD207190} & $-$11.314 &  1.3 & 2.5 &  5 &  1365\\
\object{HD223854} & 19.596 &  2.4 & 1.8 &  4 &  1401\\
\hline
\noalign{\smallskip}
\end{tabular}
\newline
$\dagger$ Weighted values.
\label{tab:5mes}
\end{table}

In Table\,\ref{tab:5mes} we present the list of targets in such conditions, their weighted average radial velocity,
the weighted rms of the measurements, the average error of the obtained radial velocities, the number of measurements, as well as their time span. 
A look at the dispersions and average values shows that none of these stars seems to present strong radial-velocity variations.
All dispersions are below 5\,m\,s$^{-1}$. Though the measurement time sampling may hide important signals, we can 
confidently exclude the existence of giant planets in short-period orbits around these stars. A statistical approach to
this issue will be taken in a separate paper. 

\subsection{Tests for stars with at least 6 measurements}

\begin{figure*}[t!]
\resizebox{5.9cm}{!}{\includegraphics[bb= 18 160 580 430]{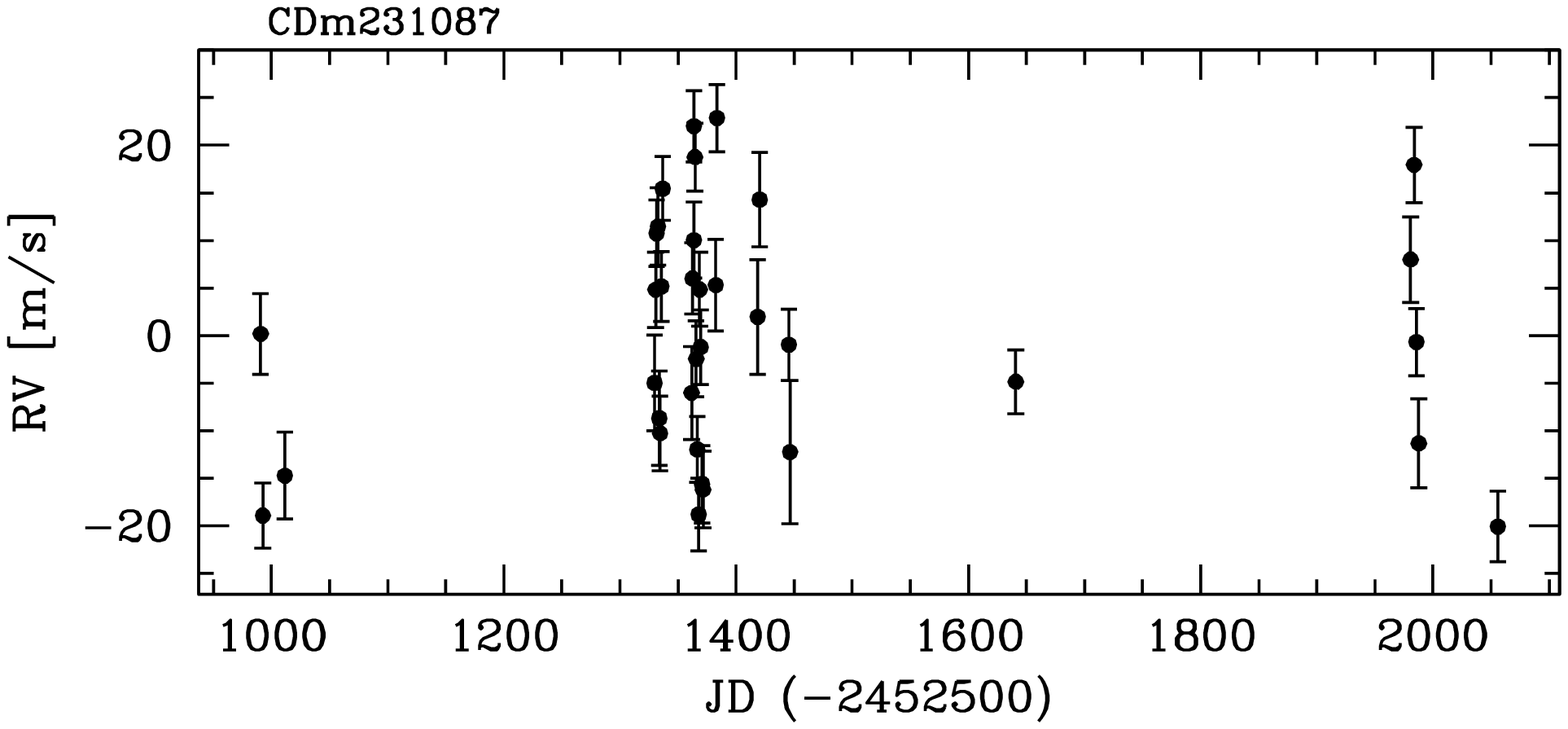}}
\resizebox{5.9cm}{!}{\includegraphics[bb= 18 160 580 430]{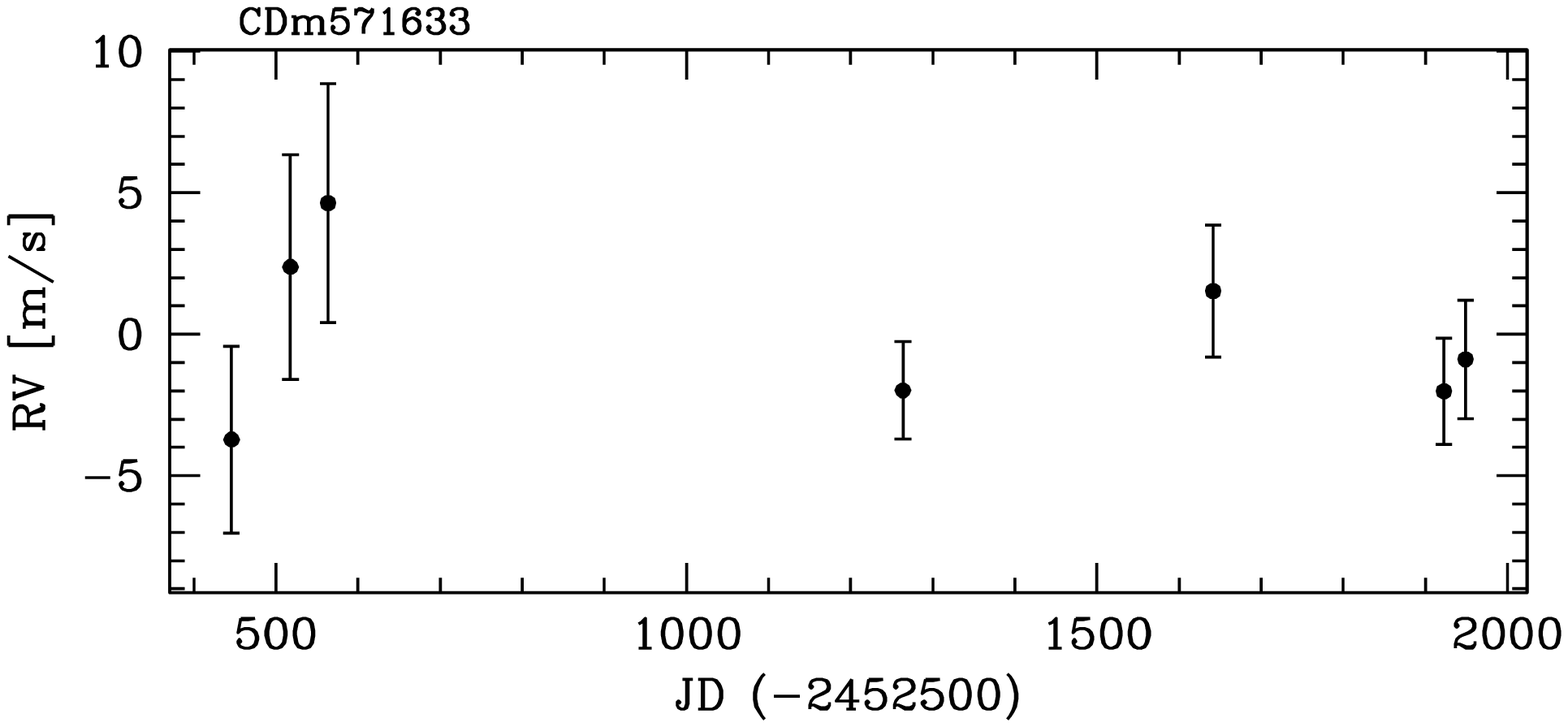}}
\resizebox{5.9cm}{!}{\includegraphics[bb= 18 160 580 430]{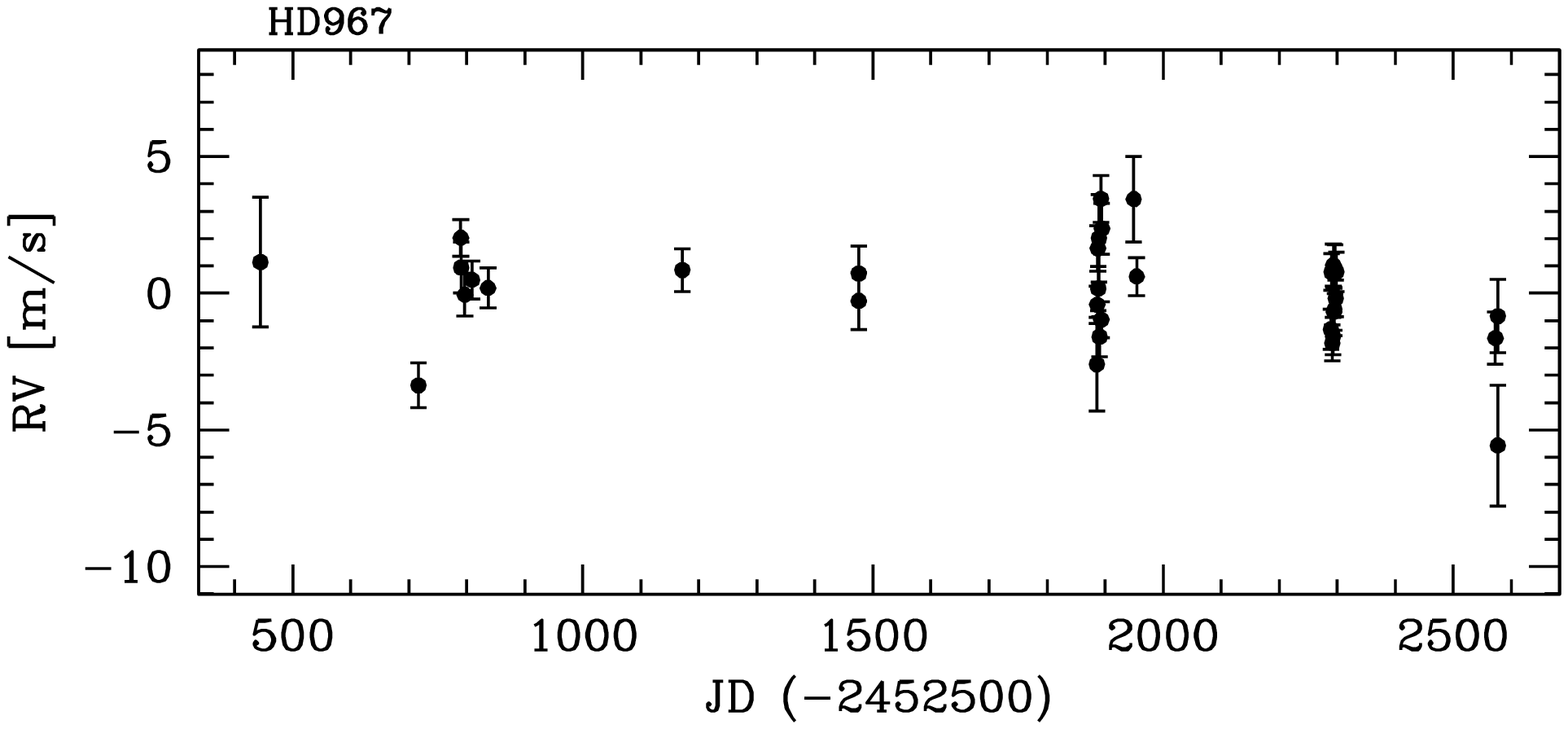}}\\
\resizebox{5.9cm}{!}{\includegraphics[bb= 18 160 580 430]{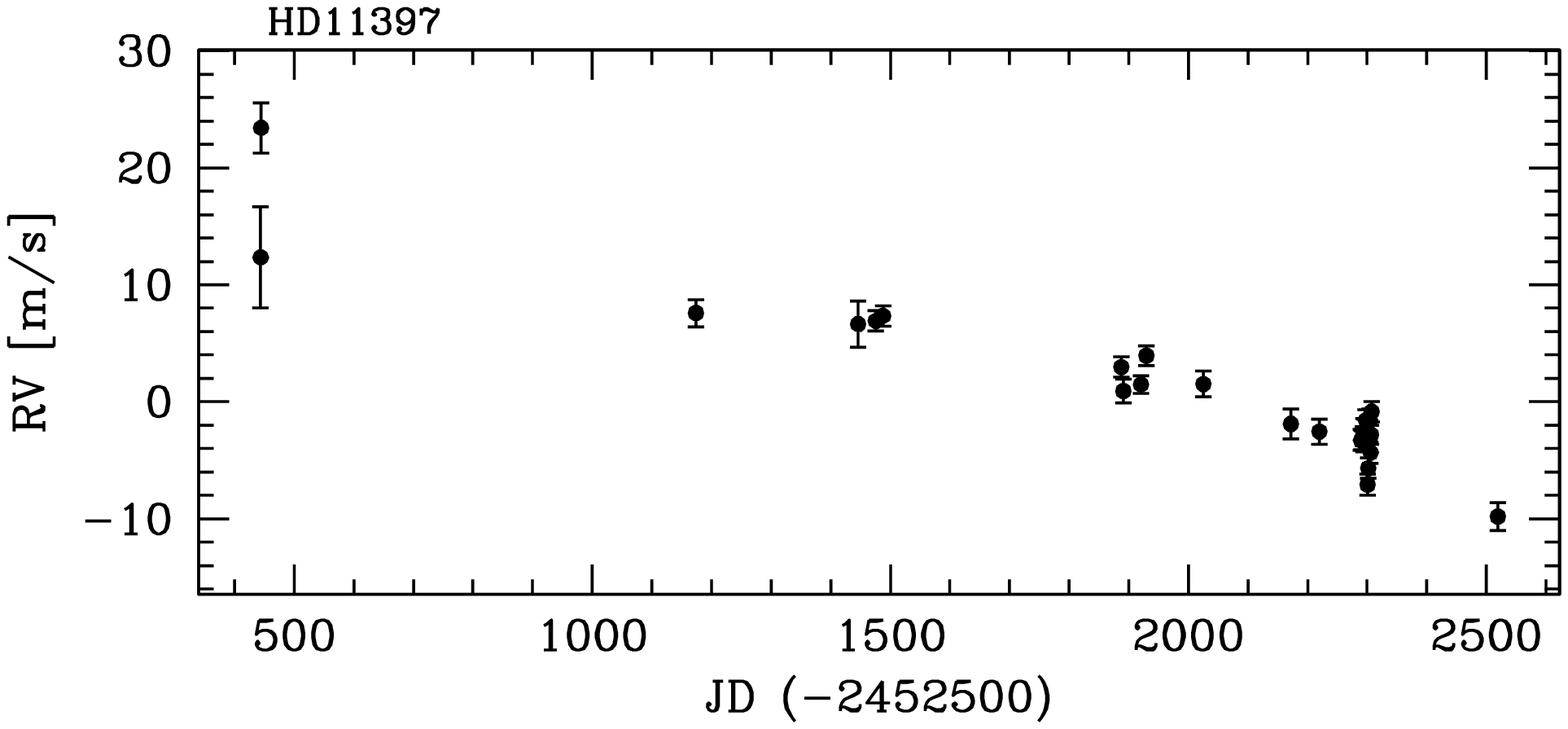}}
\resizebox{5.9cm}{!}{\includegraphics[bb= 18 160 580 430]{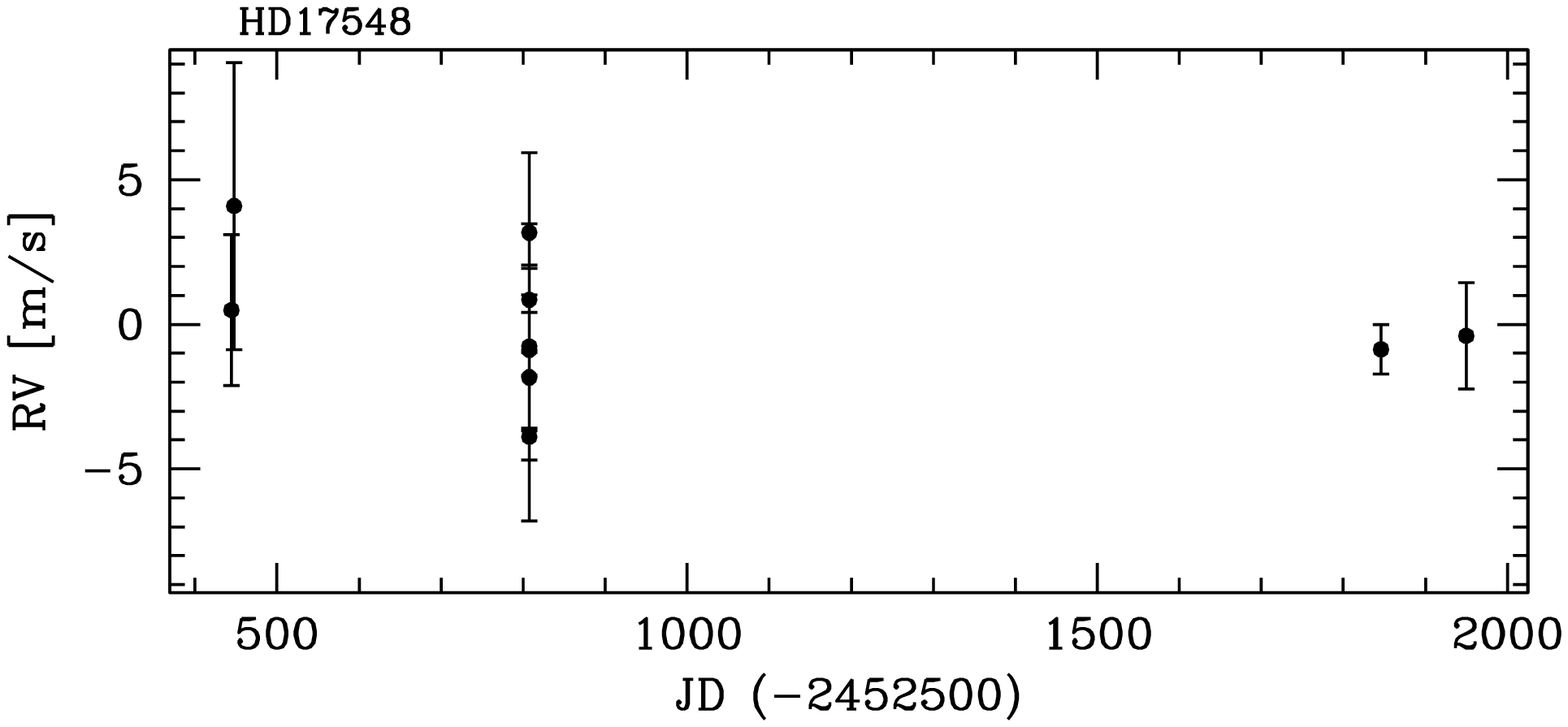}}
\resizebox{5.9cm}{!}{\includegraphics[bb= 18 160 580 430]{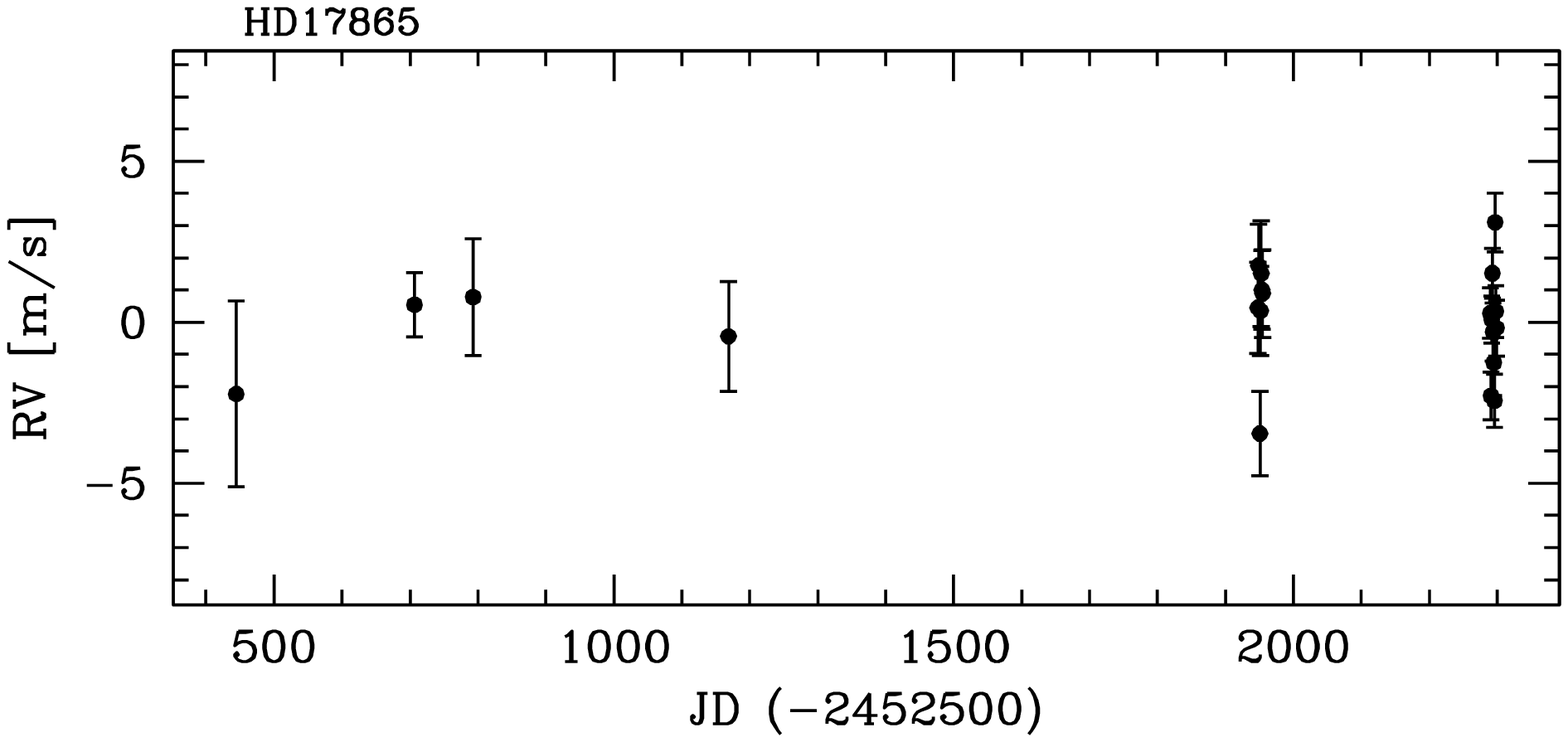}}\\
\resizebox{5.9cm}{!}{\includegraphics[bb= 18 160 580 430]{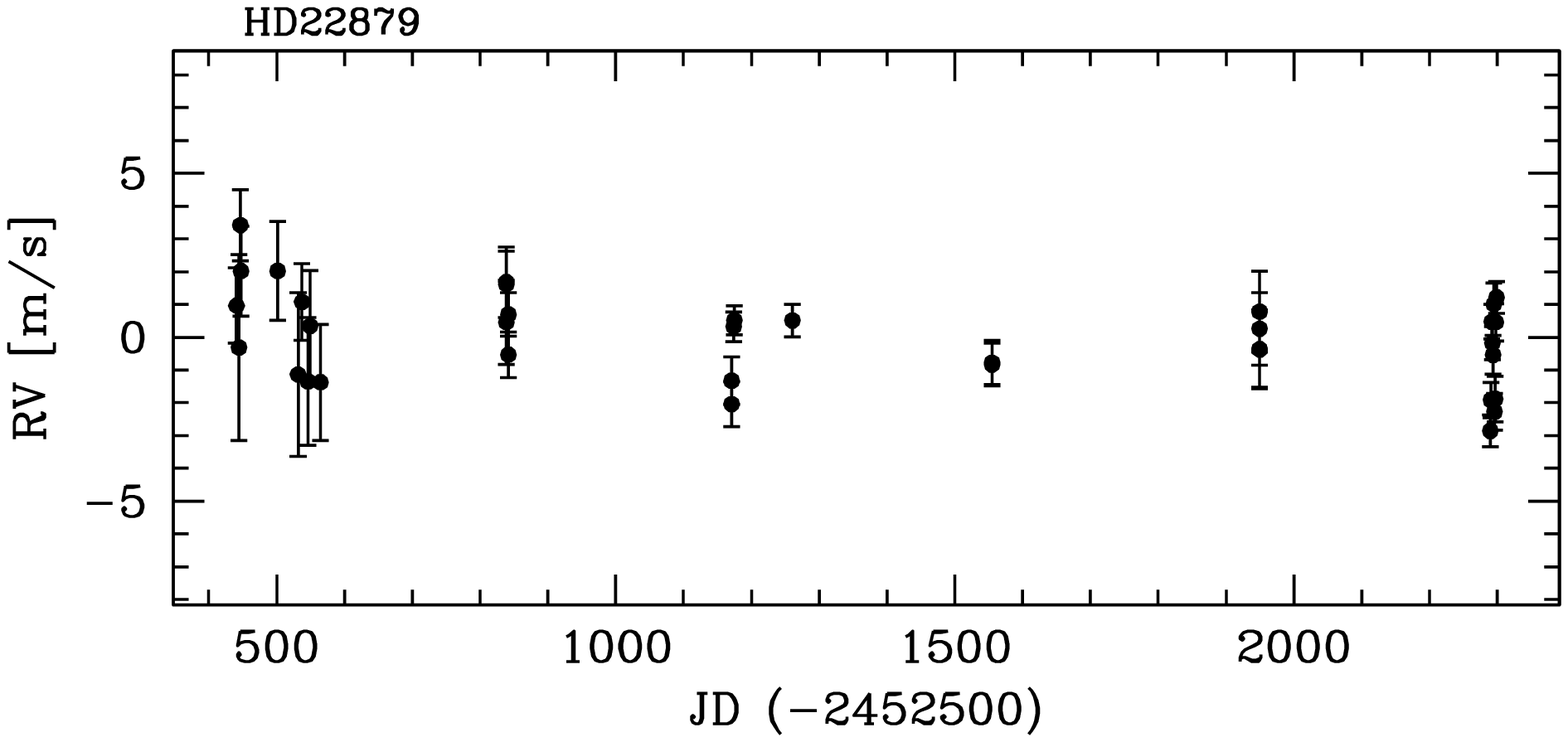}}
\resizebox{5.9cm}{!}{\includegraphics[bb= 18 160 580 430]{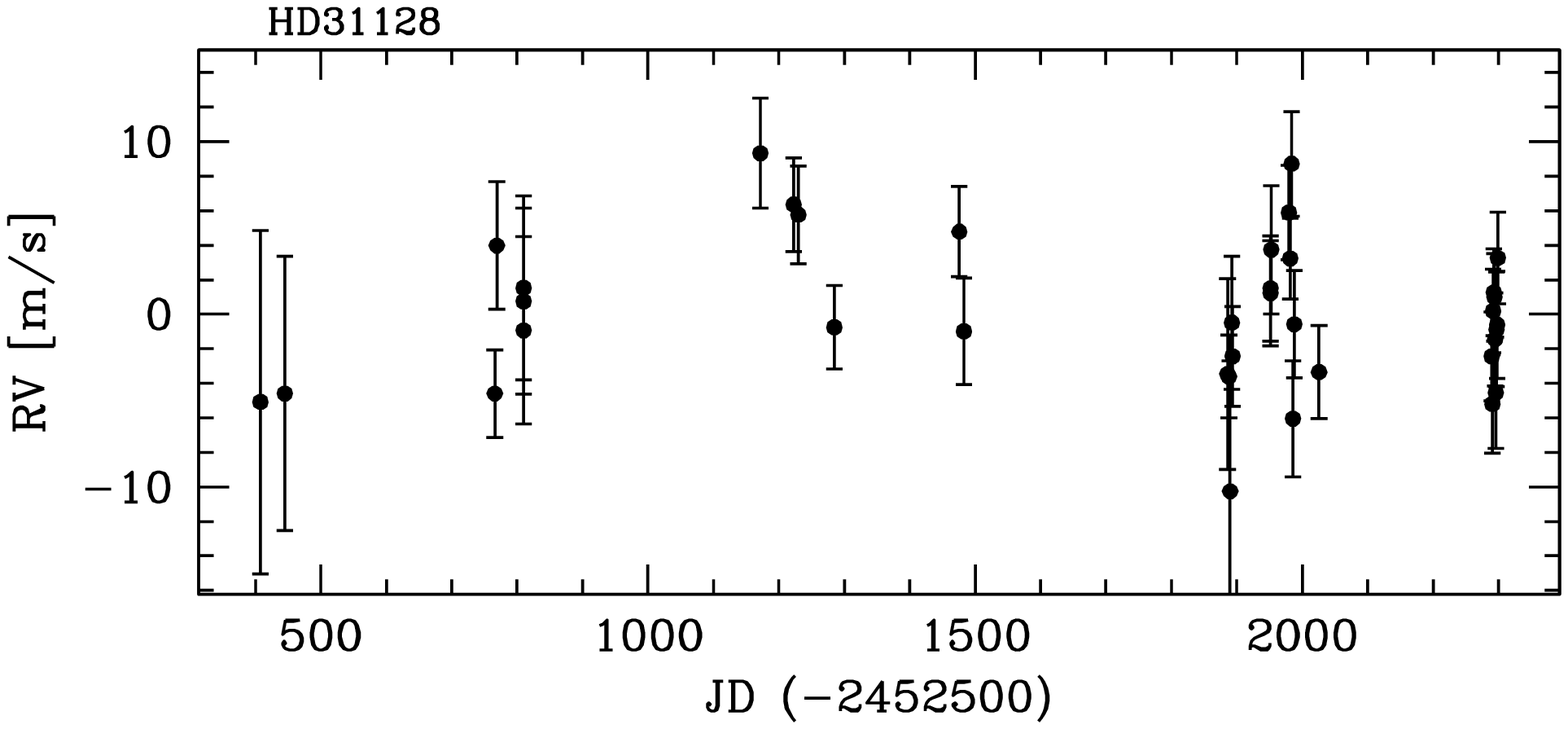}}
\resizebox{5.9cm}{!}{\includegraphics[bb= 18 160 580 430]{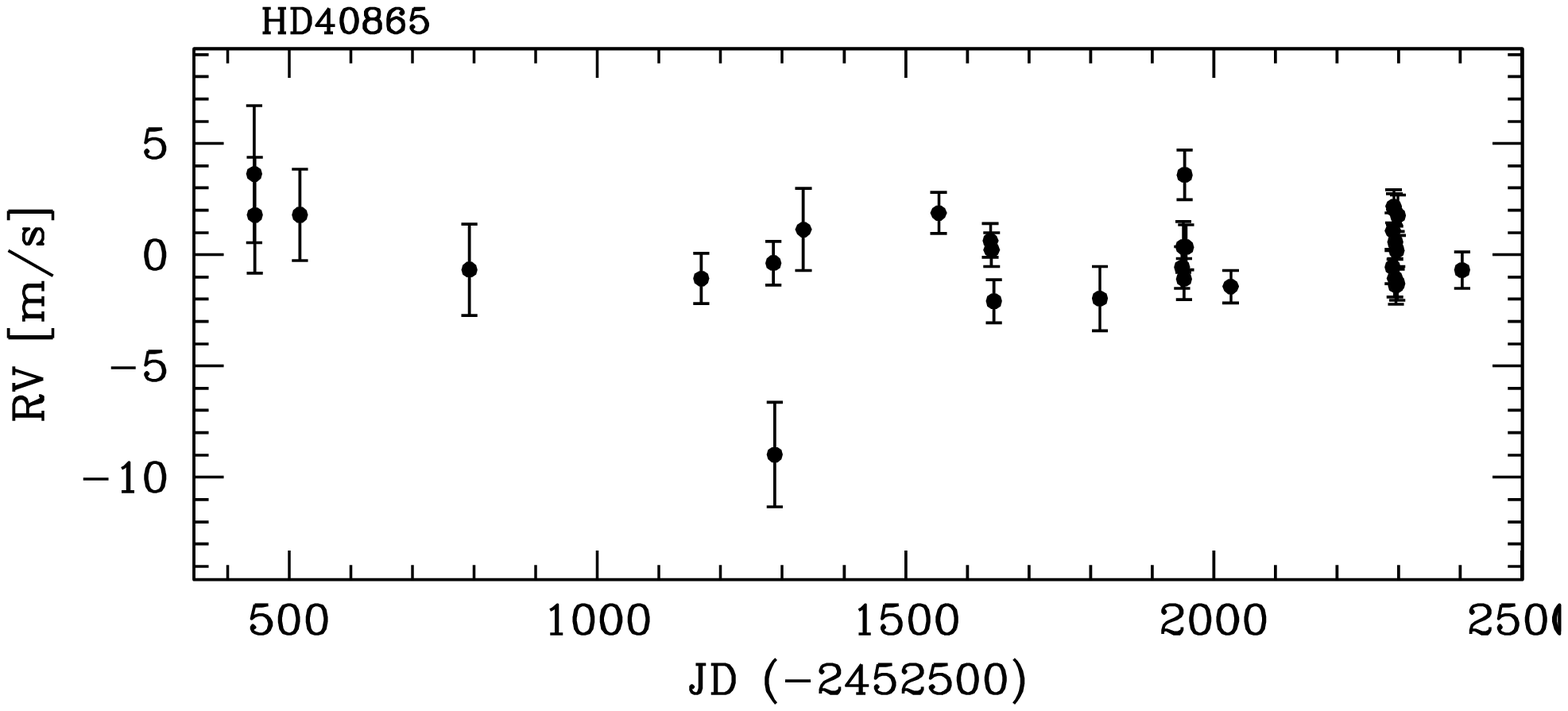}}\\
\resizebox{5.9cm}{!}{\includegraphics[bb= 18 160 580 430]{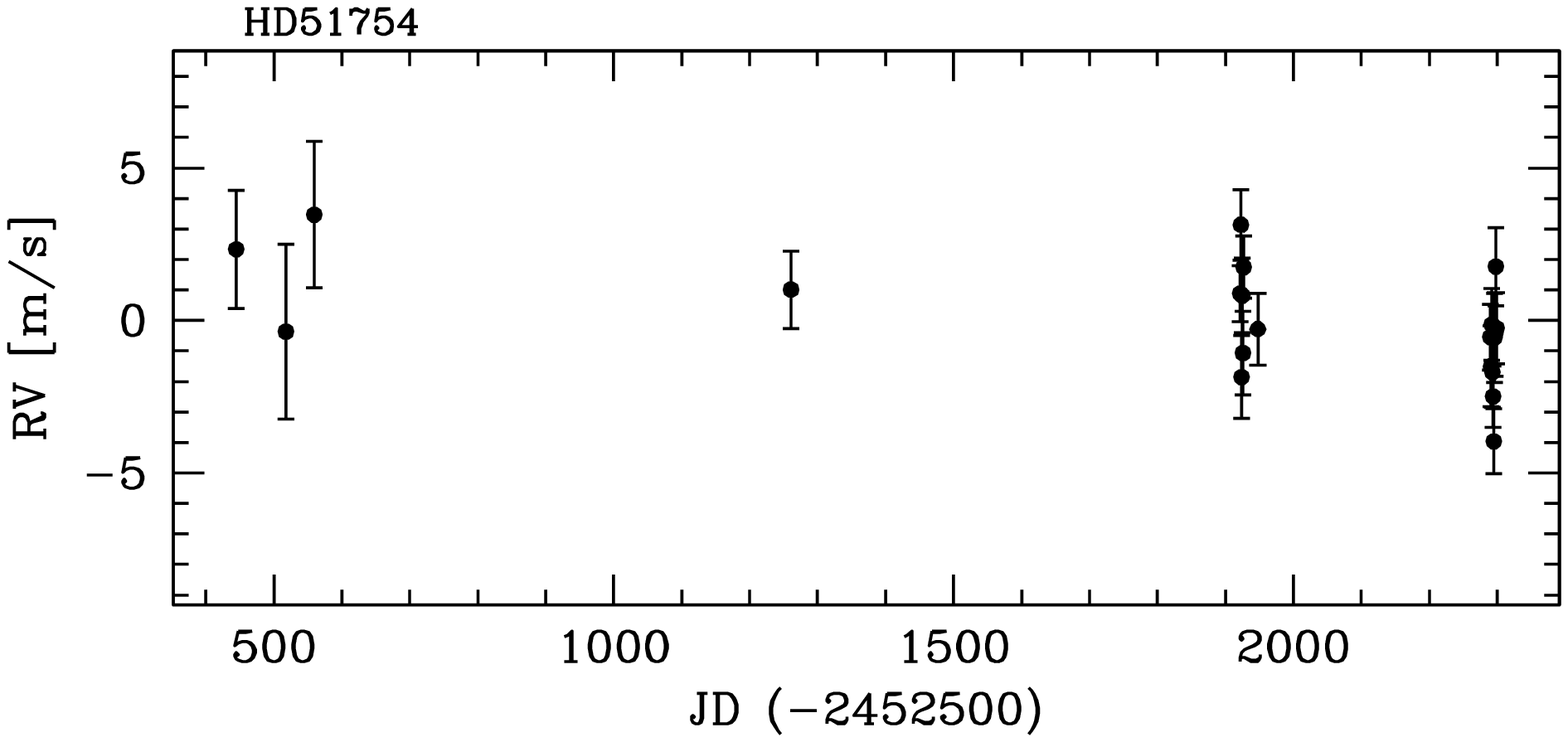}}
\resizebox{5.9cm}{!}{\includegraphics[bb= 18 160 580 430]{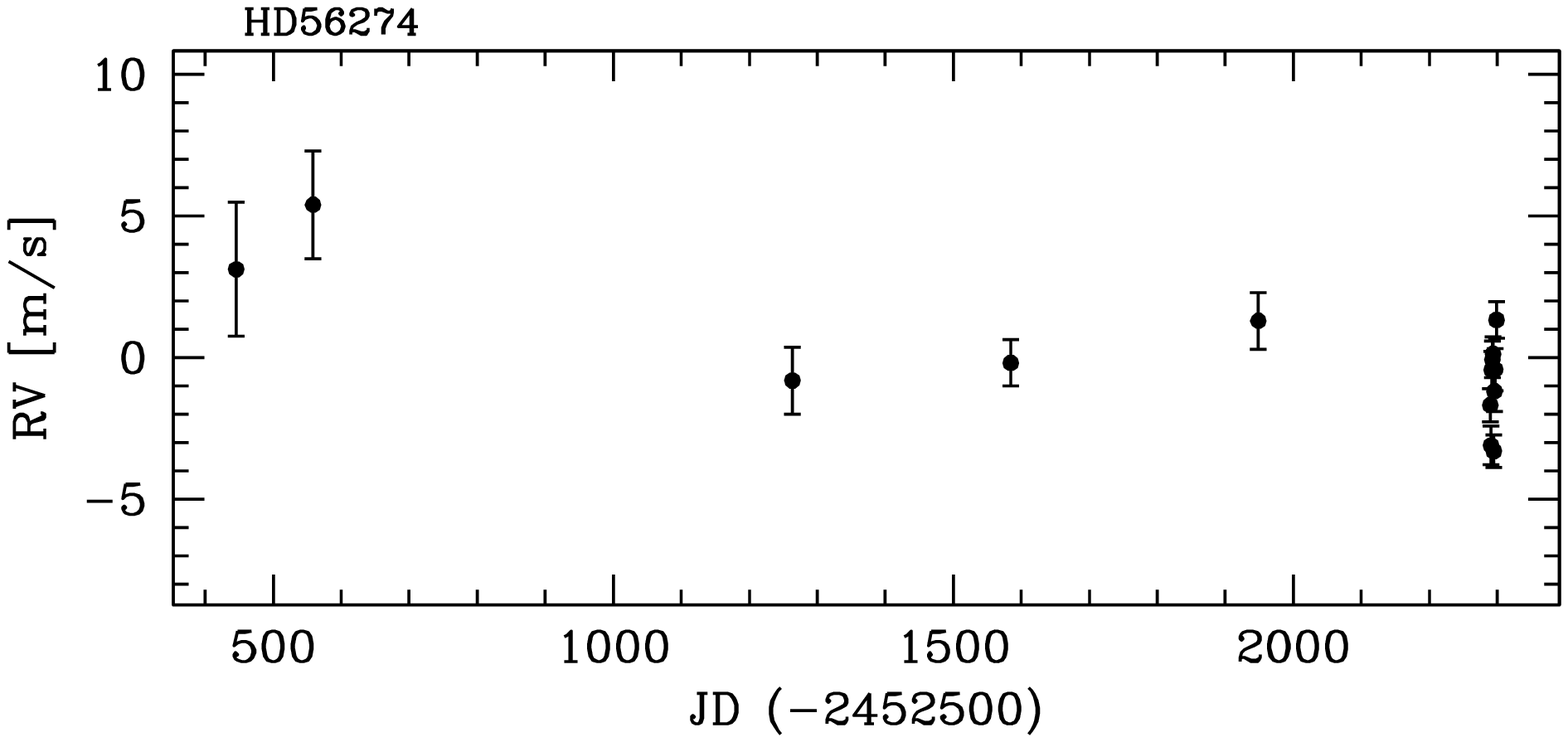}}
\resizebox{5.9cm}{!}{\includegraphics[bb= 18 160 580 430]{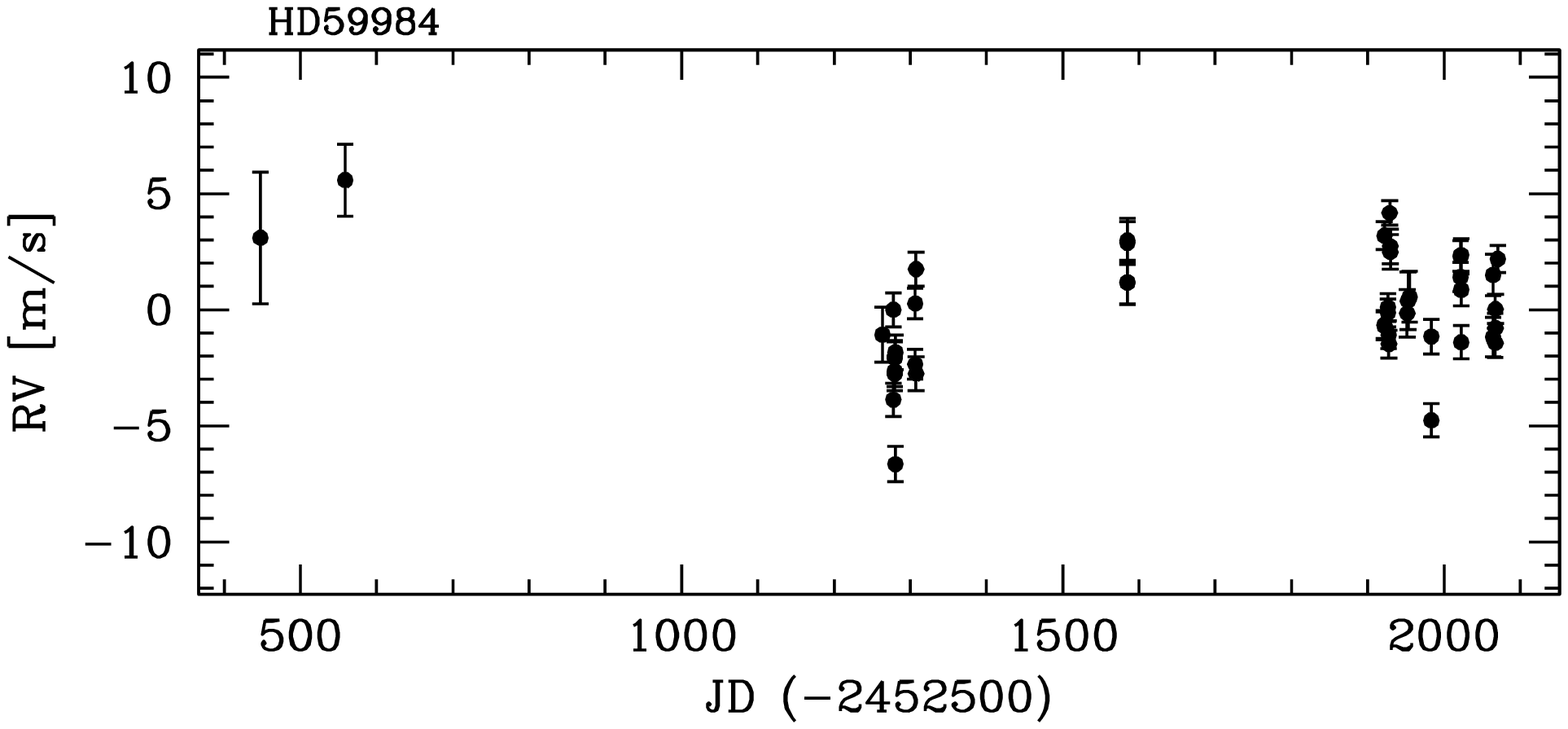}}\\
\resizebox{5.9cm}{!}{\includegraphics[bb= 18 160 580 430]{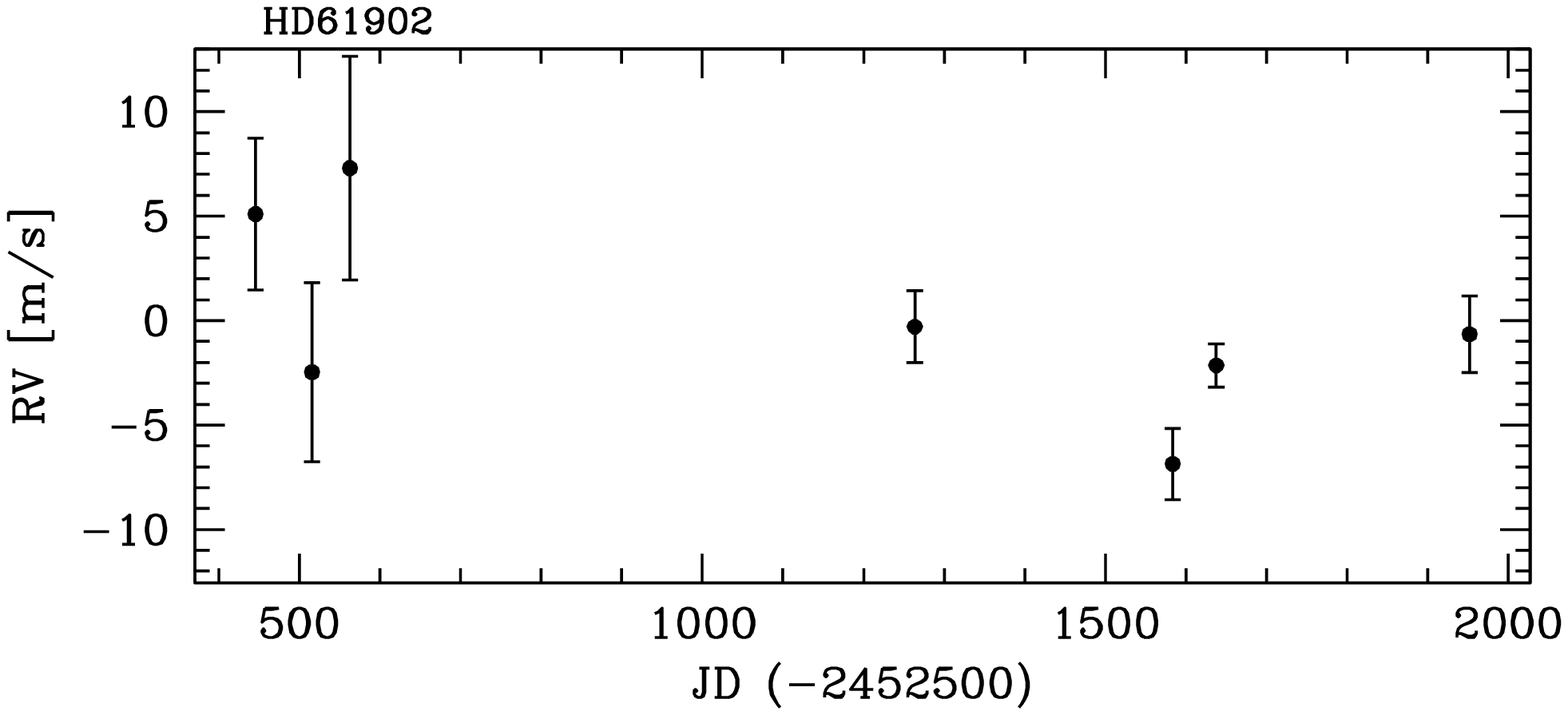}}
\resizebox{5.9cm}{!}{\includegraphics[bb= 18 160 580 430]{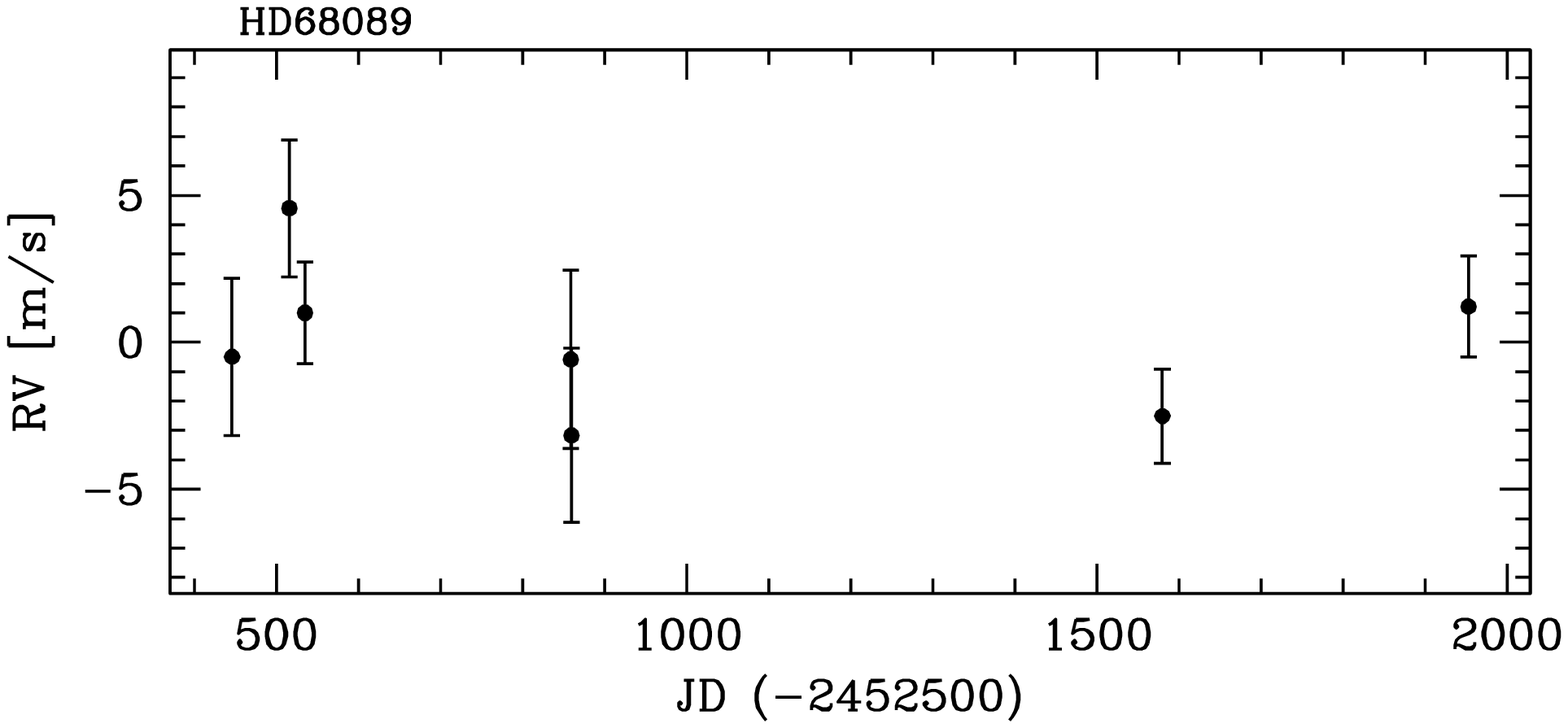}}
\resizebox{5.9cm}{!}{\includegraphics[bb= 18 160 580 430]{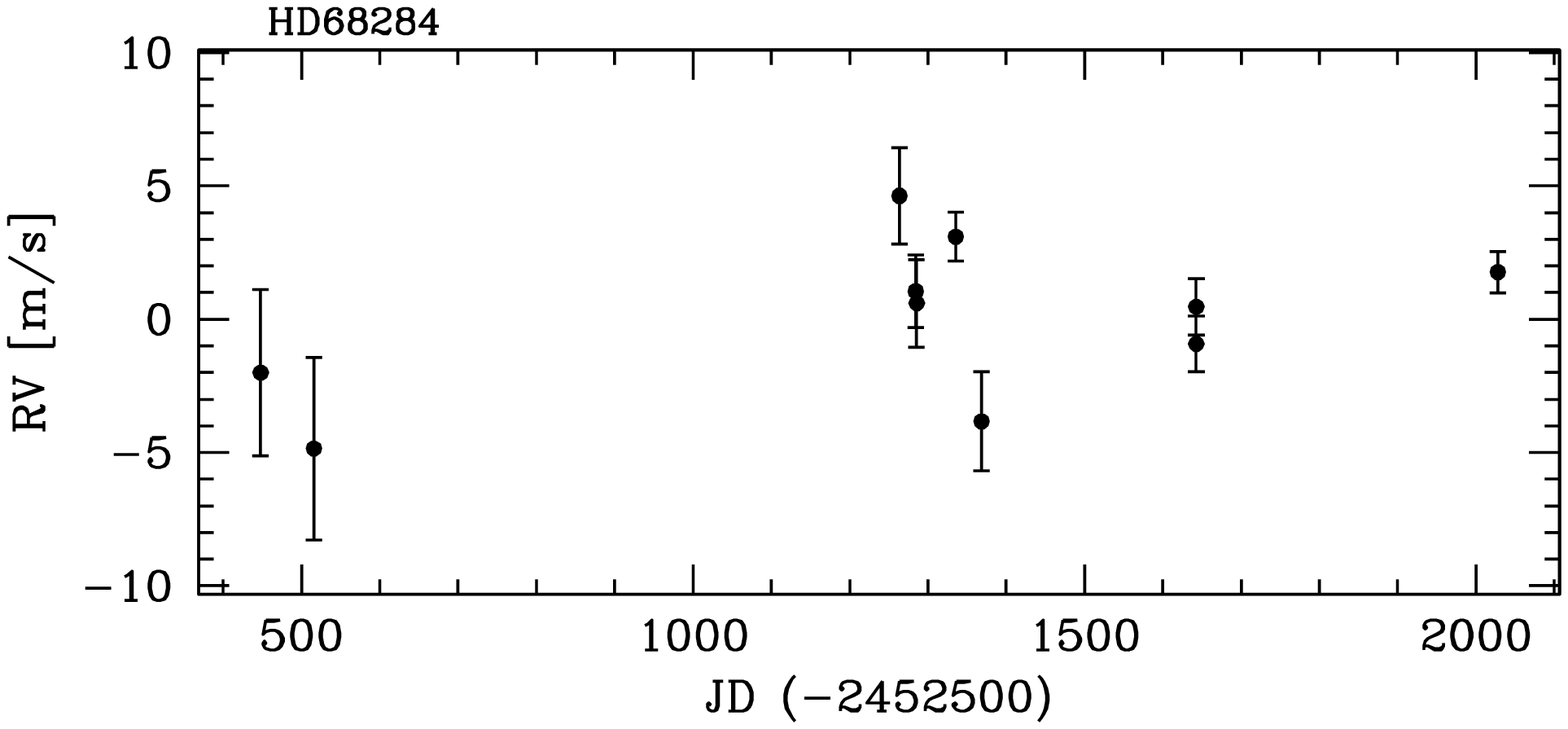}}\\
\resizebox{5.9cm}{!}{\includegraphics[bb= 18 160 580 430]{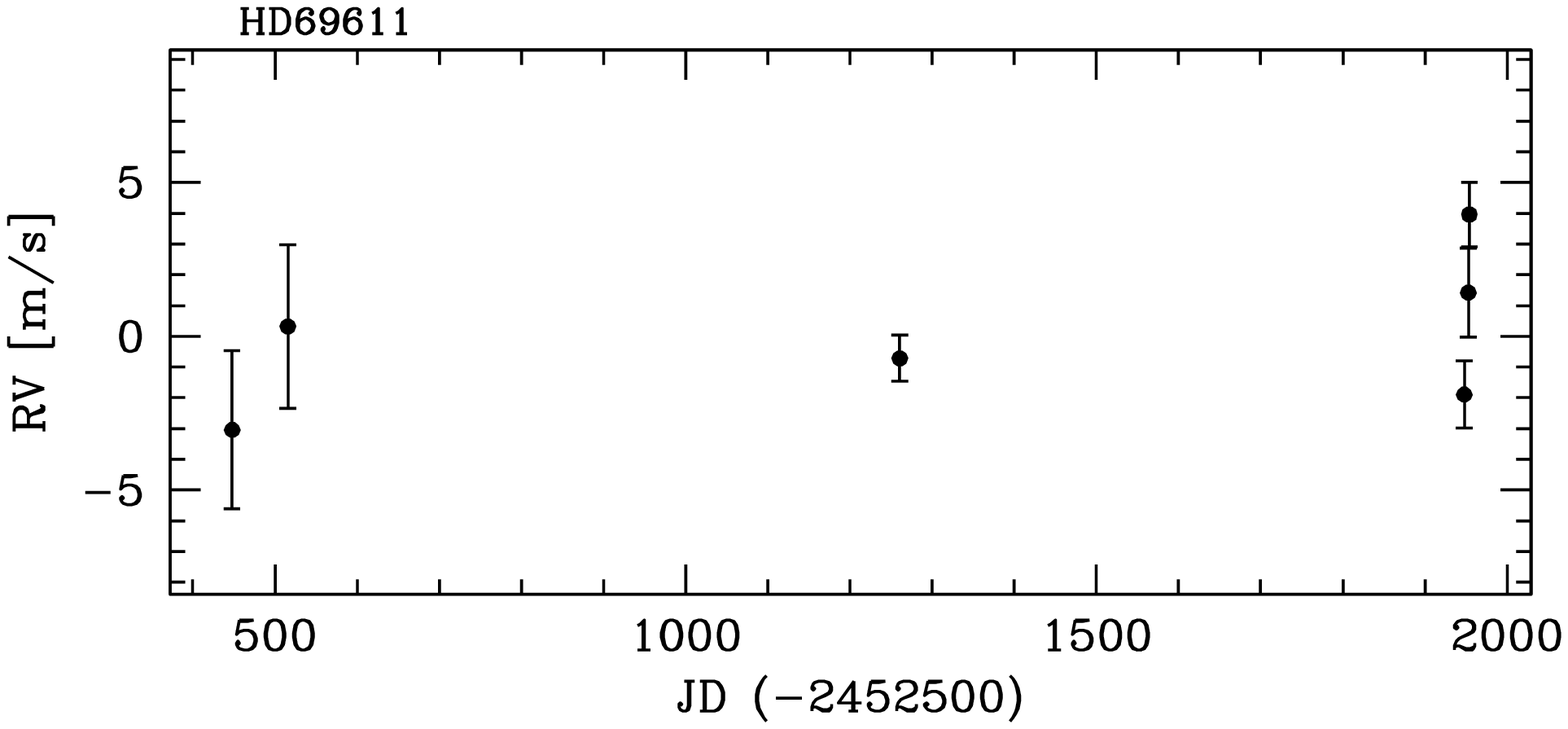}}
\resizebox{5.9cm}{!}{\includegraphics[bb= 18 160 580 430]{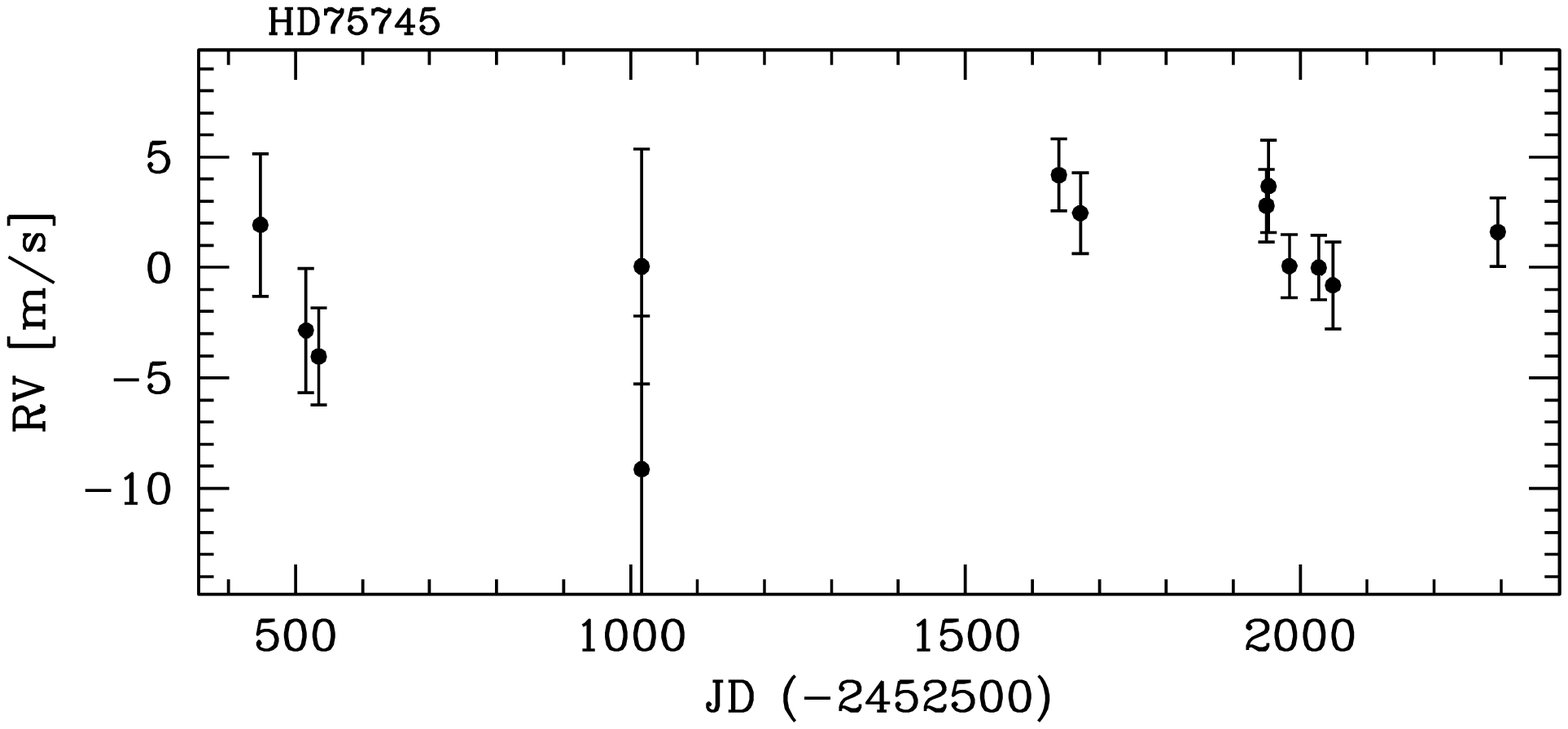}}
\resizebox{5.9cm}{!}{\includegraphics[bb= 18 160 580 430]{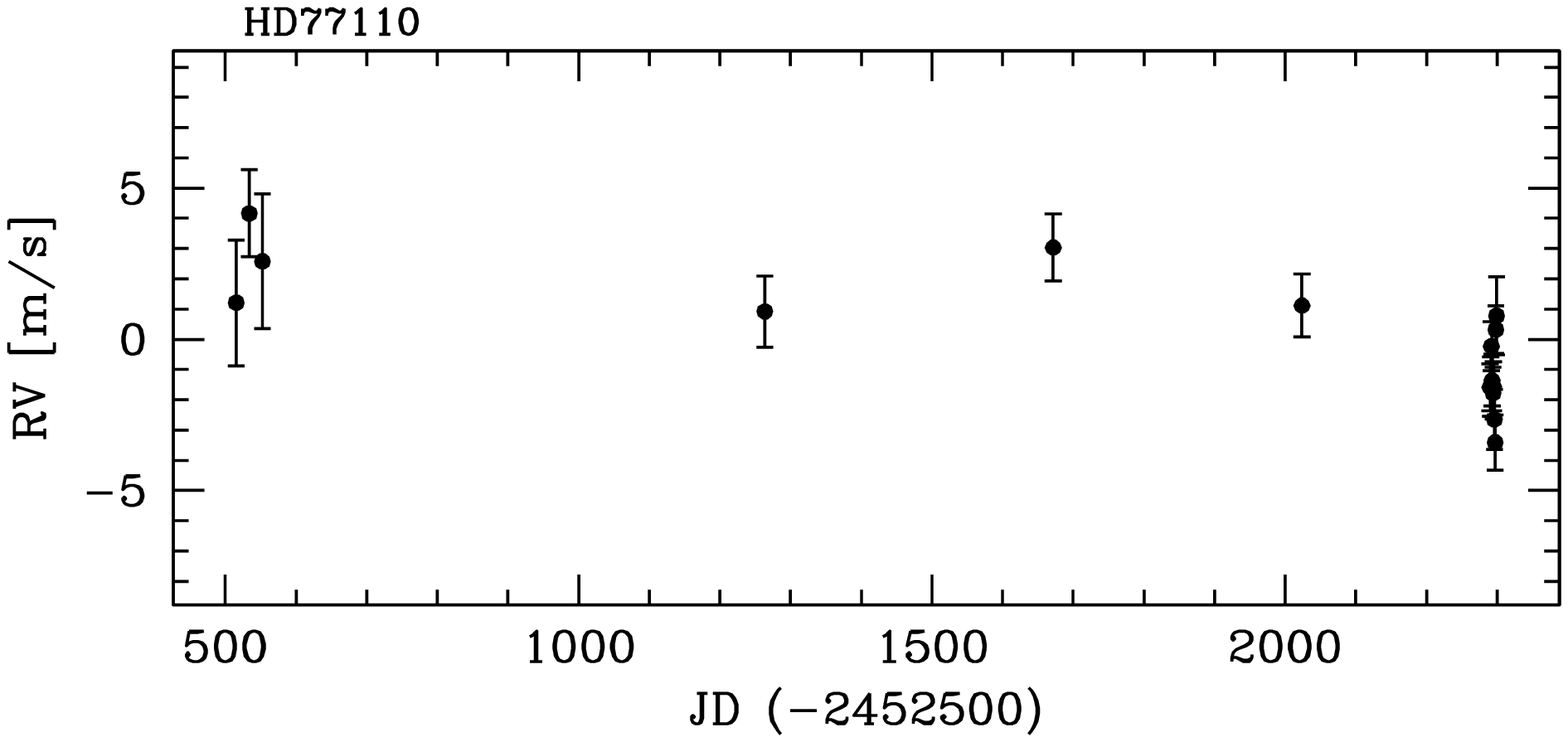}}\\
\resizebox{5.9cm}{!}{\includegraphics[bb= 18 160 580 430]{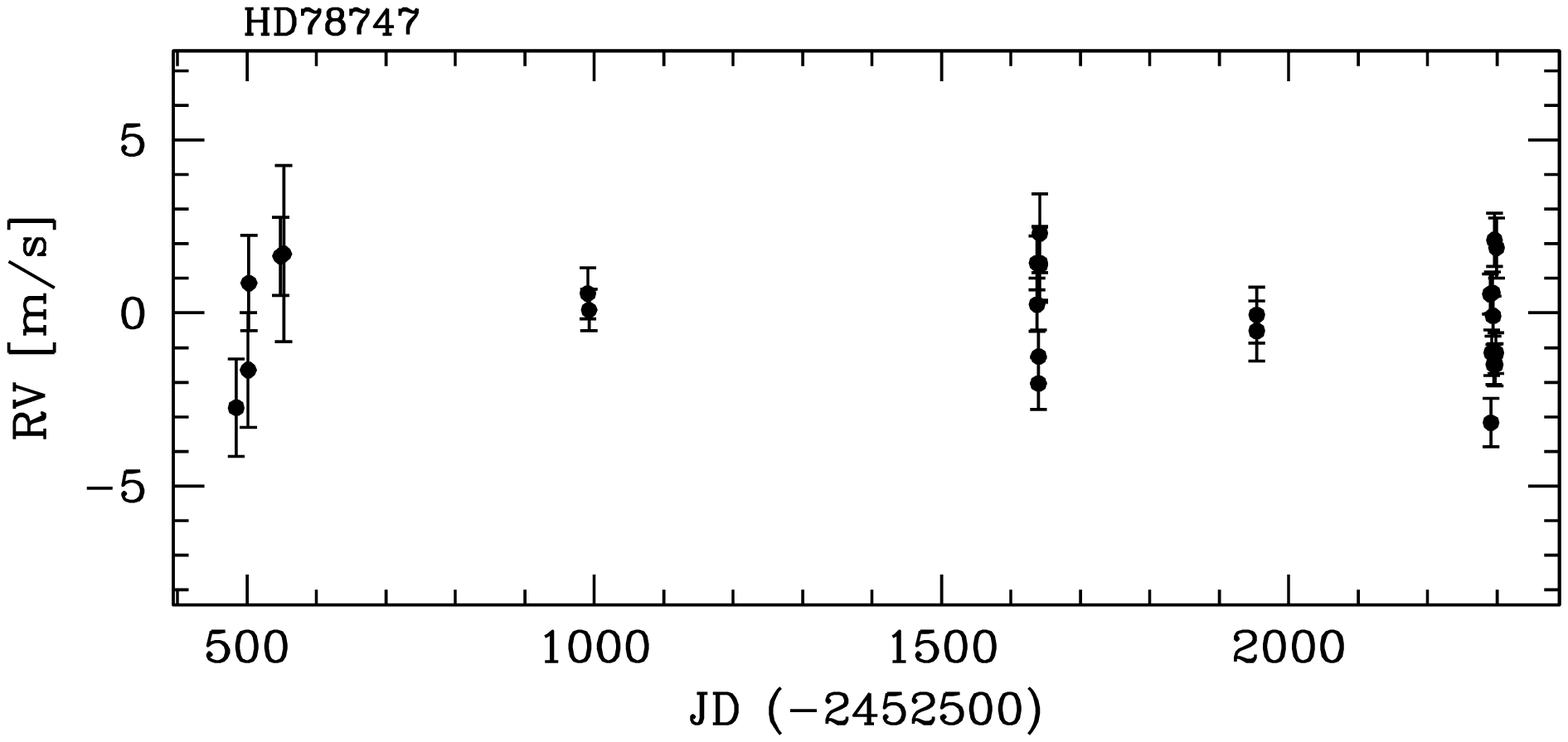}}
\resizebox{5.9cm}{!}{\includegraphics[bb= 18 160 580 430]{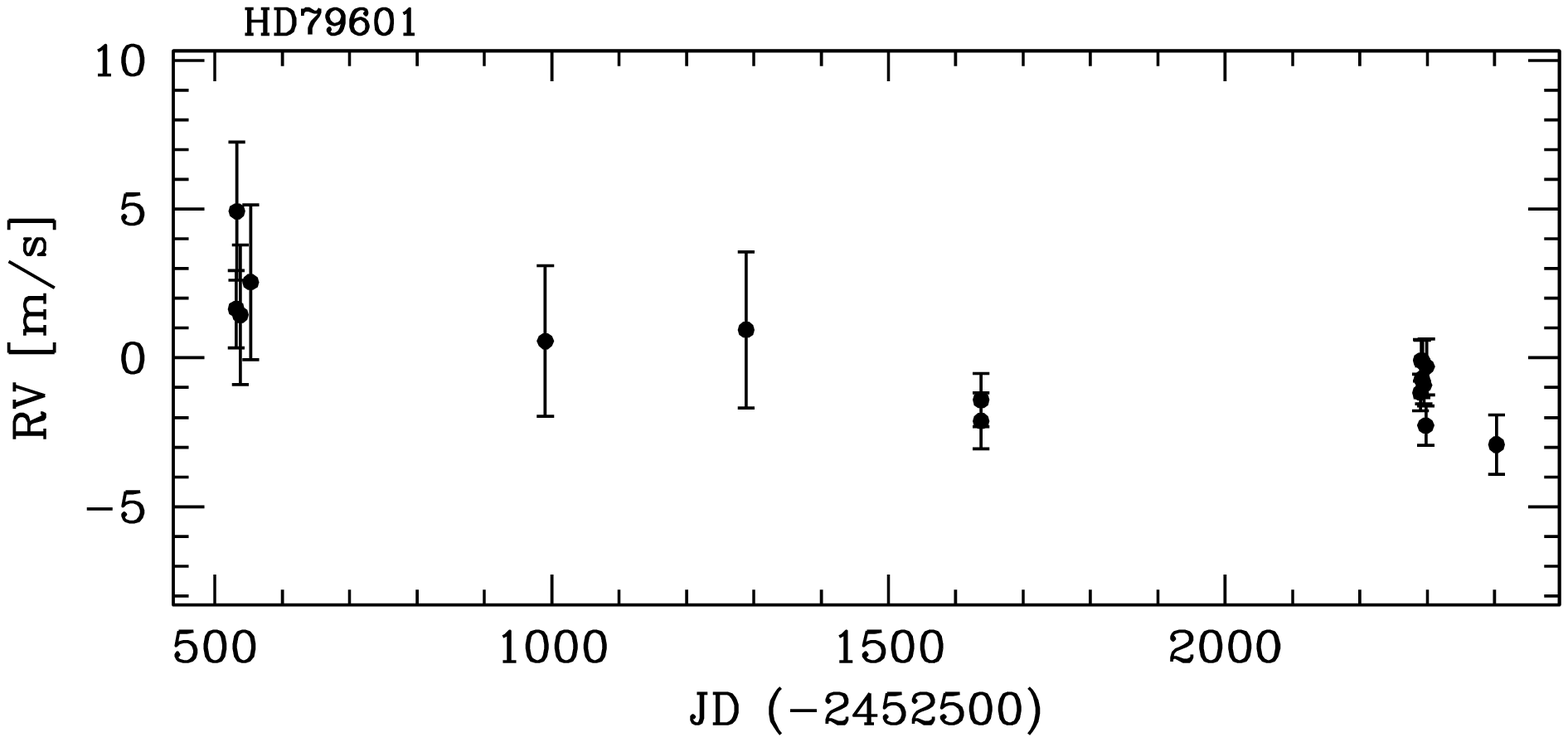}}
\resizebox{5.9cm}{!}{\includegraphics[bb= 18 160 580 430]{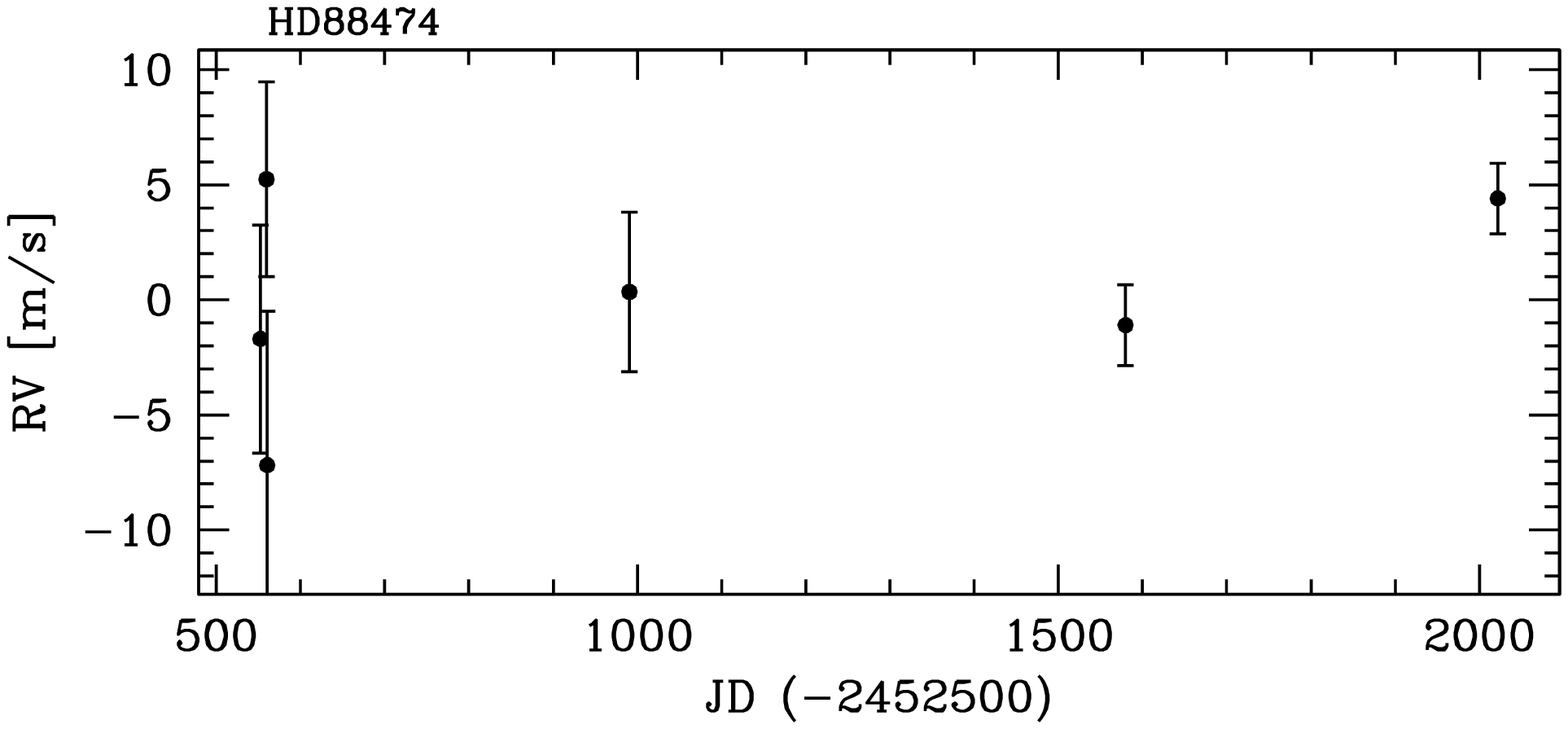}}\\
\resizebox{5.9cm}{!}{\includegraphics[bb= 18 160 580 430]{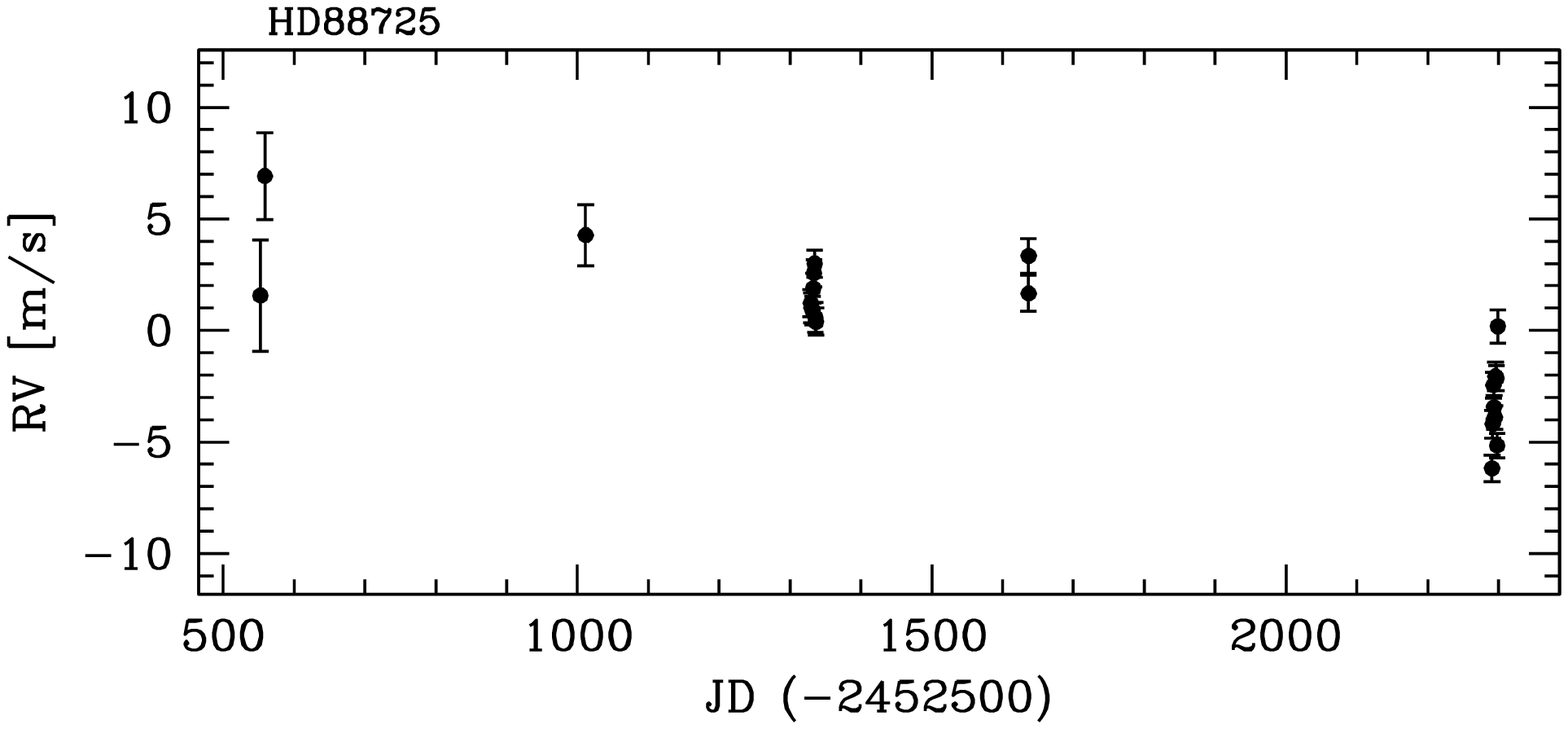}}
\resizebox{5.9cm}{!}{\includegraphics[bb= 18 160 580 430]{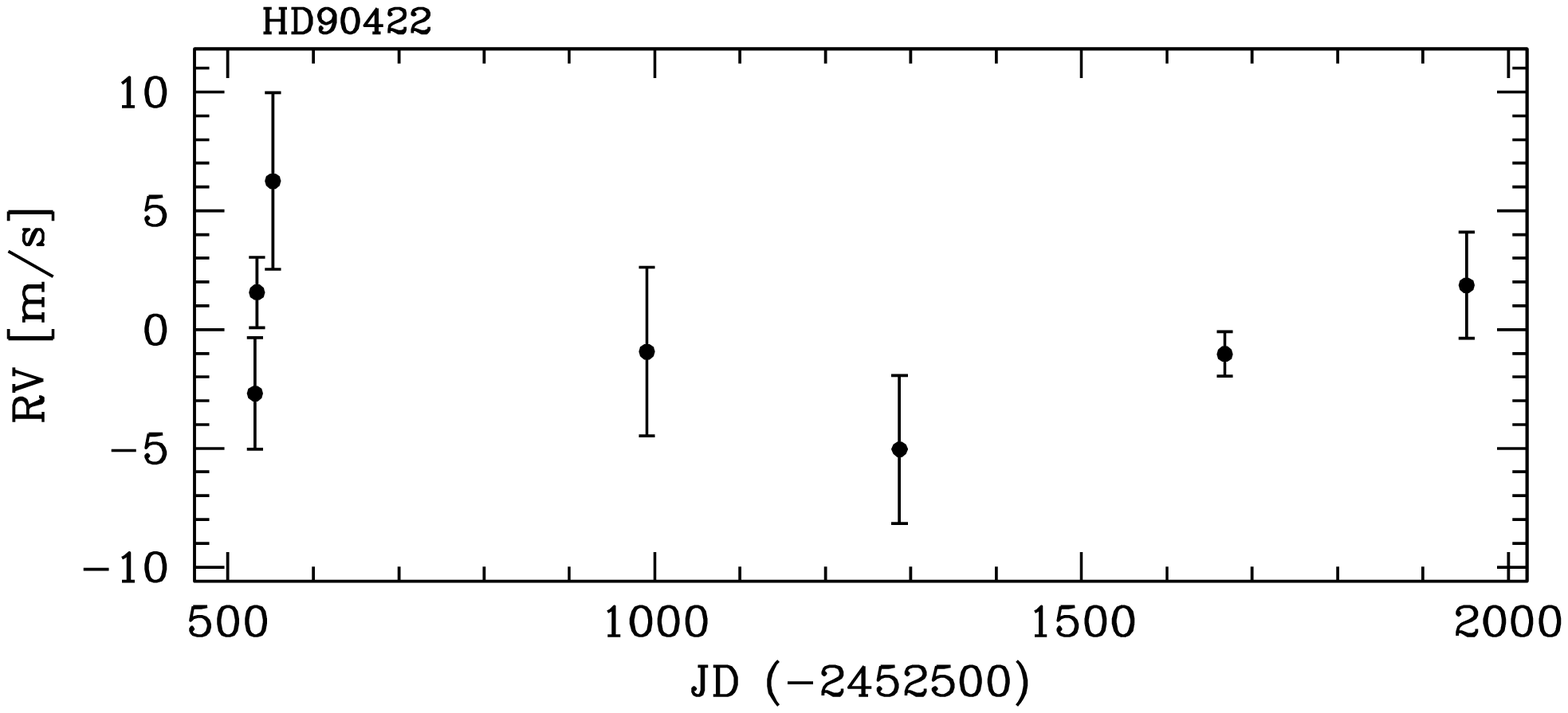}}
\resizebox{5.9cm}{!}{\includegraphics[bb= 18 160 580 430]{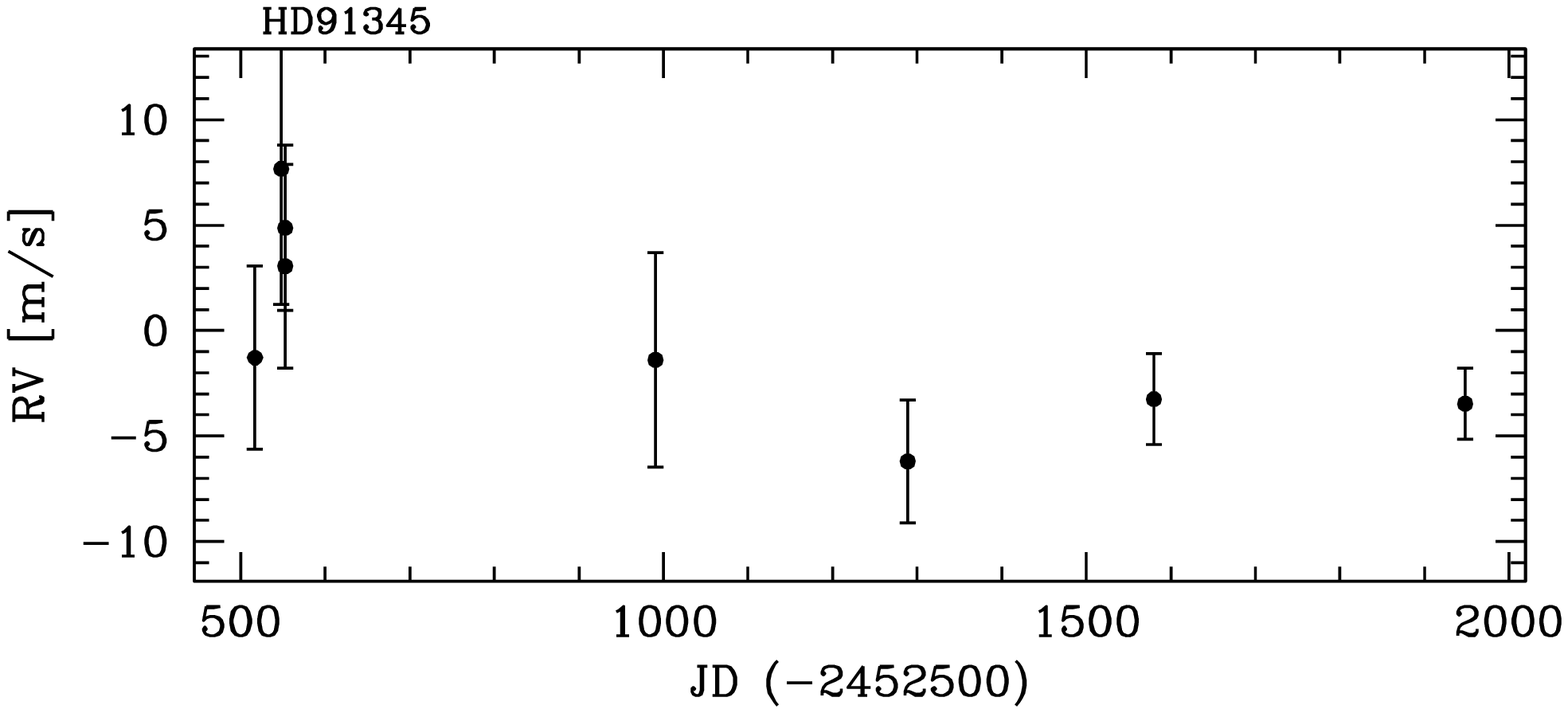}}
\caption{Radial velocity time series for stars with at least 6 radial velocity measurements.}
\label{fig:6mesa}
\end{figure*}

\begin{figure*}[t!]
\resizebox{5.9cm}{!}{\includegraphics[bb= 18 160 580 430]{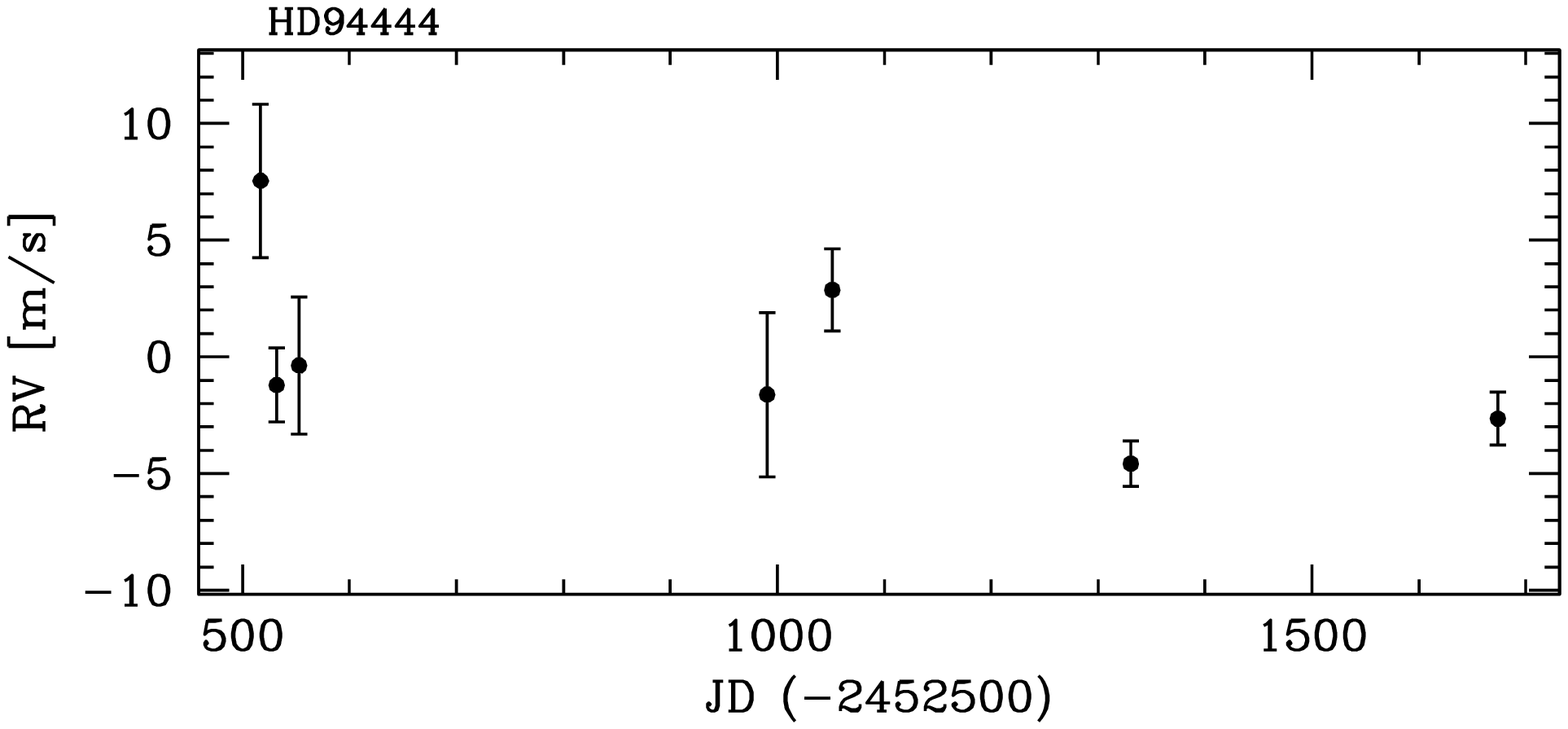}}
\resizebox{5.9cm}{!}{\includegraphics[bb= 18 160 580 430]{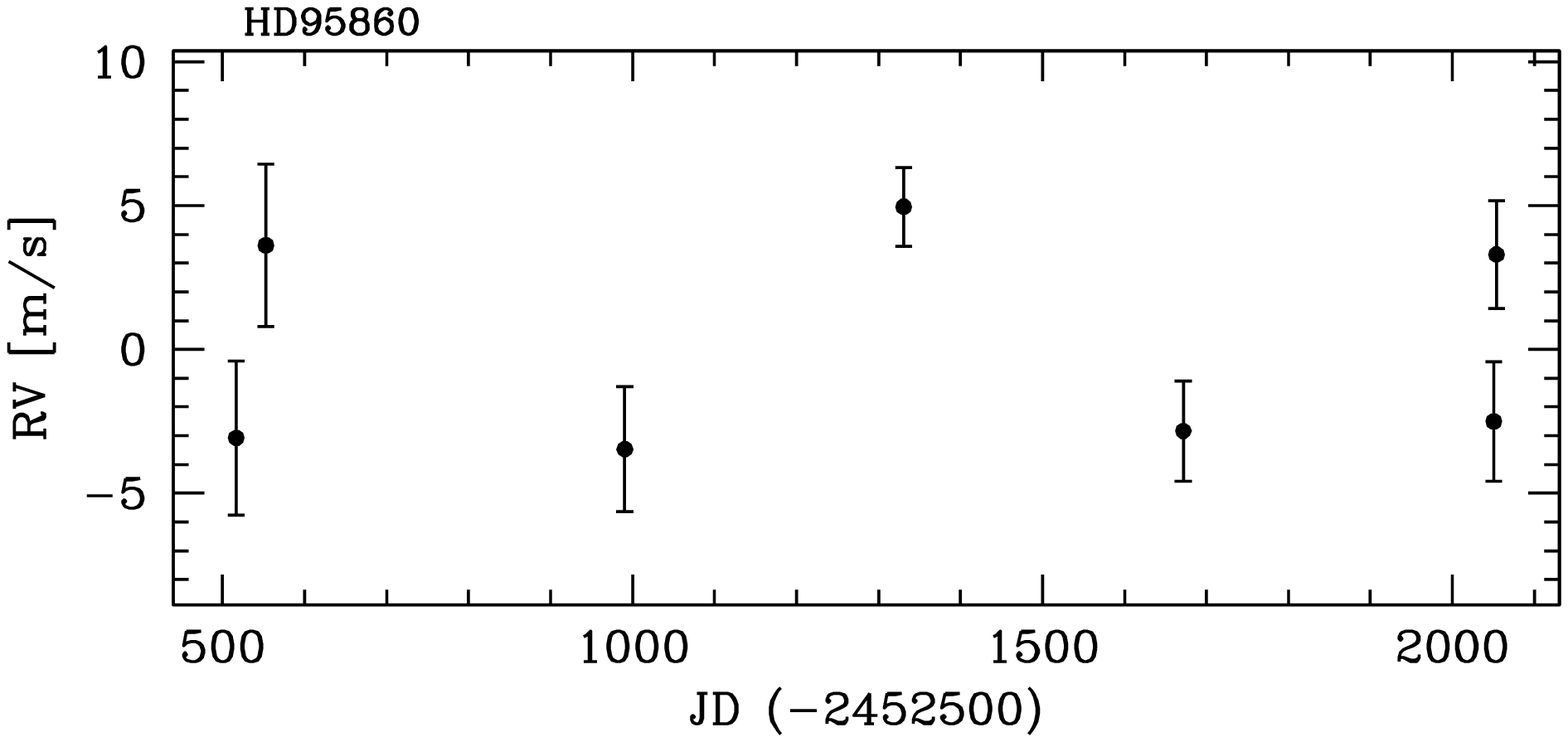}}
\resizebox{5.9cm}{!}{\includegraphics[bb= 18 160 580 430]{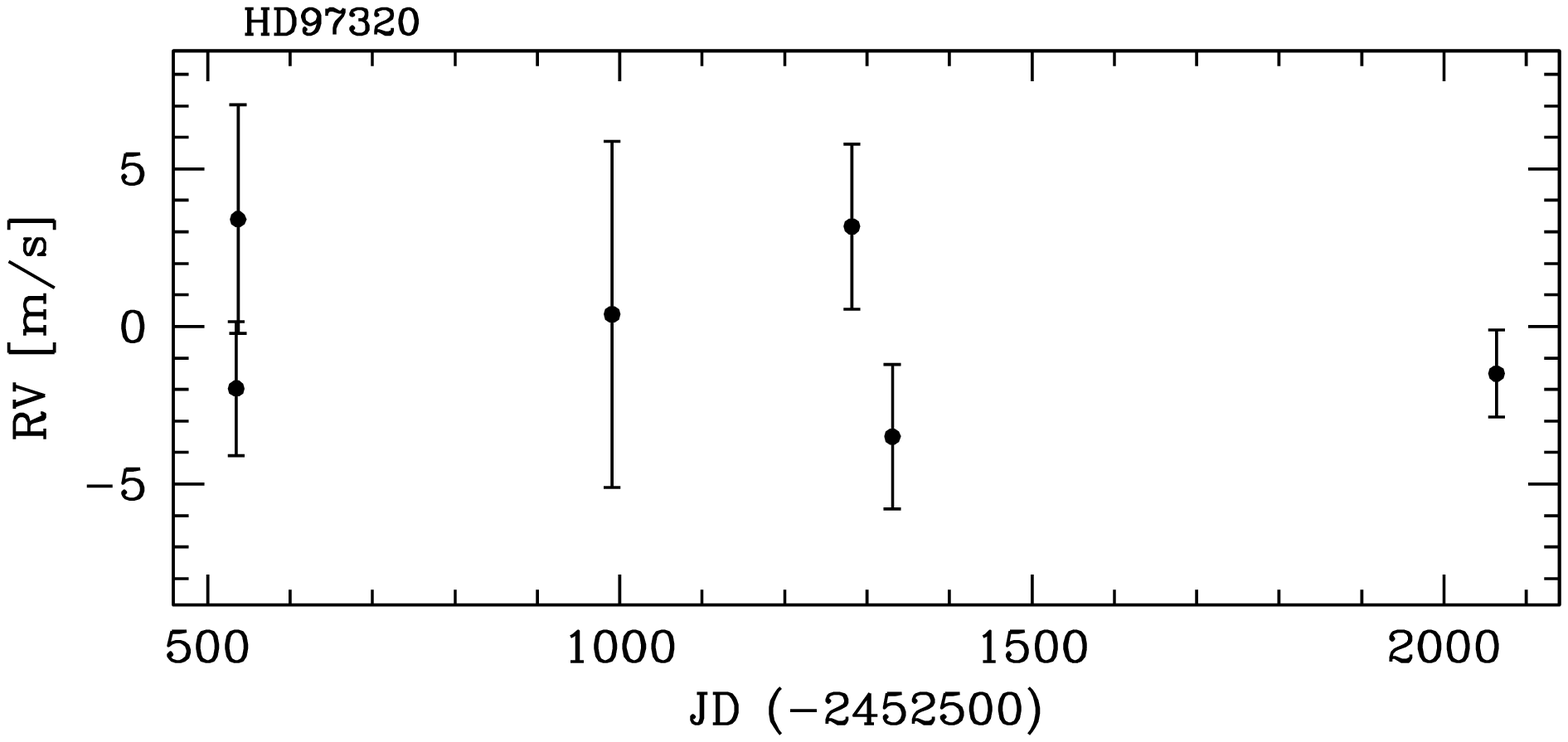}}\\
\resizebox{5.9cm}{!}{\includegraphics[bb= 18 160 580 430]{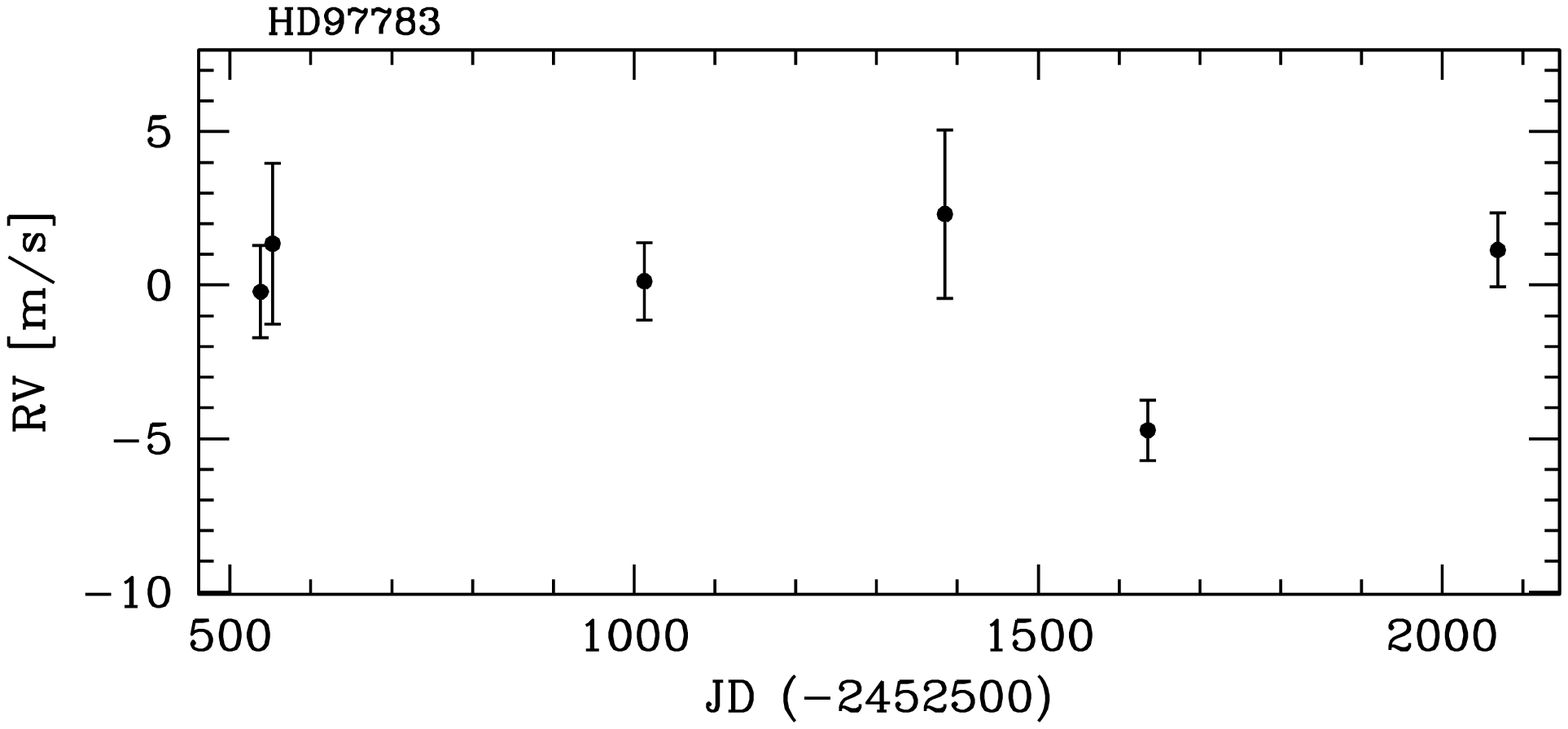}}
\resizebox{5.9cm}{!}{\includegraphics[bb= 18 160 580 430]{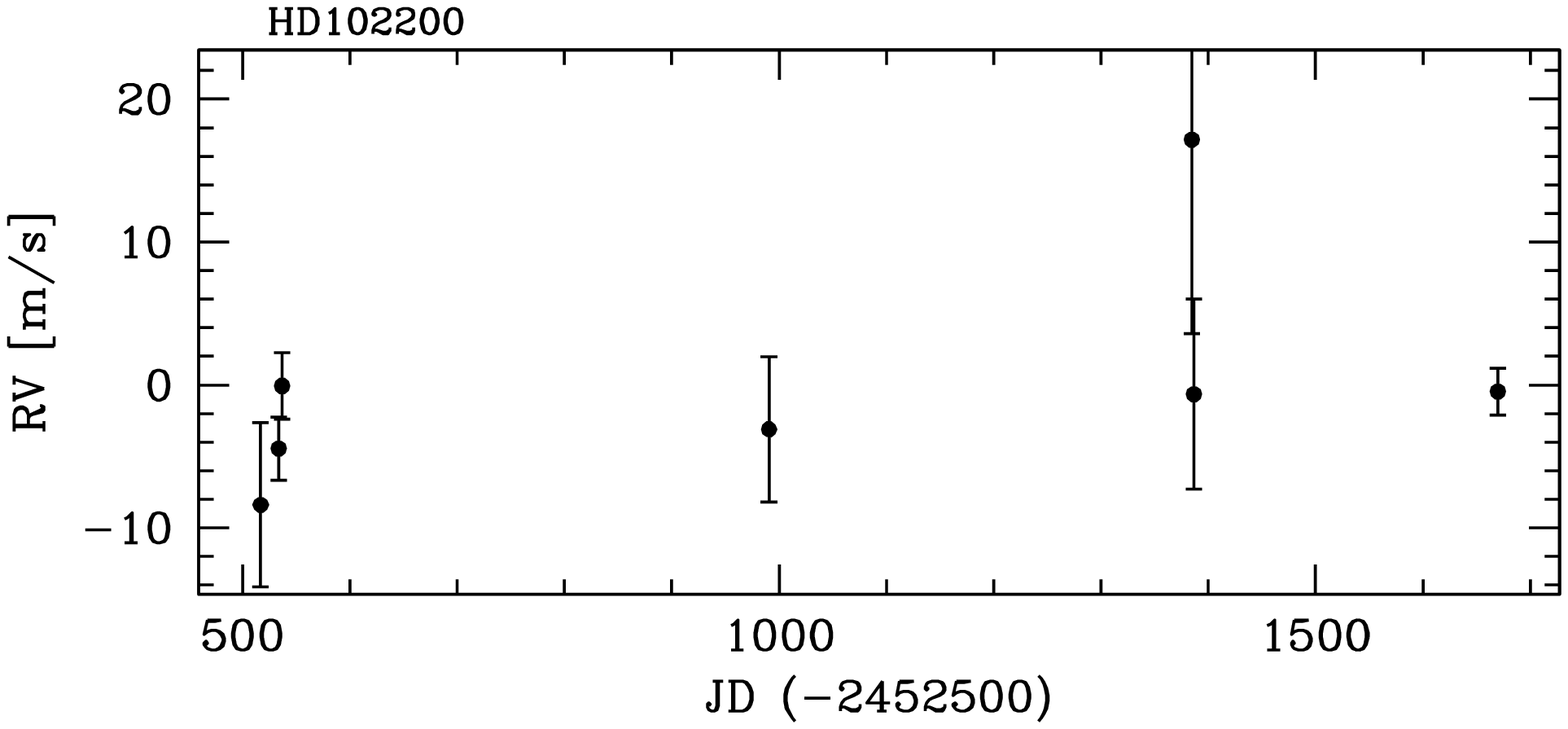}}
\resizebox{5.9cm}{!}{\includegraphics[bb= 18 160 580 430]{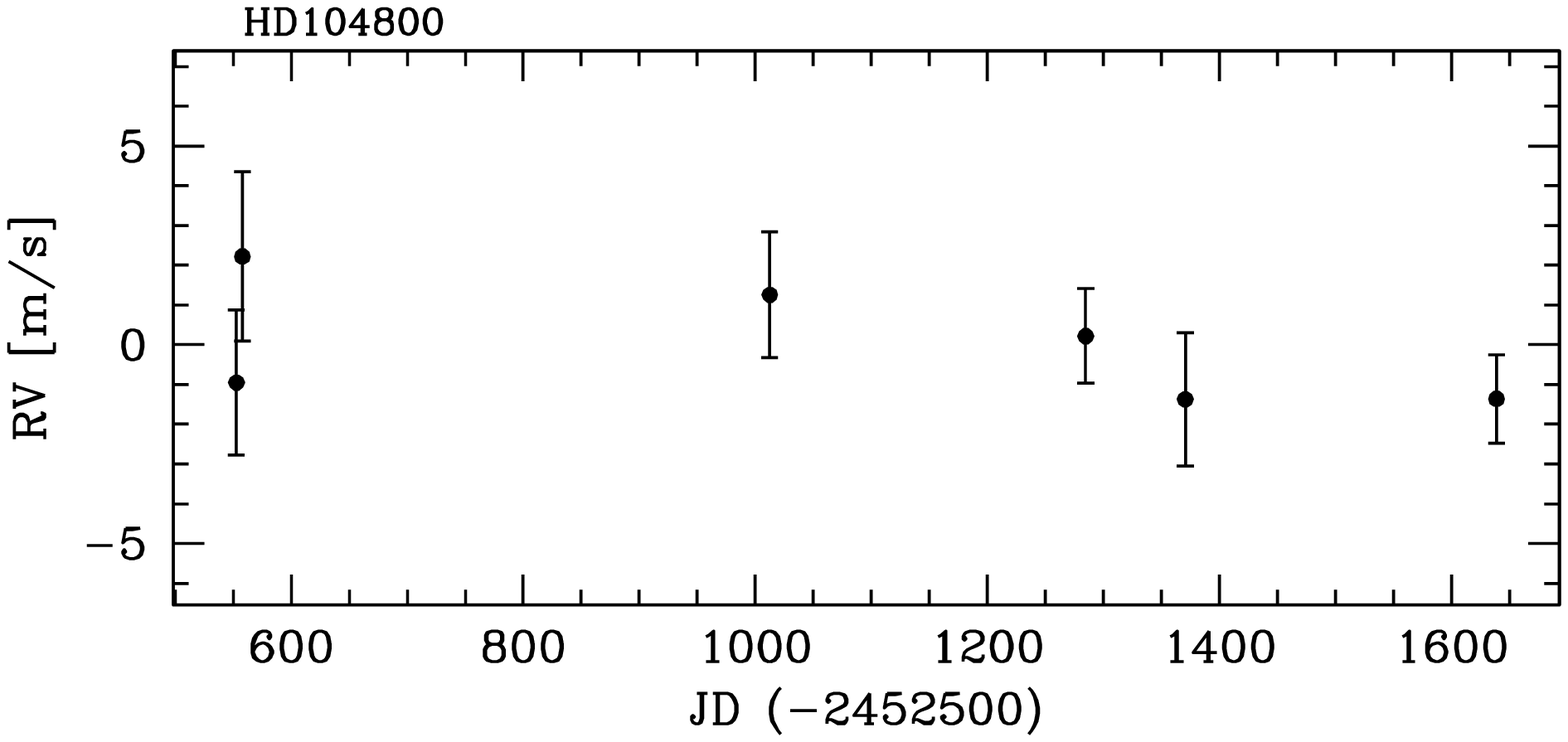}}\\
\resizebox{5.9cm}{!}{\includegraphics[bb= 18 160 580 430]{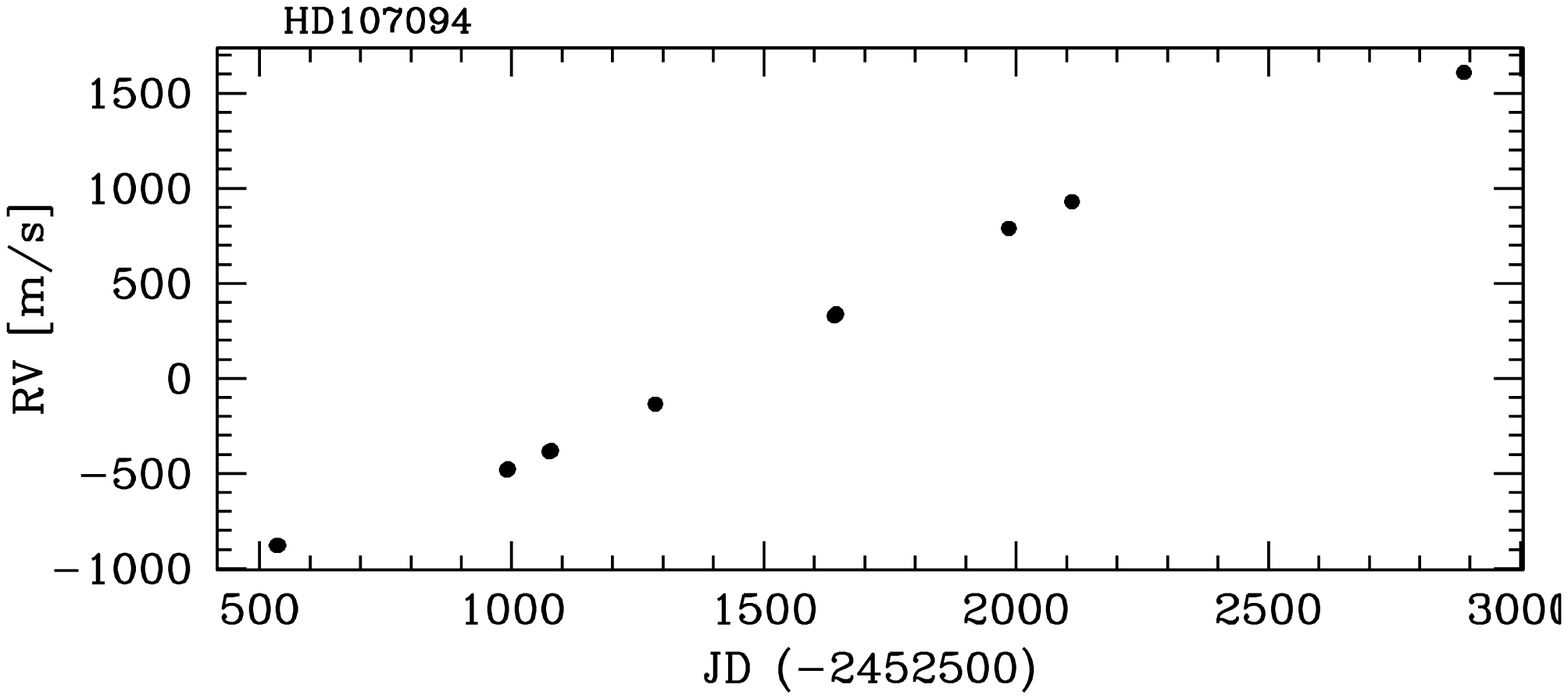}}
\resizebox{5.9cm}{!}{\includegraphics[bb= 18 160 580 430]{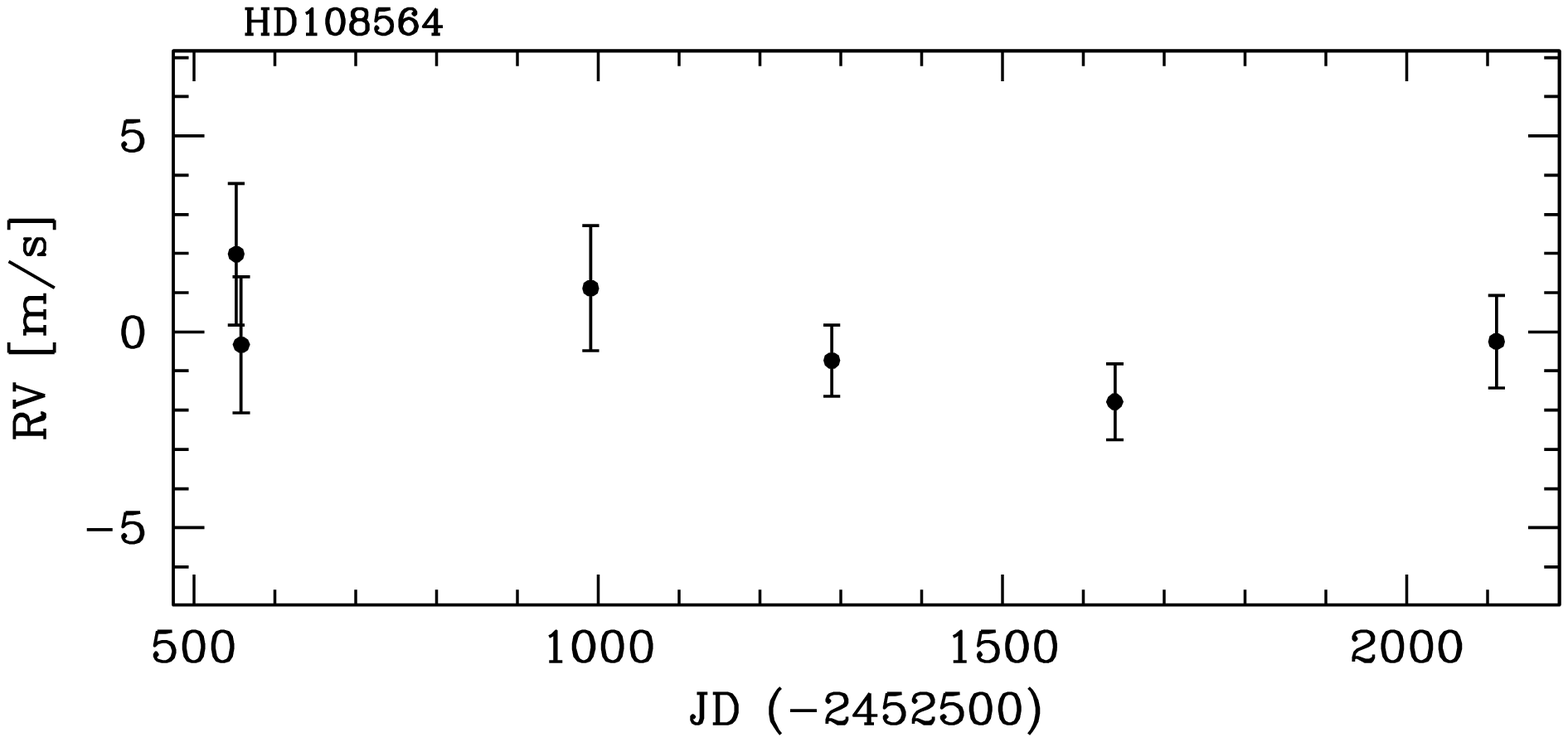}}
\resizebox{5.9cm}{!}{\includegraphics[bb= 18 160 580 430]{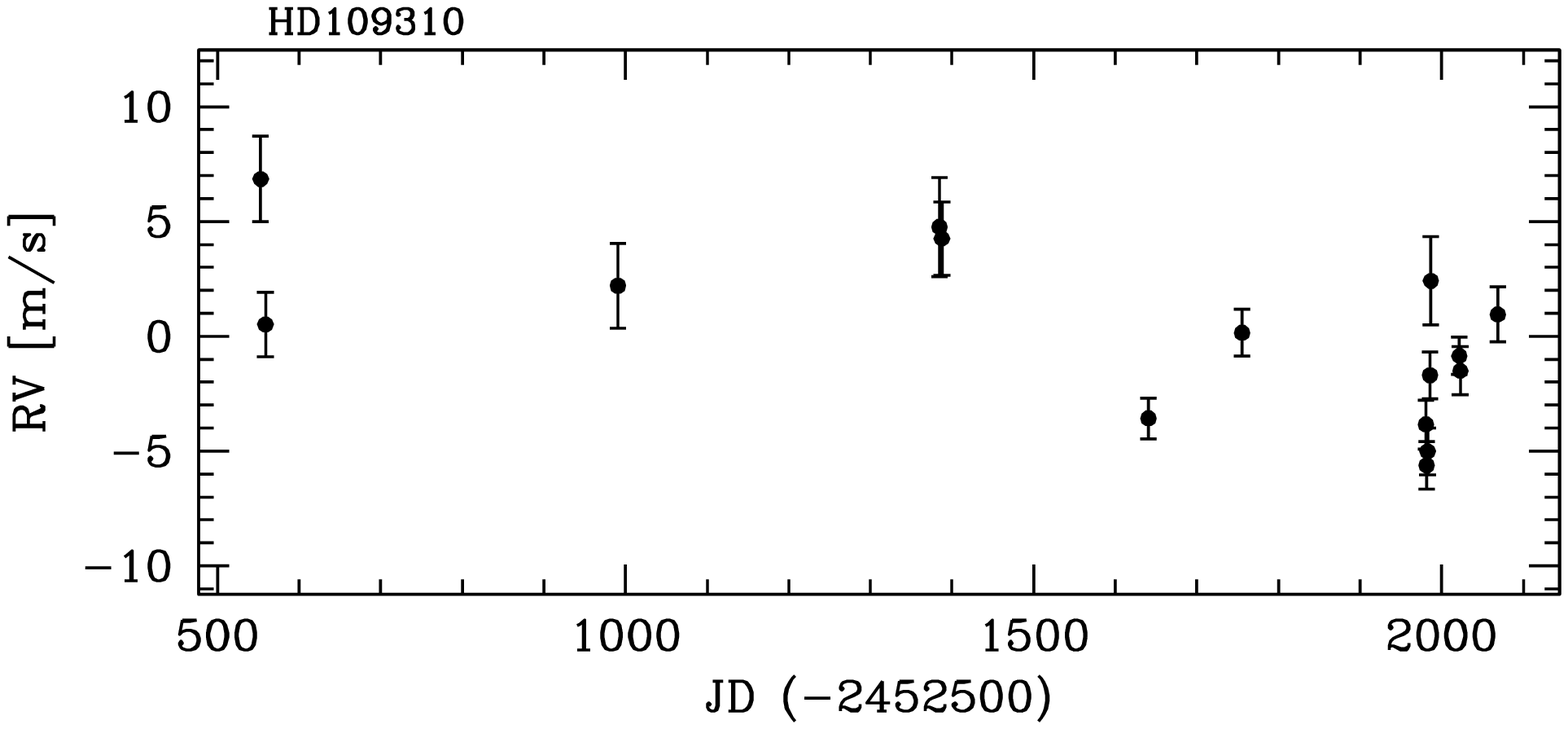}}\\
\resizebox{5.9cm}{!}{\includegraphics[bb= 18 160 580 430]{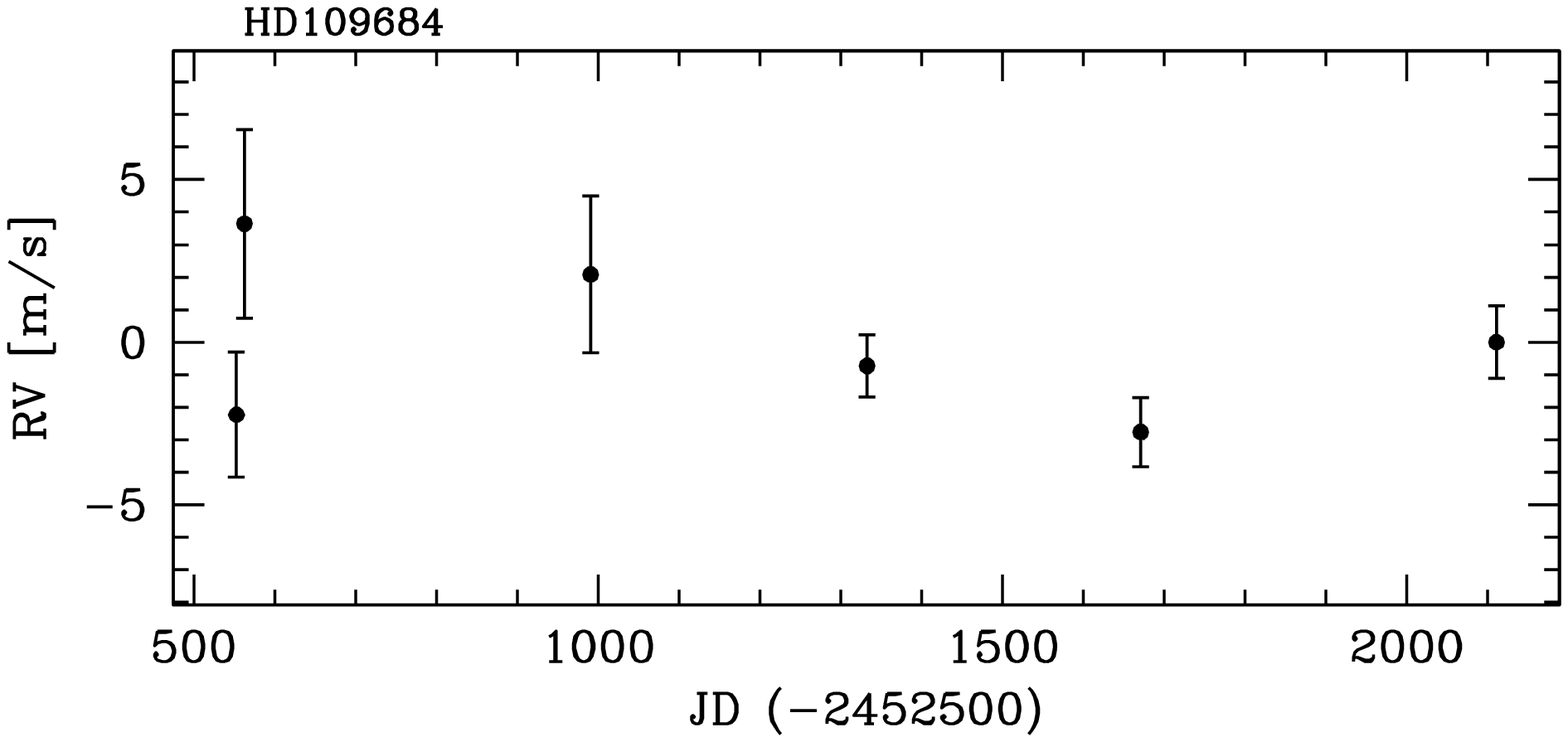}}
\resizebox{5.9cm}{!}{\includegraphics[bb= 18 160 580 430]{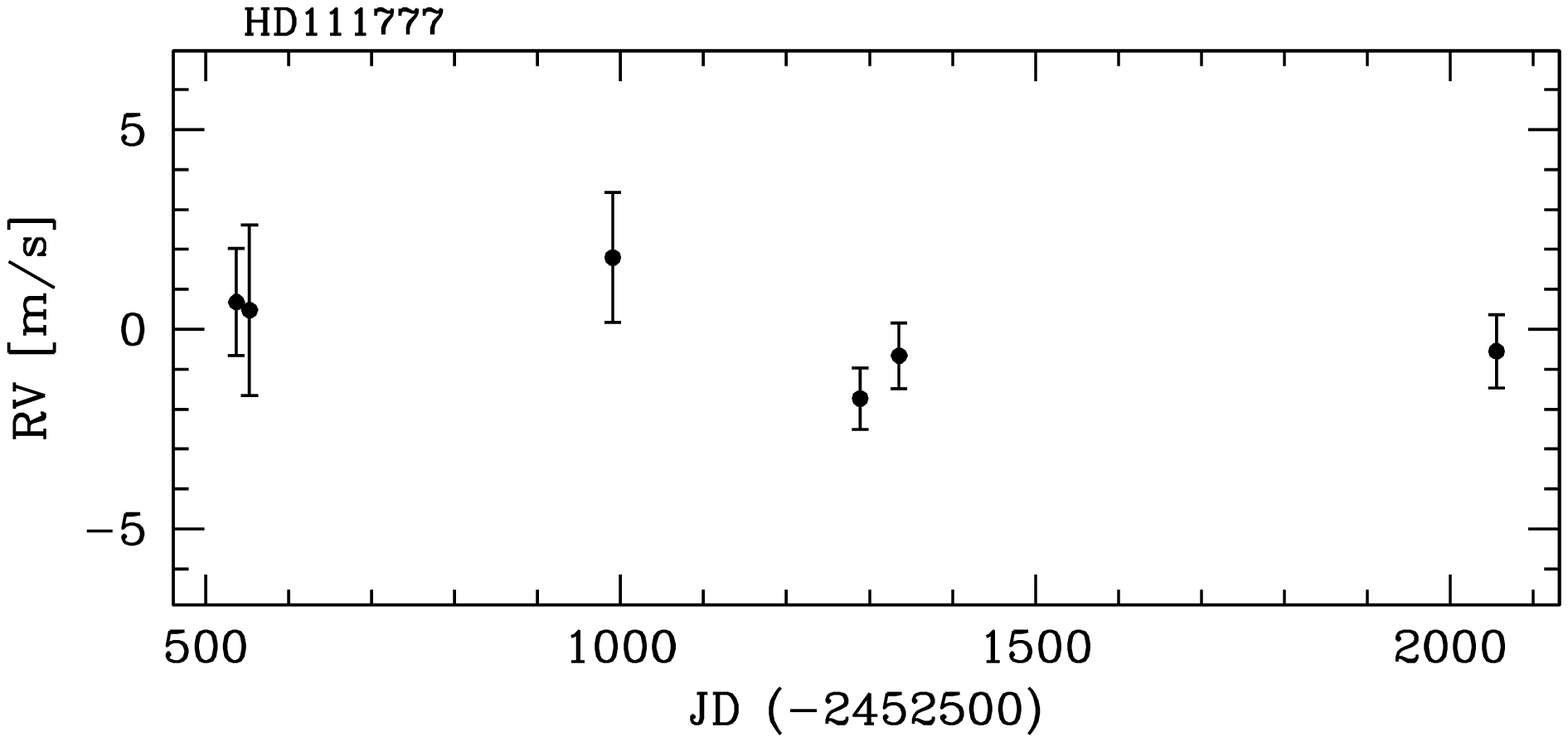}} 
\resizebox{5.9cm}{!}{\includegraphics[bb= 18 160 580 430]{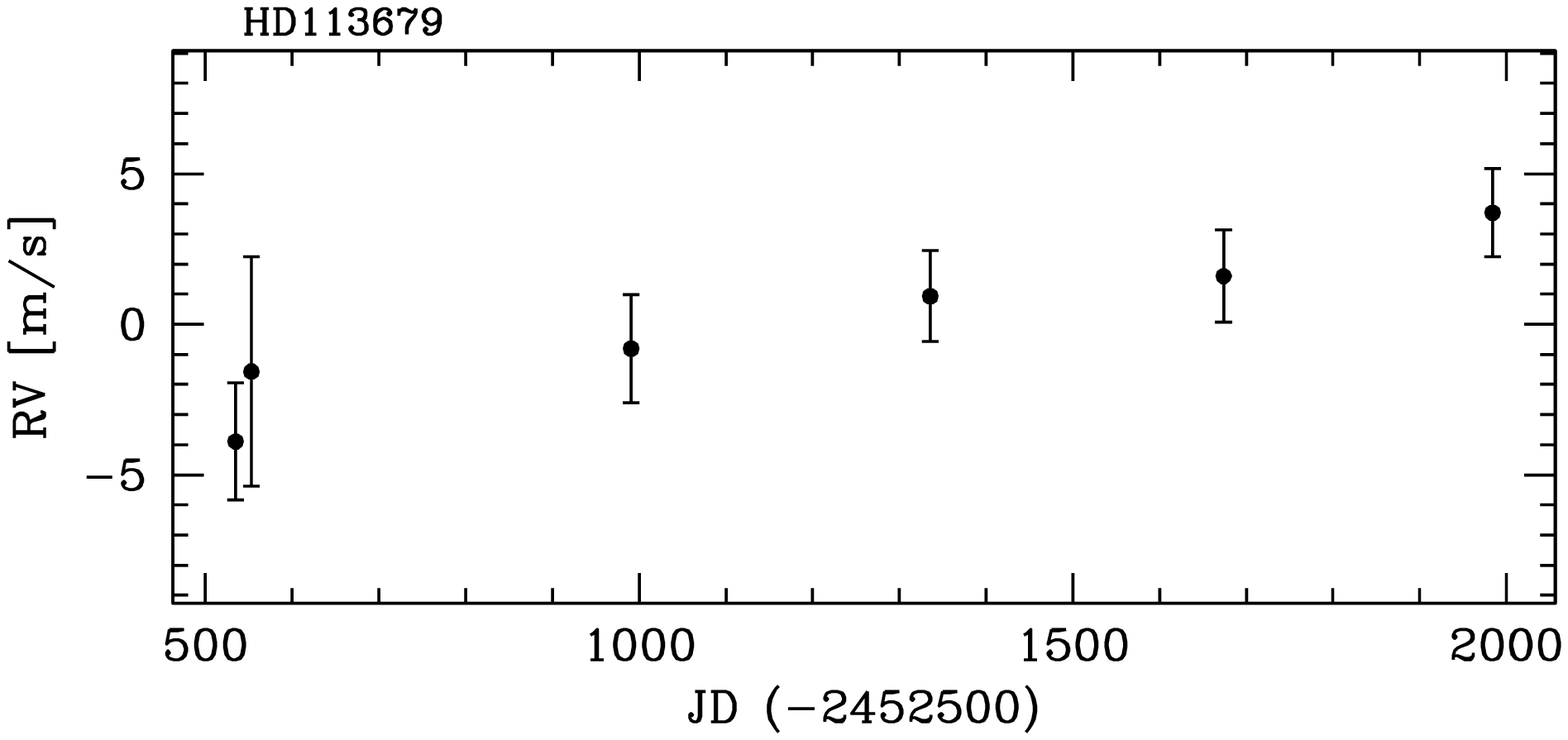}}\\
\resizebox{5.9cm}{!}{\includegraphics[bb= 18 160 580 430]{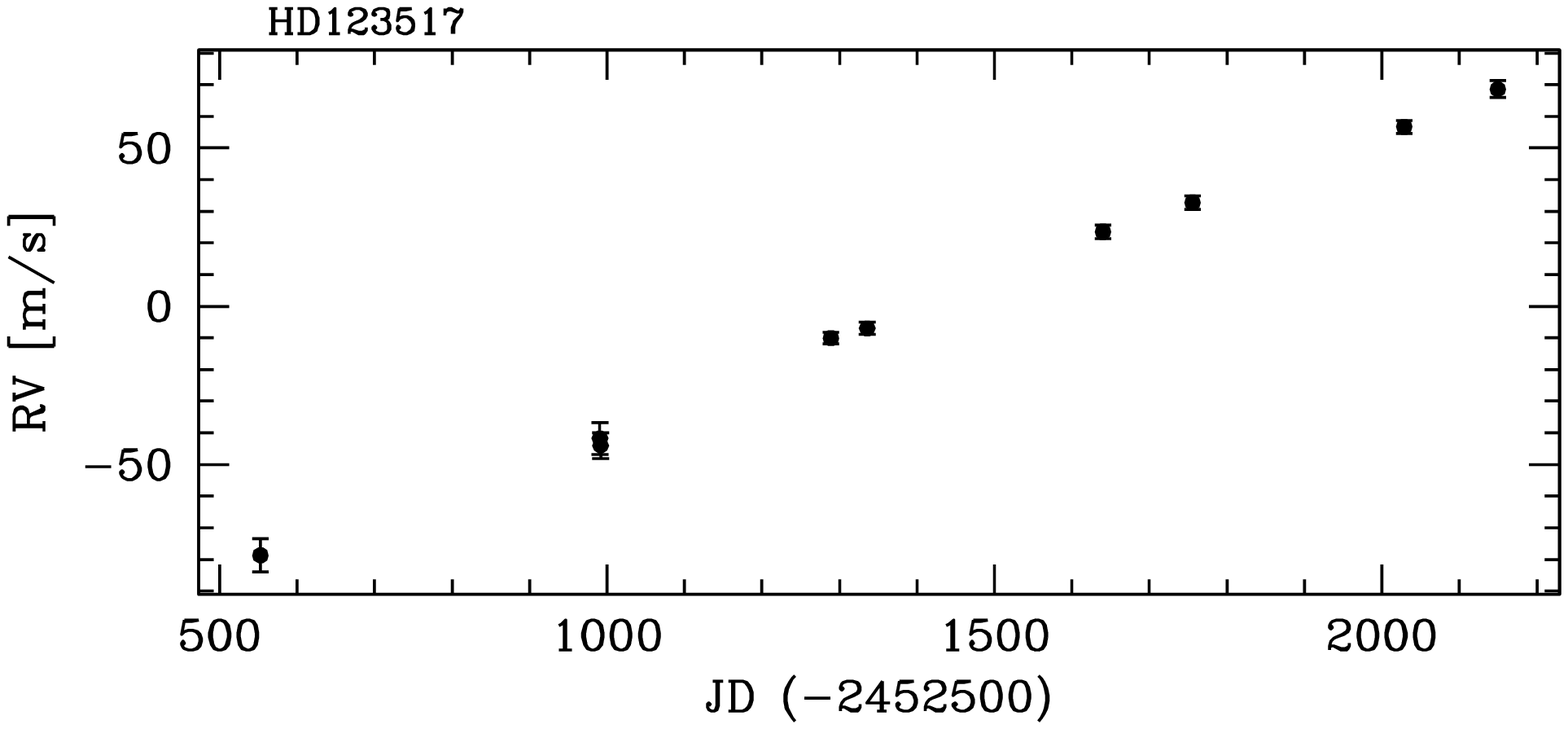}}
\resizebox{5.9cm}{!}{\includegraphics[bb= 18 160 580 430]{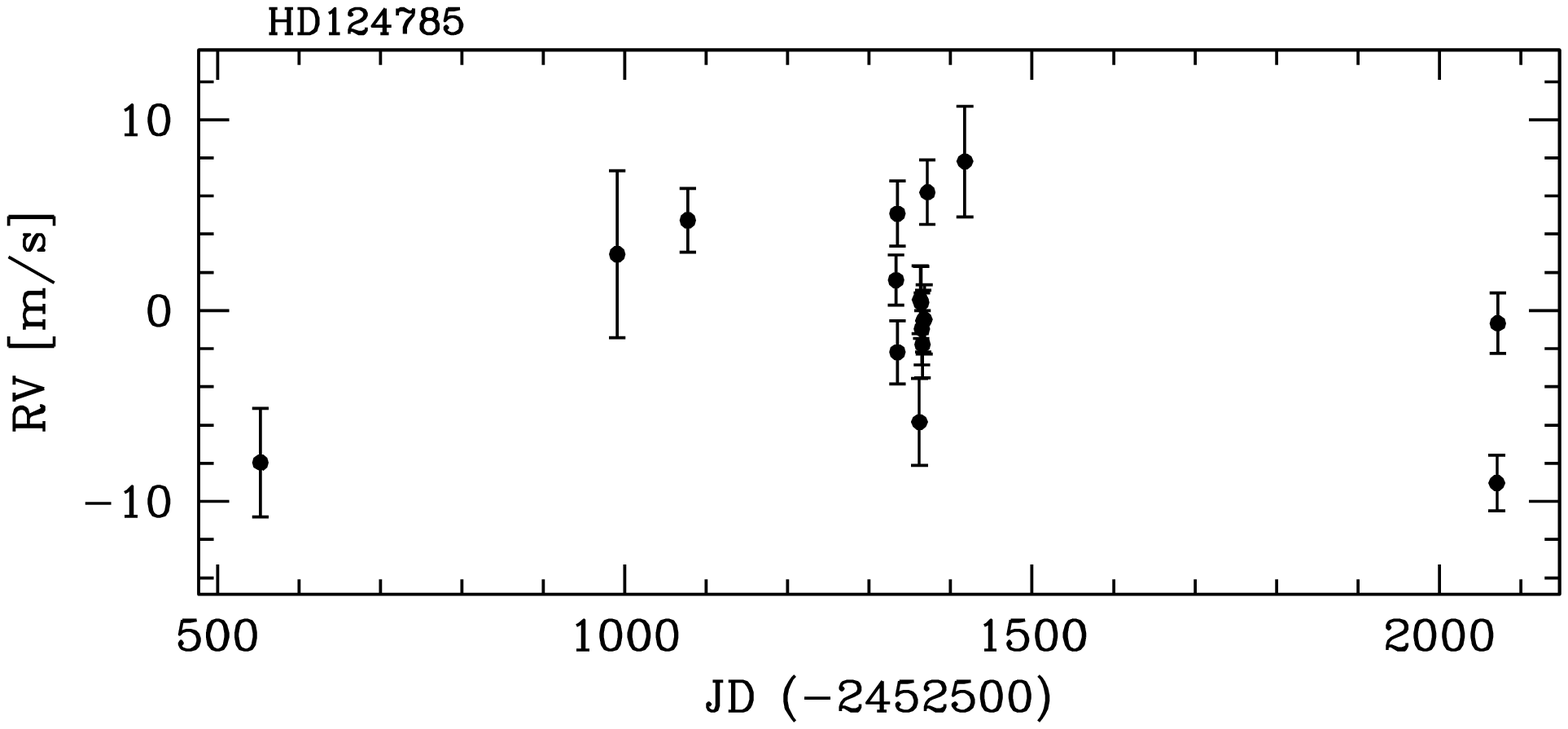}}
\resizebox{5.9cm}{!}{\includegraphics[bb= 18 160 580 430]{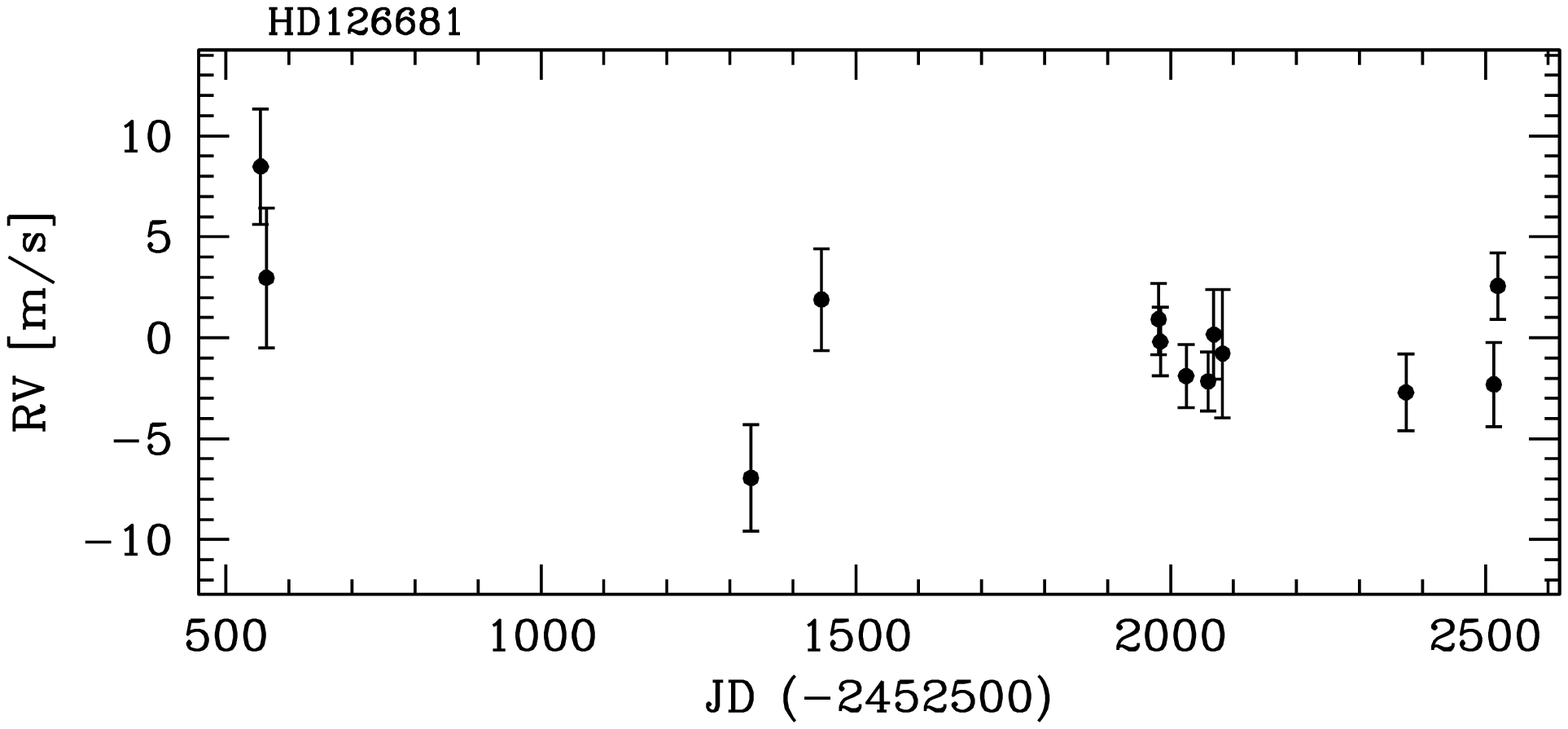}}\\
\resizebox{5.9cm}{!}{\includegraphics[bb= 18 160 580 430]{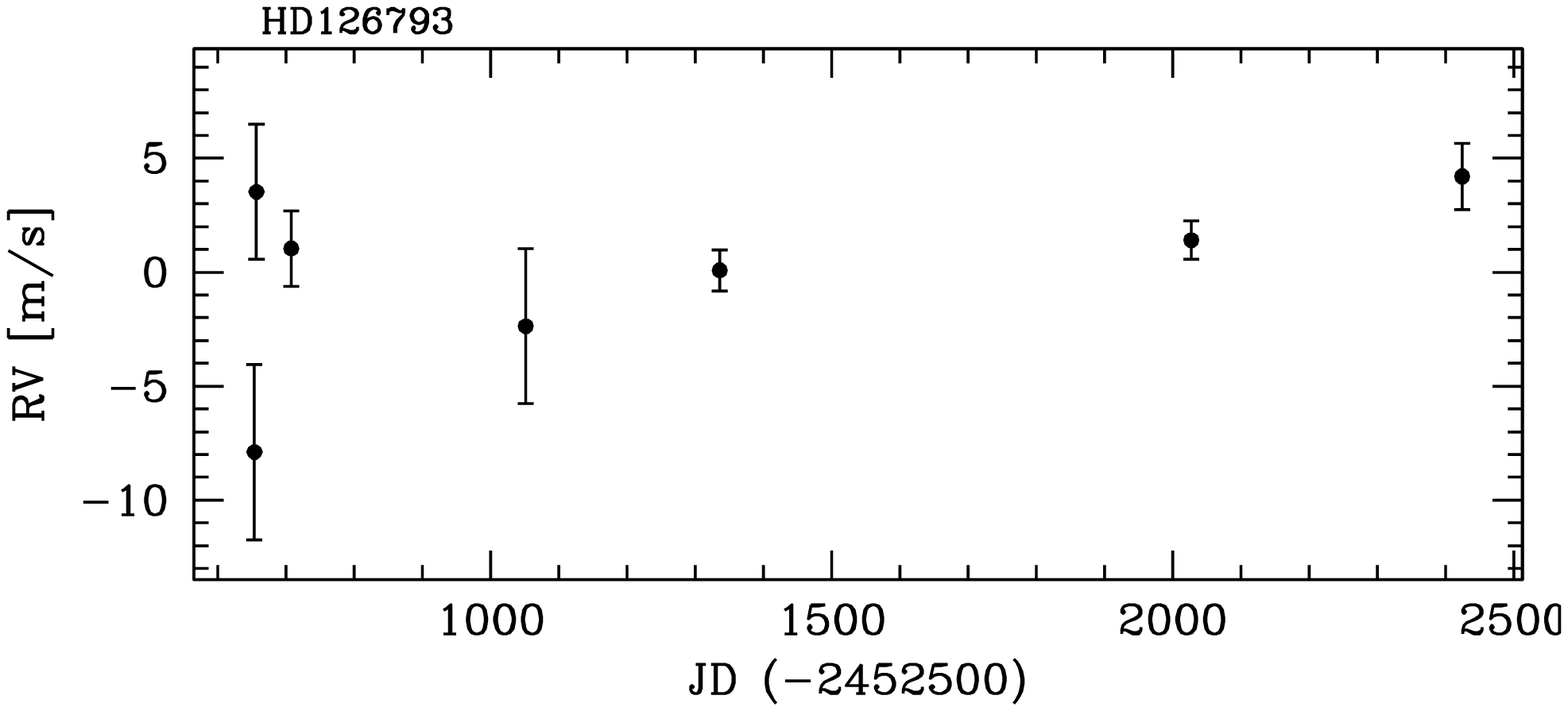}}
\resizebox{5.9cm}{!}{\includegraphics[bb= 18 160 580 430]{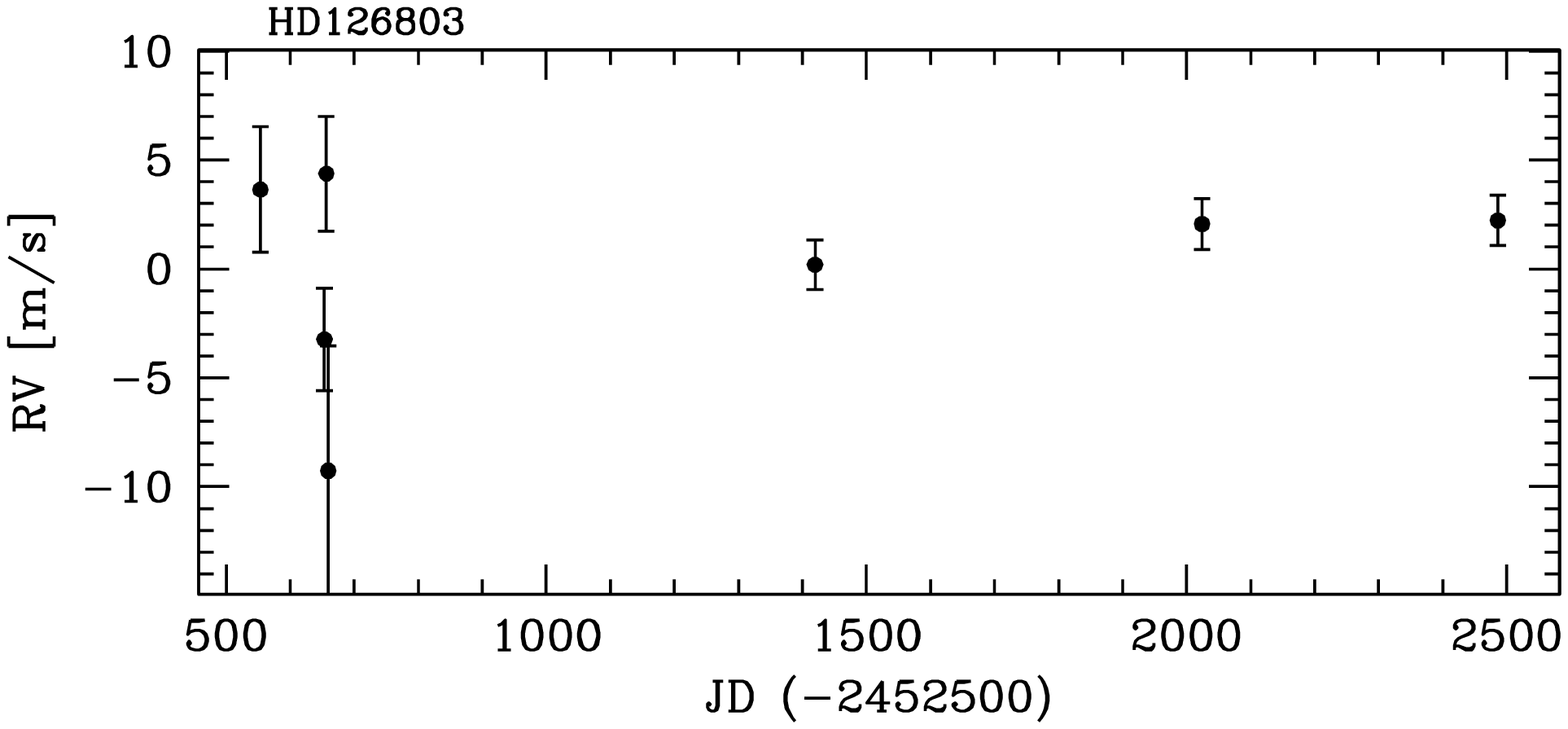}}
\resizebox{5.9cm}{!}{\includegraphics[bb= 18 160 580 430]{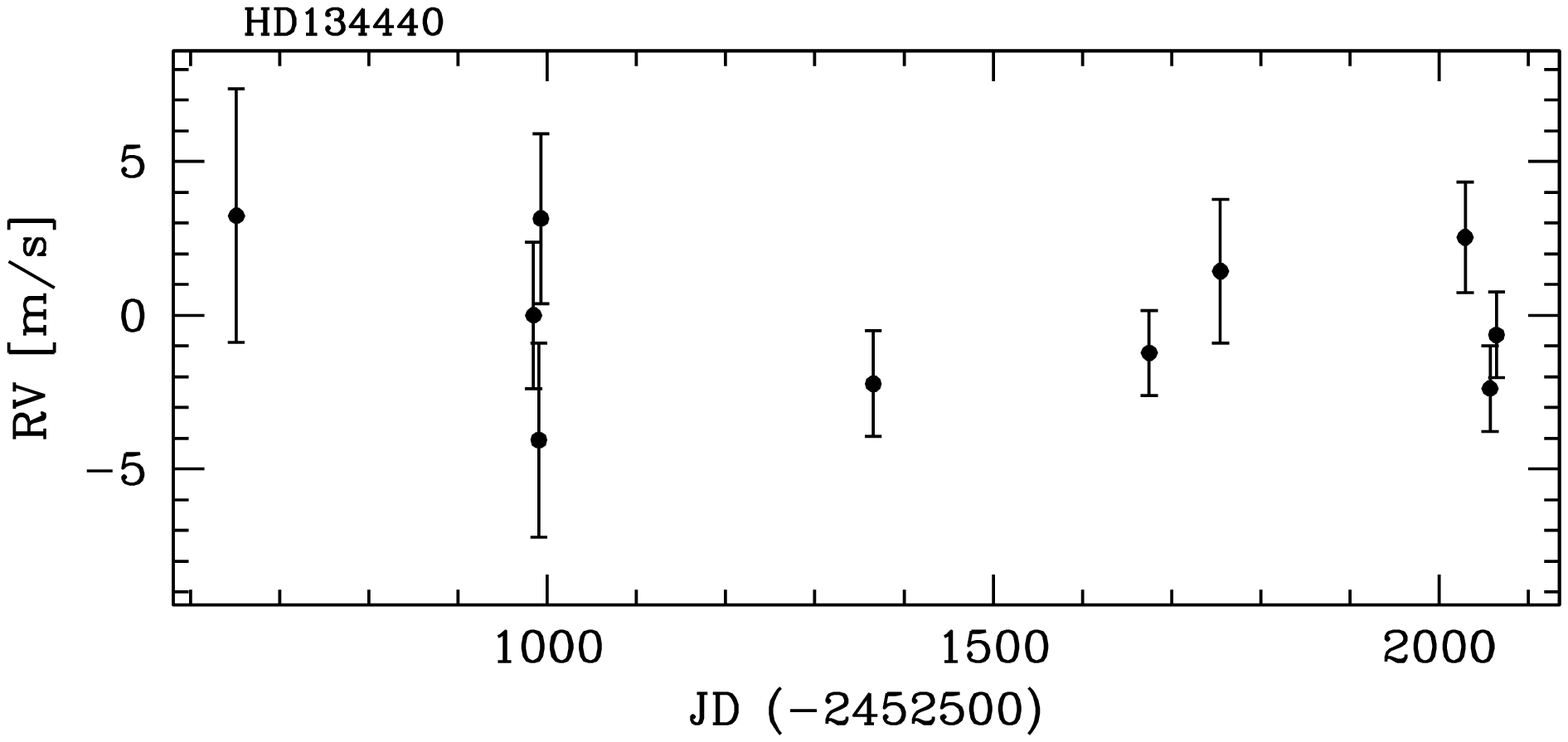}}\\
\resizebox{5.9cm}{!}{\includegraphics[bb= 18 160 580 430]{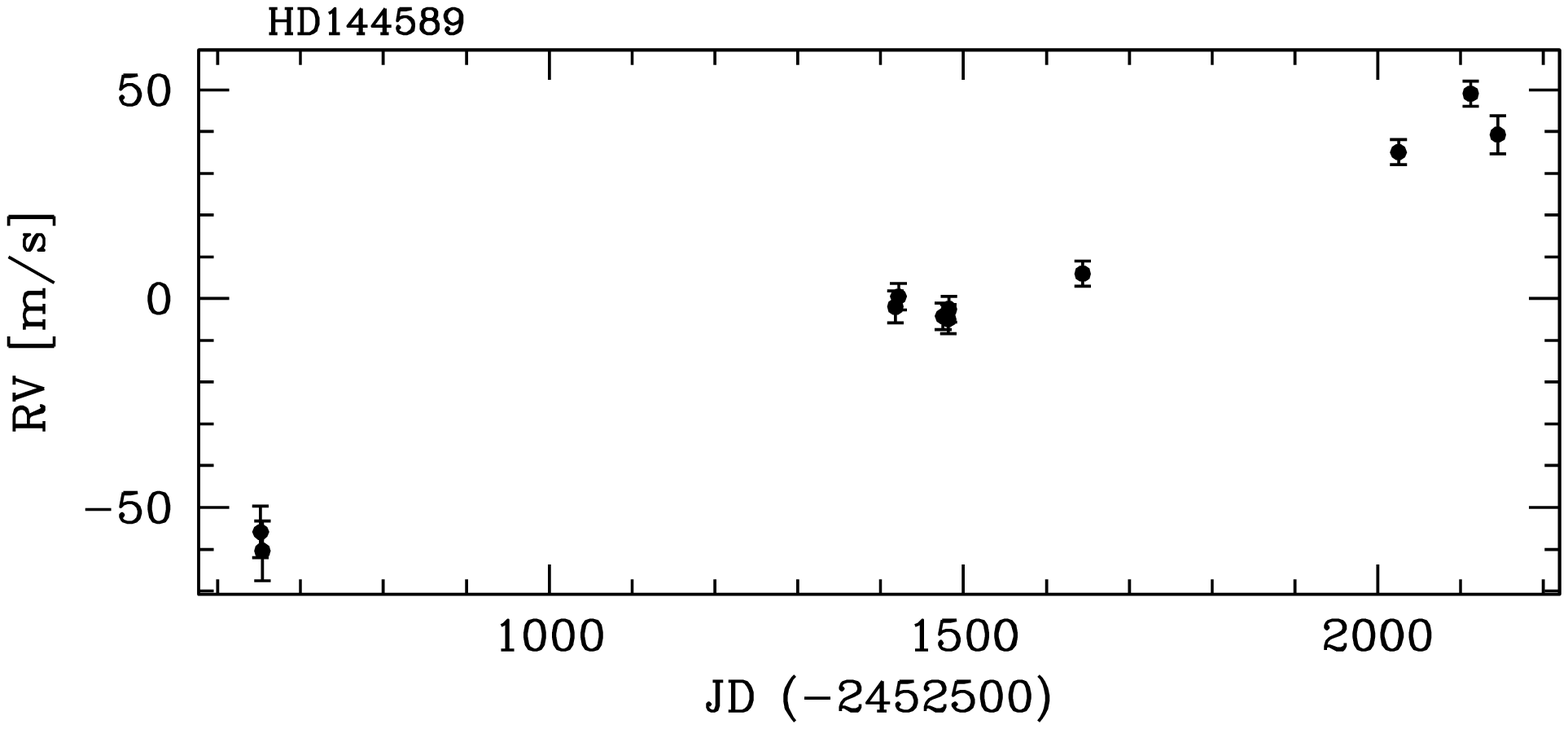}}
\resizebox{5.9cm}{!}{\includegraphics[bb= 18 160 580 430]{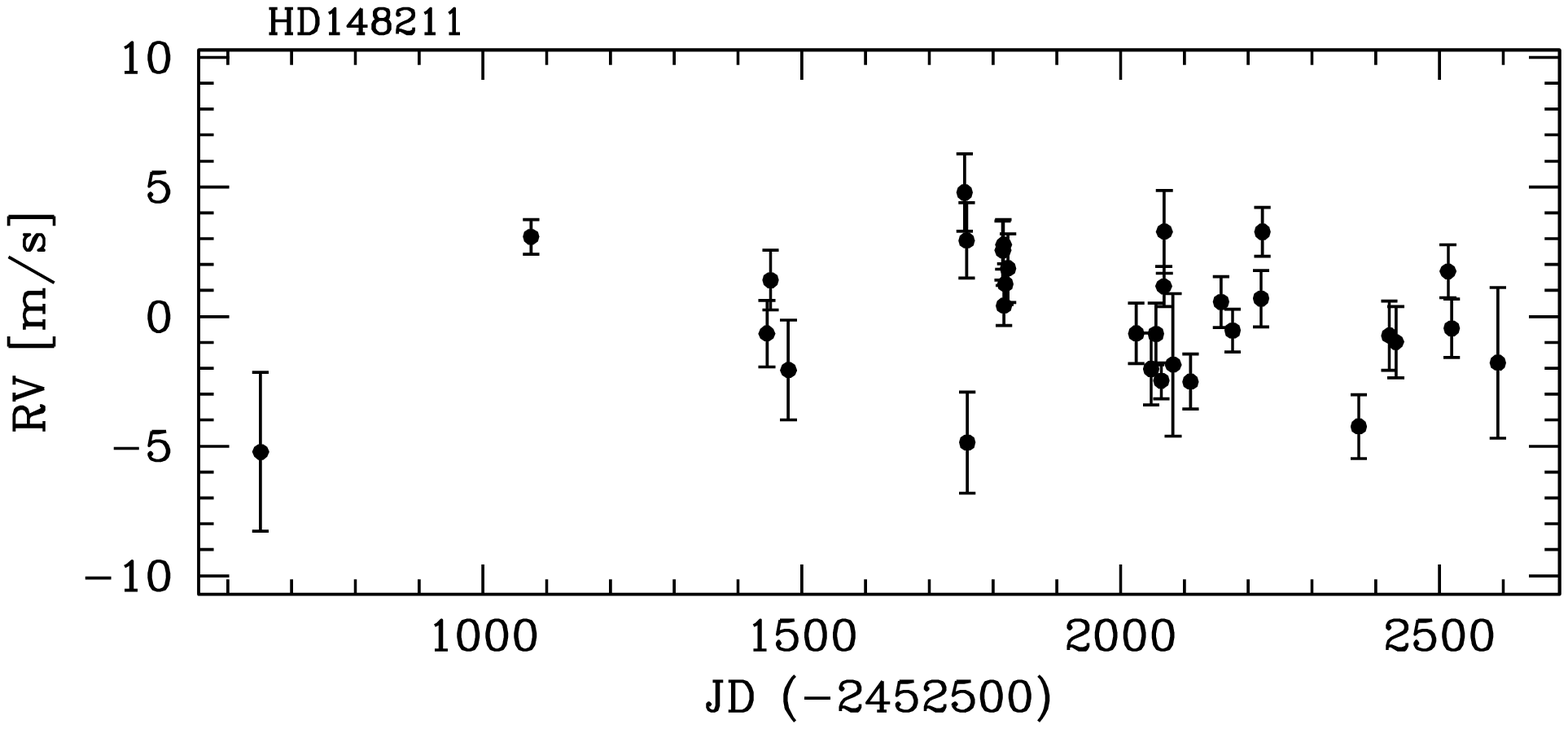}}
\resizebox{5.9cm}{!}{\includegraphics[bb= 18 160 580 430]{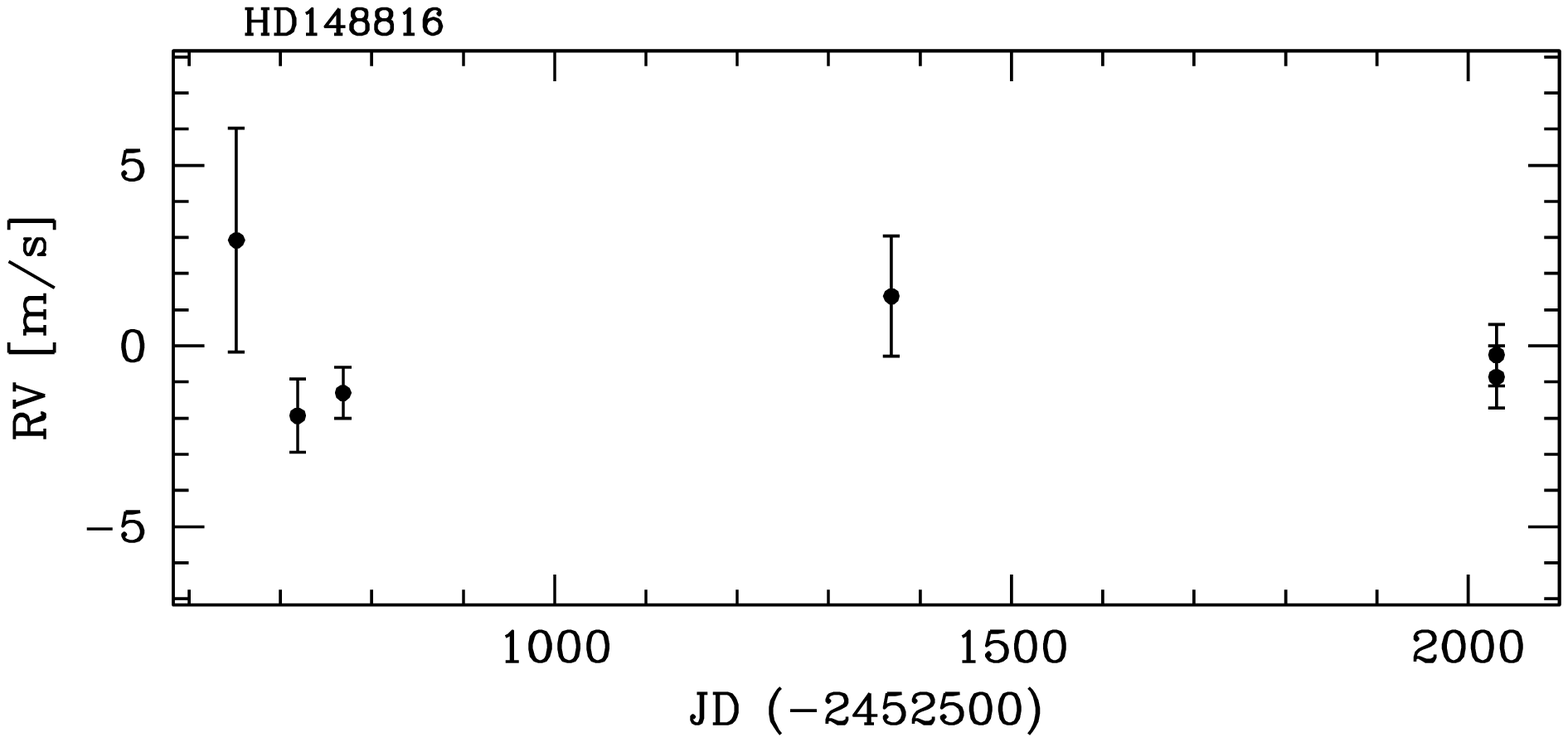}}\\
\resizebox{5.9cm}{!}{\includegraphics[bb= 18 160 580 430]{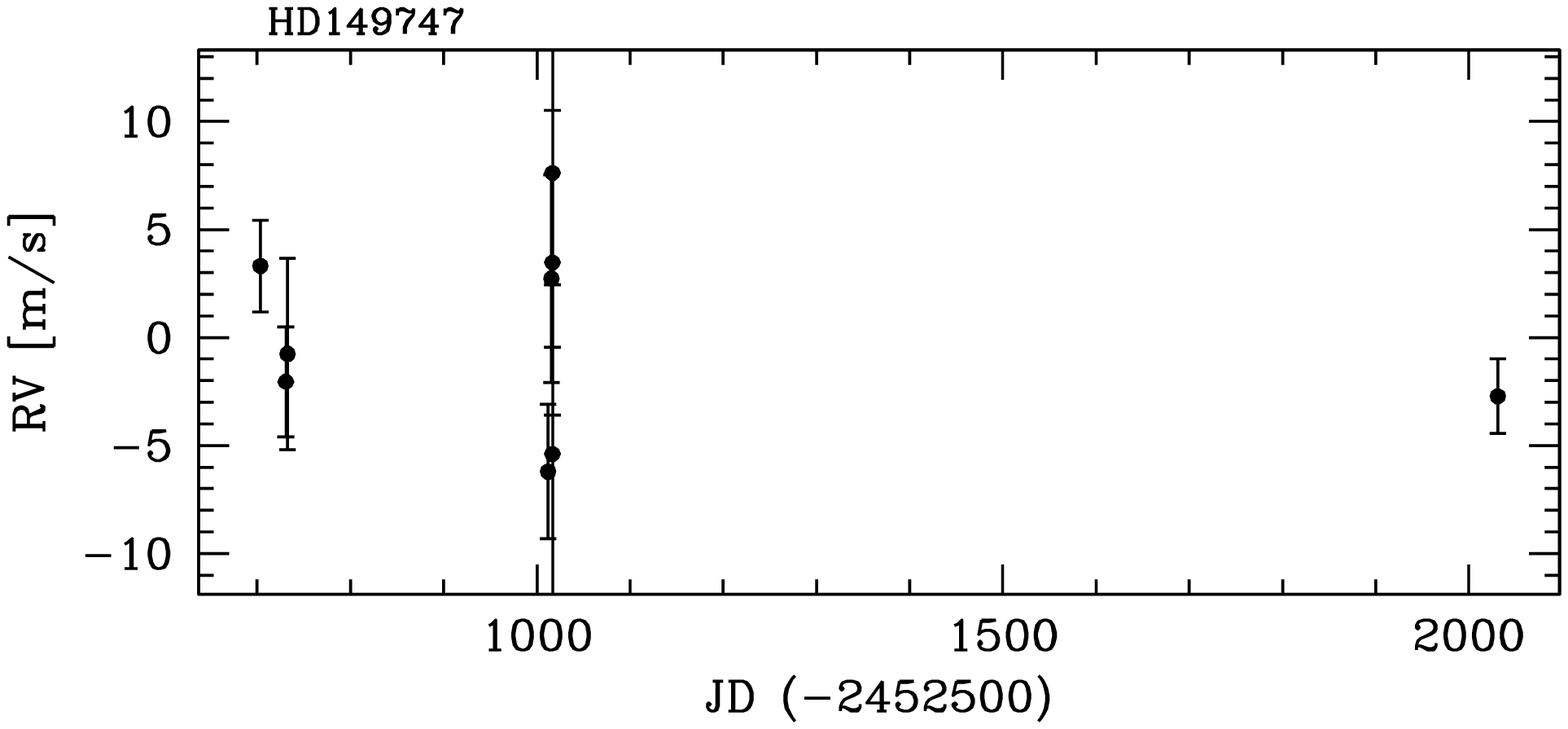}}
\resizebox{5.9cm}{!}{\includegraphics[bb= 18 160 580 430]{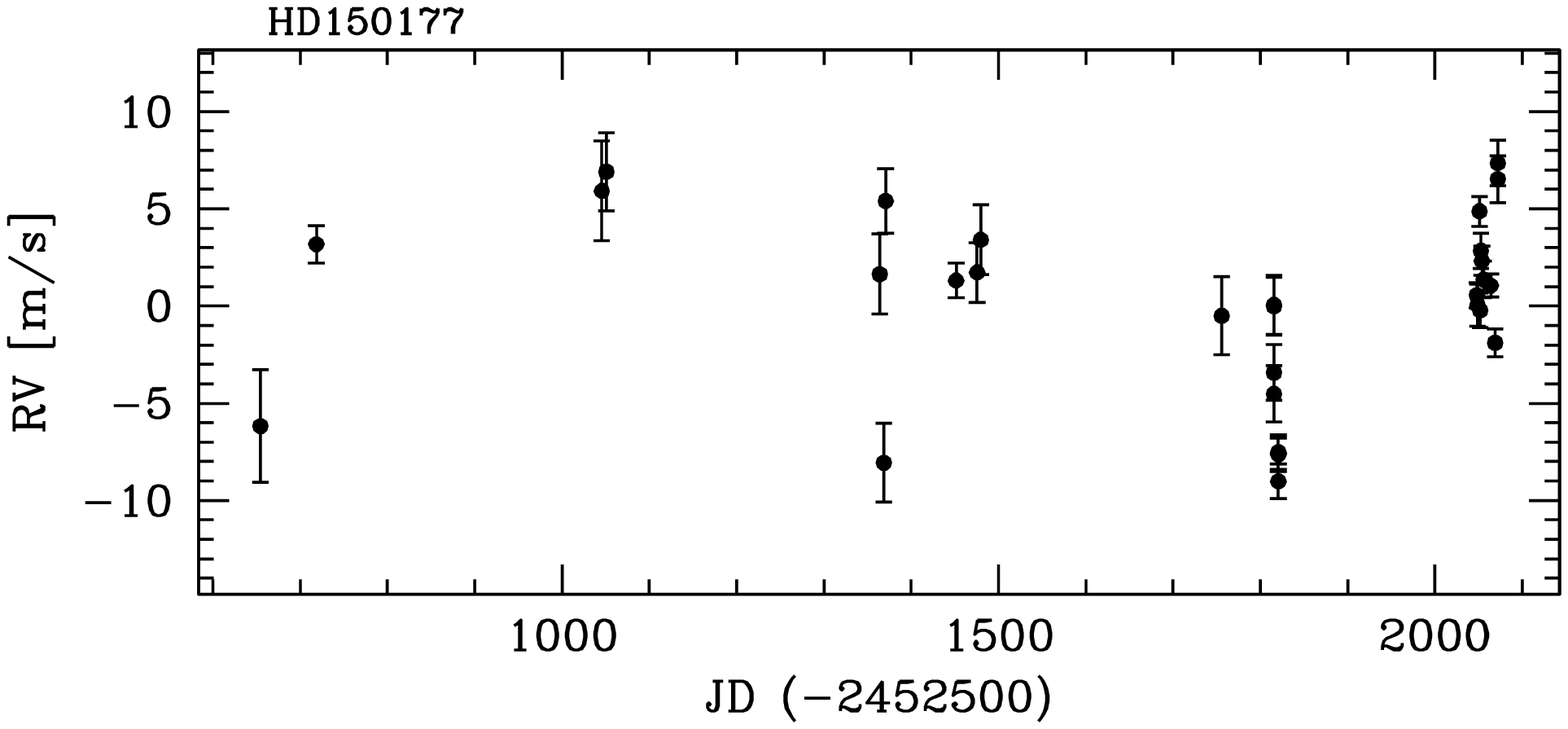}}
\resizebox{5.9cm}{!}{\includegraphics[bb= 18 160 580 430]{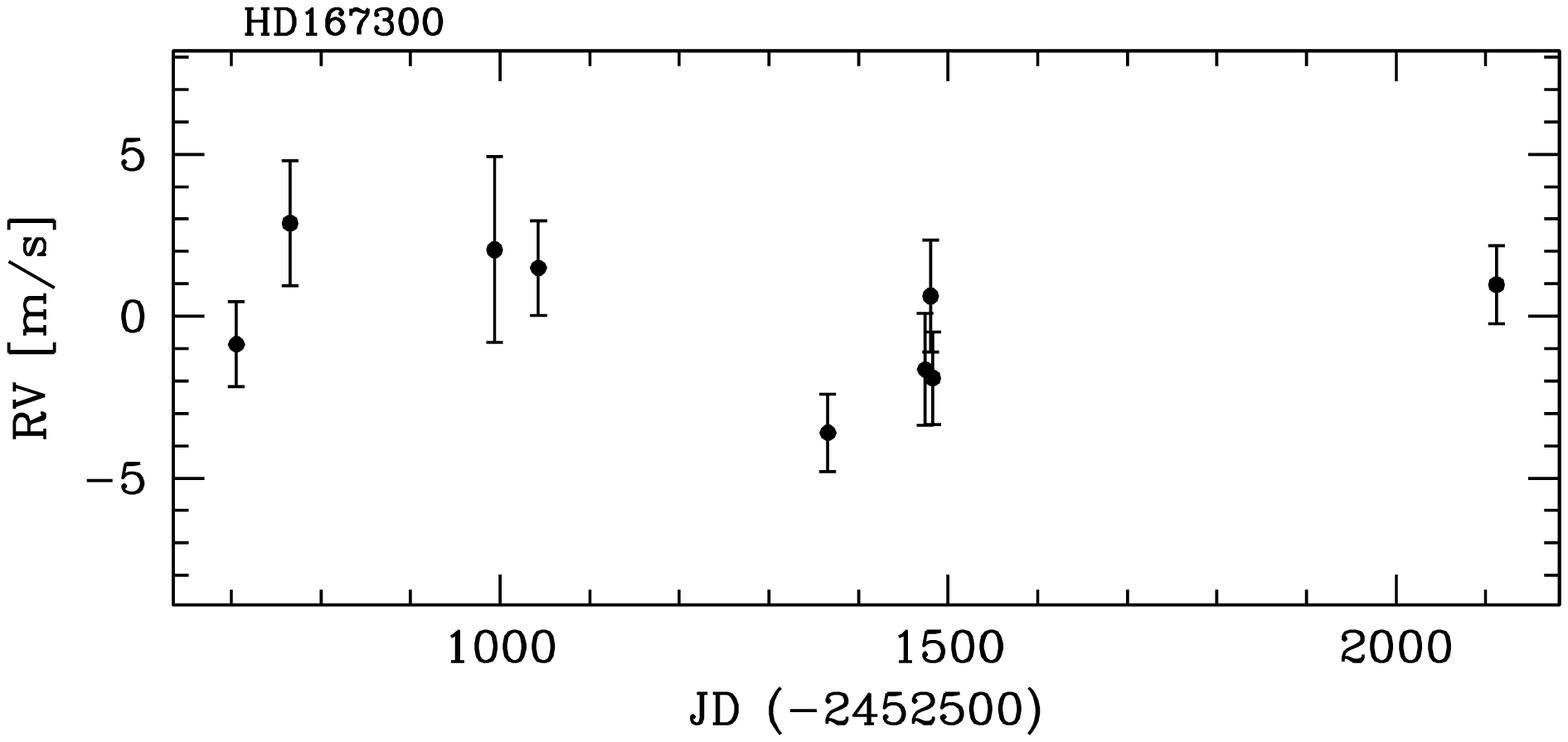}}
\caption{Continuation of Fig.\,\ref{fig:6mesa}.}
\label{fig:6mesb}
\end{figure*}

\begin{figure*}[t!]
\resizebox{5.9cm}{!}{\includegraphics[bb= 18 160 580 430]{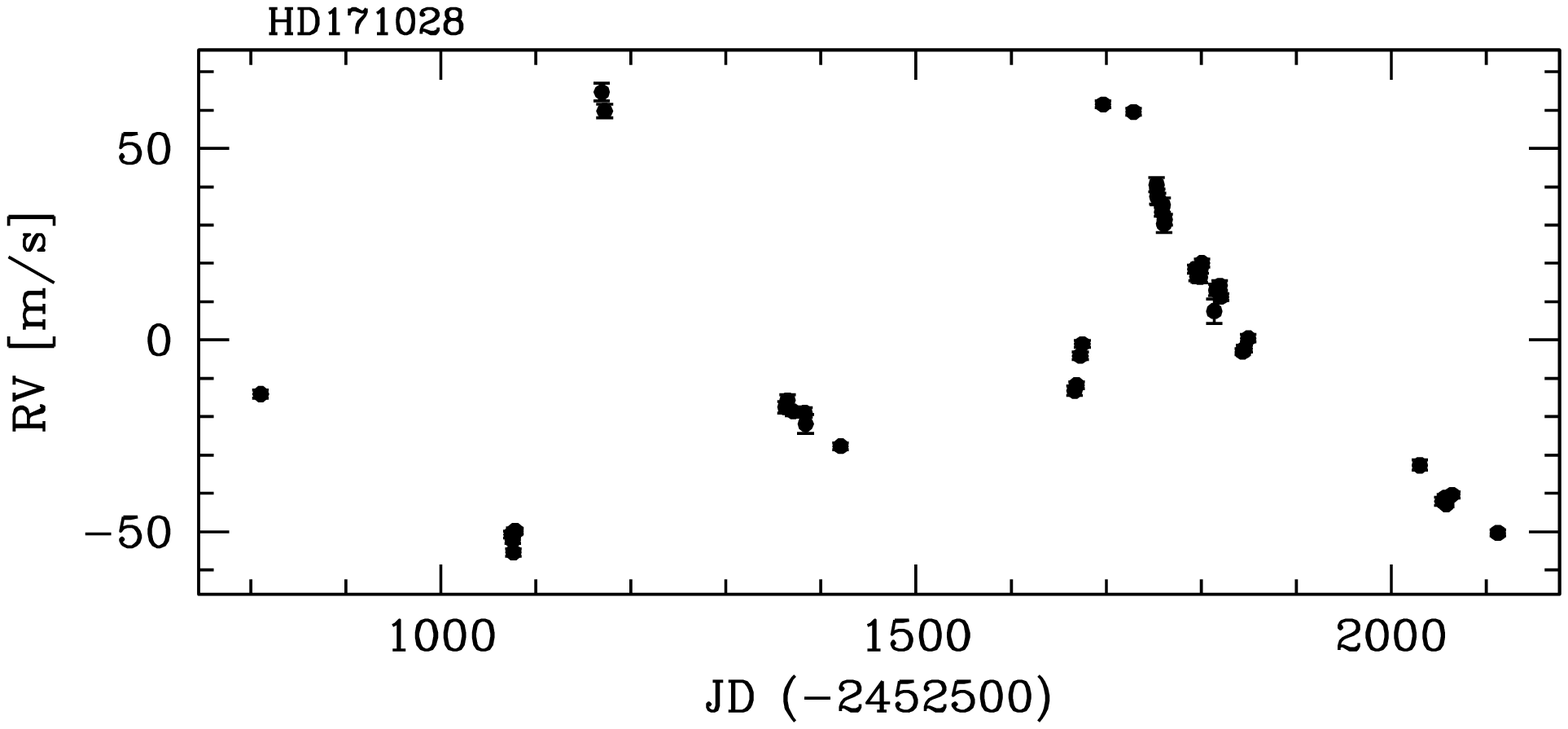}}
\resizebox{5.9cm}{!}{\includegraphics[bb= 18 160 580 430]{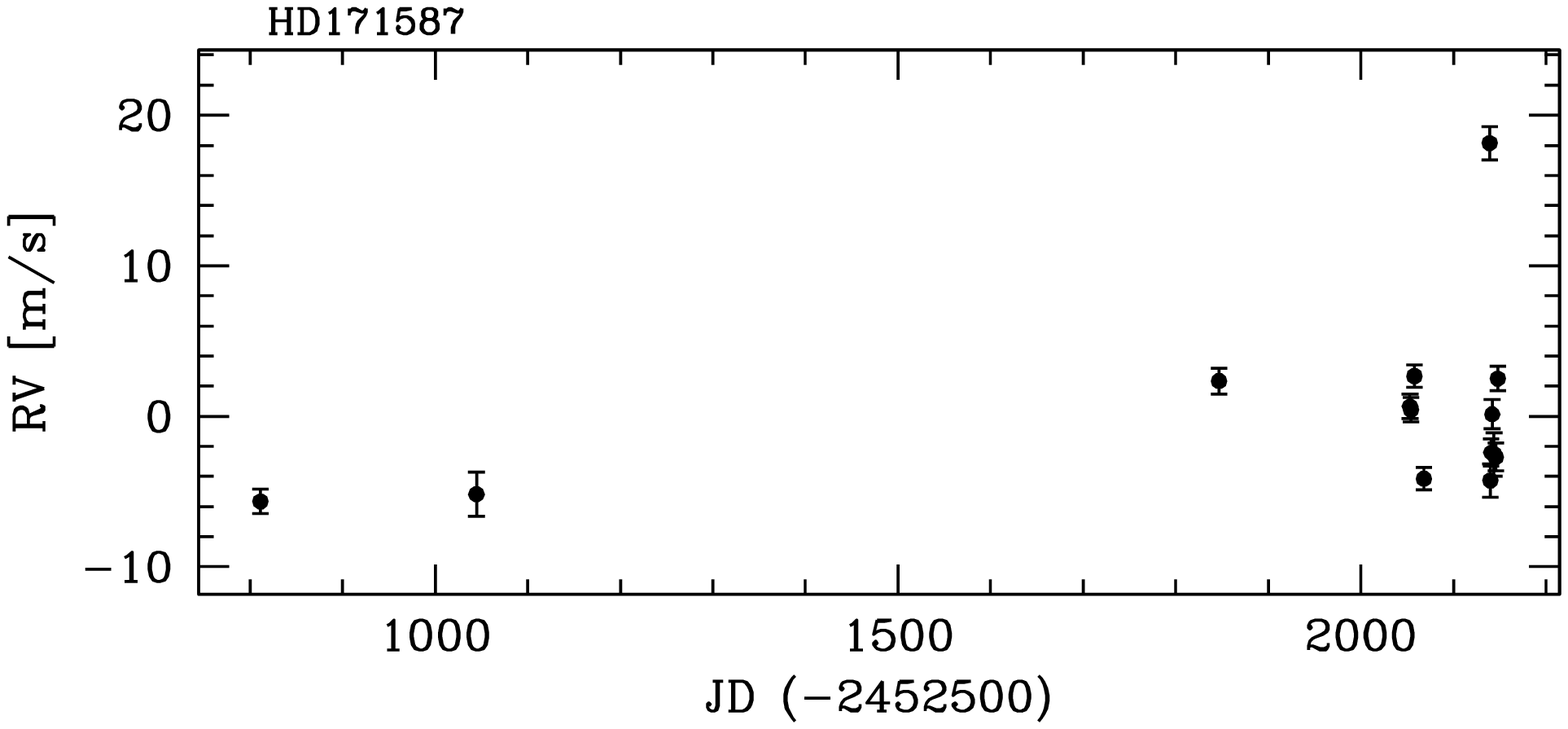}}
\resizebox{5.9cm}{!}{\includegraphics[bb= 18 160 580 430]{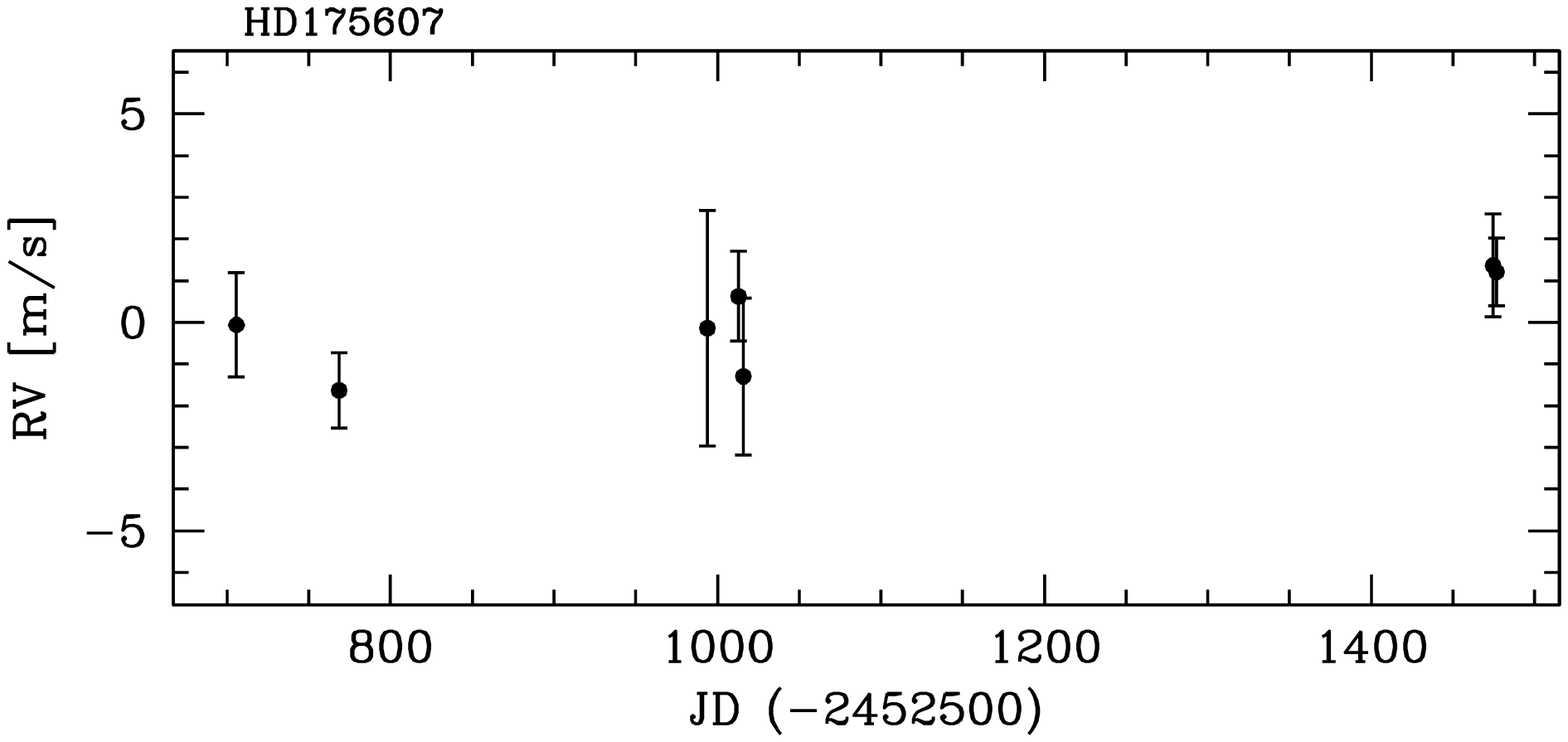}}\\
\resizebox{5.9cm}{!}{\includegraphics[bb= 18 160 580 430]{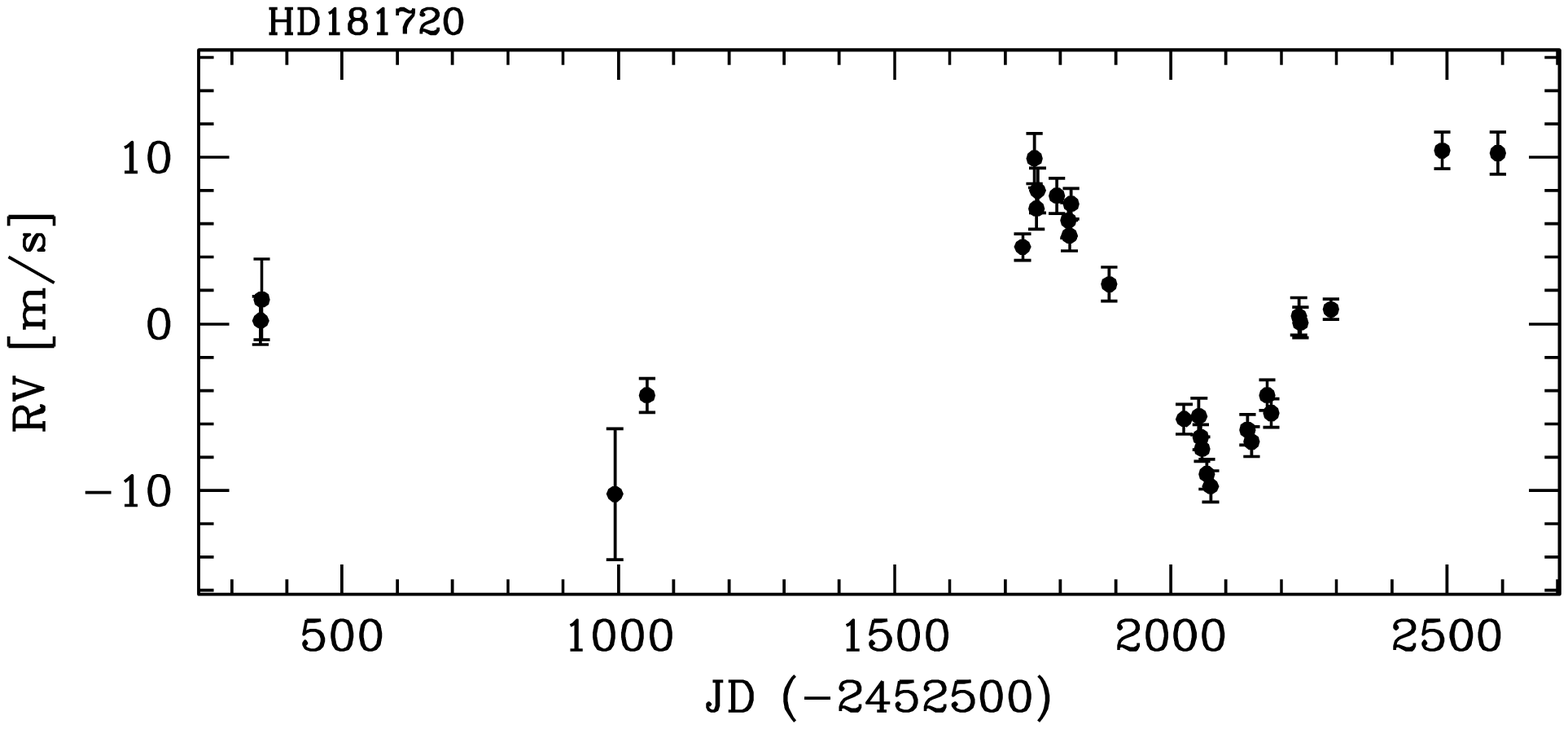}}
\resizebox{5.9cm}{!}{\includegraphics[bb= 18 160 580 430]{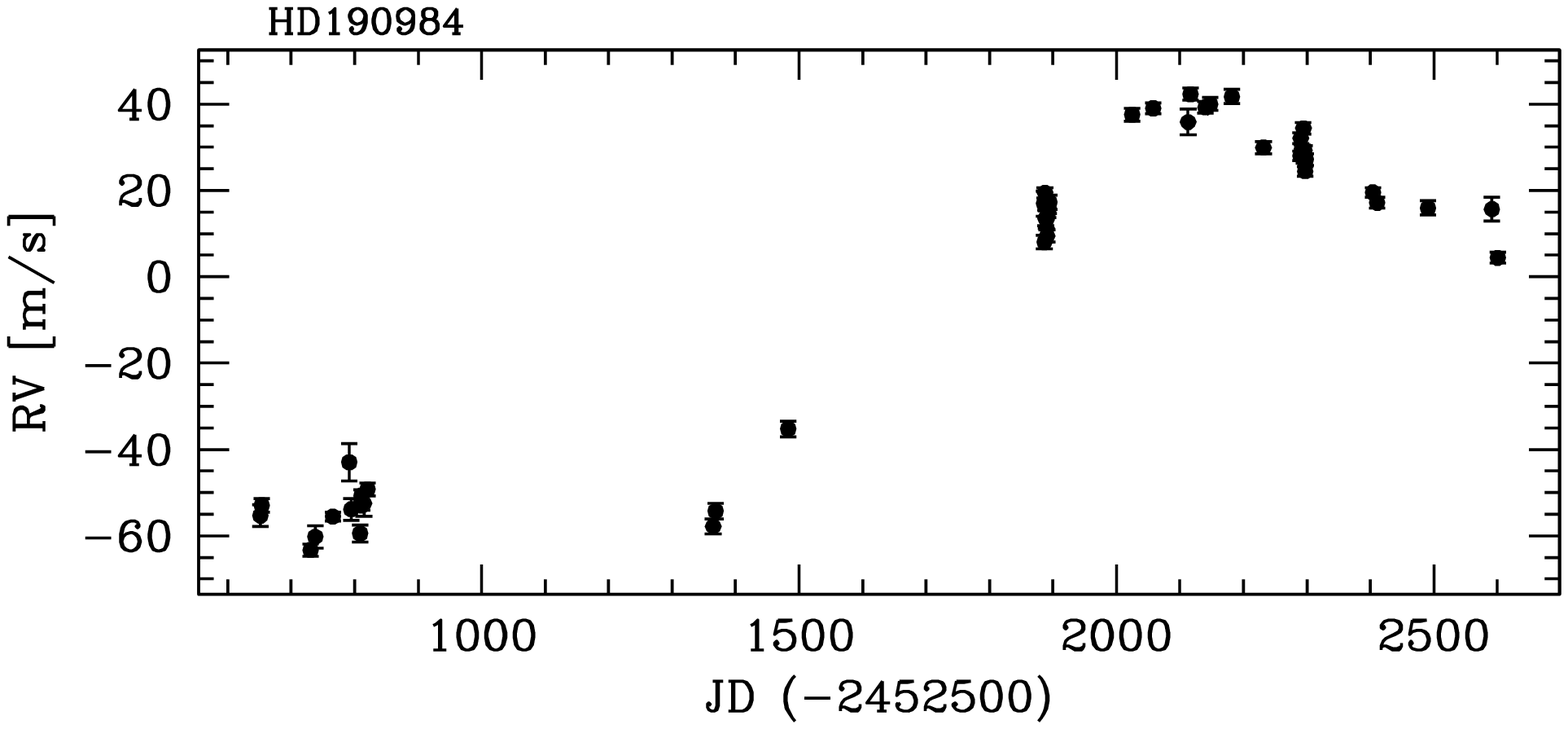}}
\resizebox{5.9cm}{!}{\includegraphics[bb= 18 160 580 430]{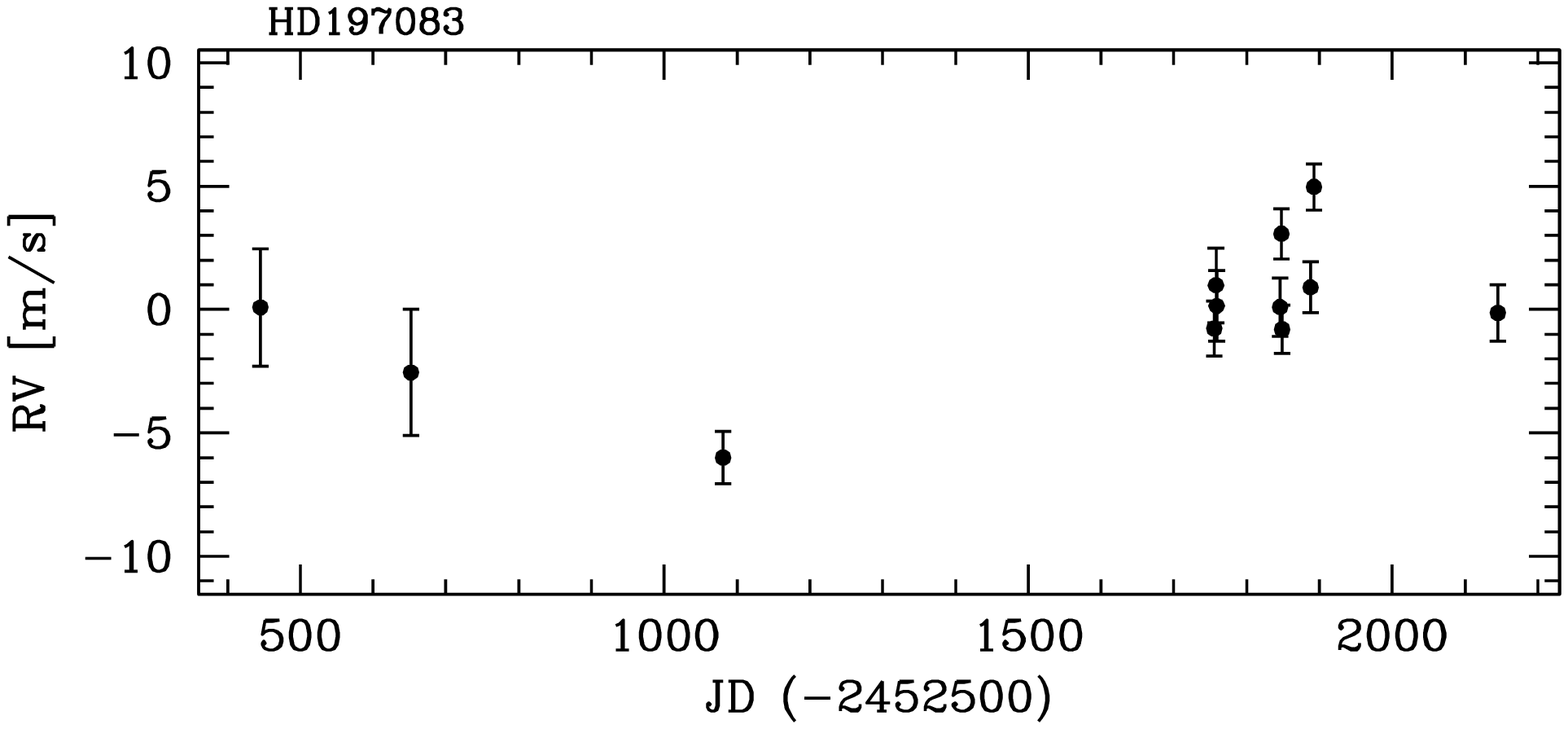}}\\
\resizebox{5.9cm}{!}{\includegraphics[bb= 18 160 580 430]{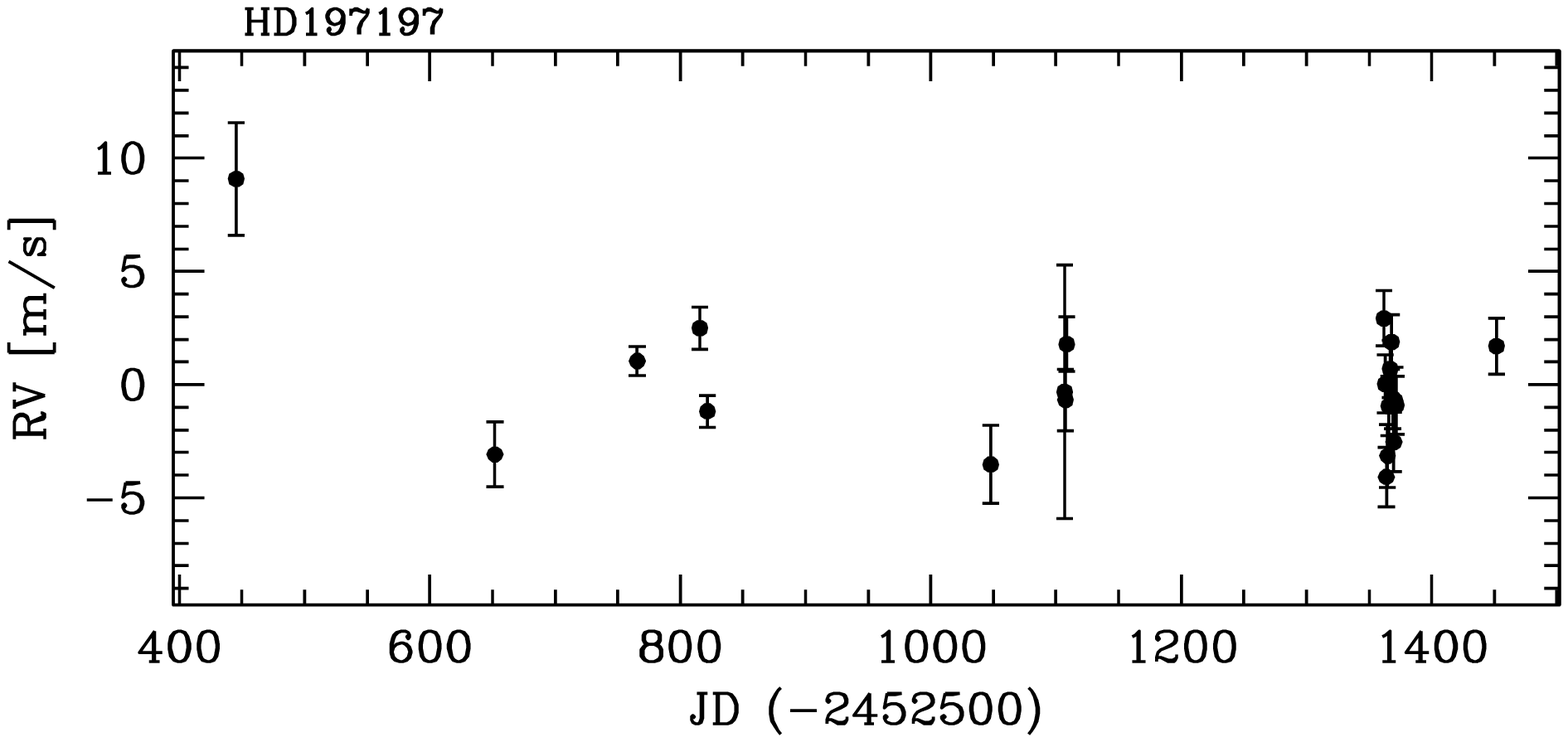}}
\resizebox{5.9cm}{!}{\includegraphics[bb= 18 160 580 430]{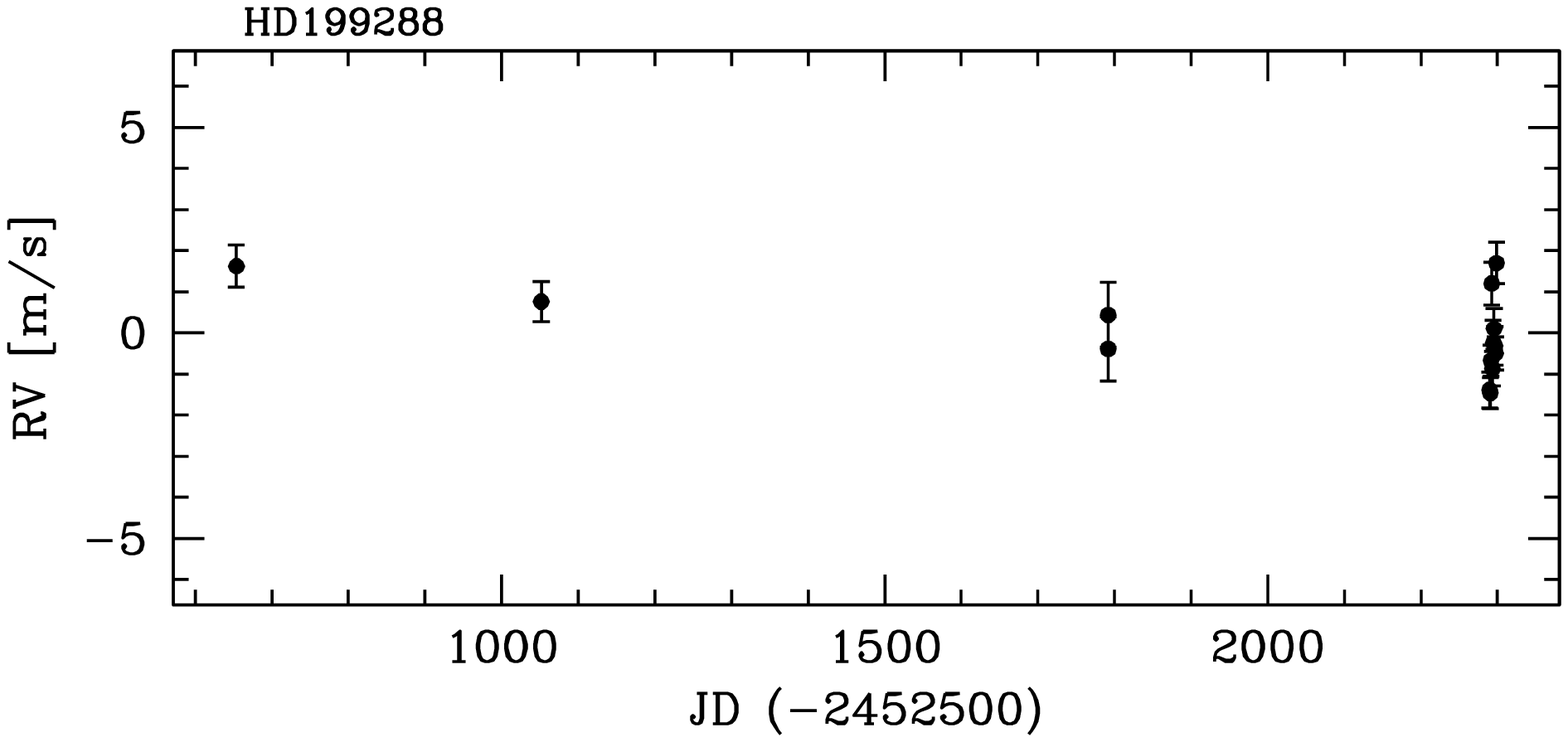}}
\resizebox{5.9cm}{!}{\includegraphics[bb= 18 160 580 430]{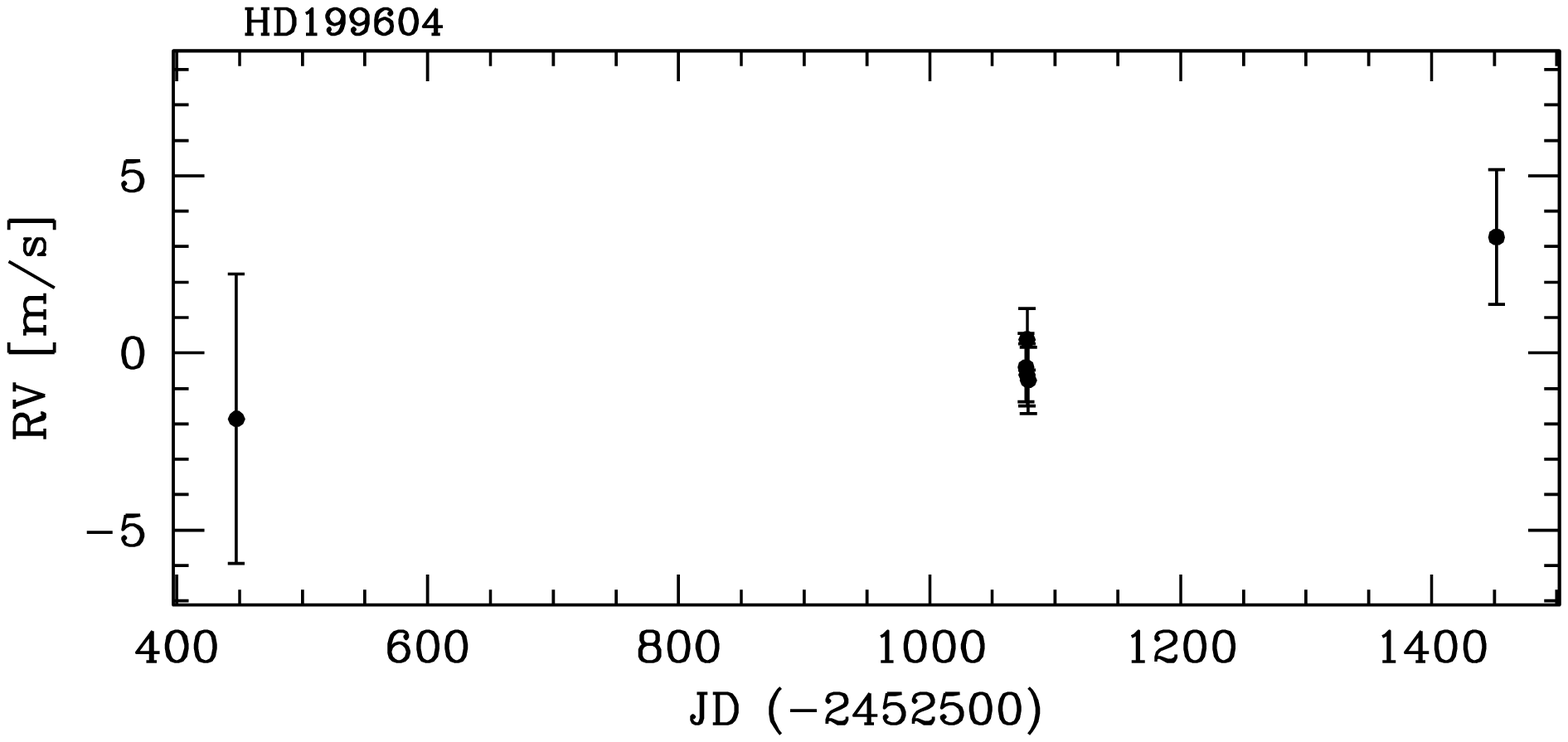}}\\
\resizebox{5.9cm}{!}{\includegraphics[bb= 18 160 580 430]{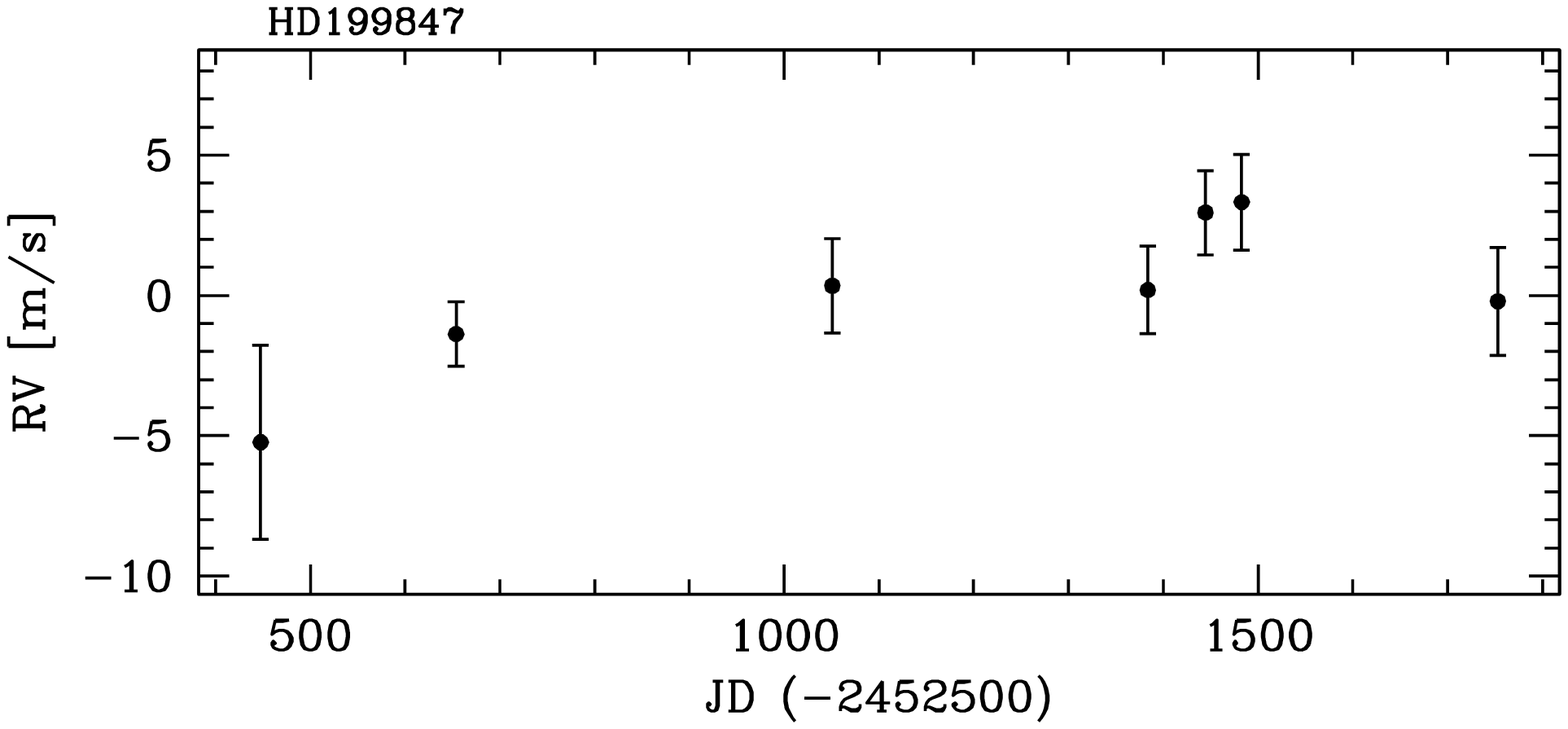}}
\resizebox{5.9cm}{!}{\includegraphics[bb= 18 160 580 430]{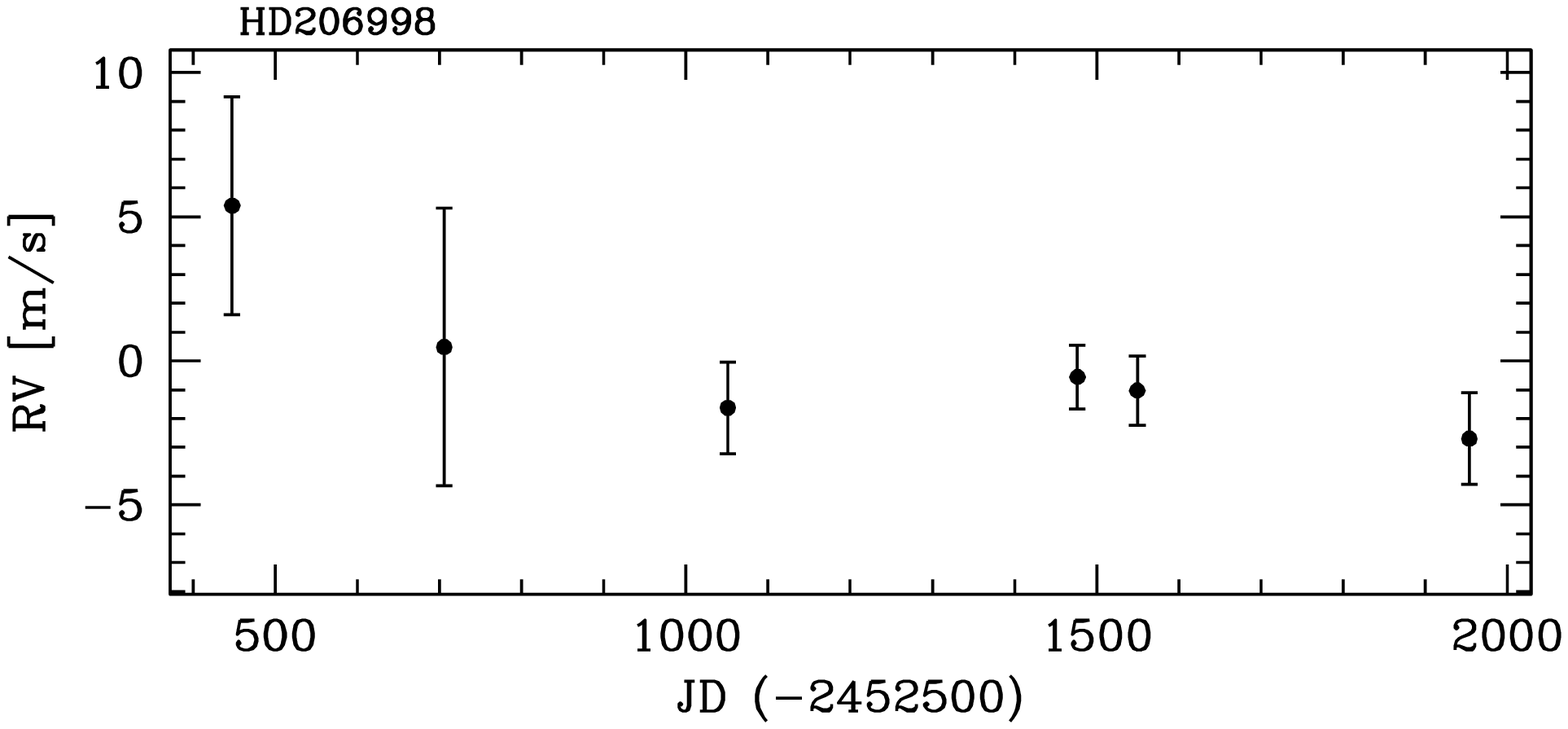}}
\resizebox{5.9cm}{!}{\includegraphics[bb= 18 160 580 430]{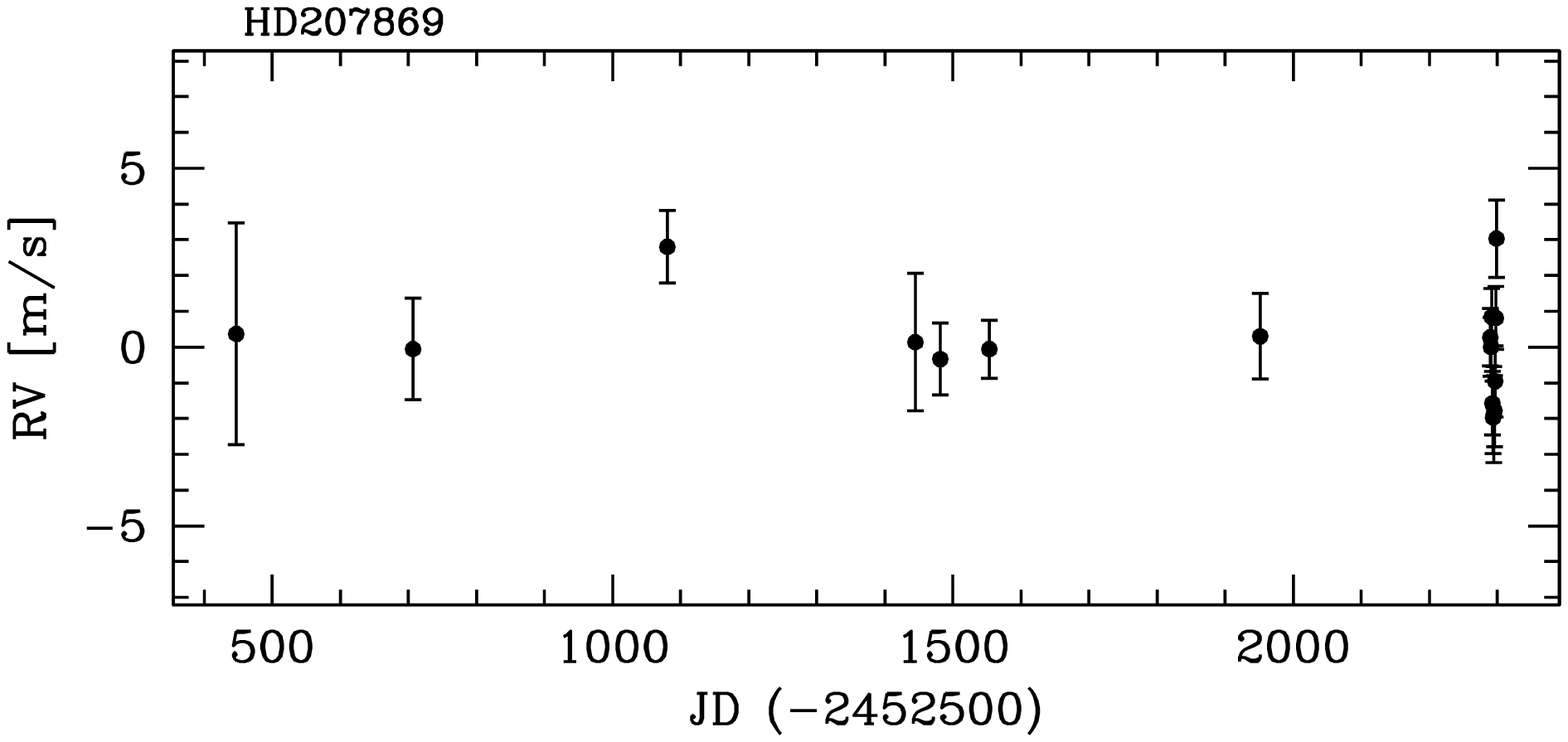}}\\
\resizebox{5.9cm}{!}{\includegraphics[bb= 18 160 580 430]{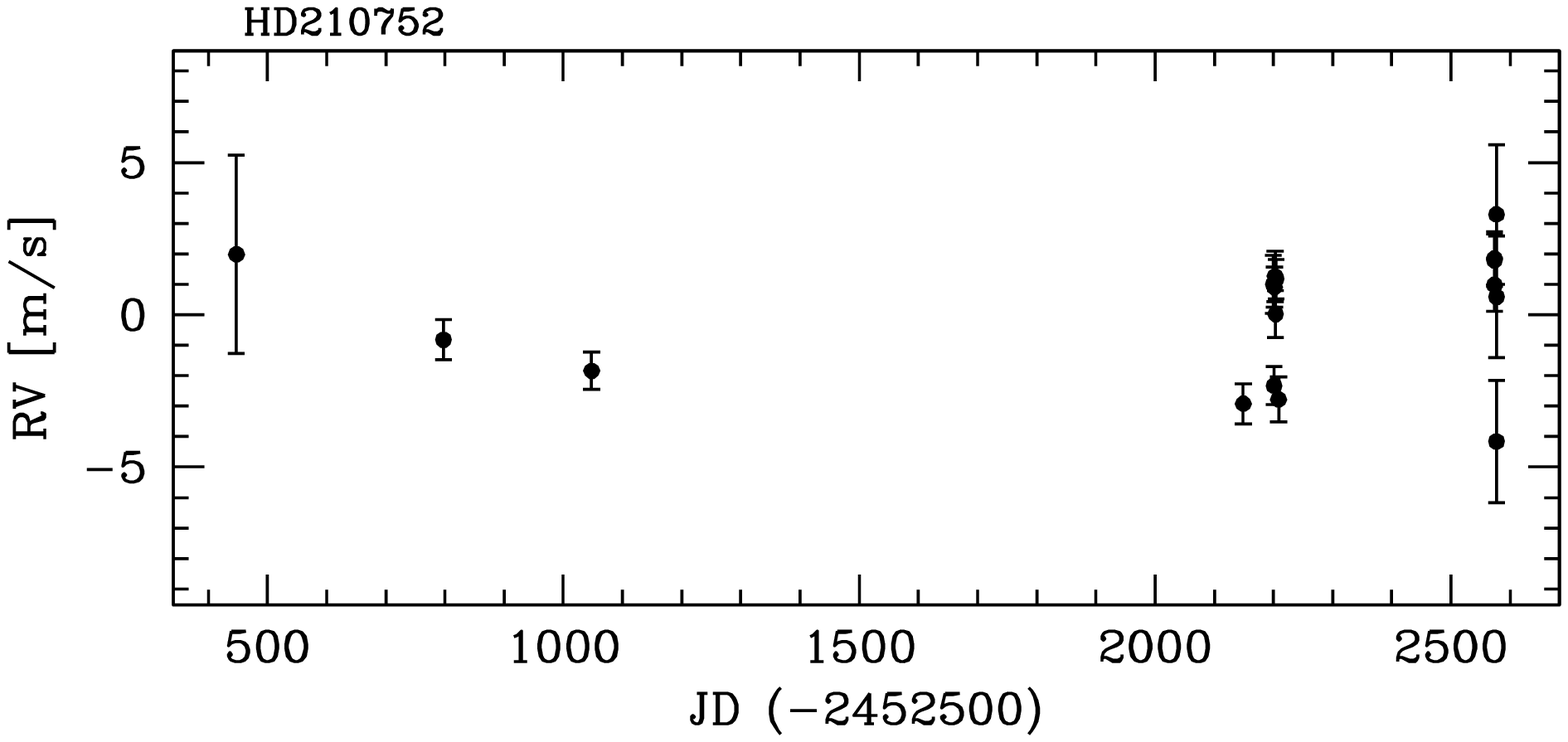}}
\resizebox{5.9cm}{!}{\includegraphics[bb= 18 160 580 430]{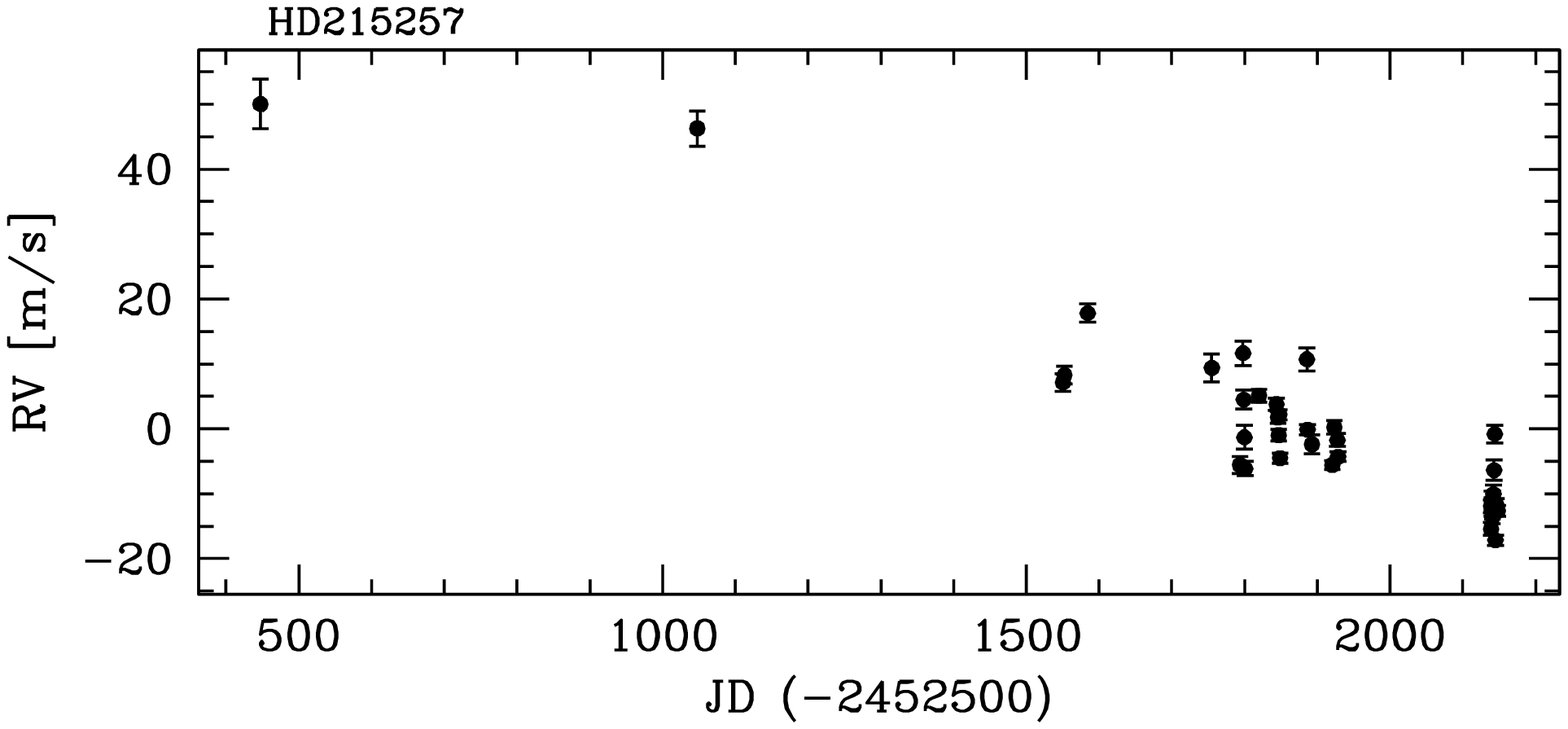}}
\resizebox{5.9cm}{!}{\includegraphics[bb= 18 160 580 430]{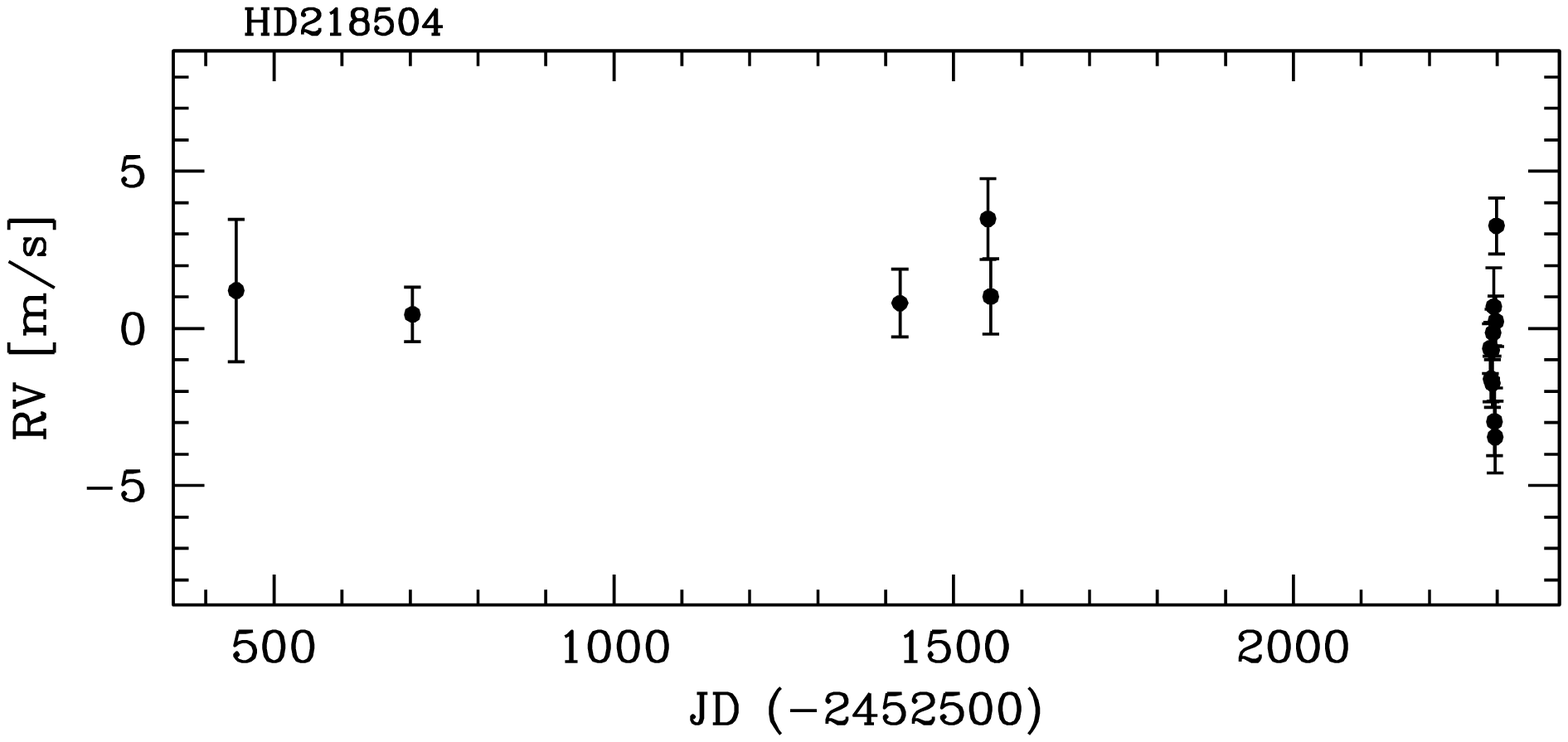}}\\
\resizebox{5.9cm}{!}{\includegraphics[bb= 18 160 580 430]{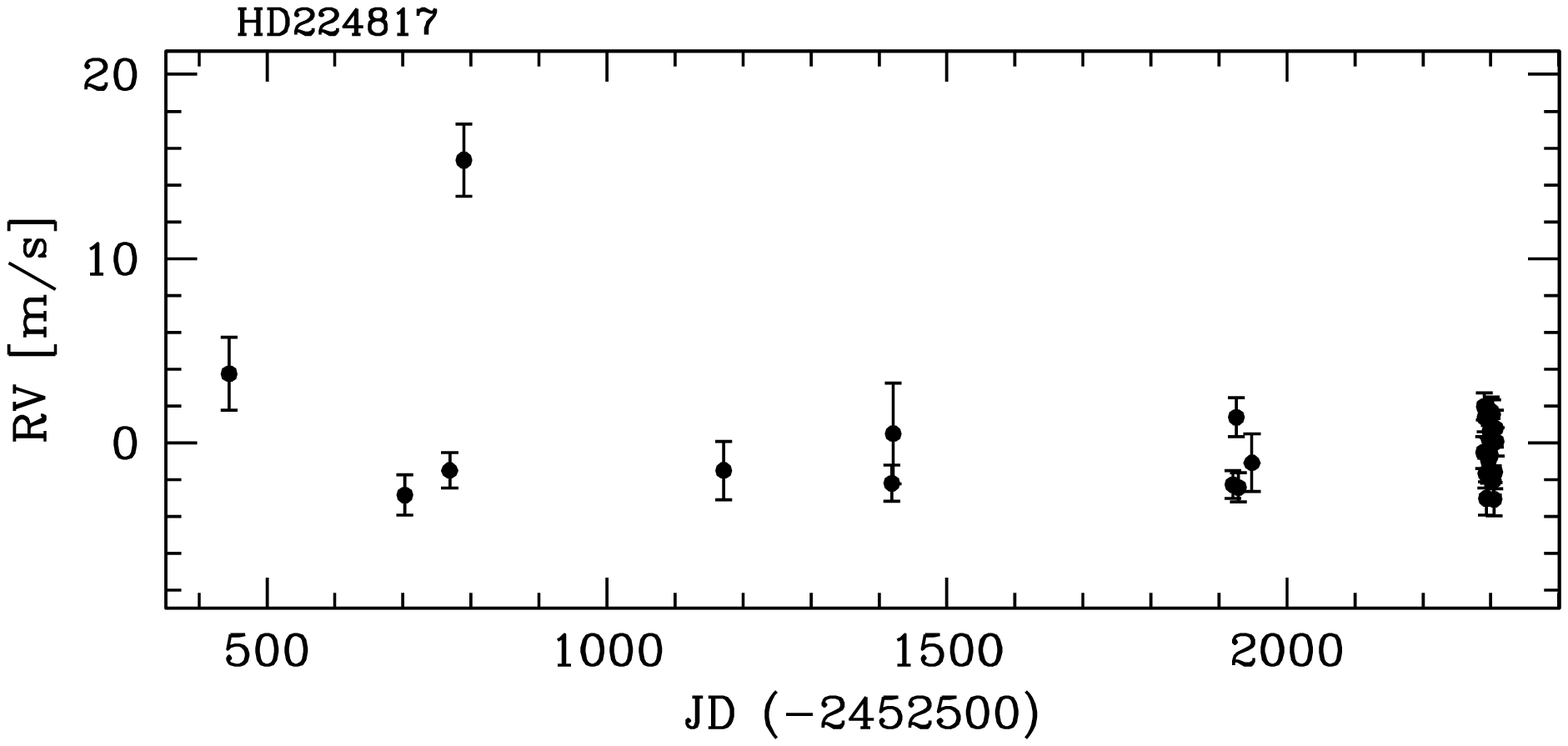}}
\caption{Continuation of Fig.\,\ref{fig:6mesa}.}
\label{fig:6mesc}
\end{figure*}

\begin{table*}[t]
\caption[]{Variability tests for stars with at least 6 measurements (CD and HD numbers up to 100\,000).}
\begin{tabular}{lcccccccccc}
\hline
\hline
Name    & N     & $\sigma_i$    & $\sigma_e$    & $P(F)$                    & $\chi_{constant}^2$      & $P(\chi_{constant}^2)$   & Slope       & $\chi^2_{slope}$      & $P(F_{slope})$    & FAP   \\
        &       & [m\,s$^{-1}$]          & [m\,s$^{-1}$]          &                           &                           &                           & [m/s/yr]    &                       &                     &       \\ 
 \hline 
\object{CD$-$2310879}	& 35	&  4.2	& 12.4	& {\boldmath $<10^{-9}$}	& 336	& {\boldmath $<10^{-9}$}	& 0.476	& 336	& $1.000$	& $0.961$\\
\object{CD$-$571633}	& 7	&  2.9	&  2.9	& $0.645$	&  5.3	& $0.505$	& -0.244	&  4.5	& $0.686$	& $0.107$\\
\object{HD967}	& 34	&  1.3	&  1.8	& $0.028$	& 56.9	& {\boldmath $0.006$}	& -0.154	& 55.4	& $0.674$	& {\boldmath$0.009$}\\
\object{HD11397}	& 33	&  1.4	&  6.3	& {\boldmath $<10^{-9}$}	& 462	& {\boldmath $<10^{-9}$}	& -4.579	& 50.1	& {\boldmath $<10^{-9}$}	& {\boldmath$<1/720$}\\
\object{HD17548}	& 10	&  2.8	&  2.3	& $0.823$	&  4.8	& $0.851$	& -0.244	&  4.2	& $0.496$	& $0.028$\\
\object{HD17865}	& 21	&  1.5	&  1.6	& $0.450$	& 27.4	& $0.123$	& -0.073	& 27.3	& $1.000$	& $0.548$\\
\object{HD22879}	& 36	&  1.3	&  1.4	& $0.468$	& 49.0	& $0.059$	& -0.297	& 40.4	& {\boldmath $<10^{-9}$}	& $0.077$\\
\object{HD31128}	& 37	&  3.7	&  4.3	& $0.245$	& 57.5	& $0.013$	& -0.547	& 54.5	& $0.035$	& $0.070$\\
\object{HD40865}	& 30	&  1.5	&  2.3	& $0.014$	& 48.3	& $0.014$	& 0.062	& 48.1	& $1.000$	& $0.019$\\
\object{HD51754}	& 21	&  1.6	&  1.9	& $0.333$	& 29.8	& $0.072$	& -0.729	& 22.4	& {\boldmath $1.3~10^{-4}$}	& $0.016$\\
\object{HD56274}	& 14	&  1.3	&  2.3	& $0.035$	& 35.8	& {\boldmath $6.3~10^{-4}$}	& -0.783	& 23.4	& {\boldmath $0.002$}	& {\boldmath$0.002$}\\
\object{HD59984}	& 45	&  1.2	&  2.4	& {\boldmath $2.0~10^{-6}$}	& 202	& {\boldmath $<10^{-9}$}	& 0.712	& 185	& {\boldmath $6.0~10^{-6}$}	& $0.019$\\
\object{HD61902}	& 7	&  2.9	&  4.8	& $0.217$	& 20.0	& {\boldmath $0.003$}	& -1.362	& 11.1	& $0.106$	& $0.056$\\
\object{HD68089}	& 7	&  2.4	&  2.6	& $0.598$	&  7.2	& $0.302$	& -0.412	&  6.6	& $0.902$	& $0.586$\\
\object{HD68284}	& 10	&  1.9	&  3.0	& $0.167$	& 21.4	& $0.011$	& 0.459	& 17.6	& $0.305$	& $0.026$\\
\object{HD69611}	& 6	&  1.8	&  2.5	& $0.392$	& 13.4	& $0.020$	& 0.774	& 10.4	& $0.609$	& $0.365$\\
\object{HD75745}	& 13	&  2.8	&  3.6	& $0.252$	& 18.9	& $0.092$	& 0.722	& 12.9	& {\boldmath $0.008$}	& $0.023$\\
\object{HD77110}	& 16	&  1.4	&  2.1	& $0.089$	& 35.5	& {\boldmath $0.002$}	& -0.964	& 15.9	& {\boldmath $2.0~10^{-6}$}	& {\boldmath$<1/720$}\\
\object{HD78747}	& 26	&  1.2	&  1.6	& $0.184$	& 40.5	& $0.026$	& -0.137	& 39.3	& $0.799$	& $0.099$\\
\object{HD79601}	& 16	&  1.6	&  2.0	& $0.282$	& 23.6	& $0.071$	& -0.569	& 11.2	& {\boldmath $4.0~10^{-6}$}	& {\boldmath$<1/720$}\\
\object{HD88474}	& 6	&  3.9	&  4.5	& $0.534$	&  9.6	& $0.088$	& 1.035	&  5.9	& $0.274$	& $0.177$\\
\object{HD88725}	& 22	&  1.2	&  3.3	& {\boldmath $1.3~10^{-5}$}	& 179	& {\boldmath $<10^{-9}$}	& -1.879	& 46.8	& {\boldmath $<10^{-9}$}	& {\boldmath$<1/720$}\\
\object{HD90422}	& 7	&  2.6	&  3.6	& $0.347$	&  8.6	& $0.199$	& -0.318	&  8.1	& $0.970$	& $0.159$\\
\object{HD91345}	& 8	&  4.0	&  4.7	& $0.476$	& 13.1	& $0.070$	& -1.574	&  4.8	& {\boldmath $0.006$}	& $0.021$\\
\object{HD94444}	& 7	&  2.3	&  4.0	& $0.181$	& 24.8	& {\boldmath $3.7~10^{-4}$}	& -1.413	& 13.7	& $0.101$	& $0.162$\\
\object{HD95860}	& 7	&  2.3	&  3.7	& $0.204$	& 20.8	& {\boldmath $0.002$}	& 0.059	& 20.3	& $0.997$	& $0.900$\\
\object{HD97320}	& 6	&  3.1	&  2.8	& $0.726$	&  5.9	& $0.318$	& -0.283	&  4.9	& $0.741$	& $0.471$\\
\object{HD97783}	& 6	&  1.9	&  2.5	& $0.443$	& 15.3	& {\boldmath $0.009$}	& -0.300	& 13.2	& $0.816$	& $0.748$\\
\hline
\end{tabular}
\label{tab:6mesa}
\end{table*}

\begin{table*}[t]
\caption[]{Variability tests for stars with at least 6 measurements (HD numbers above 100\,000).}
\begin{tabular}{lcccccccccc}
\hline
\hline
Name    & N     & $\sigma_i$    & $\sigma_e$    & $P(F)$                    & $\chi_{constant}^2$      & $P(\chi_{constant}^2)$   & Slope       & $\chi^2_{slope}$      & $P(F_{slope})$    & FAP   \\
        &       & [m\,s$^{-1}$]          & [m\,s$^{-1}$]          &                           &                           &                           & [m/s/yr]    &                       &                     &       \\ 
 \hline 
\object{HD102200}	& 7	&  5.4	&  8.1	& $0.279$	&  7.7	& $0.260$	& 0.835	&  4.5	& $0.127$	& $0.023$\\
\object{HD104800}	& 6	&  1.8	&  1.5	& $0.795$	&  3.2	& $0.663$	& -0.677	&  2.1	& $0.325$	& $0.129$\\
\object{HD107094}	& 14	&  1.8	& 599	& {\boldmath $<10^{-9}$}	& $1.8~10^{+6}$	& {\boldmath $<10^{-9}$}	& 429.034	& $1.3~10^{+4}$	& {\boldmath $<10^{-9}$}	& {\boldmath$<1/720$}\\
\object{HD108564}	& 6	&  1.6	&  1.3	& $0.791$	&  3.8	& $0.573$	& -0.436	&  2.5	& $0.321$	& $0.176$\\
\object{HD109310}	& 15	&  1.6	&  3.7	& {\boldmath $0.003$}	& 78.5	& {\boldmath $<10^{-9}$}	& -1.460	& 47.9	& {\boldmath $3.1~10^{-4}$}	& {\boldmath$0.003$}\\
\object{HD109684}	& 6	&  1.9	&  2.5	& $0.450$	&  8.0	& $0.158$	& -0.202	&  6.3	& $0.645$	& $0.170$\\
\object{HD111777}	& 6	&  1.5	&  1.2	& $0.814$	&  4.2	& $0.522$	& -0.355	&  2.9	& $0.403$	& $0.299$\\
\object{HD113679}	& 6	&  2.2	&  2.7	& $0.496$	&  9.8	& $0.082$	& 1.697	&  0.4	& {\boldmath $1.7~10^{-4}$}	& {\boldmath$<1/720$}\\
\object{HD123517}	& 9	&  3.1	& 49.5	& {\boldmath $<10^{-9}$}	& 2039	& {\boldmath $<10^{-9}$}	& 34.103	&  3.2	& {\boldmath $<10^{-9}$}	& {\boldmath$<1/720$}\\
\object{HD124785}	& 17	&  2.2	&  4.7	& {\boldmath $0.004$}	& 78.8	& {\boldmath $<10^{-9}$}	& -1.794	& 68.3	& $0.080$	& $0.100$\\
\object{HD126681}	& 13	&  2.4	&  3.7	& $0.116$	& 23.6	& $0.023$	& -0.788	& 20.1	& $0.201$	& $0.057$\\
\object{HD126793}	& 7	&  2.3	&  4.1	& $0.169$	& 14.1	& $0.029$	& 0.737	&  7.7	& $0.100$	& $0.020$\\
\object{HD126803}	& 7	&  2.6	&  4.8	& $0.143$	& 12.9	& $0.044$	& 0.414	&  8.7	& $0.252$	& $0.022$\\
\object{HD134440}	& 10	&  2.4	&  2.5	& $0.562$	&  9.7	& $0.375$	& -0.171	&  9.0	& $0.857$	& $0.083$\\
\object{HD144589}	& 11	&  4.0	& 34.7	& {\boldmath $<10^{-9}$}	& 604	& {\boldmath $<10^{-9}$}	& 25.404	& 14.2	& {\boldmath $<10^{-9}$}	& {\boldmath$0.001$}\\
\object{HD148211}	& 31	&  1.6	&  2.5	& $0.011$	& 76.0	& {\boldmath $7.0~10^{-6}$}	& -0.600	& 67.8	& {\boldmath $8.2~10^{-4}$}	& {\boldmath$0.006$}\\
\object{HD148816}	& 6	&  1.6	&  1.8	& $0.569$	&  5.7	& $0.341$	& 0.198	&  3.6	& $0.305$	& $0.070$\\
\object{HD149747}	& 9	&  4.7	&  4.6	& $0.666$	& 10.4	& $0.237$	& -0.904	&  7.9	& $0.213$	& $0.142$\\
\object{HD150177}	& 30	&  1.6	&  4.8	& {\boldmath $<10^{-9}$}	& 327	& {\boldmath $<10^{-9}$}	& -0.091	& 326	& $1.000$	& $0.313$\\
\object{HD167300}	& 9	&  1.8	&  2.1	& $0.479$	& 12.4	& $0.132$	& -0.135	& 11.9	& $0.968$	& $0.596$\\
\object{HD171028}	& 48	&  1.5	& 33.8	& {\boldmath $<10^{-9}$}	& $2.9~10^{+4}$	& {\boldmath $<10^{-9}$}	& 8.624	& $2.7~10^{+4}$	& {\boldmath $1.0~10^{-6}$}	& $0.208$\\
\object{HD171587}	& 14	&  1.3	&  6.0	& {\boldmath $2.0~10^{-6}$}	& 261	& {\boldmath $<10^{-9}$}	& 1.791	& 225	& $0.185$	& $0.175$\\
\object{HD175607}	& 7	&  1.7	&  1.2	& $0.897$	&  4.4	& $0.618$	& 1.082	&  1.5	& $0.015$	& $0.051$\\
\object{HD181720}	& 29	&  1.8	&  6.3	& {\boldmath $<10^{-9}$}	& 617	& {\boldmath $<10^{-9}$}	& -0.899	& 587	& $0.247$	& $0.107$\\
\object{HD190984}	& 46	&  2.3	& 39.1	& {\boldmath $<10^{-9}$}	& $2.0~10^{+4}$	& {\boldmath $<10^{-9}$}	& 20.284	& 2910	& {\boldmath $<10^{-9}$}	& {\boldmath$<1/720$}\\
\object{HD197083}	& 12	&  1.6	&  2.7	& $0.080$	& 45.1	& {\boldmath $5.0~10^{-6}$}	& 1.517	& 30.8	& $0.016$	& $0.278$\\
\object{HD197197}	& 21	&  1.7	&  2.9	& $0.022$	& 49.0	& {\boldmath $3.1~10^{-4}$}	& -0.609	& 46.9	& $0.721$	& $0.061$\\
\object{HD199288}	& 14	&  1.0	&  1.0	& $0.509$	& 15.7	& $0.264$	& -0.373	& 11.4	& {\boldmath $0.009$}	& $0.064$\\
\object{HD199604}	& 6	&  1.8	&  1.8	& $0.707$	&  3.6	& $0.610$	& 2.391	&  1.2	& $0.041$	& $0.055$\\
\object{HD199847}	& 7	&  2.0	&  2.9	& $0.331$	&  9.3	& $0.156$	& 1.331	&  4.5	& $0.059$	& $0.021$\\
\object{HD206998}	& 6	&  2.5	&  2.8	& $0.563$	&  5.7	& $0.332$	& -1.071	&  2.0	& $0.047$	& $0.018$\\
\object{HD207869}	& 17	&  1.4	&  1.4	& $0.613$	& 19.1	& $0.266$	& -0.305	& 17.4	& $0.339$	& $0.373$\\
\object{HD210752}	& 17	&  1.4	&  2.1	& $0.104$	& 39.0	& {\boldmath $0.001$}	& 0.378	& 34.3	& $0.126$	& $0.039$\\
\object{HD215257}	& 37	&  1.6	& 14.4	& {\boldmath $<10^{-9}$}	& 1870	& {\boldmath $<10^{-9}$}	& -15.045	& 343	& {\boldmath $<10^{-9}$}	& {\boldmath$<1/720$}\\
\object{HD218504}	& 15	&  1.3	&  2.0	& $0.122$	& 30.6	& {\boldmath $0.006$}	& -0.449	& 26.4	& $0.140$	& $0.095$\\
\object{HD224817}	& 30	&  1.4	&  3.4	& {\boldmath $6.0~10^{-6}$}	& 109	& {\boldmath $<10^{-9}$}	& -0.134	& 105	& $0.691$	& $0.090$\\
\hline
\end{tabular}
\label{tab:6mesb}
\end{table*}

For stars with 6 or more radial-velocity measurements (64 objects -- Figs.\,\ref{fig:6mesa}, \ref{fig:6mesb}, and \ref{fig:6mesc}), we performed 
a series of statistical tests, following the same methodology applied to the HARPS M-dwarf sample \citep[][]{Bonfils-2010}. First, to test whether the observed RVs ($\sigma_e$) are significantly in excess of the internal errors ($\sigma_i$), 
we performed an F-test and derived the probability, P(F), to the F-value F=$\frac{\sigma_e^2}{<\sigma_i>^2}$ \citep[see e.g. ][]{Zechmeister-2009}.
We also computed the $\chi^2_{constant}$ for the $constant$ model as well as the probability of having $\chi^2_{constant}$ given the $\sigma_i$, P($\chi^2_{constant}$).
These values are listed in Tables~\ref{tab:6mesa} and \ref{tab:6mesb}. 
For 16 out of the 64 stars, both P(F) and P($\chi^2_{constant}$) are below 1\%. These are the cases of 
\object{CD$-$2310879}, %active 		OK
\object{HD11397},  %drift		    		OK
\object{HD59984},  %				OK
\object{HD88725},  %significant drift?	OK
\object{HD107094},%drift				OK
\object{HD109310}, %significant drift?	OK
\object{HD123517}, %drift			OK
\object{HD124785}, %				OK
\object{HD144589}, %drift			OK
\object{HD150177},%				OK
 \object{HD171028},%planet			OK
 \object{HD171587}, %				NEW PLANET???
\object{HD181720}, %planet			OK
\object{HD190984}, %planet			OK
\object{HD215257}, %drift			OK
and
\object{HD224817}. %				OK

We note that the internal errors include both photon noise errors and errors from the wavelength calibration and instrumental drift correction. Furthermore,
we added an instrumental error of 0.8\,m\,s$^{-1}$ to take the estimated intrinsic precision of HARPS (guiding and internal errors) into account. Other sources of noise, 
like those coming from stellar intrinsic phenomena, are more difficult to estimate, so were not considered here.

Two tests were also done to estimate the probability that the observed radial velocities are best fitted by a linear function of type $rv=a\times\,time+b$.
First, we used an F-test to derive the probability P($F_{slope}$) that the use of an additional free parameter (2 instead of 1 in case of a constant fit) implies a statistically significant 
improvement when comparing $\chi^2_{slope}$ to $\chi^2_{constant}$. Second, we derived the false alarm probability
using a bootstrap randomization approach. In this case, we generated random RV time series by shuffling (without repetition) the original radial velocity data, 
preserving the same observing dates. 
On each obtained time series, we adjusted a slope and computed its $\chi^2$ value. The fraction of simulated 
data sets with $\chi^2$ lower than the observed one gives us the 
false-alarm probability (FAP) for the slope model. 
Thirteen (13) stars in our sample show both significant P($F_{slope}$) and FAP values:
\object{HD11397},  %DRIFT DISCUTIDO EM BAIXO	OK
\object{HD56274}, % 							OK
\object{HD77110},  % 							OK
\object{HD79601},  % 							OK
\object{HD88725},  %							OK
\object{HD107094},%DRIFT DISCUTIDO EM BAIXO	OK
\object{HD109310}, %							OK
\object{HD113679}, % 							OK
\object{HD123517}, %DRIFT DISCUTIDO EM BAIXO	OK
\object{HD144589}, %DRIFT DISCUTIDO EM BAIXO	OK
\object{HD148211},%							OK
\object{HD190984}, %PLANET DISCUTIDO EM BAIXO	OK
and
\object{HD215257}. %DRIFT DISCUTIDO EM BAIXO	OK

For stars with at least 12 measurements (37 stars), a test was also done to search for previously undetected periodic signals in the data. For this we analyzed the generalized lomb scargle periodogram. We followed the prescription used in \citet[][]{Bonfils-2010}, \citet[][]{Cumming-1999}, and \citet[][]{Zechmeister-2009b}. In brief, we created 10\,000 virtual time series by making permutations of the original radial velocity set. For each case we then computed a periodogram and located the highest power. We then derived the distribution of the power maxima. We finally attributed an FAP to the maximum power value found in the original data set by counting the fraction of the simulated power maxima that have a higher value. The resulting FAP values are listed in Table\,\ref{tab:lomb}, together with the amplitude of the period and amplitude of the highest peak in the periodogram of the real data. For stars with known planetary companions, both FAP before and after the subtraction of the Keplerian solution are shown, while for
stars presenting linear drifts the test was done only after the subtraction of the trend. This was done since a linear drift will introduce extra power at low frequencies, masking any shorter period significant peaks. Six (6) stars in our sample show the presence of significant peaks (FAP below or near 1\%):  CD$-$2310879, HD78747, HD107094, HD171028, HD181720, and HD190984.

\begin{table}[t]
\caption[]{Periodogram variability tests for stars with at least 12 measurements.}
\begin{tabular}{lccc}
\hline
\hline
Star$^\dagger$     & FAP  & Period & Amplitude \\
        & [\%]  & [days]          & [m\,s$^{-1}$]        \\ 
 \hline 
CD$-$2310879 &    {\bf 1.08} &    4.63 & 14.2\\
HD967 &  41.09 &    2.48 &  1.3\\
HD11397$^d$ &  91.19 &    2.50 &  1.2\\
HD17865 &  22.16 &    4.30 &  1.8\\
HD22879 &   7.53 &    6.11 &  1.4\\
HD31128 &  29.77 &   16.81 &  4.0\\
HD40865 &  76.46 &    8.69 &  1.3\\
HD51754 &  91.65 &   18.96 &  1.6\\
HD56274$^d$ &  90.40 &    4.78 &  1.9\\
HD59984 &   3.80 &   10.08 &  2.5\\
HD75745 &  28.95 &  307.83 &  3.0\\
HD77110$^d$ &  92.28 &    7.06 &  1.4\\
HD78747 &   {\bf 0.94} &    3.09 &  1.7\\
HD79601$^d$ &  38.62 &    9.56 &  1.1\\
HD88725$_d$ &  54.37 &    3.44 &  1.5\\
HD107094$^d$ &   {\bf $<$10$^{-4}$} & 1898.28 & 80.3\\
HD109310$^d$ &   5.93 &   77.34 &  6.0\\
HD124785 &  55.12 &   41.84 &  6.6\\
HD126681 &  17.71 & 1034.04 &  6.2\\
HD148211 &  53.78 &   63.39 &  1.9\\
HD150177 &   2.46 &  147.62 &  5.6\\
HD171028 &   {\bf $<$10$^{-4}$} &  500.55 & 47.1\\
HD171028$^p$ &   3.81 &   16.71 &  2.3\\
HD171587 &  71.19 &    4.30 &  5.1\\
HD181720 &   {\bf $<$10$^{-4}$} &  973.47 &  8.9\\
HD181720$^p$ &  50.68 &   56.25 &  1.3\\
HD190984 &   {\bf $<$10$^{-4}$} & 1948.73 & 61.6\\
HD190984$^p$ &  36.50 &   33.29 &  3.1\\
HD197083 &  21.06 &   87.60 &  7.1\\
HD197197 &   9.67 &    5.36 &  2.8\\
HD199288 &  77.67 &   59.37 &  2.0\\
HD207869 &  26.90 &   15.67 &  2.5\\
HD210752 &  25.38 &    2.60 &  1.9\\
HD215257$^d$ &  20.28 &    8.05 &  3.7\\
HD218504 &  80.32 &    4.96 &  1.9\\
HD224817 &   8.62 &    9.15 &  2.0\\
\hline
\end{tabular}
\newline
$^\dagger$ Names with suffix $d$ and $p$ denote the results after subtraction of the linear drift and Keplerian solutions, respectively.
\label{tab:lomb}
\end{table}

Finally, for all the targets a Keplerian analysis was done using $yorbit$ (S\'egransan et al., in prep.), a code that uses an hybrid method based on 
a fast linear algorithm (Levenberg-Marquardt) and genetic operators (breeding, mutations, cross-over). This approach is important since in
a periodogram an eccentric orbit is translated into not one but multiple peaks, making it difficult to identify the correct orbital parameters.

\subsection{Case-by-case analysis}

For 9 of the cases mentioned above, \object{HD11397}, \object{HD123517}, \object{HD144589}, and \object{HD215257} (clear linear drifts),  \object{HD171028}, 
\object{HD181720}, and \object{HD190984} (previously announced planets), \object{CD$-$2310879} (active star), and \object{HD107094} (candidate new planet),
the cause of the observed variation will be discussed  in Sects.\,\ref{sec:planets}, \ref{sec:drifts}, \ref{sec:active}, and \ref{sec:ambiguous}. A discussion of the remaining stars presenting 
some sort of variability according to the statistical tests is done below.

{\it\object{HD56274} (Nmes=14)}: The radial velocities of this star suggest the existence of a very low-amplitude drift ($\sim$$-$0.8\,m/s/yr), and 
our data does not allow us to reach any firm conclusions. No other significant signal is found in the data. No planet has been
detected around this star by other high-precision radial velocity surveys \citep[e.g.][]{Fischer-2005}.

{\it\object{HD59984} (Nmes=45)}: This star shows an rms that is clearly above the average photon noise error. A period search (using both a periodogram and a Keplerian analysis) did not find 
any significant periodic signal in the data. The high dispersion observed may likely be explained by the evolutionary  status of
the star. Indeed, \citet[][]{Sousa-2010} derive a surface gravity of 4.18\,dex for this T$_{eff}$=5962\,K metal-poor star ([Fe/H]=$-$0.69), indicating
that it may be slightly evolved out of the main sequence. Evolved stars are known to have higher oscillation and granulation noise \citep[e.g.][]{Mayor-2003b,Dalsgaard-2004,Dumusque-2010}.

{\it\object{HD77110}  (Nmes=16)}: This star shows a tentative very low-amplitude drift in radial velocity (amplitude $\sim$$-$1\,m/s/yr). Such trend is not discussed in the literature, 
although to our knowledge this star has not been included in any other high-precision radial-velocity survey. No significant periodic signal is found in the data.

{\it\object{HD78747}} (Nmes=26):  The analysis of the generalized Lomb Scargle periodogram (Table\,\ref{tab:lomb}) suggests the presence of a 3.1\,day period
signal in the data. It is indeed possible to find a satisfactory Keplerian fit with this period, an 
amplitude of 2.0\,m\,s$^{-1}$, and eccentricity of 0.26. In order to confirm this signal we obtained a series of 14 new radial velocities from April to July 2010. 
The analysis of the data (to be published in a separate paper) does not reveal any 3-day signal. 

{\it\object{HD79601}  (Nmes=16)}: This is another case of low-amplitude drift ($-$0.6\,m/s/yr), whose cause is not identified. No significant periodic signal was found in the data, 
even after subtracting the linear drift. No reference for a possible companion was found in the literature. %paper by Mason 1998: no binarity

{\it\object{HD88725} (Nmes=22)}: The radial velocity measurements of this star present a low amplitude linear trend ($-$1.9\,m/s/yr). The analysis of the residuals of a linear fit to the data show the presence of a peak around 3\,days. However, the FAP of this peak suggests that it is not significant. The small number of measurements precludes any further conclusions. No planet was found around this star in other high precision planet search surveys \citep[e.g.][]{Fischer-2005}.

{\it\object{HD109310} (Nmes=15)}: A low-amplitude drift is suggested from our radial velocity data ($-$1.5\,m/s/yr). The residuals of a linear fit to the data
show an rms of $\sim$2.5\,m\,s$^{-1}$, still above the average radial velocity error of 1.6\,m\,s$^{-1}$. A frequency analysis of the residuals does not reveal,
however, any significant signal. Having so few measurements does not allow us to make any further conclusions about this case.

{\it\object{HD113679} (Nmes=6)}: A long-term and low-amplitude ($\sim$1.7\,m\,s$^{-1}$) linear trend is observed for this star over a span of $\sim$1500\,days. The reduced 
number of data points obtained does not allow reaching any firm conclusions about the observed variation. We found no reference in the literature to any 
variability \citep[e.g.][]{Nordstrom-2004}, although this star was not included (to our knowledge) in any other high-precision radial-velocity survey.

{\it\object{HD124785} (Nmes=17)}: This star presents a clear excess rms in the radial velocities, when compared to the average error. As for HD59984, however,
the atmospheric parameters derived by Sousa et al. (T$_{eff}$=5867\,K, $\log{g}$=4.20) suggest that this star has already evolved off the main sequence.
This may explain the excess power observed, and is supported by not finding any significant periodicity in the data.

{\it\object{HD148211} (Nmes=31)}: The statistical tests done above suggest there is a significant drift with a very small amplitude ($-$0.6\,m\,s$^{-1}$).
This signal is visually very marginal, though. No significant periodicity is found in the data. Our statistical tests also do not indicate any significant radial-velocity excess.

\begin{figure}[t!]
\resizebox{8cm}{!}{\includegraphics[]{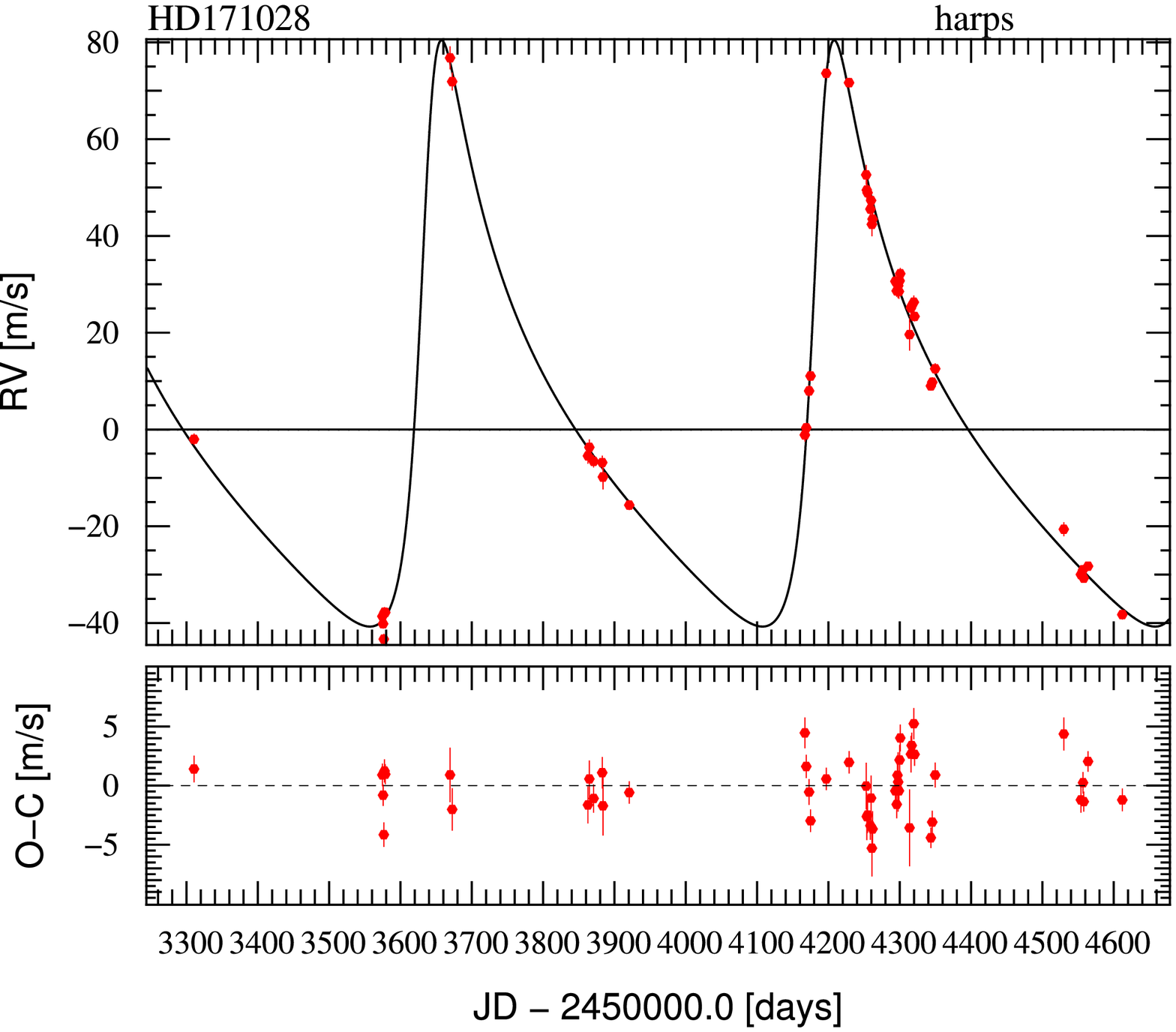}}\\
\resizebox{8cm}{!}{\includegraphics[]{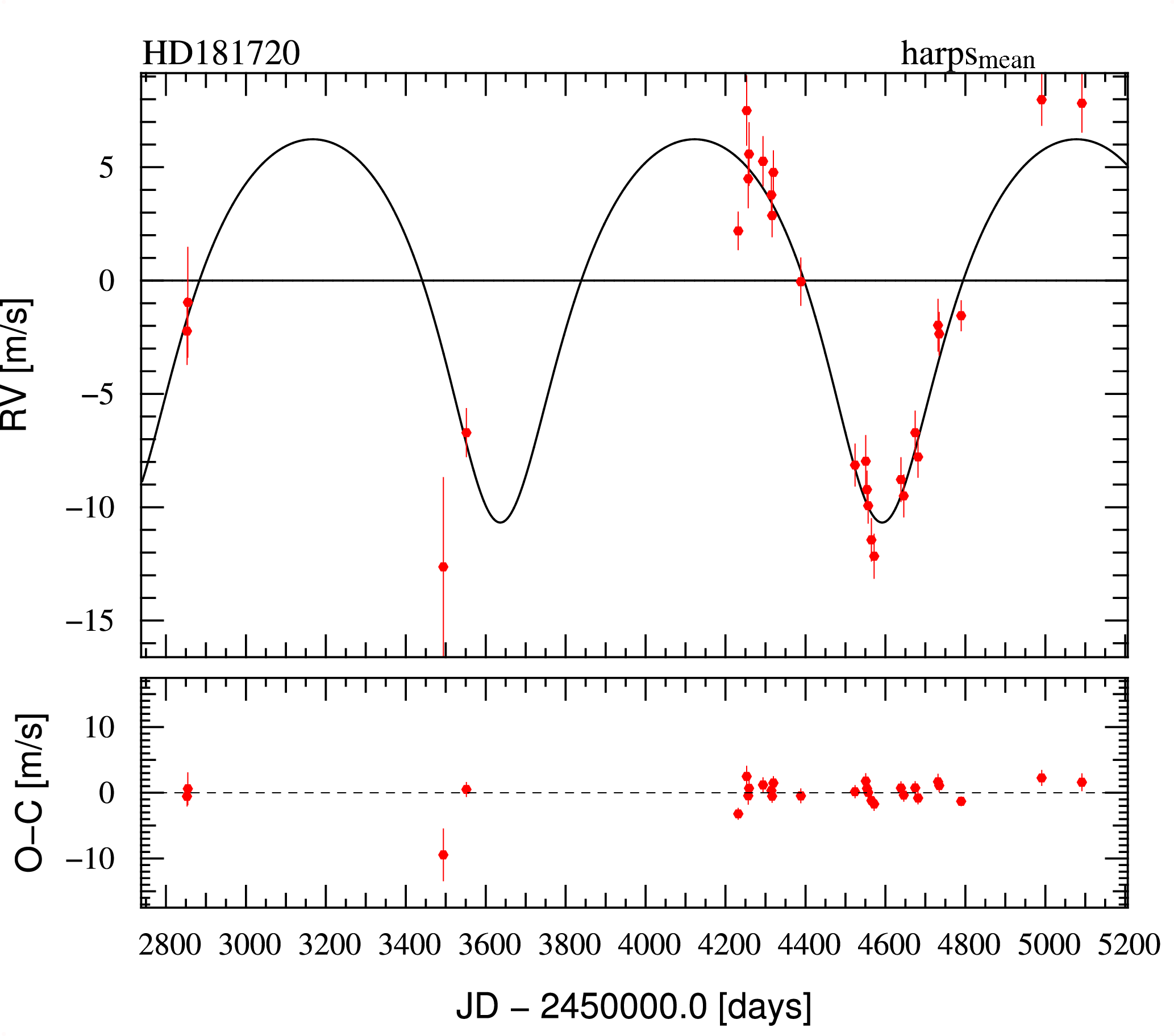}}\\
\resizebox{8cm}{!}{\includegraphics[]{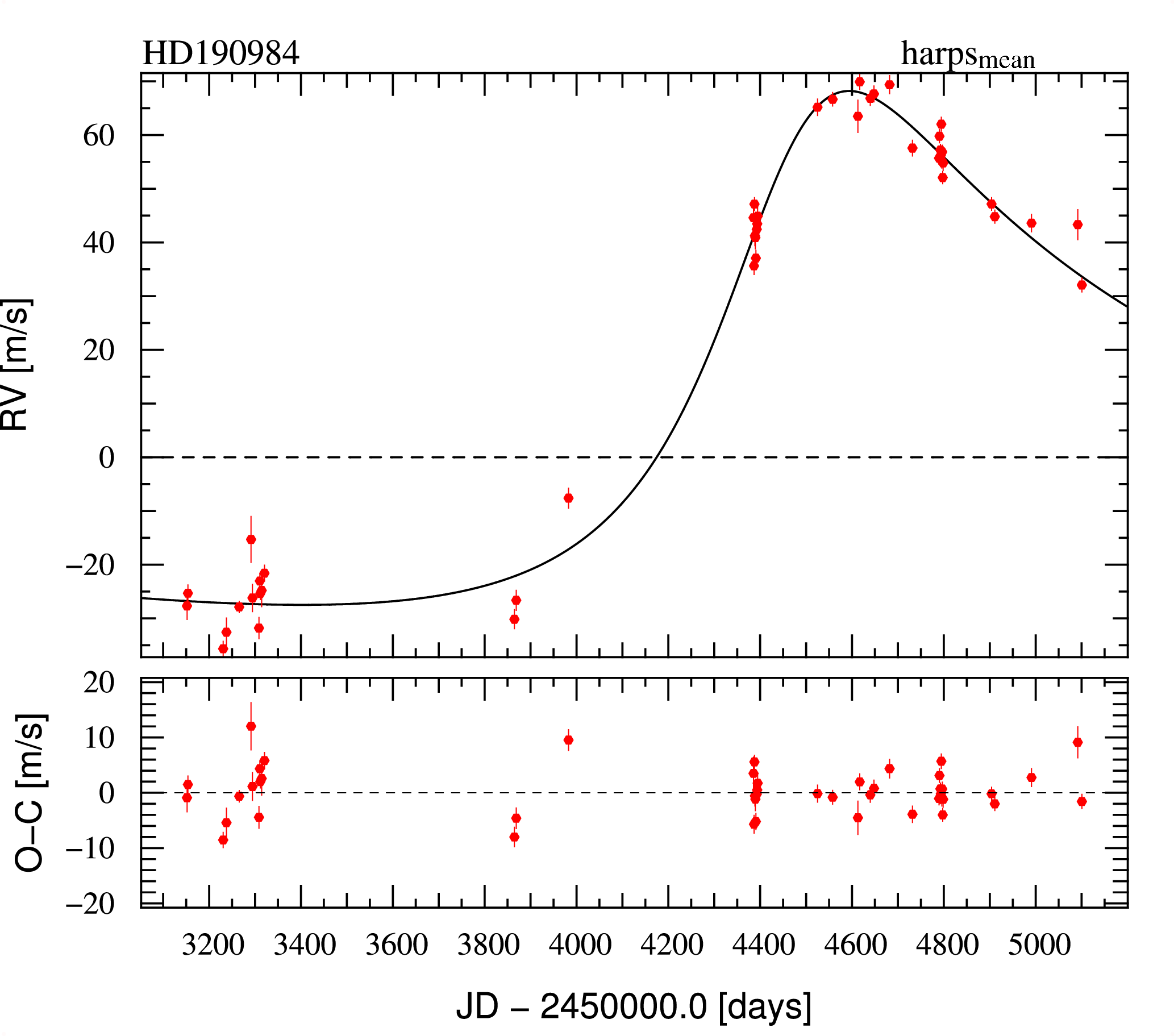}}\\
\caption{Radial velocity time series for the 3 stars with planets discussed in Sect.\,\ref{sec:planets}. }
\label{fig:orbits}
\end{figure}

{\it\object{HD150177} (Nmes=30)}: A high level of residuals is seen in the data for this star. No significant periodic signal(s) was found though. The early spectral type and 
evolutionary status suggested by the precise atmospheric parameters derived by Sousa et al. (T$_{eff}$=6216\,K, $\log{g}$=4.18) indicate that the observed noise may be due to 
stellar oscillation and granulation phenomena.

{\it\object{HD171587}} (Nmes=14): 
The radial velocity measurements of this star show a clear radial-velocity rms excess with respect to the measurement errors. The analysis of the data show that, excluding the
point at JD=54638 ($\sim$20\,m\,s$^{-1}$ above the average of the other measurements), a clear periodic signal appears that is best fit with 
a Keplerian function with P=4.49\,days, K=4.0\,m\,s$^{-1}$, and e=0.57. The residuals of this fit are only 0.56\,m\,s$^{-1}$ (compared to the average error of 1.3\,m\,s$^{-1}$), making us 
suspect that there are not enough points to obtain a reliable fit. This signal would be compatible with the presence of a 7.1 Earth mass planet orbiting 
this 0.76\,M$_{\odot}$ star (T$_{eff}$=5412\,K, $\log{g}$=4.59, [Fe/H]=$-$0.64). The analysis of the \ion{Ca}{ii} H and K lines suggests that this star 
is chromospherically very active ($\log{R'_{HK}}$=$-$4.75). Interestingly, this activity value indicates 
a rotational period of 21\,days \citep[][]{Noyes-1984}, $\sim$5 times longer than the observed signal. A periodogram of the individual S$_{MW}$ values shows a peak around 23.5\,days, with a significance level of 93\%. This peak may be the signature of the rotational period of the star. We note, however, that 
the Keplerian fit described is only based on 13 points (excluding the above-mentioned date), and the time sampling is very wide (individual points are very separated in time). 
A series of additional data points were obtained in 5 consecutive nights in July 2010 under ESO program 085.C-0063 (to be published in a separate paper). The results do not show any clear 4.5\,day periodicity. 

{\it\object{HD224817} (Nmes=30)}: According to our statistical tests, this star shows an rms above the expected value when taking the average internal errors into account.
However, this is largely due to the only point at JD=53289, without which no significant variation is seen.

\subsubsection{Previously announced planets}
\label{sec:planets}

Three stars in our sample were previously announced as harboring long-period giant planets \citep[][]{Santos-2007,Santos-2010b}, see Fig.\,\ref{fig:orbits}. For one of these, new HARPS radial velocities were obtained after the discovery paper. 

\begin{table*}[t]
\label{tab:orbits}
\caption[]{Elements of the fitted orbits for known planets.}
\begin{tabular}{lllll}
\hline
\hline
\noalign{\smallskip}
			&  \object{HD\,171028} & \object{HD\,181720}$\dagger$	    & \object{HD190984}	& \\
\hline
$P$             	& 550$\pm$3             & 956$\pm$14			    & 4885$\pm$1600	& [d]\\
$T$             	& 2\,454\,187$\pm$1    & 2\,453\,631$\pm$30      & 2\,449\,572$\pm$1600	& [d]\\
$a$			& 1.32                          & 1.78				    & 5.5                                 & [AU]\\
$e$             	& 0.59$\pm$0.01       & 0.26$\pm$0.06		    & 0.57$\pm$0.10           &  \\
$V_r$           	& 13.641$\pm$0.001& $-$45.3352$\pm$0.0004 & 20.269$\pm$0.004    & [km\,s$^{-1}$]\\
$\omega$        	& 304$\pm$1              & 177$\pm$12			   & 318$\pm$5		& [degr] \\ 
$K_1$           	& 60.6$\pm$1.0          & 8.4$\pm$0.4			   & 48$\pm$1			& [m\,s$^{-1}$] \\
$f_1(m)$        	& 6.61\,10$^{-9}$	 &  0.053\,10$^{-9}$		   & 30.95				& [M$_{\odot}$]\\ 
$\sigma(O-C)$	& 2.3                            & 1.37				   & 3.44				& [m\,s$^{-1}$]  \\    
$N$             	& 48                              & 28					   & 47				&  \\
$m_2\,\sin{i}$  	& 1.98                           & 0.37				   & 3.1				 & [M$_{\mathrm{Jup}}$]\\
\noalign{\smallskip}
\hline
\end{tabular}
\newline
$\dagger$ Orbital parameters from \citet[][]{Santos-2010b}
\end{table*}

{\it \object{HD171028}}: This star was the first in our sample to be announced to have an orbiting giant planet \citep[][]{Santos-2007}. Since the announcement paper, 29 new radial velocity measurements
were obtained (in a total of 48), allowing us to put stronger constraints to the orbital parameters of HD171028b:
period of 550\,days, eccentricity of 0.59, and minimum mass of 1.98\,M$_{Jup}$ \citep[supposing a stellar mass of 1.01\,M$_{\odot}$ --][]{Sousa-2010}. In Table\,\ref{tab:orbits} we present our solution. A Lomb Scargle Periodogram of the residuals reveal the presence of a non significant peak at $\sim$17\,day period. This may be related to the rotational period of the star. We tried to fit a Keplerian function to the residuals of the Keplerian fit, but no satisfactory solution was found. 

{\it \object{HD181720}} and {\it \object{HD190984}}: giant planets were found around these two stars and recently announced in \citet[][]{Santos-2010b}.
The orbital parameters for the two planets are listed in Table\,\ref{tab:orbits}. We refer to the discovery paper for more details.

\subsubsection{High-amplitude linear drifts}
\label{sec:drifts}

A few stars in our sample present very clear long-term trends with total 
amplitudes of at least 10 to 20\,m\,s$^{-1}$. These are compatible with the existence of long period giant planets or 
brown dwarfs. However, the insufficient time coverage of the data does not allow making any conclusion about the origin 
of the observed signal.

{\it \object{HD11397}}: The 33 radial velocities of this star show the signature of a clear long-term trend with a total amplitude
of 33\,m\,s$^{-1}$ over 2075\,days. When excluding the first two measurements, the trend shows a slight curvature. The data could be well fit
by a linear slope, together with a low-amplitude long-period signal. However, the lower quality of the first two data points
(photon noise error of 2-4\,m\,s$^{-1}$ and obtained with short exposure times not allowing supression of stellar oscillation noise) keeps us
from deeper insight into the observed signal. No significant short-period signal is found in the data. A quadratic fit that excludes the first two 
points shows a low rms of 1.27\,m\,s$^{-1}$, not strongly above the average photon noise (0.90\,m\,s$^{-1}$).
More data are needed to clarify this case. 

{\it \object{HD123517}}:  The 9 radial velocity measurements of this star reveal a clear long-term linear drift ($\sim$150\,m\,s$^{-1}$ 
over $\sim$1600\,days). The residuals of a linear fit to the data (1.5\,m\,s$^{-1}$) are smaller than the average photon noise error of
the measurements (2.3\,m\,s$^{-1}$; this is one of the faintest stars in our sample with mv=9.6), suggesting that no detectable short period signal 
is hidden in the data.  More data are needed to confirm the source of the observed radial velocity variation.

{\it \object{HD144589}}:  We obtained 11 radial velocity measurements of this star, spanning a total of $\sim$1500\,days. The data shows the presence of a clear linear trend, with a fitted slope of 25\,m\,s$^{-1}$\,yr$^{-1}$. The residuals of a linear fit to the data (4.0\,m\,s$^{-1}$) are slightly above the average photon noise of the measurements (3.4\,m\,s$^{-1}$). The stellar atmospheric parameters derived for this star (T$_{eff}$=6372\,K, $\log{g}$=4.28, [Fe/H]=$-$0.05) by \citet[][]{Sousa-2010} suggest that this is one of the hottest stars in our sample and likely slightly evolved off the main sequence. Stellar oscillations, along with the relatively high projected rotational velocity (5.6\,km\,s$^{-1}$ -- derived from the HARPS cross-correlation function), may explain the excess observed scatter.

{\it \object{HD215257}}:  The 37 radial velocities of this star (average error of 1.6\,m\,s$^{-1}$) indicate the presence of
a low amplitude linear drift ($-$15\,m\,s$^{-1}$\,yr$^{-1}$ over 1700\,days). A signal with a similar slope was also found by \citet[][]{Sozzetti-2009} using lower precision data from their Keck-HIRES survey. The high residuals (3.98\,m\,s$^{-1}$) around the linear trend indicate an extra signal in the data. No reliable Keplerian fit was found, though, and no significant peak is found in a generalized Lomb Scargle periodogram. Since the star is an early-G/late-F dwarf with T$_{eff}$=6052\,K, $\log{g}$=4.46, and [Fe/H]=$-$0.63 \citep[][]{Sousa-2010}, at least part of this signal may come from stellar oscillation and granulation ``noise'' \citep[more important in early type stars --][]{Mayor-2003b,Dumusque-2010}. More data are needed to confirm that this is the only source of the observed rms.

\subsubsection{Activity-induced periodic signals}
\label{sec:active}

\begin{figure}[t!]
\resizebox{\hsize}{!}{\includegraphics[]{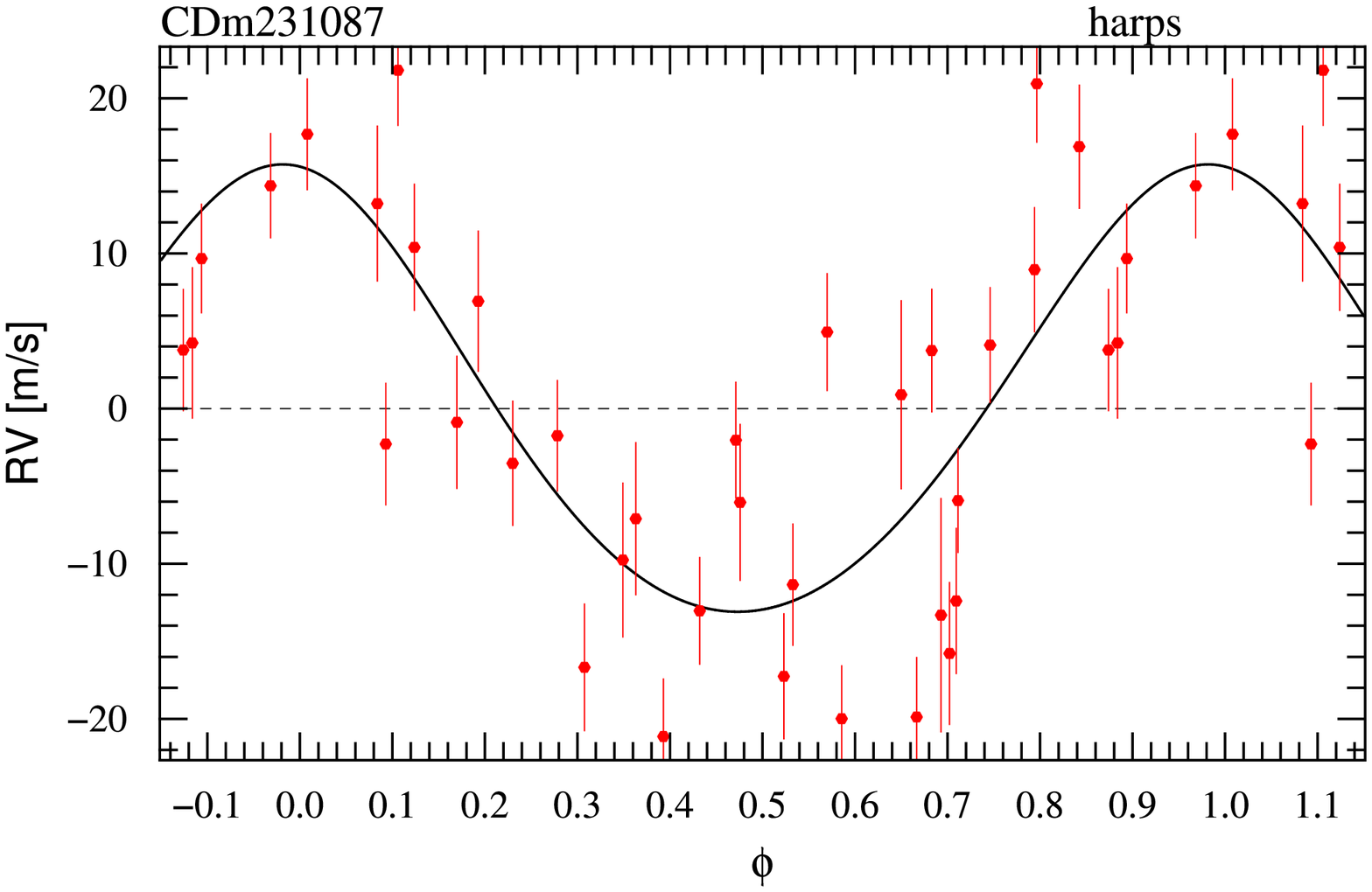}}
\resizebox{\hsize}{!}{\includegraphics[bb= 25 147 580 441]{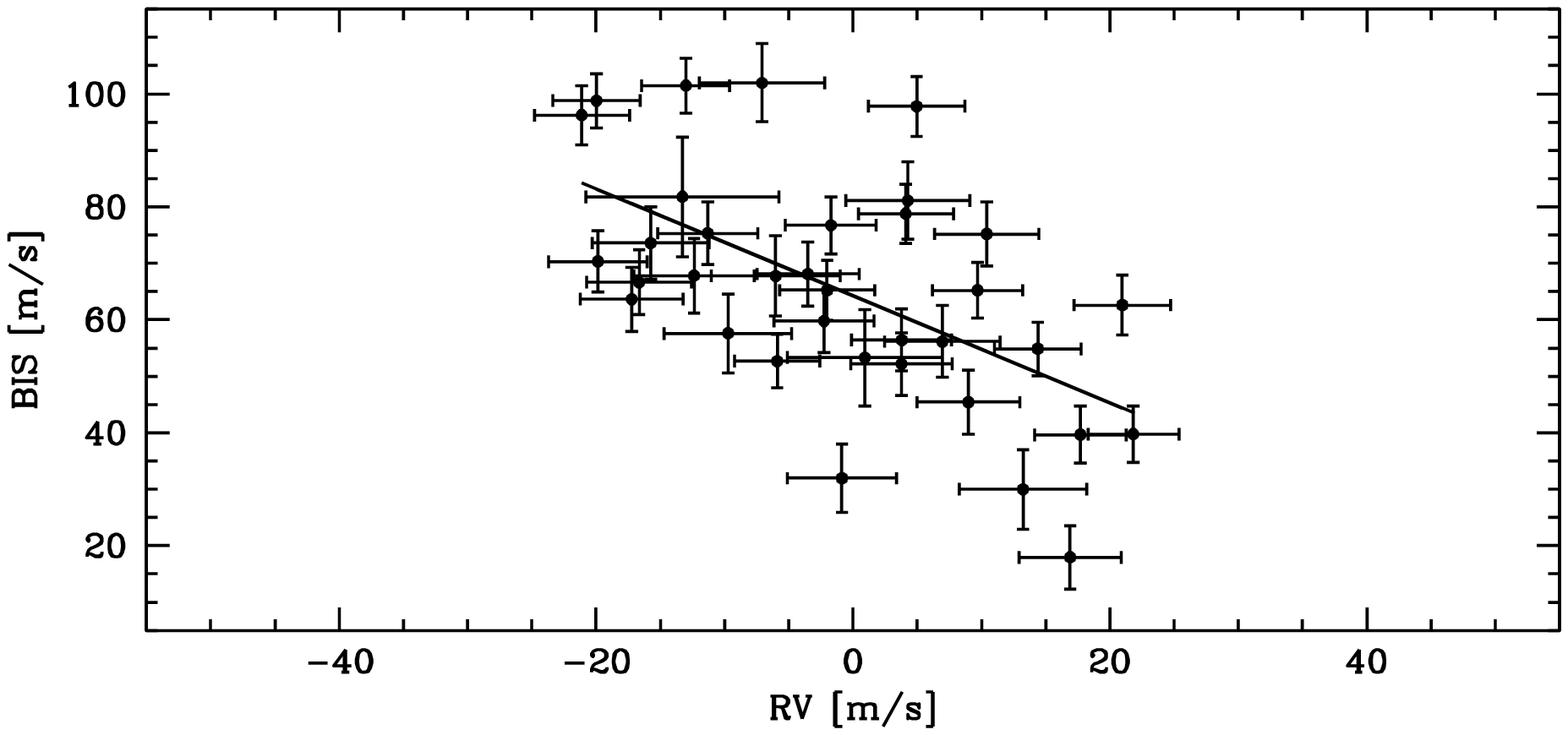}}
\caption{{\it Top}: Phase folded radial-velocity measurements of \object{CD-2310879} with a period of  4.626\,days. The line represents the best Keplerian function fit. {\it Bottom}: Radial velocities as a function of the HARPS cross-correlation bisector inverse slope. The line represents a linear fit to the data.}
\label{fig:cd231087}
\end{figure}

{\it \object{CD$-$23\,10879}}: The 35 radial-velocity measurements of this F dwarf (T$_{eff}$=6788\,K, $\log{g}$=4.67, [Fe/H]=$-$0.24) present an rms of 12.9\,m\,s$^{-1}$, clearly above the median photon noise (3.9\,m\,s$^{-1}$). A Lomb Scargle periodogram shows a significant peak around 4.5\,days, well fit by a Keplerian function (see Fig.\,\ref{fig:cd231087}) with P=4.626\,days, eccentricity of 0.092, and semi-amplitude of 14.4\,m\,s$^{-1}$ (residuals of 3.96\,m\,s$^{-1}$, similar to the photon noise). This signal is compatible with the existence of a planet with 42 times the mass of the Earth orbiting this 1.20\,M$_{\odot}$ star. However, a Lomb Scargle periodogram of the bisector inverse slope (BIS) of the HARPS cross-correlation function shows a similar peak at $\sim$4.5\,days. A clear negative correlation is also seen between BIS and the radial velocities (slope=$-$0.94$\pm$0.23 -- Fig.\,\ref{fig:cd231087}), typical of the signal expected from activity-induced radial-velocity variations \citep[see case of HD\,166435 --][]{Queloz-2000}. The high value of the activity index  obtained from the HARPS spectra, $\log{R'_{HK}}$=$-$4.60, suggests a rotational period around 4\,days, similar to the observed radial-velocity periodicity \citep[][]{Noyes-1984}. We thus conclude that the best explanation for the observed radial-velocity variation of this star is given by active regions modulated
by stellar rotation.

\subsubsection{Ambiguous cases: new candidate planets}
\label{sec:ambiguous}

\begin{figure}[t!]
\resizebox{8cm}{!}{\includegraphics[]{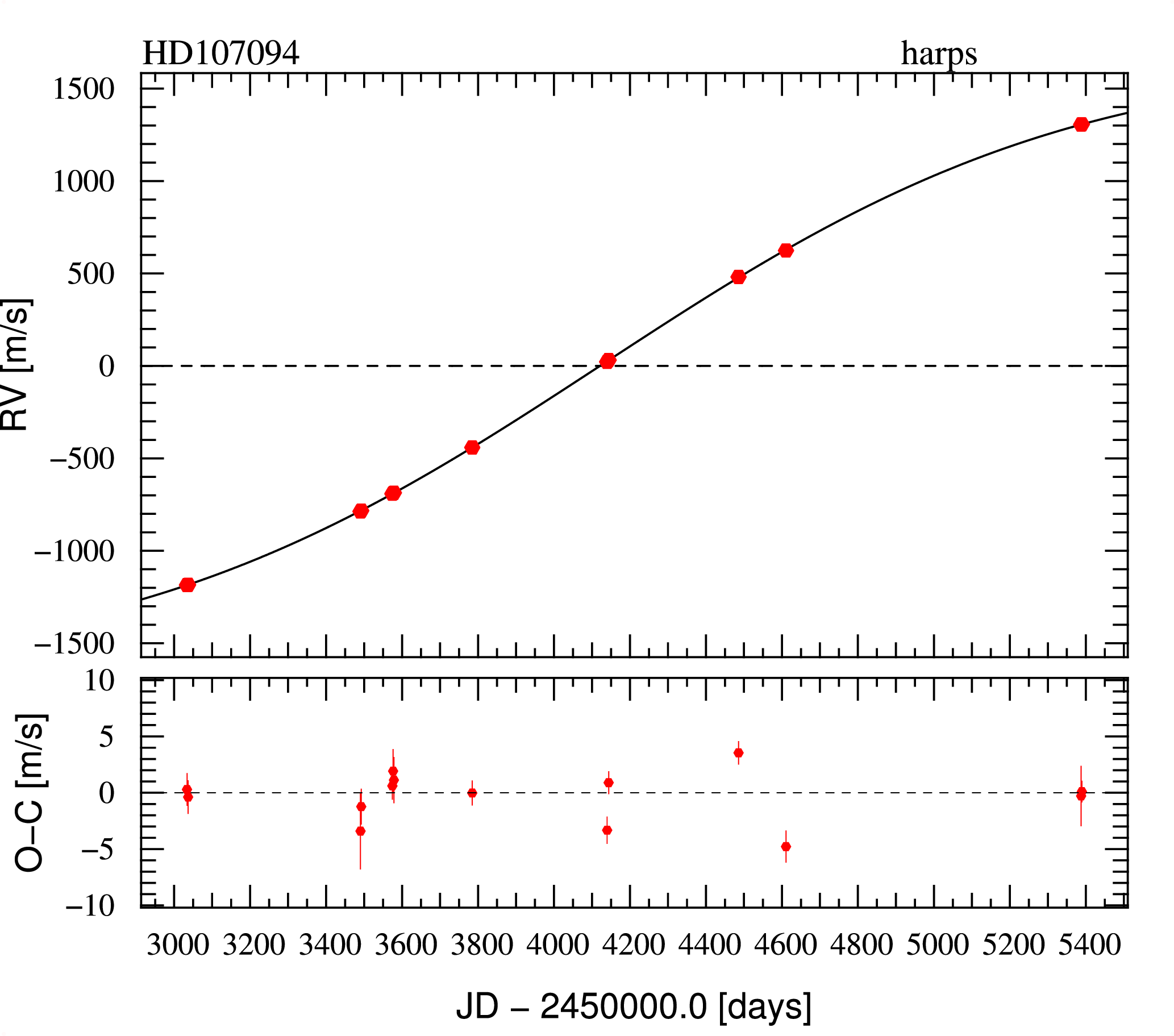}}
\resizebox{8cm}{!}{\includegraphics[]{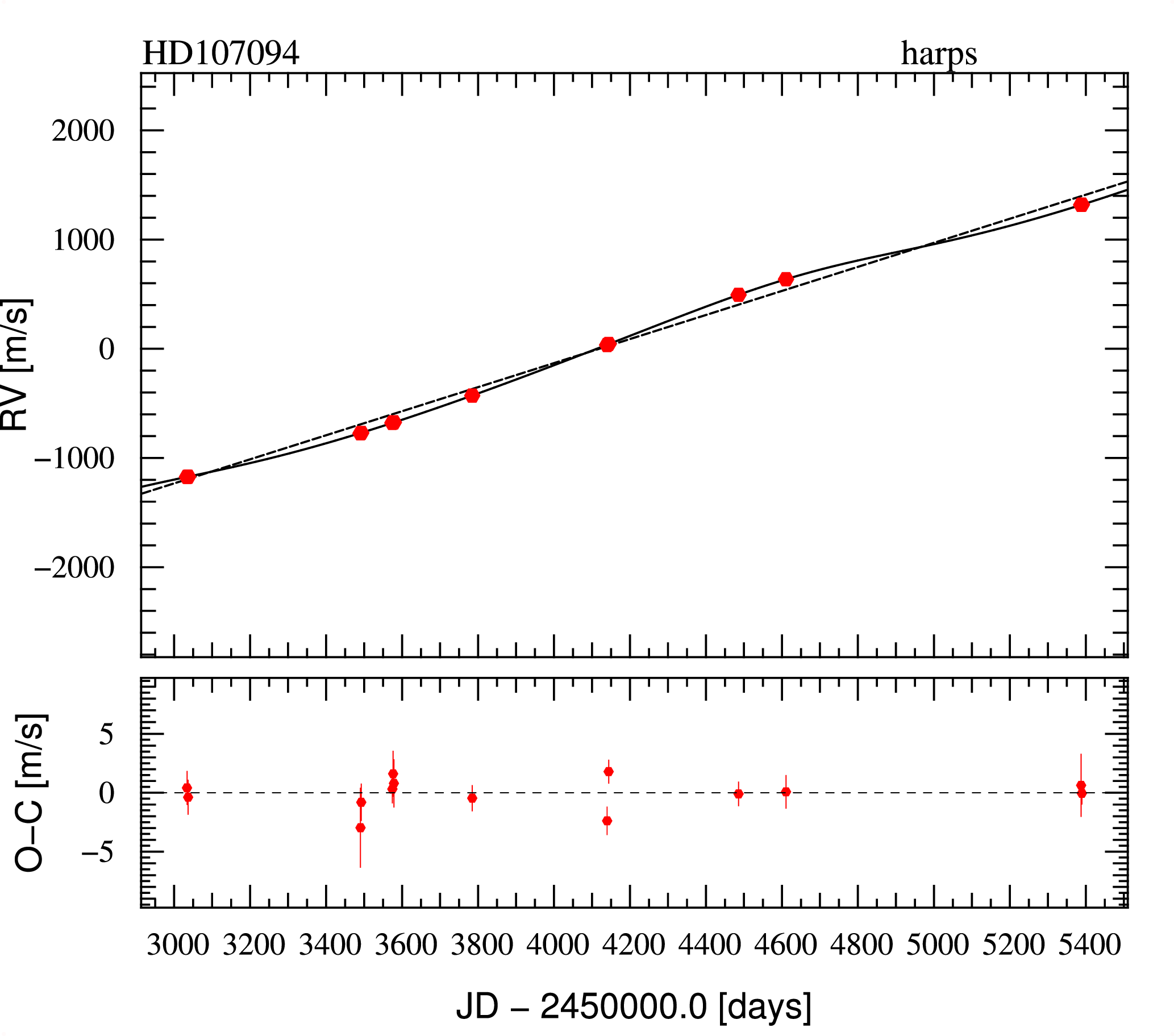}}
\resizebox{8cm}{!}{\includegraphics[]{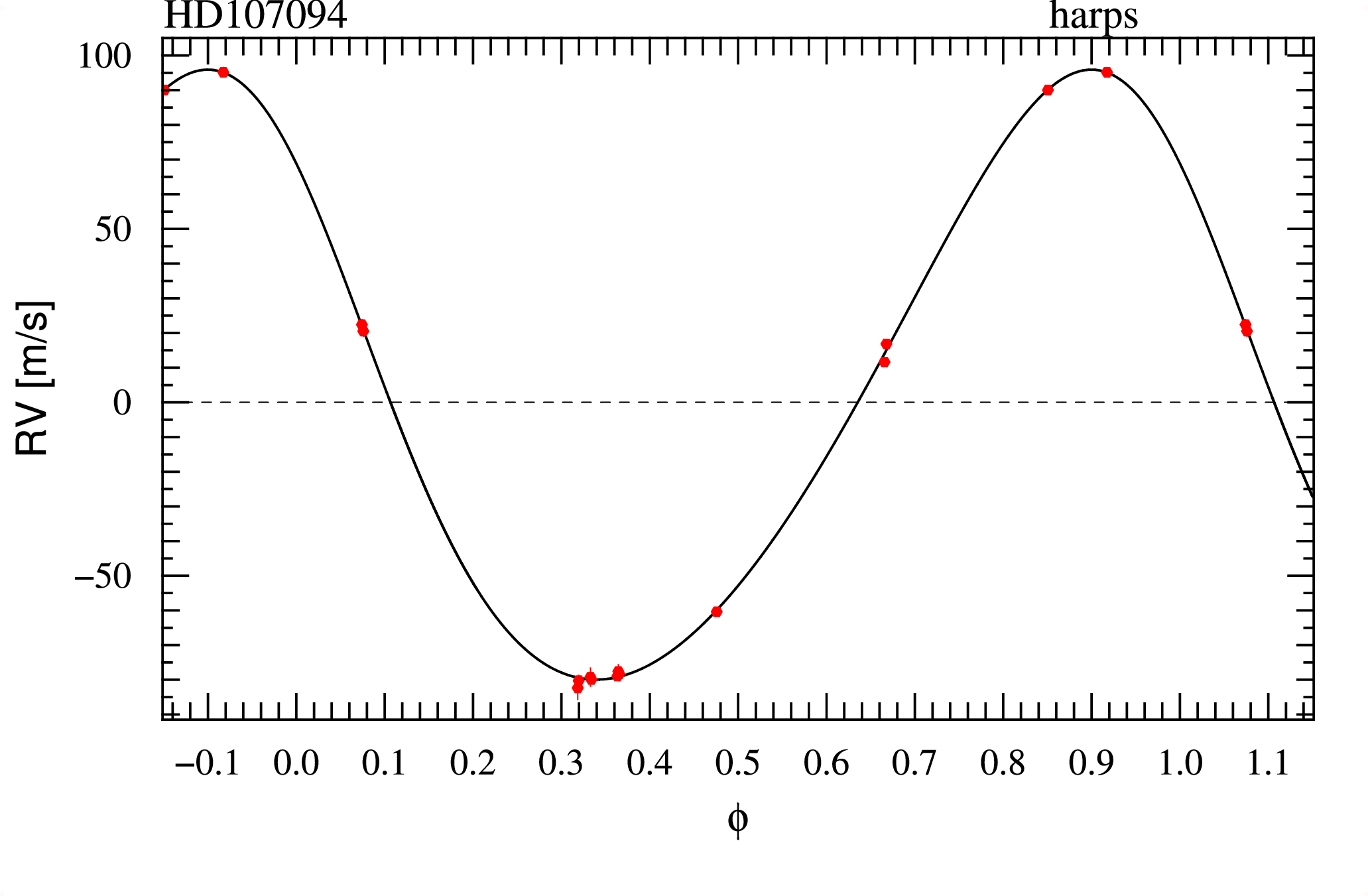}}
\caption{{\it Top}: Radial velocity measurements of HD107094 and best 1-Keplerian fit to the data. {\it Middle}: same as above but after fitting a linear trend, together with a Keplerian function.
In both cases the residuals are shown. {\it Bottom}: phase-folded radial velocity measurements with the period of the candidate planet.}
\label{fig:HD107094}
\end{figure}

\begin{table}[t]
\label{tab:ambiguous}
\caption[]{Elements of the fitted orbit for HD107094.}
\begin{tabular}{llll}
\hline
\hline
\noalign{\smallskip}
%			&  \object{HD\,107094} 	& \\
%\hline
$P$             	& 1870$\pm$34       	& [d]\\
$T$             	& 2\,454\,765$\pm$29    & [d]\\
$a$			& 2.76                            	    & [AU]\\
$e$             	& 0.13$\pm$0.02          &  \\
$V_r$           	& 15.33$\pm$0.05     & [km\,s$^{-1}$]\\
$\omega$        	& 46$\pm$3             	   & [degr] \\ 
$K_1$           	& 88$\pm$5          	   & [m\,s$^{-1}$] \\
$f_1(m)$        	& 128.3\,10$^{-9}$		   & [M$_{\odot}$]\\ 
$\sigma(O-C)$	& 1.10                         			   & [m\,s$^{-1}$]  \\    
$N$             	& 14                          		   &  \\
$m_2\,\sin{i}$  	& 4.5                          		    & [M$_{\mathrm{Jup}}$]\\
\noalign{\smallskip}
\hline
\end{tabular}
\end{table}

% Planet + Drift?

{\it HD107094}:
A total of 14 radial velocities were obtained for this star, showing a clear long-term (quasi-linear) radial-velocity signal. Twelve of these measurements were obtained 
within the HARPS GTO program, and a last two in July 2010 under ESO program 085.C-0063. In a first analysis,
the total amplitude ($\sim$2.5\,km\,s$^{-1}$ over 2355\,days) suggests there is a stellar or sub-stellar companion orbiting this 0.8 solar mass star. 
A single Keplerian fit to the data shows that it is well fit by a long-period Keplerian function, though the orbital parameters are difficult to constrain (see Fig.\,\ref{fig:HD107094}). 
The residuals of this fit, on the order of 2.1\,m\,s$^{-1}$, are still high, and they called attention to a different possible orbital solution to the system.
Indeed, a fit of the linear drift together with a Keplerian function reveals the presence of a possible shorter period solution, with P$\sim$1870\,days, K=88\,m\,s$^{-1}$,
and e=0.13 (Table\,\ref{tab:ambiguous}). This signal, first detected using only the GTO data, is compatible with the presence of a 4.5 Jupiter mass 
companion. The residuals of this fit (1.10\,m\,s$^{-1}$) are significantly better than in the previous case.
The fitted linear trend (which, to our knowledge, has not been previously reported in the literature) has a slope of 402$\pm$1\,m/s/yr.
More data are now needed to disentangle the two possible solutions, but we are likely seeing a new giant planet. Interestingly, 
with [Fe/H]=$-$0.51 \citep[][]{Sousa-2010}, HD107094 is close to the high-metallicity limit of our sample.

{\it HD197083}: 
Although the statistical analysis done for this star does not indicate any strong signal, a visual inspection of the radial velocities 
hints at the presence of a long-term periodic signal (P$\sim$1500-2000\,days), see Fig.\,\ref{fig:6mesc}. The number of points covering the first $\sim$1000\,days is very reduced (3) with high error bars. More data are thus clearly needed to confirm this signal.

\section{Discussion and conclusions}
\label{sec:conclusions}

In this paper we present the overall results of our HARPS program to search for planets orbiting a sample of metal-poor stars.
In a total of 104 stars, 3 giant planets were confirmed, and one more very promising giant planet candidate is
presented. These discoveries significantly increase the number of known giant planets orbiting metal-poor stars \citep[see also case of HD\,155358 --][]{Cochran-2007}. Several binary stars were also found in the sample.  

A proper statistical analysis of our results requires not only identifying the planetary signals in our time series but also assessing their detection limit; i.e. what planets, as a function of mass and period, can be rejected considering our observations. Although this is the topic of a forthcoming paper, partial figures can already be given. 
Excluding the 16 stars discussed in Sect.\,\ref{sec:cleaning} (targets not suitable for a high-precision radial-velocity planet search), 
our sample is constituted of 64/87 stars with at least 6/3 radial velocity measurements.
We have detected at least 3 giant planets with masses greater than 0.3 Jupiter masses and with periods ranging from a few hundred to a few thousand days. Of course, time series with only a few measurements ($\sim$3-6) cannot always rule out a Jupiter-mass companion on a short or moderate period orbit. Moreover, we have a possible fourth detection and several more stars with a linear RV drift indicative of a long-period companion. Three (3) detections over 87 targets, or 3.4$^{+3.2}_{-1.0}$\%\footnote{The 1-sigma error bars were obtained using a binomial distribution.}, is therefore a lower estimate to the frequency of Jupiter-mass planets orbiting metal-poor stars. If we only consider stars with [Fe/H] between $-$0.40 and $-$0.60\,dex (an interval where the stars with planets are in our sample), 
then only 34 stars with at least 3 measurements exist, and the percentages increase to 11.3$^{+4.9}_{-5.3}$\%.

We can additionally note that no planet has been detected around the 32 stars with [Fe/H]$<-0.60$\,dex and at least 6 measurements. For these stars it is reasonable to assume completeness for Jupiter-mass planets on short or moderate period orbits ($<100$\,days). Those stars have metallicities from $-$1.4 to $-$0.6\,dex, with a median value of $-0.69$\,dex. A null detection for that subsample implies a maximum occurrence of Jupiter-mass planets (on short or moderate orbital periods) of $\sim$5\%. 

 \citet[][]{Santos-2001,Santos-2004b} have shown that the frequency of giant planets orbiting solar type dwarfs is a strong 
 function of the stellar metallicity. This trend, confirmed by several other studies \citep[e.g.][]{Fischer-2005,Johnson-2010}, makes these facts
surprising. {The above studies have also shown that the frequency of giant planets orbiting stars with
metallicity near $-$0.5\,dex is $\sim$3\% \citep[see also review by ][]{Udry-2007}, strongly below the $\sim$11\% mentioned above (though 
both results are still likely compatible within the error bars)}. 
We have to take into account, however, that the exquisite precision of HARPS has increased the planet detection
rate when compared to the results of the lower precision programs on which the existence of the metallicity-giant
planet correlation has been based \citep[][]{Santos-2004b,Fischer-2005}. The present results, although based on small
number statistics, may hint at a much higher frequency of giant planets around solar type stars (of all metallicities).

The stars with giant planets discovered in our survey (HD\,171028, HD\,181720, HD\,190984, and likely HD\,107094),
all have metallicity values on the high-metallicity side of our sample ($-$0.48, $-$0.53, $-$0.49, and $-$0.51, respectively), see Fig.\,\ref{fig:param}.
This suggests that even for metal-poor stars the frequency of giant planets is a rising function of the stellar metallicity, a
result that does not confirm former suspicions that the metallicity-giant planet correlation could be flat for [Fe/H] values below solar \citep[][]{Santos-2004b}.
This result further suggests that the core accretion model of planet formation is still at play in these low metallicity
values \citep[][]{Ida-2004b,Mordasini-2009a}. Since no planets were discovered in our sample at [Fe/H] significantly below $-$0.5\,dex,
and only one giant planet was found around a stars with lower metallicity \citep[$-$0.68\,dex --][]{Cochran-2007}, 
this could also hint that we are close to the metallicity limit below which no giant planets can be formed. More data are, however,
needed to settle this issue.

As mentioned in Sect.\,\ref{sec:sample}, it is interesting to observe that two of the stars discussed in Sect.\,\ref{sec:drifts}, the ones that present long-term radial-velocity linear drifts (HD\,123517 and HD\,144589), have the highest metallicity values ($+$0.09 and $-$0.05\,dex) among all the objects in our sample (see also Fig.\,\ref{fig:param}). Both present relatively low-amplitude linear trends, suggestive of the presence of long-period giant planets. If such detections are confirmed, this result would be in line with the metallicity-giant planet correlation. 

Interestingly, no short period planet was found in our data. All the 3 (or 4) giant planets discovered orbit in long-period orbits. 
This suggests that giant planets in short-period orbits are not common around stars with low metallicity, in agreement with the
results of \citet[][]{Sozzetti-2009}. The lack of such detections is, however, expected, since the frequency of hot
jupiters in radial velocity surveys is only $\sim$1\% \citep[][]{Udry-2007}. Only 87 stars were effectively monitored in
the present survey, precluding a significant statistical meaning for this lack of detection.  

Theoretical studies \citep[e.g.][]{Mordasini-2009a}, backed up by recent observations \citep[e.g.][]{Sousa-2008}, suggest that the frequency of low-mass 
planets may increase for stars in the metallicity range of our sample when compared to solar metallicity objects. That no Neptune or super-Earth mass planet has been
discovered up to now in our sample does not contradict this conclusion. Instead, the observing
strategy used (data sampling and exposure times) was simply not adequate for the purpose of finding
such low-mass objects. A program with HARPS is presently being done to fill this gap.

\begin{acknowledgements}
We would like to thank the referee, Michael Endl, for the positive and constructive report. We acknowledge the support by the European Research Council/European Community under the FP7 through Starting Grant agreement number 239953. NCS also acknowledges the support from 
Funda\c{c}\~ao para a Ci\^encia e a Tecnologia (FCT) through program Ci\^encia\,2007 funded by FCT/MCTES (Portugal) and POPH/FSE (EC), and in 
the form of grants reference PTDC/CTE-AST/098528/2008 and PTDC/CTE-AST/098604/2008. SGS is supported by grant SFRH/BPD/47611/2008 from
FCT/MCTES.
\end{acknowledgements}

\bibliographystyle{aa}
\bibliography{santos_bibliography}

\begin{thebibliography}{53}
\expandafter\ifx\csname natexlab\endcsname\relax\def\natexlab#1{#1}\fi

\bibitem[{{Bonfils} {et~al.}(2010){Bonfils}, {Delfosse}, {Udry}, {Forveille},
  {Mayor}, {Perrier}, {Bouchy}, {Gillon}, {Lovis}, {Pepe}, {Queloz}, {Santos},
  {S\'egransan}, \& {Bertaux}}]{Bonfils-2010}
{Bonfils}, X., {Delfosse}, X., {Udry}, S., {et~al.} 2010, A\&A

\bibitem[{{Bonfils} {et~al.}(2005){Bonfils}, {Forveille}, {Delfosse}, {Udry},
  {Mayor}, {Perrier}, {Bouchy}, {Pepe}, {Queloz}, \& {Bertaux}}]{Bonfils-2005b}
{Bonfils}, X., {Forveille}, T., {Delfosse}, X., {et~al.} 2005, A\&A, 443, L15

\bibitem[{{Boss}(1997)}]{Boss-1997}
{Boss}, A.~P. 1997, Science, 276, 1836

\bibitem[{{Boss}(2002)}]{Boss-2002}
{Boss}, A.~P. 2002, ApJ, 567, L149

\bibitem[{{Boss}(2006)}]{Boss-2006}
{Boss}, A.~P. 2006, ApJL, 644, L79

\bibitem[{{Bouchy} {et~al.}(2005){Bouchy}, {Bazot}, {Santos}, {Vauclair}, \&
  {Sosnowska}}]{Bouchy-2005c}
{Bouchy}, F., {Bazot}, M., {Santos}, N.~C., {Vauclair}, S., \& {Sosnowska}, D.
  2005, A\&A, 440, 609

\bibitem[{{Christensen-Dalsgaard}(2004)}]{Dalsgaard-2004}
{Christensen-Dalsgaard}, J. 2004, Sol. Phys., 220, 137

\bibitem[{{Cochran} {et~al.}(2007){Cochran}, {Endl}, {Wittenmyer}, \&
  {Bean}}]{Cochran-2007}
{Cochran}, D.~C., {Endl}, M., {Wittenmyer}, R.~A., \& {Bean}, J.~L. 2007, ApJ,
  in press

\bibitem[{{Cumming} {et~al.}(1999){Cumming}, {Marcy}, \&
  {Butler}}]{Cumming-1999}
{Cumming}, A., {Marcy}, G.~W., \& {Butler}, R.~P. 1999, ApJ, 526, 890

\bibitem[{{Da Silva} {et~al.}(2006){Da Silva}, {Udry}, {Bouchy}, {Mayor},
  {Moutou}, {Pont}, {Queloz}, {Santos}, {S{\'e}gransan}, \&
  {Zucker}}]{DaSilva-2006}
{Da Silva}, R., {Udry}, S., {Bouchy}, F., {et~al.} 2006, A\&A, 446, 717

\bibitem[{{Dumusque} {et~al.}(2010){Dumusque}, {Udry}, {Lovis}, {Santos}, \&
  {Monteiro}}]{Dumusque-2010}
{Dumusque}, X., {Udry}, S., {Lovis}, C., {Santos}, N., \& {Monteiro}, M. 2010,
  A\&A

\bibitem[{{ESA}(1997)}]{ESA-1997}
{ESA}. 1997, The Hipparcos and Tycho Catalogues

\bibitem[{{Fischer} {et~al.}(2005){Fischer}, {Laughlin}, {Butler}, {Marcy},
  {Johnson}, {Henry}, {Valenti}, {Vogt}, {Ammons}, {Robinson}, {Spear},
  {Strader}, {Driscoll}, {Fuller}, {Johnson}, {Manrao}, {McCarthy},
  {Mu{\~n}oz}, {Tah}, {Wright}, {Ida}, {Sato}, {Toyota}, \&
  {Minniti}}]{Fischer-2005b}
{Fischer}, D.~A., {Laughlin}, G., {Butler}, P., {et~al.} 2005, ApJ, 620, 481

\bibitem[{{Fischer} \& {Valenti}(2005)}]{Fischer-2005}
{Fischer}, D.~A. \& {Valenti}, J. 2005, ApJ, 622, 1102

\bibitem[{{Gonzalez}(1997)}]{Gonzalez-1997}
{Gonzalez}, G. 1997, MNRAS, 285, 403

\bibitem[{{Gratton} {et~al.}(2000){Gratton}, {Sneden}, {Carretta}, \&
  {Bragaglia}}]{Gratton-2000}
{Gratton}, R.~G., {Sneden}, C., {Carretta}, E., \& {Bragaglia}, A. 2000, A\&A,
  354, 169

\bibitem[{{Ida} \& {Lin}(2004{\natexlab{a}})}]{Ida-2004a}
{Ida}, S. \& {Lin}, D.~N.~C. 2004{\natexlab{a}}, ApJ, 604, 388

\bibitem[{{Ida} \& {Lin}(2004{\natexlab{b}})}]{Ida-2004b}
{Ida}, S. \& {Lin}, D.~N.~C. 2004{\natexlab{b}}, ApJ, 616, 567

\bibitem[{{Johnson} {et~al.}(2010){Johnson}, {Aller}, {Howard}, \&
  {Crepp}}]{Johnson-2010}
{Johnson}, J.~A., {Aller}, K.~M., {Howard}, A.~W., \& {Crepp}, J.~R. 2010,
  ArXiv e-prints

\bibitem[{{Johnson} {et~al.}(2007){Johnson}, {Butler}, {Marcy}, {Fischer},
  {Vogt}, {Wright}, \& {Peek}}]{Johnson-2007}
{Johnson}, J.~A., {Butler}, R.~P., {Marcy}, G.~W., {et~al.} 2007, ApJ, 670, 833

\bibitem[{{Latham} {et~al.}(2002){Latham}, {Stefanik}, {Torres}, {Davis},
  {Mazeh}, {Carney}, {Laird}, \& {Morse}}]{Latham-2002}
{Latham}, D.~W., {Stefanik}, R.~P., {Torres}, G., {et~al.} 2002, AJ, 124, 1144

\bibitem[{{Laughlin} {et~al.}(2004){Laughlin}, {Bodenheimer}, \&
  {Adams}}]{Laughlin-2004}
{Laughlin}, G., {Bodenheimer}, P., \& {Adams}, F.~C. 2004, ApJ, 612, L73

\bibitem[{{Lovis} \& {Mayor}(2007)}]{Lovis-2007}
{Lovis}, C. \& {Mayor}, M. 2007, A\&A, 472, 657

\bibitem[{{Lovis} {et~al.}(2006){Lovis}, {Mayor}, {Pepe}, {Alibert}, {Benz},
  {Bouchy}, {Correia}, {Laskar}, {Mordasini}, {Queloz}, {Santos}, {Udry},
  {Bertaux}, \& {Sivan}}]{Lovis-2006}
{Lovis}, C., {Mayor}, M., {Pepe}, F., {et~al.} 2006, Nature, 441, 305

\bibitem[{{Matsuo} {et~al.}(2007){Matsuo}, {Shibai}, {Ootsubo}, \&
  {Tamura}}]{Matsuo-2007}
{Matsuo}, T., {Shibai}, H., {Ootsubo}, T., \& {Tamura}, M. 2007, ApJ, 662, 1282

\bibitem[{{Mayer} {et~al.}(2002){Mayer}, {Quinn}, {Wadsley}, \&
  {Stadel}}]{Mayer-2002}
{Mayer}, L., {Quinn}, T., {Wadsley}, J., \& {Stadel}, J. 2002, Science, 298,
  1756

\bibitem[{{Mayor} {et~al.}(2009){Mayor}, {Bonfils}, {Forveille}, {Delfosse},
  {Udry}, {Bertaux}, {Beust}, {Bouchy}, {Lovis}, {Pepe}, {Perrier}, {Queloz},
  \& {Santos}}]{Mayor-2009}
{Mayor}, M., {Bonfils}, X., {Forveille}, T., {et~al.} 2009, ArXiv e-prints

\bibitem[{{Mayor} {et~al.}(2003){Mayor}, {Pepe}, {Queloz}, {Bouchy},
  {Rupprecht}, {Lo Curto}, {Avila}, {Benz}, {Bertaux}, {Bonfils}, {dall},
  {Dekker}, {Delabre}, {Eckert}, {Fleury}, {Gilliotte}, {Gojak}, {Guzman},
  {Kohler}, {Lizon}, {Longinotti}, {Lovis}, {Megevand}, {Pasquini}, {Reyes},
  {Sivan}, {Sosnowska}, {Soto}, {Udry}, {van Kesteren}, {Weber}, \&
  {Weilenmann}}]{Mayor-2003b}
{Mayor}, M., {Pepe}, F., {Queloz}, D., {et~al.} 2003, The Messenger, 114, 20

\bibitem[{{Mordasini} {et~al.}(2009{\natexlab{a}}){Mordasini}, {Alibert}, \&
  {Benz}}]{Mordasini-2009a}
{Mordasini}, C., {Alibert}, Y., \& {Benz}, W. 2009{\natexlab{a}}, A\&A, 501,
  1139

\bibitem[{{Mordasini} {et~al.}(2009{\natexlab{b}}){Mordasini}, {Alibert},
  {Benz}, \& {Naef}}]{Mordasini-2009b}
{Mordasini}, C., {Alibert}, Y., {Benz}, W., \& {Naef}, D. 2009{\natexlab{b}},
  A\&A, 501, 1161

\bibitem[{{Nordstr{\"o}m} {et~al.}(2004){Nordstr{\"o}m}, {Mayor}, {Andersen},
  {Holmberg}, {Pont}, {J{\o}rgensen}, {Olsen}, {Udry}, \&
  {Mowlavi}}]{Nordstrom-2004}
{Nordstr{\"o}m}, B., {Mayor}, M., {Andersen}, J., {et~al.} 2004, A\&A, 418, 989

\bibitem[{{Noyes} {et~al.}(1984){Noyes}, {Hartmann}, {Baliunas}, {Duncan}, \&
  {Vaughan}}]{Noyes-1984}
{Noyes}, R.~W., {Hartmann}, L.~W., {Baliunas}, S.~L., {Duncan}, D.~K., \&
  {Vaughan}, A.~H. 1984, ApJ, 279, 763

\bibitem[{{Paulson} {et~al.}(2002){Paulson}, {Saar}, {Cochran}, \&
  {Hatzes}}]{Paulson-2002}
{Paulson}, D.~B., {Saar}, S.~H., {Cochran}, W.~D., \& {Hatzes}, A.~P. 2002, AJ,
  124, 572

\bibitem[{{Pollack} {et~al.}(1996){Pollack}, {Hubickyj}, {Bodenheimer},
  {Lissauer}, {Podolak}, \& {Greenzweig}}]{Pollack-1996}
{Pollack}, J., {Hubickyj}, O., {Bodenheimer}, P., {et~al.} 1996, Icarus, 124,
  62

\bibitem[{{Pourbaix} {et~al.}(2004){Pourbaix}, {Tokovinin}, {Batten}, {Fekel},
  {Hartkopf}, {Levato}, {Morrell}, {Torres}, \& {Udry}}]{Pourbaix-2004}
{Pourbaix}, D., {Tokovinin}, A.~A., {Batten}, A.~H., {et~al.} 2004, A\&A, 424,
  727

\bibitem[{{Queloz} {et~al.}(2000){Queloz}, {Mayor}, {Weber}, {Bl{\' e}cha},
  {Burnet}, {Confino}, {Naef}, {Pepe}, {Santos}, \& {Udry}}]{Queloz-2000}
{Queloz}, D., {Mayor}, M., {Weber}, L., {et~al.} 2000, A\&A, 354, 99

\bibitem[{{Saar} \& {Donahue}(1997)}]{Saar-1997}
{Saar}, S.~H. \& {Donahue}, R.~A. 1997, ApJ, 485, 319

\bibitem[{{Santos} {et~al.}(2004{\natexlab{a}}){Santos}, {Bouchy}, {Mayor},
  {Pepe}, {Queloz}, {Udry}, {Lovis}, {Bazot}, {Benz}, {Bertaux}, {Lo Curto},
  {Delfosse}, {Mordasini}, {Naef}, {Sivan}, \& {Vauclair}}]{Santos-2004a}
{Santos}, N.~C., {Bouchy}, F., {Mayor}, M., {et~al.} 2004{\natexlab{a}}, A\&A,
  426, L19

\bibitem[{{Santos} {et~al.}(2001){Santos}, {Israelian}, \&
  {Mayor}}]{Santos-2001}
{Santos}, N.~C., {Israelian}, G., \& {Mayor}, M. 2001, A\&A, 373, 1019

\bibitem[{{Santos} {et~al.}(2004{\natexlab{b}}){Santos}, {Israelian}, \&
  {Mayor}}]{Santos-2004b}
{Santos}, N.~C., {Israelian}, G., \& {Mayor}, M. 2004{\natexlab{b}}, A\&A, 415,
  1153

\bibitem[{{Santos} {et~al.}(2010){Santos}, {Mayor}, {Benz}, {Bouchy},
  {Figueira}, {Lo Curto}, {Lovis}, {Melo}, {Moutou}, {Naef}, {Pepe}, {Queloz},
  {Sousa}, \& {Udry}}]{Santos-2010b}
{Santos}, N.~C., {Mayor}, M., {Benz}, W., {et~al.} 2010, A\&A, 512, A47+

\bibitem[{{Santos} {et~al.}(2007){Santos}, {Mayor}, {Bouchy}, {Pepe}, {Queloz},
  \& {Udry}}]{Santos-2007}
{Santos}, N.~C., {Mayor}, M., {Bouchy}, F., {et~al.} 2007, A\&A, 474, 647

\bibitem[{{Santos} {et~al.}(2000){Santos}, {Mayor}, {Naef}, {Pepe}, {Queloz},
  {Udry}, \& {Blecha}}]{Santos-2000a}
{Santos}, N.~C., {Mayor}, M., {Naef}, D., {et~al.} 2000, A\&A, 361, 265

\bibitem[{{Santos} {et~al.}(2002){Santos}, {Mayor}, {Naef}, {Pepe}, {Queloz},
  {Udry}, {Burnet}, {Clausen}, {Helt}, {Olsen}, \& {Pritchard}}]{Santos-2002a}
{Santos}, N.~C., {Mayor}, M., {Naef}, D., {et~al.} 2002, A\&A, 392, 215

\bibitem[{{Setiawan} {et~al.}(2005){Setiawan}, {Rodmann}, {da Silva}, {Hatzes},
  {Pasquini}, {von der L{\"u}he}, {de Medeiros}, {D{\"o}llinger}, \&
  {Girardi}}]{Setiawan-2005}
{Setiawan}, J., {Rodmann}, J., {da Silva}, L., {et~al.} 2005, A\&A, 437, L31

\bibitem[{{Sousa} {et~al.}(2010){Sousa}, {Santos}, {Israelian}, {Mayor},
  {Silva}, \& {Udry}}]{Sousa-2010}
{Sousa}, S.~G., {Santos}, N.~C., {Israelian}, G., {et~al.} 2010, A\&A, in press

\bibitem[{{Sousa} {et~al.}(2008){Sousa}, {Santos}, {Mayor}, {Udry},
  {Casagrande}, {Israelian}, {Pepe}, {Queloz}, \& {Monteiro}}]{Sousa-2008}
{Sousa}, S.~G., {Santos}, N.~C., {Mayor}, M., {et~al.} 2008, A\&A, 487, 373

\bibitem[{{Sozzetti} {et~al.}(2009){Sozzetti}, {Torres}, {Latham}, {Stefanik},
  {Korzennik}, {Boss}, {Carney}, \& {Laird}}]{Sozzetti-2009}
{Sozzetti}, A., {Torres}, G., {Latham}, D.~W., {et~al.} 2009, \apj, 697, 544

\bibitem[{{Tinney} {et~al.}(2003){Tinney}, {Butler}, {Marcy}, {Jones}, {Penny},
  {McCarthy}, {Carter}, \& {Bond}}]{Tinney-2003}
{Tinney}, C.~G., {Butler}, R.~P., {Marcy}, G.~W., {et~al.} 2003, ApJ, 587, 423

\bibitem[{{Udry} \& {Santos}(2007)}]{Udry-2007}
{Udry}, S. \& {Santos}, N. 2007, ARAA, 45, 397

\bibitem[{{Zacharias} {et~al.}(2004){Zacharias}, {Monet}, {Levine}, {Urban},
  {Gaume}, \& {Wycoff}}]{Zacharias-2004}
{Zacharias}, N., {Monet}, D.~G., {Levine}, S.~E., {et~al.} 2004, in Bulletin of
  the American Astronomical Society, Vol.~36, 1418

\bibitem[{{Zechmeister} \& {K{\"u}rster}(2009)}]{Zechmeister-2009b}
{Zechmeister}, M. \& {K{\"u}rster}, M. 2009, A\&A, 496, 577

\bibitem[{{Zechmeister} {et~al.}(2009){Zechmeister}, {K{\"u}rster}, \&
  {Endl}}]{Zechmeister-2009}
{Zechmeister}, M., {K{\"u}rster}, M., \& {Endl}, M. 2009, A\&A, 505, 859

\end{thebibliography}

\end{document}